\documentclass[AMA,Times1COL,fleqn]{WileyNJDv5} 

\usepackage{amsmath}
\usepackage{amsfonts}
\usepackage{amsthm}
\usepackage{mathtools}
\usepackage{subfig}
\usepackage{lineno}
\usepackage{textcomp}
\usepackage{algorithm,algpseudocode,algorithmicx}
\usepackage{mathrsfs}
\usepackage{booktabs}

\usepackage{multicol}
\usepackage{xcolor}

\usepackage{tabularx}

\usepackage{amssymb}
\usepackage{empheq}
\usepackage{siunitx}

\usepackage{float}
\usepackage{array,multirow}
\usepackage{color}
\usepackage{tikz}

\usepackage{pgfplots}
\pgfplotsset{compat=1.16}

\definecolor{lightgray}{gray}{0.80}

\newcommand{\Bezier}{B\'ezier~}

\definecolor{grey1}{rgb}{0.5, 0.5, 0.5}
\definecolor{green1}{rgb}{0.13, 0.55, 0.13} 
\definecolor{blue1}{rgb}{0.26, 0.41, 0.88} 
\definecolor{red1}{rgb}{0.8600, 0.0800, 0.2400}
\definecolor{yellow1}{rgb}{0.93, 0.75, 0.125}
\definecolor{purple1}{rgb}{0.40, 0.20, 0.60}
\definecolor{lightblue1}{rgb}{0.3010, 0.7450, 0.9330}
\definecolor{bordeaux1}{rgb}{0.6350, 0.0780, 0.1840}
\definecolor{brown1}{rgb}{0.65, 0.16, 0.16}
\definecolor{pink1}{rgb}{1.0, 0.08, 0.58}
\definecolor{orange1}{rgb}{1, 0.65, 0.0}

\definecolor{burntorange}{rgb}{0.8702,0.4791176,0}

\newcommand{\PreserveBackslash}[1]{\let\temp=\\#1\let\\=\temp}
\newcolumntype{C}[1]{>{\PreserveBackslash\centering}p{#1}}
\newcolumntype{R}[1]{>{\PreserveBackslash\raggedleft}p{#1}}
\newcolumntype{L}[1]{>{\PreserveBackslash\raggedright}p{#1}}

\usepackage{color, soul}
\definecolor{blue2}{RGB}{125, 249, 255}

\DeclareRobustCommand{\reviewerI}[1]{{\sethlcolor{pink}\hl{#1}}}
\DeclareRobustCommand{\reviewerII}[1]{{\sethlcolor{blue2}\hl{#1}}}

\DeclareRobustCommand{\changed}[1]{{\sethlcolor{blue2}\hl{#1}}}

\soulregister\reviewerI{1}
\soulregister\reviewerII{1}
\soulregister\changed{1}

\soulregister\ref7
\soulregister\pageref7
\soulregister\eqref7
\soulregister\cite7

\makeatletter
\renewcommand*\env@matrix[1][*\c@MaxMatrixCols c]{%
  \hskip -\arraycolsep
  \let\@ifnextchar\new@ifnextchar
  \array{#1}}
\makeatother

\theoremstyle{plain}

\newtheorem{configuration}[theorem]{Configuration}

\theoremstyle{definition}

\newcommand{\vect}[1]{\boldsymbol{#1}} 									
\newcommand{\mat}[1]{\mathbf{#1}} 											
\newcommand{\List}[1]{\left\{\right.#1\left.\right\}}					
\newcommand{\normalvect}{\boldsymbol{\nu}}    

\newcommand{\domain}{\Omega}														
\newcommand{\pardomain}{\hat{\Omega}}										
\newcommand{\ndofs}{N}																	
\newcommand{\trialf}{B}																	
\newcommand{\testf}{\tilde{B}}													
\newcommand{\dual}{\bar{\trialf}}												
\newcommand{\adual}{\hat{\trialf}}                      

\newcommand{\strainOp}{\mathscr{E}}

\newcommand{\detJ}{C}                                   

\newcommand{\n}{n}																				
\newcommand{\m}{m}																	

\newcommand{\sspace}[2]{\mathbb{S}^{#1}_{#2}}					
\newcommand{\Span}[1]{\text{span} \left( #1 \right)}					

\def\genbox#1#2#3#4#5#6{
    \leavevmode\raise#4bp\hbox to#5bp{\vrule height#5bp depth0bp width0bp
    \pdfliteral{q .5 w \csname #2COLOR\endcsname\space RG
                       \csname #3PDF\endcsname{#5}{#6} S Q
             \ifx1#1 q \csname #2COLOR\endcsname\space rg 
                       \csname #3PDF\endcsname{#5}{#6} f Q\fi}\hss}}

\usepackage{needspace}
\usepackage{adjustbox}

\articletype{Research Article}%

\received{Date Month Year}
\revised{Date Month Year}
\accepted{Date Month Year}

\journal{International Journal for Numerical Methods in Engineering}
\volume{00}
\copyyear{2024}
\startpage{1}

\begin{document}

\title{A locking-free isogeometric thin shell formulation based on higher order accurate diagonalized strain projection via approximate dual splines} 

\author[1,3]{Thi-Hoa Nguyen}

\author[2,3]{Ren\'e R. Hiemstra}

\author[3]{Dominik Schillinger}

\authormark{Nguyen \textsc{et al.}}
\titlemark{A locking-free isogeometric thin shell formulation based on higher order accurate diagonalized strain projection via approximate dual splines}

\address[1]{\orgdiv{Geophysical Institute and Bergen Offshore Wind Centre}, \orgname{University of Bergen}, \orgaddress{ \country{Norway}}}

\address[2]{\orgdiv{Department of Mechanical Engineering}, \orgname{Eindhoven University of Technology}, \orgaddress{ \country{Netherlands}}}

\address[3]{\orgdiv{Institute for Mechanics, Computational Mechanics Group}, \orgname{Technical University of Darmstadt}, \orgaddress{ \country{Germany}}}

\corres{\email{hoa.nguyen@uib.no}}

\abstract[Abstract]{We present a novel isogeometric discretization approach for the Kirchhoff-Love shell formulation based on the Hellinger-Reissner variational principle. 
For mitigating membrane locking, we discretize the independent strains with spline basis functions that are one degree lower than those used for the displacements. To enable computationally efficient condensation of the independent strains, we first discretize the variations of the independent strains with approximate dual splines to obtain a projection matrix that is close to a diagonal matrix. We then diagonalize this strain projection matrix via row-sum lumping.  Due to this diagonalization, the static condensation of the independent strain fields becomes computationally inexpensive, as no matrix needs to be inverted. At the same time, our approach maintains higher-order accuracy at optimal rates of convergence. We illustrate the numerical properties and the performance of our approach through numerical benchmarks, including a curved Euler-Bernoulli beam, 
the classical shell obstacle course, and geometrically nonlinear cylindrical shell problems.}


\maketitle

\section{Introduction}
	





Isogeometric analysis (IGA), first introduced by Hughes et al. \cite{hughes_isogeometric_2005} in 2005, is a framework that improves the integration of computer-aided design (CAD) and finite element analysis (FEA). 
The fundamental idea is to employ the same higher-order smooth spline basis functions for both the geometry representation in CAD and the approximation of the field solutions in FEA \cite{Haberleitner:17.1}. 
Another benefit of IGA is that it can be directly applied for the discretization of variational formulations that require higher-order smoothness. A salient example is Kirchhoff-Love shell theory, whose Galerkin formulation requires $C^1$-continuity of the trial and test functions. 
A major challenge for finite element schemes that discretize formulations of thin structures such as shells is numerical locking. 
Primary locking phenomena are membrane and transverse shear locking in beam, plate, and shell elements \cite{Stolarski1982,Stolarski1983,Bischoff:04.1} and 
volumetric locking due to incompressibility in solid elements \cite{hughes_finite_2003}. 
Membrane and transverse shear locking are due to artificial bending stiffness that arises as a result of the coupling of the bending response with the membrane response and transverse shear response, respectively, in the structural model.
Locking phenomena negatively affect accuracy and convergence, see e.g. \cite{Stolarski1983,Bischoff:04.1,Echter:10.1,nguyen2022leveraging,hiemstra2023_1}. 

Locking-preventing discretization technology has been developed for more than 50 years, first within classical FEA and then also within IGA. 
For the former, prominent locking-preventing approaches are 
strain modification methods such as 
the B-bar method \cite{belytschko_locking_1985,Hughes1977,Hughes1980,Recio2006}, 
the assumed natural strain (ANS) method \cite{Simo1986,Bucalem1993,Bathe1985}, 
and the enhanced assumed strain (EAS) method \cite{Andelfinger1993,Bischoff1999,Alves2005};
the reduced and selective reduced integration techniques \cite{Zienkiewicz1971,Malkus1978,Noor1981,Schwarze2009}, often combined with hourglass control \cite{Flanagan1981,Belytschko1984,Reese2007}; 
and the utilization of mixed formulations, often based on the Hu-Washizu variational principle \cite{Malkus1978,Noor1981,Kim1998,Wagner_mixed_2008} 
or the Hellinger-Reissner variational principle \cite{Bischoff1999,Stolarski1986,Lee2012}. 
Moreover, the effect of locking phenomena can be reduced by increasing the polynomial degree of the basis functions \cite{Ashwell1976,Rank1998}, which, however, does not completely eliminate locking, see e.g. \cite{nguyen2022leveraging,Rank1998}. 
In the context of IGA, these approaches have been transferred and extended, for instance, for 
rod elements \cite{Greco2017}, 
beam elements \cite{Echter:10.1,Bouclier2012,BeiraodaVeiga2012,bieber_locking_mixedForm2018,Zhang2018}, 
plate elements \cite{bieber_locking_mixedForm2018,Elguedj2007,Adam2015,Adam2015b}, 
shell elements \cite{bieber_locking_mixedForm2018,Adam2015b,Benson:10.1,Bouclier:15.1,Caseiro2015,Greco_ANS_2018,Leonetti2018,Antolin2019,Echter:13.1,Kikis2019,Kim2022}, 
and solid elements \cite{Elguedj2007,Bouclier2013,Cardoso2012,Taylor2011}. 
For isogeometric discretizations, 
there exist other reduced quadrature rules that require a minimal number of quadrature points while preserving accuracy and optimal convergence \cite{hiemstra2017optimal,schillinger2014reduced}. 
In \cite{Zou_quadrature_locking_2021,Zou_quadrature_locking_2022}, the authors employed a specific quadrature rule based on the Greville abscissa to alleviate membrane and shear locking for Reissner-Mindlin and Kirchhoff-Love shell elements. 
Recently, Casquero and co-workers introduced the so-called CAS element based on an assumed-strain approach for quadratic spline functions, where they interpolate the strain fields corresponding to locking with linear Lagrange polynomials. 
The CAS element idea has been applied for removing membrane and shear locking in linear rod formulations \cite{Golestanian_casEle_TimoRod_2023,Casquero_casEle_KirchhoffRod_2022}, membrane locking in linear Kirchhoff-Love shell formulations \cite{Casquero_casEle_KLshell_2023}, and volumetric locking in nearly-incompressible linear elasticity \cite{Casquero_casEle_volumetricLocking_2023}. 
For Reissner-Mindlin shell formulations based on the Hellinger-Reissner principle, Zou and co-workers \cite{zou_dual_locking_2020} introduced an efficient technique to condense out the strain variables in the mixed formulation. To this end, they suggested to employ the \Bezier dual spline basis \cite{zou_mortar_2018,Miao_dual_2018}, such that the condensation does not require any matrix inversion. 
Closely related to this local strain condensation approach are the mixed isogeometric solid-shell element that was introduced by Bouclier and co-workers \cite{Bouclier2013}, which employs a local quasi-interpolation of the strain variables and enables local strain condensation without matrix inversion; 
and the blended mixed formulation by Greco and co-workers \cite{Greco2017,Greco_ANS_2018} that is based on a local $L^2$ projection of the strain variables \cite{Thomas2015}.

In this work, inspired by the strain projection technique introduced in \cite{zou_dual_locking_2020}, 
we propose a locking-free isogeometric mixed formulation for Kirchhoff-Love shells based on the Hellinger-Reissner principle. 
It consists of the following key components. First, to eliminate membrane locking, we follow the widely used idea to discretize the independent strain fields with B-splines of one degree lower than those used for the displacements \cite{Echter:10.1,nguyen2022leveraging,Bouclier:15.1}. 
We choose the multivariate spline spaces for each independent strain component based on those discussed in \cite{Guo2021}. 
Second, to condense out the independent strain fields, we employ the higher-order diagonalization technique that was applied in \cite{nguyen_mass_lumping2023,hiemstra2023} for mass lumping in explicit dynamics.
In particular, we discretize all trial functions and variations of the displacement field with smooth B-splines, but discretize the variations of the independent strain fields with the corresponding modified approximate dual functions \cite{nguyen_mass_lumping2023,hiemstra2023}. 
Row-sum lumping of the projection matrix yields a diagonal matrix, eliminating the need for matrix inversion in the global condensation procedure while preserving higher-order accuracy and optimal convergence. 
We note that similar to the projection technique introduced in \cite{zou_dual_locking_2020}, we discretize the trial functions and variations of the independent strain fields with different function spaces and hence obtain a Petrov-Galerkin formulation for the strain projection.

The structure of our article is as follows: in Section \ref{sec:approx_dual}, we briefly review some of the properties and the construction of the approximate dual functions employed in this work. 
In Section \ref{sec:KLshell},
we briefly review the kinematics of the Kirchhoff-Love shell model. 
In Section \ref{sec:mixed_form}, we review the mixed variational formulation of the Kirchhoff-Love shell problem based on the Hellinger-Reissner principle in a continuous setting. 
We then discretize the independent strain and displacement fields in Section \ref{sec:discretization} and recall the resulting matrix equations. 
In Section \ref{sec:strain_condensation}, we specify our approach to obtain a fast and robust diagonalized strain projection, using approximate dual functions and row-sum lumping. In addition, we analyze the corresponding significant reduction of the computational cost in terms of the static strain condensation and the solution of the displacement-based system in comparison to the consistent mixed Galerkin formulation via theoretical estimates and computing times. 
In Section \ref{sec:results}, we present numerical benchmarks of beam and shell problems, demonstrating via spectral analysis and convergence
studies that our approach eliminates membrane locking and achieves higher-order accuracy and convergence. 
In Section \ref{sec:conclusions}, we summarize our results and draw conclusions.

\section{Approximate dual spline basis}\label{sec:approx_dual}

In this section, we review some of the construction and key properties of approximate dual spline functions introduced in \cite{chui_wavelet_2004}. In addition, we illustrate that employing these approximate dual functions to discretize the test functions in an $L^2$ projection preserves the optimal accuracy and convergence, also when the resulting system matrix is diagonalized via row-sum lumping.

\subsection{Construction and computation}

Splines are piecewise polynomials that are joined smoothly with a prescribed continuity. 
Spline functions can be represented by B-splines \cite{boor_practical_2001, schumaker_spline_2007}. 
Let $\List{\trialf_i, \; i=1, \ldots , N}$ denote a set of B-splines of polynomial degree $p$ defined on an open knot vector that corresponds to a partitioning of the interval $\pardomain \subset \mathbb{R}$. 
These functions are a basis for the complete spline space:
\begin{align}
	\sspace{}{}(\pardomain) = \Span{\trialf_i, \; i=1, \ldots , N}.
\end{align}

Let $\langle \cdot , \cdot \rangle $ denote the standard $L^2$ inner product on $\pardomain$. A set of functions $\List{\dual_i, \; i=1, \ldots , N}$ is called an $L^2$ dual basis (or bi-orthogonal basis) to the set of B-splines if the following property holds:
\begin{align}
	\langle \dual_i, \, \trialf_j \rangle = \delta_{ij}, \quad \text{for } i,j \in 1, \ldots , N,
\end{align}
where $ \delta_{ij}$ denotes the Kronecker delta. 
If $\dual_i \in \sspace{}{}$, then dual functions are uniquely defined and can be computed using:
\begin{align}
	\dual_i(x) = \sum_{j=1}^{N} \left( \mat{G}^{-1} \right)_{ij} \trialf_j(x), \quad \text{where }   G_{ij} =  \langle \trialf_i, \, \trialf_j \rangle.
	\label{eq:dual}
\end{align}
The dual functions are linearly independent and also form a basis for $\sspace{}{}$. However, they are globally supported, which limits their practical usability as test functions. 

We loosen the restriction of duality and, instead, search for an alternative basis $\List{\adual_i, \; i=1, \ldots , N}$ for the space of splines that achieves duality only in an approximate sense while maintaining local support. Let $\mat{S} \in \mathbb{R}^{N \times N}$ denote a symmetric positive definite matrix. We call the set of functions:
\begin{align}
	\adual_i(x) = \sum_{j=1}^{N} S_{ij} \trialf_j(x), \quad i  =  1, \ldots , N.
\end{align}
an approximate dual basis to the set of B-splines if:
\begin{align}
	\langle f, \, \adual_i \rangle = \langle f, \, \dual_i \rangle \quad \forall f \in \mathcal{P}^p,
	\label{eq:quasiinterpolant}
\end{align}
where $\mathcal{P}^p$ denotes the space of piecewise polynomials of degree $p \geq 0$. 
The functions $\List{\adual_i, \; i=1, \ldots , N}$ are called \emph{approximate dual functions} by virtue of \eqref{eq:quasiinterpolant}. They span the space of splines and are linearly independent since $\mat{S}$ is invertible.

Notably, the representation is non-unique. A particularly useful approximate dual basis has been developed in \cite{chui_wavelet_2004}. The approximate dual functions presented therein have minimum compact support, in the sense that there exists no other basis that satisfies the approximate duality with smaller supports. This translates into $\mat{S}$ having a banded structure with typically $2p+1$ diagonals. Importantly, the matrix $\mat{S}$ is an approximate left and right inverse of the Gramian matrix $\mat{G}$, that is,
\begin{align}
	\mat{S} \approx \mat{G}^{-1}.
	\label{eq:approx_inv}
\end{align}
The matrix is computed by a recursive algorithm, based on fundamental properties of B-splines such as the nestedness of spaces under knot insertion. We refer to the article\cite{chui_wavelet_2004} for additional details. For application in isogeometric analysis, we refer to \cite{nguyen_mass_lumping2023,hiemstra2023,Dornisch2017}. In the remainder of this paper, we shall base our construction on this approach.

\begin{remark}
    The approximation \eqref{eq:approx_inv} can be iteratively improved based on a predictor-multicorrector scheme, see \cite{nguyen_mass_lumping2023}.
\end{remark}

\subsection{Geometric mapping}
\label{sec22}

\begin{figure}[ht!]
    \centering
    \captionsetup[subfloat]{labelfont=scriptsize,textfont=scriptsize}
    \subfloat[B-splines]{{\includegraphics[width=0.45\textwidth]{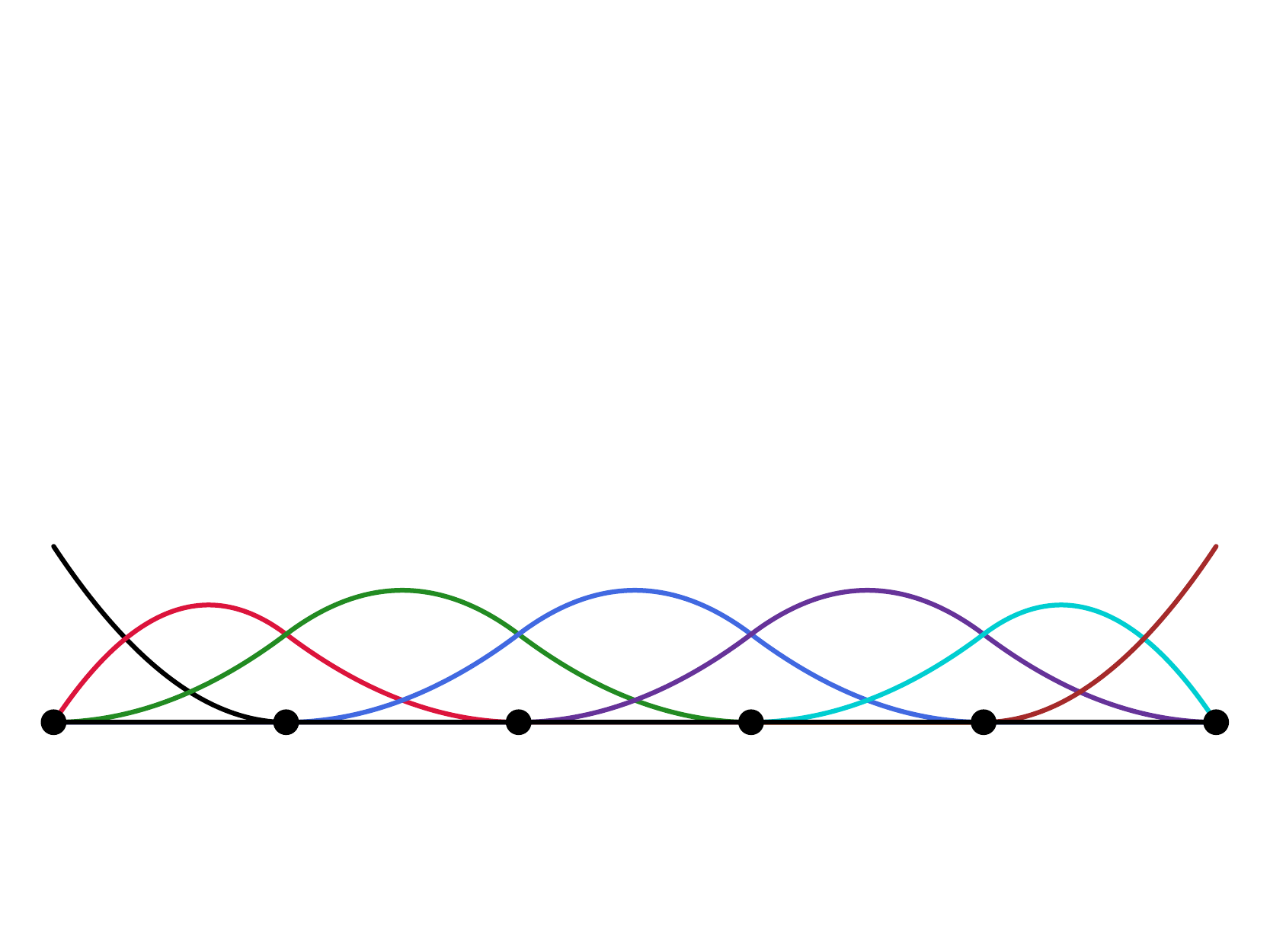} }}
    \subfloat[Modified approximate dual functions]{{\includegraphics[width=0.45\textwidth]{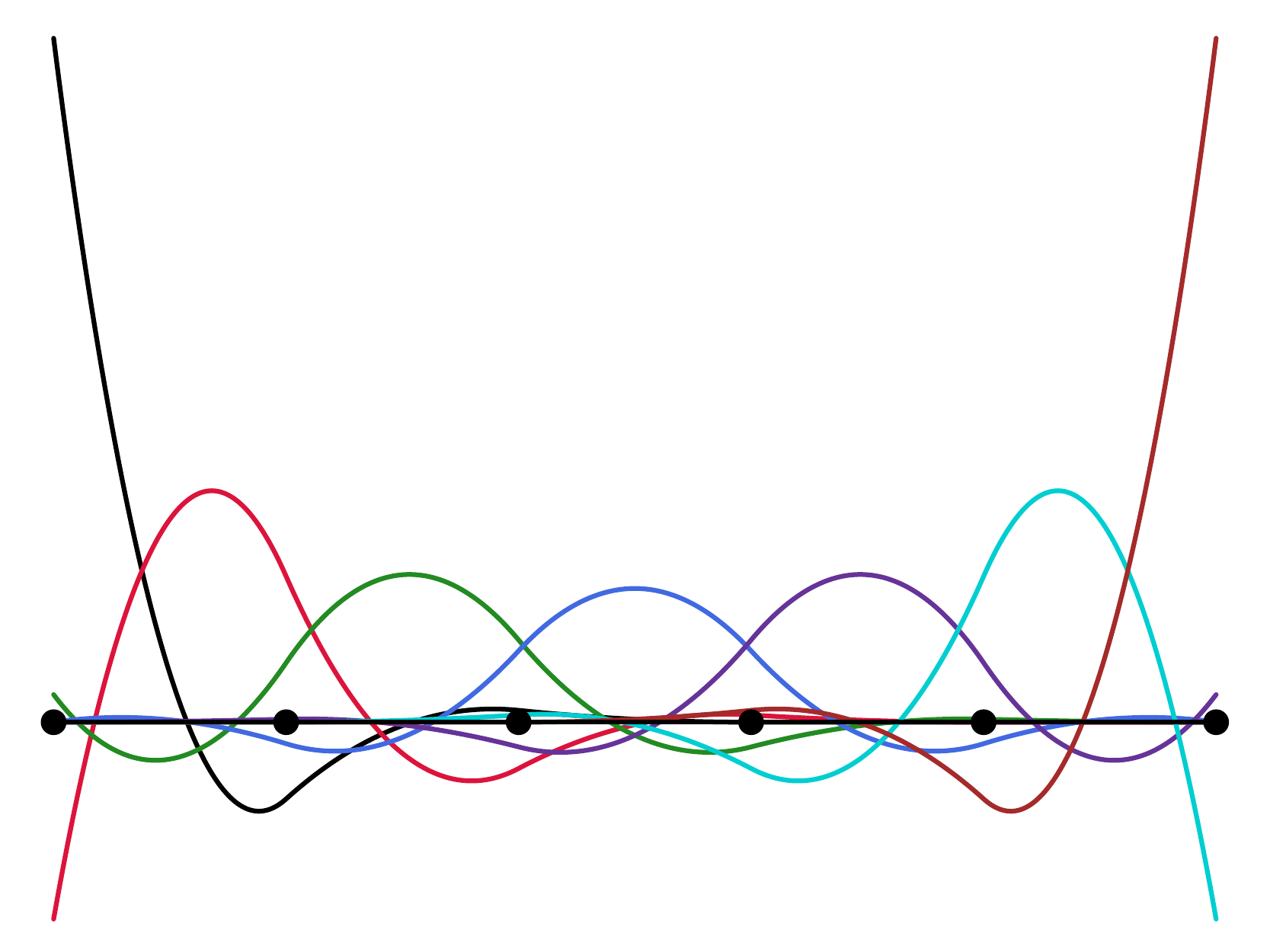} }}

    \subfloat[Jacobian determinant function for a quarter circle.]{{\includegraphics[width=0.45\textwidth]{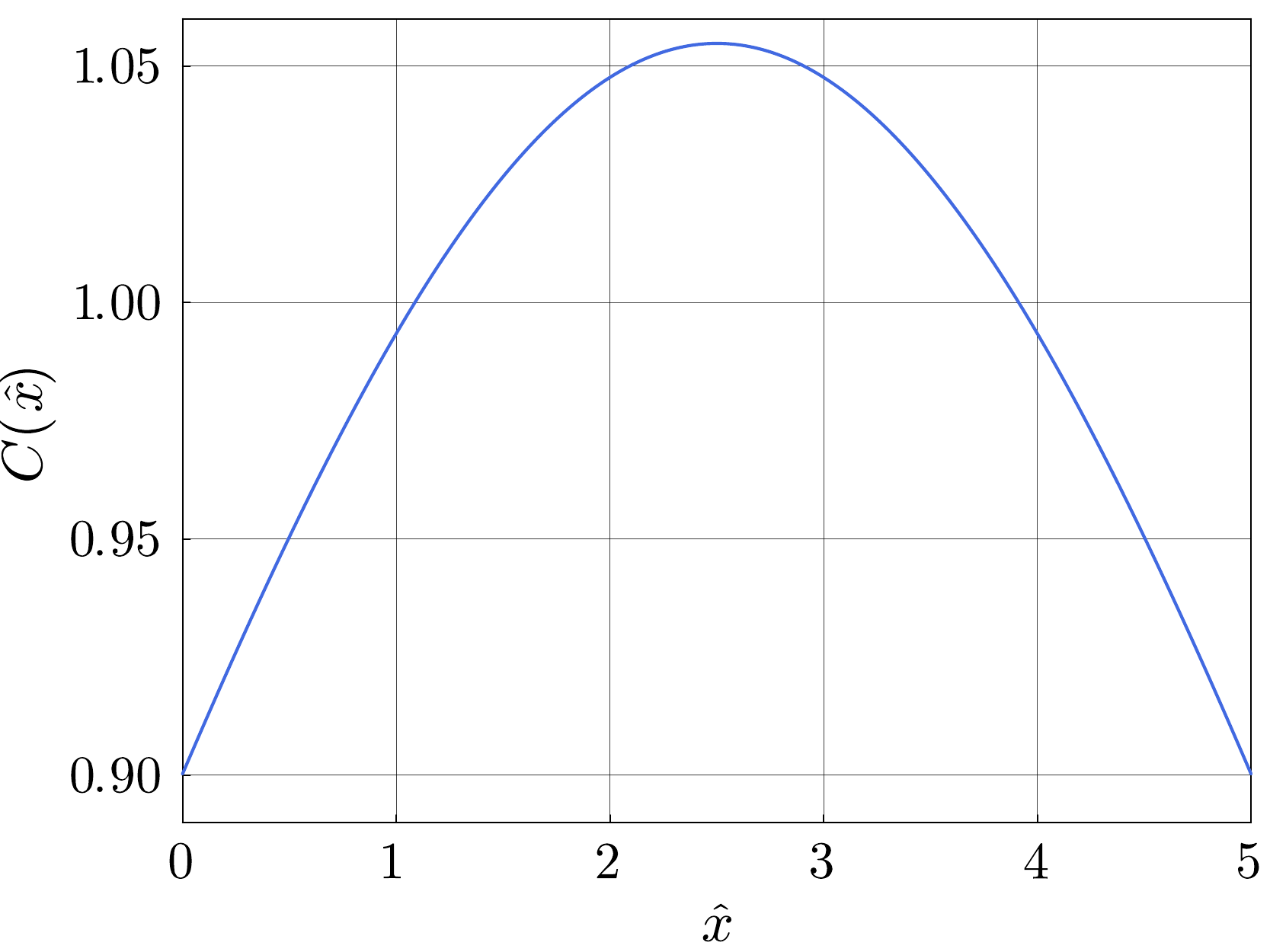} }}

    \vspace{0.3cm}
    \caption{Standard B-splines and modified approximate dual functions for a 1D patch of quadratic B-splines and a non-constant Jacobian determinant (from \cite{nguyen_mass_lumping2023}).}
    \label{fig:trial_test_funcs}
\end{figure}

For general non-affine geometric mappings, the approximation \eqref{eq:approx_inv}, and thus the approximate duality, is not preserved when mapped from the parametric domain $\pardomain$ to the physical domain $\domain$. 
To tackle this problem, we followed \cite{mika2021matrix} and introduced in \cite{nguyen_mass_lumping2023} a modification of
the original approximate dual functions, $\adual_i$, by multiplying with the inverse of 
the determinant of the Jacobian matrix of the mapping, $\detJ(\hat{\mat{x}})$. 
The modified approximate dual functions are:
\begin{align}
    \testf_i(\hat{\vect{x}}) \, := \, \frac{\adual_i(\hat{\vect{x}})}{\detJ(\hat{\vect{x}})} \, , \qquad \, i=1,\,\ldots,\,\ndofs \,, \quad  \hat{\vect{x}} \in \pardomain \,. \label{eq:test_func}
\end{align}
As shown in \cite{nguyen_mass_lumping2023}, $\testf_i$ satisfy the approximate duality in the sense of \eqref{eq:approx_inv} in the physical domain. 
$\testf_i$ are linearly independent due to the linear independence of the approximate dual functions $\adual_i$ and preserve their local support. 
Their regularity, however, depends on the smoothness of the Jacobian $\detJ(\hat{\vect{x}})$. For practical scenarios, we can assume that the underlying geometric mapping is sufficiently smooth and invertible such that the Jacobian matrix and its inverse are well-defined. 

For illustration, we consider a quarter circle with a unit radius represented by a non-uniform rational B-spline (NURBS) curve. In Figure \ref{fig:trial_test_funcs}, we plot the modified approximate dual functions \eqref{eq:test_func} next to the corresponding quadratic $C^1$-continuous B-splines. We also plot the corresponding non-constant function of the Jacobian determinant. 
We observe that the modified approximate dual functions have local support and preserve the $C^{p-1}$-continuity of the corresponding B-spline functions.

\subsection{$L^2$ projection with row-sum lumped matrix}\label{sec:l2projection}

\begin{figure}[ht!]
    \centering
    \captionsetup[subfloat]{labelfont=scriptsize,textfont=scriptsize}
    \subfloat[$p=2$]{{\def\svgwidth{0.45\textwidth}
    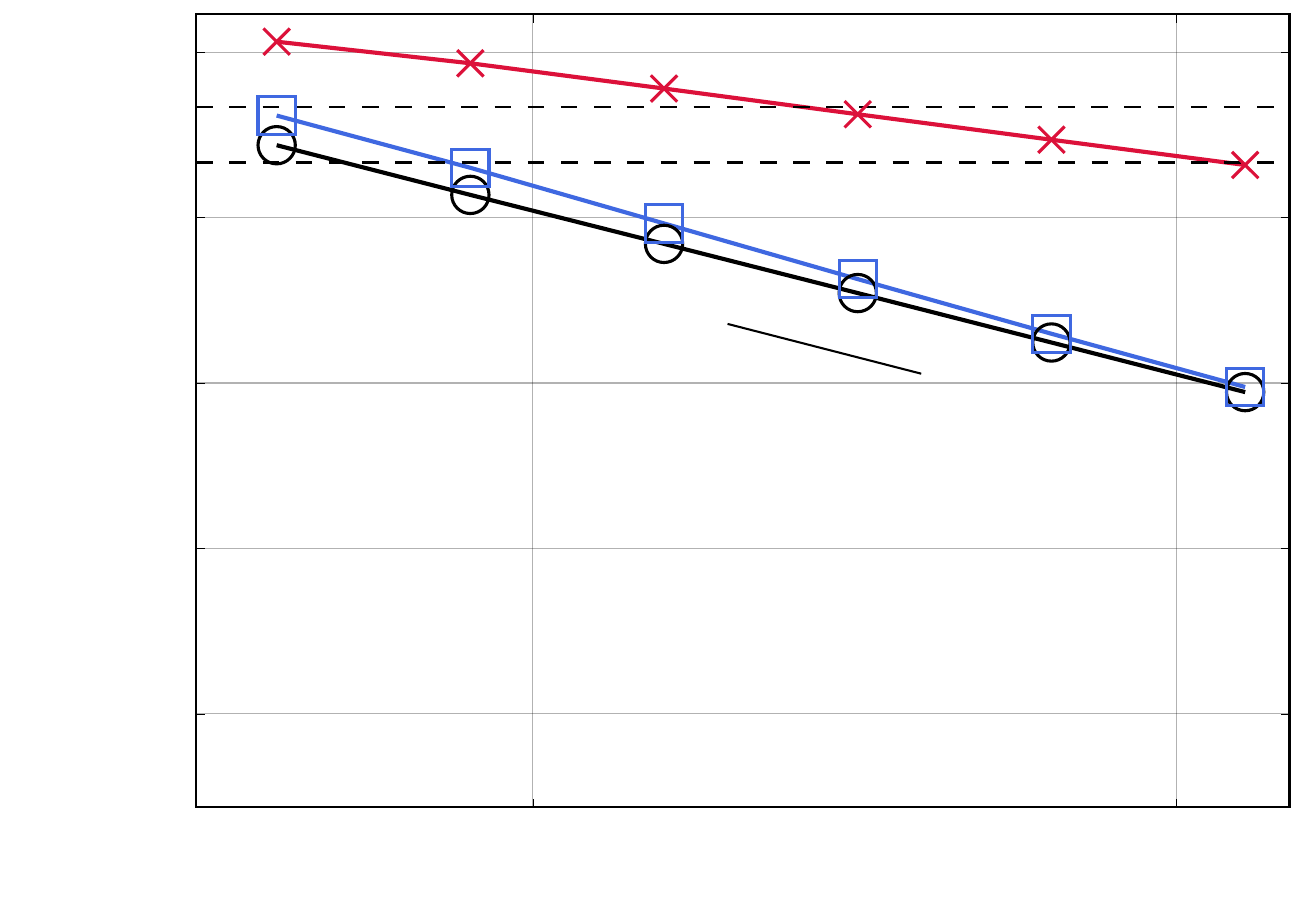 }}
    \subfloat[$p=3$]{{\def\svgwidth{0.45\textwidth}
    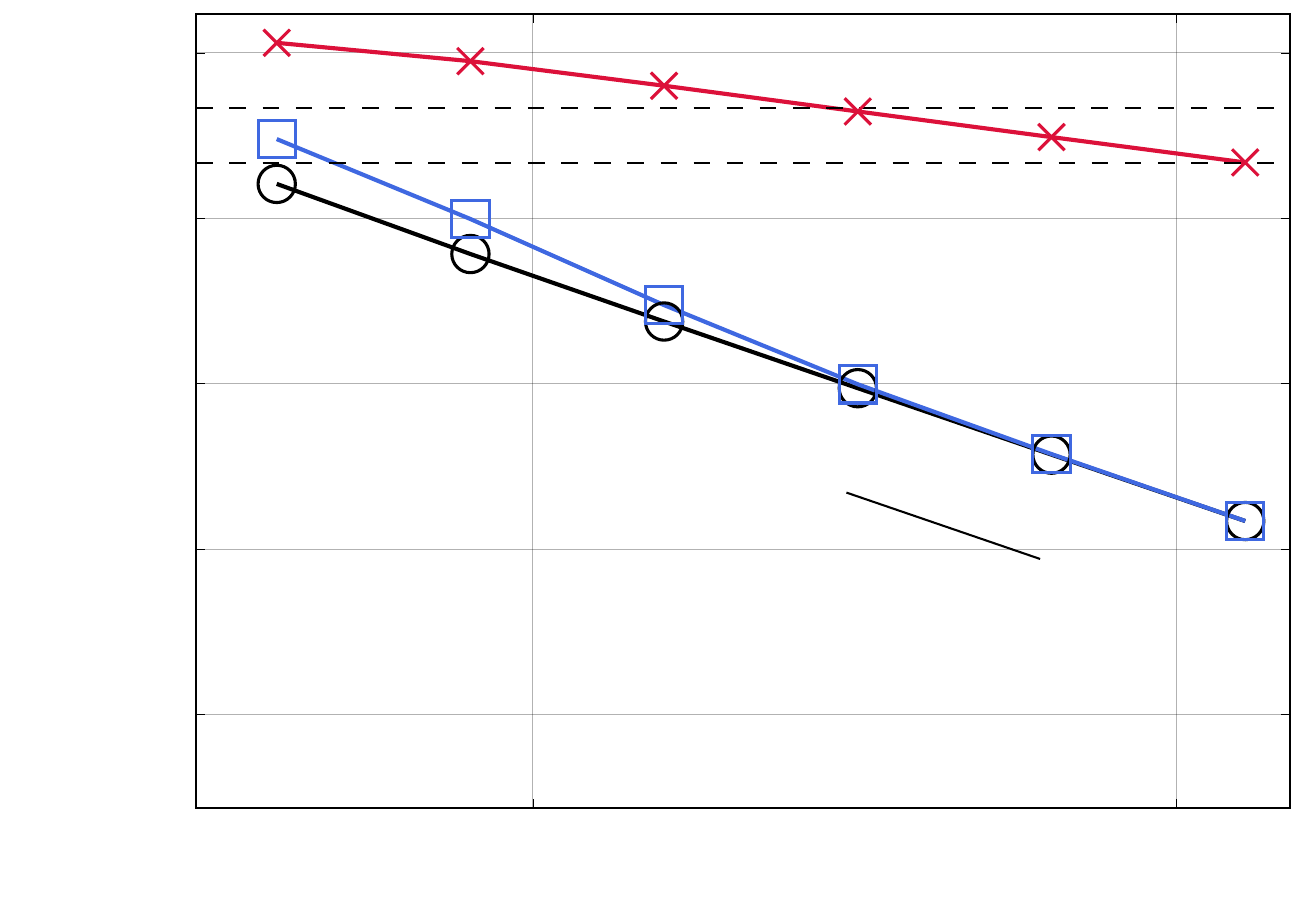 }}

    \subfloat[$p=4$]{{\def\svgwidth{0.45\textwidth}
    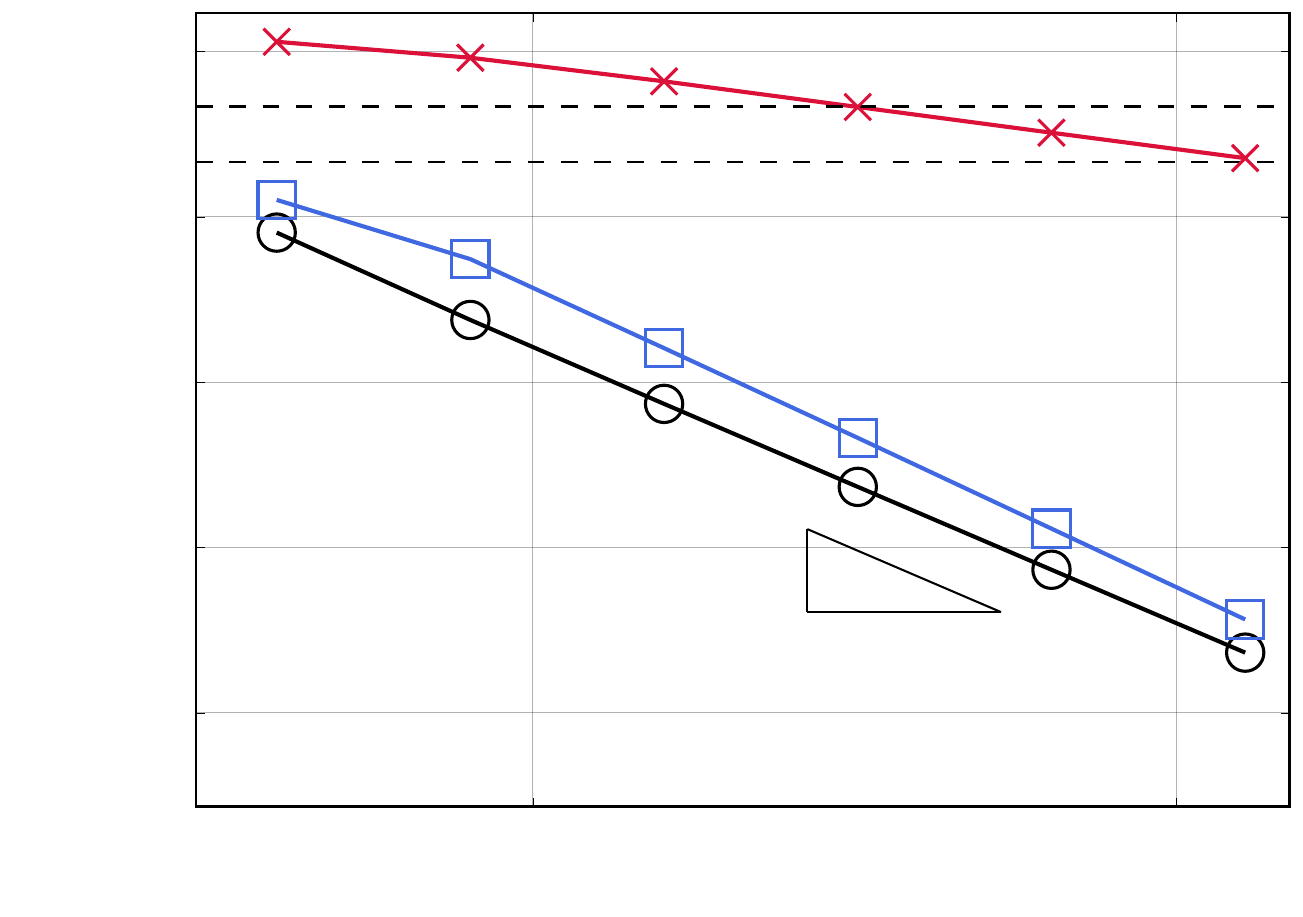 }}
    \subfloat[$p=5$]{{\def\svgwidth{0.45\textwidth}
    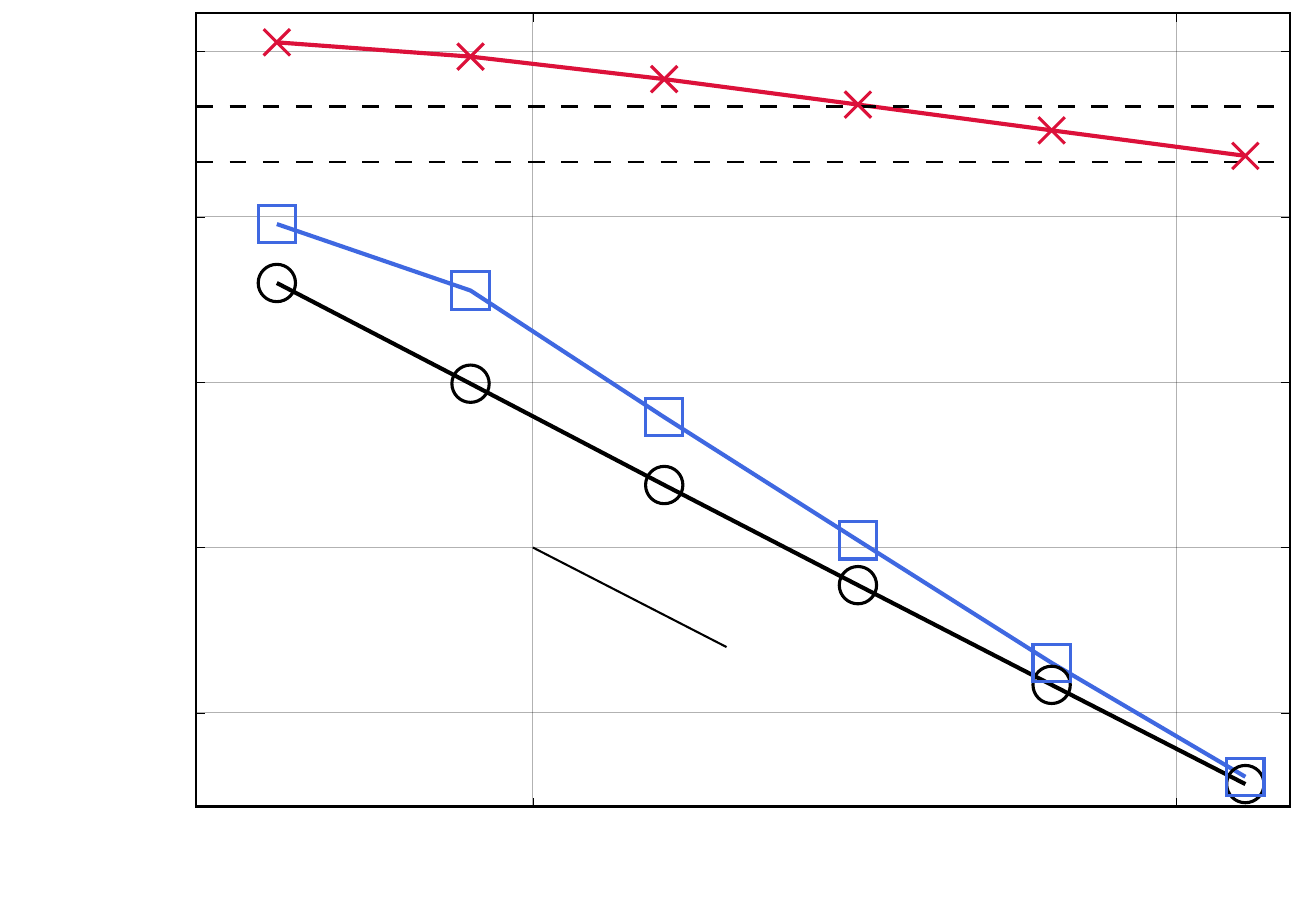 }}
    
    \vspace{0.2cm}
    \begin{tikzpicture}
    \filldraw[black,line width=1pt, solid] (0.0,0) -- (0.6,0);
    \filldraw[black,line width=1pt] (0.3,0) [fill=none] circle (3pt);
    \filldraw[black,line width=1pt] (0.7,0) node[right]{\scriptsize Consistent projection};
    \filldraw[red1,line width=1pt, solid] (5.0,0) -- (5.6,0);
    \filldraw[red1,line width=1pt] (5.0,0) node[right]{$\boldsymbol{\bigtimes}$};
    \filldraw[red1,line width=1pt] (5.7,0) node[right]{\scriptsize Lumped projection, B-splines};
    \filldraw[blue1,line width=1pt, solid] (11.0,0.05) -- (11.7,0.05);
    \filldraw[blue1,line width=1pt] (11.2,-0.08) [fill=none] rectangle ++(0.25,0.25);
    \filldraw[blue1,line width=1pt] (11.8,0) node[right]{\scriptsize Lumped projection, approximate duals};
\end{tikzpicture}
    
    \caption{Accuracy and convergence for an $L^2$ projection test case (trigonometric function on a quarter circle).
    } \label{fig:l2projection}
\end{figure}

In view of our mixed approach discussed in the following sections, 
we are interested in the accuracy of the $L^2$ projection employing the modified approximate dual functions as test functions when the matrix is diagonalized via row-sum lumping. To this end, we consider the following test problem, which consists of the $L^2$ projection of an arbitrary function $f$ onto a space $\mathcal{S}$: \\
\noindent
Find $u \in \mathcal{S}$ such that:
\begin{align}\label{eq:l2-projection}
    \langle w , u \rangle  = \langle w , f \rangle \,, \quad \forall \, w \in \mathcal{V}\,,
\end{align}
where
$u$ denotes the unknown trial function that is the projection of the function $f$, and $w$ denotes the test function. 
In the case where the two function spaces $\mathcal{S}$ and $\mathcal{V}$ are the same, expression \eqref{eq:l2-projection} is a Galerkin formulation. 
When $\mathcal{S}$ and $\mathcal{V}$ are two different function spaces, expression \eqref{eq:l2-projection} is a Petrov-Galerkin formulation.

We discretize $u$ and $w$ using smooth B-splines of degree $p$ and the corresponding modified approximate dual functions \eqref{eq:test_func}, respectively, such that:
\begin{align}
    u^h = \sum_{i=1}^{n} \, \trialf_i \, \hat{u}^h_i \,, \qquad
    w^h = \sum_{i=1}^{n} \, \testf_i \, \hat{w}^h_i \,,
\end{align}
The resulting matrix equation is:
\begin{align}
    \mat{M} \, \vect{\hat{u}}^h \,=\, \vect{F} \,, 
\end{align}
where $M_{ij} = \langle \testf_i ,\, \trialf_j \rangle$ and $F_i = \langle \testf_i ,\, f \rangle$.
Here, $\vect{\hat{u}}^h$ is the vector of unknown coefficients $\hat{u}^h_i$, $i=1,\ldots,n$. 
As shown in \cite{hiemstra2023}, row-sum lumping of the matrix $\mat{M}$ yields the identity matrix due to the definition and construction of the approximate dual functions. Hence, we directly obtain the solution $\vect{\hat{u}}^h \,=\, \vect{F}$ \emph{without any matrix inversion}. 
For explicit dynamics of beam and shell models, the authors showed in \cite{nguyen_mass_lumping2023,hiemstra2023} that when an isogeometric discretization uses the modified approximate dual functions as test functions, employing such a row-sum lumped mass matrix does not compromise higher-order accuracy. 

We now illustrate, via the $L^2$ projection of the specific function $f(x) = \sin(x) \, \cos(x)$ on a quarter circle with a unit radius, exactly represented by a NURBS-curve, that row-sum lumping of the matrix $\mat{M}$ does not affect higher-order accuracy and optimal convergence, given that the test functions are discretized with modified approximate dual functions. 
In Figure \ref{fig:l2projection}, 
we plot the convergence of the relative $L^2$ error in this projection computed with quadratic, cubic, quartic, and quintic discretizations and three different methods: (a) a Galerkin formulation with B-splines as both trial and test functions that uses the inverse of a consistent (black) projection matrix; (b) a Galerkin formulation with B-splines as both trial and test functions that uses the inverse of a row-sum lumped projection matrix (red); and (c) a Petrov-Galerkin formulation with B-splines as trial functions and approximate dual splines as test functions that uses a row-sum lumped projection matrix (blue). 
We perform uniform mesh refinement in the angular direction with 4, 8, 16, 32, 64, and 128 \Bezier elements.

We observe that the Petrov-Galerkin approach with approximate dual test functions leads to results that are several orders of magnitude more accurate than the Galerkin approach using a lumped projection matrix. 
It achieves the same optimal convergence rate as the reference Galerkin method using the inverse of a consistent projection matrix but at a slightly higher error level. 
We also see that the accuracy of the projection based on a Galerkin formulation is significantly affected by row-sum lumping, limiting the convergence rate at second order irrespective of the polynomial degree of the spline basis.

\section{The Kirchhoff-Love shell model}\label{sec:KLshell}

In this section, we briefly review the Kirchhoff-Love shell model \cite{Bischoff:04.1,Kiendl_shell_2009,reddy_shell2006}, which we employ in this work.

\begin{figure}[ht!]
	\begin{center}
    \vspace{-0.5cm}
		\def\svgwidth{0.85\textwidth}
		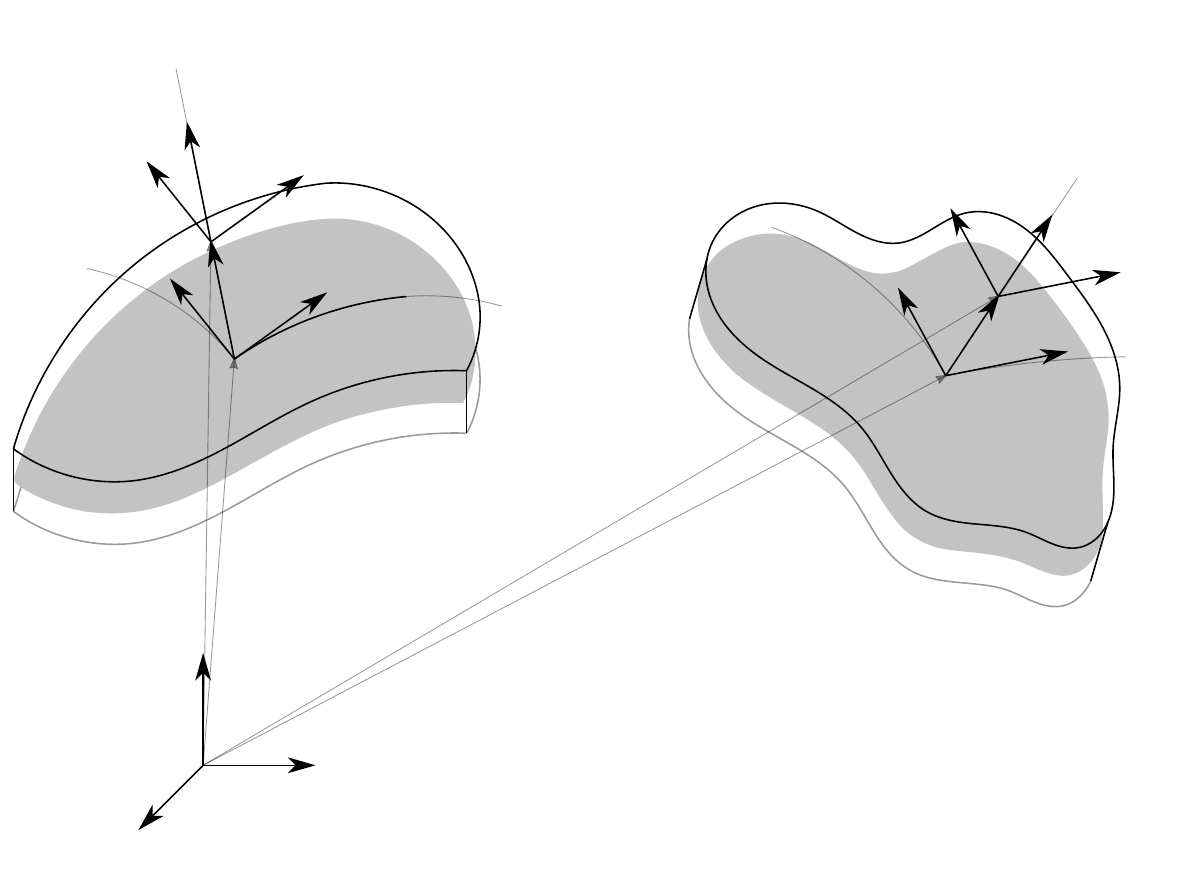
    \end{center}
    \caption{A schematic of shell kinematics.}
    \label{fig:shell}
\end{figure}

Consider a shell-like body in $\mathbb{R}^3$, illustrated in Figure \ref{fig:shell}, that is, locally, described by curvilinear coordinates $\theta = (\theta^1, \theta^2, \theta^3)$. In the following, we adopt the Einstein summation convention and the convention for Greek and Latin letters, i.e. $i=1,2,3$ and $\alpha=1,2$. 
The reference and current configurations of the shell-like body are described by the position vector $\vect{X}(\theta)$ and $\vect{x}(\theta)$, respectively, as follows:
\begin{align}
\label{eq1}
  \vect{X}(\theta) = \theta^\alpha \vect{A}_\alpha + \theta^3 \vect{A}_3 \,, \qquad
  \vect{x}(\theta) = \theta^\alpha \vect{a}_\alpha + \theta^3 \vect{a}_3 \,.
\end{align}
The two are related through the deformation:
\begin{align}
	\vect{u}(\theta) = \vect{x}(\theta) - \vect{X}(\theta)\,.
\end{align}
Here, $\theta^3$ is the coordinate in the thickness direction $-d/2 \leq \theta^3 \leq d/2$, where $d$ denotes the shell thickness. 
$\vect{A}_i$ and $\vect{a}_i$ denote the covariant base vectors of the midsurface in the reference configuration and the current configuration, respectively. 
$\vect{A}_i$ and $\vect{a}_i$ are defined in terms of the position vector of the midsurface $\bar{\vect{X}}$ and $\bar{\vect{x}}$ as follows:
\begin{align}
  & \vect{A}_\alpha = \bar{\vect{X}}_{,\alpha}\,, \qquad \vect{A}_3 = \frac{\vect{A}_1 \times \vect{A}_2}{|\vect{A}_1 \times \vect{A}_2|} \,,\\
  & \vect{a}_\alpha = \bar{\vect{x}}_{,\alpha}\,, \qquad \vect{a}_3 = \frac{\vect{a}_1 \times \vect{a}_2}{|\vect{a}_1 \times \vect{a}_2|} \,, \label{eq:base-vector-current-config}
\end{align}
where $(\cdot)_{,\alpha}$ denotes $\partial (\cdot) / \partial \theta^\alpha$. 
The covariant base vectors at any point in the shell continuum in the reference and current configurations are:
\begin{align}
  \vect{G}_{\alpha} = \vect{X}_{,\alpha}  \quad \text{and} \quad  \vect{g}_{\alpha} = \vect{x}_{,\alpha}\,,
\end{align}
respectively. 
The corresponding normal vectors are denoted by $\vect{G}_{3}$ and $\vect{g}_{3}$, respectively. 
Contravariant base vectors, $\vect{G}^{\alpha}$ and $\vect{g}^{\alpha}$, can be computed directly from the first fundamental form.

The Green-Lagrange strain tensor is defined as $\mat{E} = E_{\alpha \beta} \, \vect{G}^\alpha \otimes \vect{G}^\beta$.
For Kirchhoff-Love shells, the non-zero strain components $E_{\alpha \beta}$ can be represented by the sum of a constant membrane action and a linearly varying bending action due to the change of curvature: 
\begin{align}
	E_{\alpha \beta} = \varepsilon_{\alpha \beta} + \theta^3 \kappa_{\alpha \beta}.
\end{align}
The coefficients $\varepsilon_{\alpha \beta}$ and $\kappa_{\alpha \beta}$, which can be computed from the first and second fundamental form, see e.g. \cite{Bischoff:04.1,Kiendl_shell_2009,reddy_shell2006}, are:
\begin{align}\label{eq:membrane_bending_strains}
  \varepsilon_{\alpha \beta} = \frac{1}{2} \left(\vect{a}_\alpha \cdot \vect{a}_\beta - \vect{A}_\alpha \cdot \vect{A}_\beta \right)  \quad \text{and} \quad
  \kappa_{\alpha \beta} = \vect{A}_{\alpha,\beta} \cdot \vect{A}_3 - \vect{a}_{\alpha,\beta} \cdot \vect{a}_3 \,.
\end{align}
In Voigt notation, the strains due to membrane and bending action are: are:
\begin{align}
  \vect{\varepsilon} = \begin{bmatrix}
    \varepsilon_{11} \\ \varepsilon_{22} \\ 2\varepsilon_{12}
  \end{bmatrix} \,, \quad  \vect{\kappa} = \begin{bmatrix}
    \kappa_{11} \\ \kappa_{22} \\ 2\kappa_{12}
  \end{bmatrix} \,,
\end{align}
respectively. 
The first variation of the strain components are:
\begin{align}
  & \delta \varepsilon_{\alpha \beta} = \frac{1}{2} \left(\delta \vect{a}_\alpha \cdot \vect{a}_\beta + \vect{a}_\alpha \cdot \delta \vect{a}_\beta \right) \,, \label{eq:variation-epsilon}\\
  & \delta \kappa_{\alpha \beta} = - \left(\delta \vect{a}_{\alpha,\beta} \cdot \vect{a}_3 + \vect{a}_{\alpha,\beta} \cdot \delta \vect{a}_3 \right)\,, \label{eq:variation-kappa}
\end{align}
and the second variation are:
\begin{align}
  & \Delta \delta \varepsilon_{\alpha \beta} = \frac{1}{2} \left(\delta \vect{a}_\alpha \cdot \Delta \vect{a}_\beta + \Delta \vect{a}_\alpha \cdot \delta \vect{a}_\beta \right) \,, \label{eq:2ndvariation-epsilon} \\
  & \Delta \delta \kappa_{\alpha \beta} = - \left(\delta \vect{a}_{\alpha,\beta} \cdot \Delta \vect{a}_3 + \Delta \vect{a}_{\alpha,\beta} \cdot \delta \vect{a}_3 + \vect{a}_{\alpha,\beta} \cdot \Delta \delta \vect{a}_3 \right)\,. \label{eq:2ndvariation-kappa}
\end{align}

The components of the second Piola Kirchhoff stress resultants $\vect{n}$ and $\vect{m}$ (which are energetically conjugate to the Green-Lagrange strains) are:
\begin{align}\label{eq:constitutive}
	\vect{n} 	= d \, \mat{C} \; \vect{\varepsilon} \,, \qquad
   \vect{m} = \frac{d^3}{12} \mat{C} \; \vect{\kappa} \,,
\end{align}
where $\mat{C}$ denotes the material matrix that is:
\begin{align}\label{eq:material-matrix}
  \mat{C} = \frac{E}{1-\nu^2} \, \begin{bmatrix} 
    1 & \nu & 0 \\ \nu & 1 & 0 \\ 0 & 0 & \frac{1-\nu}{2}
\end{bmatrix}\,.
\end{align}
where we assume the Saint Venant-Kirchhoff material model, i.e. a linear relationship. 
Here, $E$ and $\nu$ are the Young's modulus and Poisson's ratio. 
We note that the strains and stress resultants in \eqref{eq:constitutive} are defined in a local Cartesian coordinate system.

\section{Mixed variational formulation}\label{sec:mixed_form}

In this section, we briefly recall the Hu-Washizu and the Hellinger-Reissner variational principles of elastostatics, see e.g. \cite{felippa1994survey},
which form the starting point for the developments in the remainder of this work.

\subsection{The Hu-Washizu principle}

We recall the three-field Hu-Washizu functional \cite{fraeijs1951diffusion,hu1954some,washizu1955variational}:
\begin{align} \label{eq:hu-washizu-func}
    \Pi_{HW}(\vect{u}, \mat{E}, \mat{S}) \;  = & \;
        \frac{1}{2} \int_{\Omega} \mat{E}^T \, (\mat{C} \ \mat{E}) \; \mathrm{d} \Omega 
         - \int_{\Omega} \vect{u}^T \vect{F}_b \; \mathrm{d} \Omega 
       - \int_{\Gamma_t} \vect{u}^T \, \vect{F}_t \; \mathrm{d} \Gamma \nonumber \\
      & - \int_{\Omega} \mat{S}^T (\mat{E} - \strainOp(\vect{u})) \; \mathrm{d} \Omega
        - \int_{\Gamma_u} (\mat{S} \,^T \normalvect) \, (\vect{u} - \vect{u}_0) \; \mathrm{d} \Gamma \,,
\end{align}
defined for an elastic body with domain $\Omega$, where we write all strain and stress quantities in Voigt notation. 

The variables $\vect{u}$, $\mat{E}$, and $\mat{S}$ denote \textit{independent} displacement, strain, and stress fields, which are also sometimes denoted as \textit{master} fields in this context \cite{felippa1994survey}. To distinguish the dependent strain fields from the independent strain fields, we introduce the strain-displacement operator $\strainOp(\vect{u})$, which acts on the displacement field to produce the dependent strain fields via the kinematic relations. The resulting dependent fields are also sometimes denoted as \textit{slave} fields in this context \cite{felippa1994survey}. 

Furthermore, $\mat{C}$ is the material matrix, relating stresses and strains, $\vect{F}_b$ is the body force acting on the elastic body $\Omega$, $\vect{F_t}$ the traction acting on the traction boundary $\Gamma_t$, $\normalvect$ is a matrix that acts on a stress vector in Voigt notation to produce the traction vector based on the outward-facing unit normal to the displacement boundary $\Gamma_u$, and $\vect{u}_0$ the prescribed displacement on $\Gamma_u$. 

We observe in \eqref{eq:hu-washizu-func} that its last two terms contain the residuals of the kinematic strain-displacement relation and the displacement boundary conditions. The corresponding relations are imposed variationally by multiplication with energetically conjugate stress and traction fields that act as Lagrange multipliers.

The Hu-Washizu variational theorem then states that:
\begin{align}
    \Pi_{HW}(\vect{u}, \mat{E}, \mat{S}) \; = \; \text{Stationary} \,.
\end{align}

\subsection{The Hellinger-Reissner principle}

A Hellinger-Reissner functional \cite{hellinger1907allgemeinen,reissner1950variational} can be derived from the Hu-Washizu functional \eqref{eq:hu-washizu-func} by
assuming that the constitutive law and the Dirichlet boundary conditions are strongly satisfied \cite{zou_dual_locking_2020}, i.e.:
\begin{align}
    & \mat{S} = \mat{C} \ \mat{E} \quad \text{ on } \Omega \,, \\
    & \vect{u} = \vect{u}_0 \quad \text{ on } \Gamma_u \,,
\end{align}
The Hellinger-Reissner functional follows as:
\begin{align} \label{HR}
    \Pi_{HR}(\vect{u}, \mat{E}) \; & = \;
        \int_{\Omega} \mat{E}^T \, \left(\mat{C} \, \strainOp(\vect{u})\right) - \frac{1}{2} \, \mat{E}^T \, (\mat{C} \ \mat{E}) - \vect{u}^T \vect{F}_b \; \mathrm{d} \Omega - \int_{\Gamma_t} \vect{u}^T \, \vect{F}_t \; \mathrm{d} \Gamma \,.
\end{align}
Our $\Pi_{HR}(\vect{u}, \mat{E})$ is also known as the modified Hellinger-Reissner functional, since it considers the strains, instead of the stresses, as the independent master field \cite{zou_dual_locking_2020}. We note that to distinguish the dependent strain field derived from the displacements, we again use the strain-displacement operator $\strainOp(\vect{u})$, which acts on the displacement field to produce the dependent strain fields. 

The Hellinger-Reissner variational theorem states that:
\begin{align}
    \Pi_{HR}(\vect{u}, \mat{E}) \; = \; \text{Stationary} \,,
\end{align}
i.e. the variation of the Hellinger-Reissner functional equals zero, that is:
\begin{align}
    \delta \Pi_{HR}(\vect{u},\delta \vect{u}, \mat{E}, \delta \mat{E}) 
    = \int_\Omega \delta \strainOp(\vect{u})^T \, (\mat{C} \ \mat{E}) + \delta \mat{E}^T \, \left[\mat{C} \, \left( \strainOp(\vect{u}) - \mat{E} \right) \right] \; \mathrm{d} \Omega
    - \int_\Omega \delta \vect{u}^T \, \vect{F}_b \; \mathrm{d} \Omega - \int_{\Gamma_t} \delta \vect{u}^T \, \vect{F}_t \; \mathrm{d} \Gamma  = 0 \,.
\end{align}
These nonlinear equations must be solved via linearization and the Newton-Raphson method. The linearization yields \cite{zou_dual_locking_2020}:
\begin{align}\label{eq:HW-linearization}
    L\left[\delta \Pi_{HR}(\vect{u},\delta \vect{u}, \mat{E}, \delta \mat{E})\right] = 
    \delta \Pi_{HR}(\vect{u},\delta \vect{u}, \mat{E}, \delta \mat{E}) + 
    \Delta \delta \Pi_{HR}(\Delta \vect{u},\delta \vect{u}, \Delta \mat{E}, \delta \mat{E}) = 0 \,,
\end{align}
where
\begin{align}
    \Delta \delta \Pi_{HR}(\Delta \vect{u},\delta \vect{u}, \Delta \mat{E}, \delta \mat{E}) 
    & = \underbrace{\int_\Omega - \delta \mat{E}^T \, \left(\mat{C} \, \Delta \mat{E}\right) + \delta \mat{E}^T \, \left[\mat{C} \, \Delta \strainOp(\vect{u})\right] + \delta \strainOp(\vect{u})^T \, \left(\mat{C} \, \Delta \mat{E}\right) \; \mathrm{d} \Omega}_{\text{material stiffness}} \nonumber \\
    & + \underbrace{\int_\Omega \Delta \delta \strainOp(\vect{u})^T \, (\mat{C} \ \mat{E}) + \Delta \delta \mat{E}^T \, \left[\mat{C} \, \left(\strainOp(\vect{u}) - \mat{E}\right) \right] \; \mathrm{d} \Omega}_{\text{geometric stiffness}}
\end{align}

\subsection{Weak formulation of Kirchhoff-Love shells in mixed format}

We now cast the Kirchhoff shell formulation reviewed in the previous section into Hellinger-Reissner format. To this end, we consider the membrane and bending strains as independent master fields. 
We now distinguish between the independent strain field
\begin{align}\label{eq:strain-var-KL}
    \mat{E} = \vect{e} + \theta^3 \, \vect{k} \,.
\end{align}
which consists of the independent membrane and bending strain fields, $\vect{e}$ and $\vect{k}$, and the strain-displacement operator
\begin{align}\label{eq:strain-operator-KL}
    \strainOp(\vect{u}) = \vect{\varepsilon} + \theta^3 \, \vect{\kappa} \,.
\end{align}
which consists of the displacement-dependent membrane and bending fields $\vect{\varepsilon}$ and $\vect{\kappa}$. We note that the latter are strongly related to the displacements via the shell kinematics reviewed in \eqref{eq1} to \eqref{eq:2ndvariation-kappa}.

We now substitute \eqref{eq:strain-var-KL} and \eqref{eq:strain-operator-KL} (in Voigt notation) into the Hellinger-Reissner variational formulation \eqref{HR}, use the material matrix \eqref{eq:material-matrix} and analytically integrate through the thickness. We then obtain the following mixed variational formulation of the Kirchhoff-Love shell problem:
\begin{align}\label{eq:HR-mixed-form-KLa}
    & d \int_{\bar{\Omega}} \delta \vect{e}^T \mat{C}\left(\vect{\varepsilon} - \vect{e} \right) + \delta \vect{\varepsilon}^T \mat{C} \, \vect{e} \; \mathrm{d} \bar{\Omega} \,+\,
    \frac{d^3}{12} \int_{\bar{\Omega}} \delta \vect{k}^T \mat{C}\left(\vect{\kappa} - \vect{k} \right) + \delta \vect{\kappa}^T \mat{C} \, \vect{k} \; \mathrm{d} \bar{\Omega} 
    - d \int_{\bar{\Omega}} \, \delta \vect{u}^T \, \vect{F}_b \; \mathrm{d} \bar{\Omega} - d \int_{\bar{\Gamma}_t} \, \delta \vect{u}^T \vect{F}_t \; \mathrm{d} \bar{\Gamma} \, = 0 \,,
\end{align}
where $\bar{\Omega}$ and $\bar{\Gamma}_t$ denote the middle surface of the shell body and its traction boundary, respectively.

\begin{remark}
    In \eqref{eq:HR-mixed-form-KLa}, we assume that the shell is subjected merely to the body force $\vect{F}_b$ and the traction $\vect{F}_t$ at the traction boundary (see e.g. \cite{Guo2021}). 
    For ease of notation, we neglect here other Neumann boundary conditions, for instance, due to bending moments. 
    The Dirichlet boundary conditions are enforced strongly via prescribing the displacement of the corresponding control points, see e.g. \cite{Kiendl_shell_2009}.
\end{remark}

Taking into account the independence of the variations of the different fields, we can write this equation in the following three separate variational statements:
\begin{subequations}\label{eq:HR-mixed-form-KL}
    \begin{empheq}[left=\empheqlbrace]{align}
       d \int_{\bar{\Omega}} \delta \vect{\varepsilon}^T \mat{C} \, \vect{e} \; \mathrm{d} \bar{\Omega}
        + \frac{d^3}{12} \int_{\bar{\Omega}} \delta \vect{\kappa}^T \mat{C} \, \vect{k} \; \mathrm{d} \bar{\Omega} \; & = \; d \int_{\bar{\Omega}} \, \delta \vect{u}^T \, \vect{F}_b \; \mathrm{d} \bar{\Omega} + d \int_{\bar{\Gamma}_t} \, \delta \vect{u}^T \vect{F}_t \; \mathrm{d} \bar{\Gamma} \\
         d \int_{\bar{\Omega}} \delta \vect{e}^T \mat{C}\left(\vect{\varepsilon} - \vect{e} \right) \; \mathrm{d} \bar{\Omega} \; & = \;  0 \\
        \frac{d^3}{12} \int_{\bar{\Omega}} \delta \vect{k}^T \mat{C}\left(\vect{\kappa} - \vect{k} \right) \; \mathrm{d} \bar{\Omega}  \; & = \; 0 \,.
    \end{empheq}
\end{subequations}
It is straightforward to identify the last two equations in \eqref{eq:HR-mixed-form-KL} as the projection of the displacement-dependent membrane and bending strain fields, $\vect{\varepsilon}$ and $\vect{\kappa}$, on the independent membrane and bending strain fields, such that the remaining difference between the corresponding fields is minimized with respect to the strain energy norm. The last two equations of the system \eqref{eq:HR-mixed-form-KL} are therefore equivalent to the projection \eqref{eq:l2-projection}, with the exception of the two different norms.

Following \eqref{eq:HW-linearization}, we can write the linearization of the system \eqref{eq:HR-mixed-form-KL} as follows:
\begin{subequations}\label{eq:HR-mixed-form-KL-linearized}
    \begin{empheq}[left=\empheqlbrace]{align}
        d \,& \int_{\bar{\Omega}} \Delta \delta \vect{\varepsilon}^T \mat{C} \, \vect{e} \; \mathrm{d} \bar{\Omega} + \frac{d^3}{12} \, \int_{\bar{\Omega}} \Delta \delta \vect{\kappa}^T \mat{C} \, \vect{k} \; \mathrm{d} \bar{\Omega} \,+\, 
        d \,\int_{\bar{\Omega}} \delta \vect{\varepsilon}^T \mat{C} \, \Delta \vect{e} \; \mathrm{d} \bar{\Omega} \,+\,
        \frac{d^3}{12} \int_{\bar{\Omega}} \delta \vect{\kappa}^T \mat{C} \, \Delta  \vect{k} \; \mathrm{d} \bar{\Omega} \nonumber  \\
        & \qquad \qquad = d \, \int_{\bar{\Omega}} \, \delta \vect{u}^T \, \vect{F}_b \; \mathrm{d} \bar{\Omega} + d \, \int_{\bar{\Gamma}_t} \, \delta \vect{u}^T \vect{F}_t \; \mathrm{d} \bar{\Gamma}
        - d \, \int_{\bar{\Omega}} \delta \vect{\varepsilon}^T \mat{C} \, \vect{e} \; \mathrm{d} \bar{\Omega}
        - \frac{d^3}{12} \int_{\bar{\Omega}} \delta \vect{\kappa}^T \mat{C} \, \vect{k} \; \mathrm{d} \bar{\Omega} \\
        d \, & \int_{\bar{\Omega}} \delta \vect{e}^T \mat{C} \, \Delta \vect{\varepsilon} - \delta \vect{e}^T \mat{C} \, \Delta \vect{e} \; \mathrm{d} \bar{\Omega} \,=\, - d \, \int_{\bar{\Omega}} \delta \vect{e}^T \mat{C}\left(\vect{\varepsilon} - \vect{e} \right) \; \mathrm{d} \bar{\Omega} \\
        \frac{d^3}{12} \, & \int_{\bar{\Omega}} \delta \vect{k}^T \mat{C} \, \Delta \vect{\kappa} - \delta \vect{k}^T \mat{C} \, \Delta \vect{k} \; \mathrm{d} \bar{\Omega} \,=\, - \frac{d^3}{12} \int_{\bar{\Omega}} \delta \vect{k}^T \mat{C}\left(\vect{\kappa} - \vect{k} \right) \; \mathrm{d} \bar{\Omega} \,.
    \end{empheq}
\end{subequations}
which is the starting point to set up a Newton-Raphson procedure in a straightforward way.

\begin{remark}\label{remark_strain_projection}
    For Kirchhoff-Love shells, where only membrane locking occurs, we note that an alternative is to merely consider the membrane strains, $\vect{e}$, as unknown auxiliary variables \cite{Guo2021}. 
    This is equivalent to assuming $\vect{k} = \vect{\kappa}$ for all points in $\Omega$. Equation \eqref{eq:HR-mixed-form-KLa} then becomes:
    \begin{align}
        & d \int_{\bar{\Omega}} \delta \vect{e}^T \mat{C}\left(\vect{\varepsilon} - \vect{e} \right) + \delta \vect{\varepsilon}^T \mat{C} \, \vect{e} \; \mathrm{d} \bar{\Omega} \,+\,
        \frac{d^3}{12} \int_{\bar{\Omega}} \delta \vect{\kappa}^T \mat{C} \, \vect{\kappa} \; \mathrm{d} \bar{\Omega} 
        - d \int_{\bar{\Omega}} \, \delta \vect{u}^T \, \vect{F}_b \; \mathrm{d} \bar{\Omega} - d \int_{\bar{\Gamma}_t} \, \delta \vect{u}^T \vect{F}_t \; \mathrm{d} \bar{\Gamma} \, = 0 \,,
    \end{align}
    which can be again expressed in the following separate variational statements:
    \begin{subequations}
        \begin{empheq}[left=\empheqlbrace]{align}
           d \int_{\bar{\Omega}} \delta \vect{\varepsilon}^T \mat{C} \, \vect{e} \; \mathrm{d} \bar{\Omega}
            + \frac{d^3}{12} \int_{\bar{\Omega}} \delta \vect{\kappa}^T \mat{C} \, \vect{\kappa} \; \mathrm{d} \bar{\Omega} \; & = \; d \int_{\bar{\Omega}} \, \delta \vect{u}^T \, \vect{F}_b \; \mathrm{d} \bar{\Omega} + d \int_{\bar{\Gamma}_t} \, \delta \vect{u}^T \vect{F}_t \; \mathrm{d} \bar{\Gamma} \\
             d \int_{\bar{\Omega}} \delta \vect{e}^T \mat{C}\left(\vect{\varepsilon} - \vect{e} \right) \; \mathrm{d} \bar{\Omega} \; & = \;  0 \,.
        \end{empheq}
    \end{subequations}
    We note that this is no longer a second-order problem as in the case of \eqref{eq:HR-mixed-form-KLa}, but a fourth-order problem. 
\end{remark}

\section{Isogeometric discretization and matrix formulation}\label{sec:discretization}

To eliminate membrane locking in the Kirchhoff-Love shell formulation, we follow the well-established strategy of discretizing the independent strains with spline basis functions that are one degree lower than those used for the displacements \cite{Echter:10.1,nguyen2022leveraging,Bouclier:15.1,Guo2021}. Following the discrete spline spaces chosen in Guo and co-workers \cite{Guo2021}, we discretize 
the displacement vector $\vect{u}$, 
the membrane strains $\vect{e}$, and 
the bending strains $\vect{k}$ as follows: 
\begin{align}
    & \vect{u}^h = \begin{bmatrix} u_1^h \\ u_2^h \\ u_3^h \end{bmatrix} 
    = \sum_{i=1}^{n} \, \begin{bmatrix} 
        B_i^{p,p} \, \hat{u}_{1,i}^h \\ 
        B_i^{p,p} \, \hat{u}_{2,i}^h \\ 
        B_i^{p,p} \, \hat{u}_{3,i}^h 
    \end{bmatrix}
    = \sum_{i=1}^{n} \, B_i^{p,p} \mat{I} \, \begin{bmatrix} 
        \hat{u}_{1,i}^h \\ \hat{u}_{2,i}^h \\ \hat{u}_{3,i}^h 
    \end{bmatrix}
    = \sum_{i=1}^{n} \, \mat{B}_i \, \vect{\hat{u}}_i^h \,, \label{eq:discrete-trial-displacement}
\end{align}
\begin{align}
    & \vect{e}^h = \begin{bmatrix} e_{11}^h \\ e_{22}^h \\ 2e_{12}^h \end{bmatrix} = 
    \sum_{i=1}^{m} \, \begin{bmatrix}
        N_i^{p-1,p} \, \hat{e}_{11,i}^h \\
        N_i^{p,p-1} \, \hat{e}_{22,i}^h \\
        N_i^{p-1,p-1} \, 2\hat{e}_{12,i}^h
    \end{bmatrix} = 
    \sum_{i=1}^{m} \,\begin{bmatrix}N_i^{p-1,p} & 0 & 0 \\ 0 & N_i^{p,p-1} & 0 \\ 0 & 0 & N_i^{p-1,p-1} \end{bmatrix} \, \begin{bmatrix} \hat{e}_{11,i}^h \\ \hat{e}_{22,i}^h \\ 2\hat{e}_{12,i}^h \end{bmatrix} = 
    \sum_{i=1}^{m} \, \mat{N}_i \, \vect{\hat{e}}_i^h \,. \label{eq:discrete-trial-strain} \\
    & \vect{k}^h = \begin{bmatrix} k_{11}^h \\ k_{22}^h \\ 2k_{12}^h \end{bmatrix} 
    = \sum_{i=1}^{m} \, \begin{bmatrix}
        N_i^{p-1,p} \, \hat{k}_{11,i}^h \\
        N_i^{p,p-1} \, \hat{k}_{22,i}^h \\
        N_i^{p-1,p-1} \, 2\hat{k}_{12,i}^h
    \end{bmatrix} = 
    \sum_{i=1}^{m} \,\begin{bmatrix}N_i^{p-1,p} & 0 & 0 \\ 0 & N_i^{p,p-1} & 0 \\ 0 & 0 & N_i^{p-1,p-1} \end{bmatrix} \, \begin{bmatrix} \hat{k}_{11,i}^h \\ \hat{k}_{22,i}^h \\ 2\hat{k}_{12,i}^h \end{bmatrix}
    = \sum_{i=1}^{m} \, \mat{N}_i \, \vect{\hat{k}}_i^h \,. \label{eq:discrete-trial-bending}
\end{align}

Here, the basis $B_i^{p,p}$, $i=1,\ldots,\n$, consists of spline basis functions of degree $p$ in both $\theta^1$ and $\theta^2$ directions, and the basis 
$N_i^{p-1,p}$, $i=1,\ldots,\m$, consists of spline functions of degree $p-1$ and $p$ in $\theta^1$ and $\theta^2$ directions, respectively. 
The notation follows analogously for the spline basis $N_i^{p,p-1}$ and the spline basis $N_i^{p-1,p-1}$.
The unknown coefficients can be summarized in subvectors $\vect{\hat{u}}_i^h$, $\vect{\hat{e}}_i^h$, and $\vect{\hat{k}}_i^h$ for each basis function, and later assembled into corresponding global vectors $\vect{\hat{u}}^h$, $\vect{\hat{e}}^h$, and $\vect{\hat{k}}^h$.

One can discretize the variations of the displacements, membrane and bending strains using the same spline spaces of the corresponding independent master field, i.e.: 
\begin{align}
    & \delta \vect{u}^h = \sum_{i=1}^{n} \, \mat{B}_i \, \delta \vect{\hat{u}}^h_i \,, \\
    & \delta \vect{e}^h = \sum_{i=1}^{m} \, \mat{N}_i \, \delta \vect{\hat{e}}_i^h \,, \label{eq:discrete-test-strain-galerkin}\\
    & \delta \vect{k}^h = \sum_{i=1}^{m} \, \mat{N}_i \, \delta \vect{\hat{k}}_i^h \,, \label{eq:discrete-test-bending-galerkin}
\end{align}
Inserting the interpolation of the displacements, membrane and bending strains, and their variations, into the linearized equations \eqref{eq:HR-mixed-form-KL-linearized} leads to the following linear system solved in each Newton-Raphson iteration:
\begin{align}\label{eq:matrix-eq-shell}
    \begin{bmatrix}
        \mat{K}_{11}^{\text{geom}} & \mat{K}_{12} & \mat{K}_{13} \\
        \mat{K}_{21} & \mat{K}_{22} & \mat{0} \\
        \mat{K}_{31} & \mat{0} & \mat{K}_{33} \\
    \end{bmatrix} \, \begin{bmatrix}
        \Delta \vect{\hat{u}}^h \\ \Delta \vect{\hat{e}}^h \\ \Delta \vect{\hat{k}}^h
    \end{bmatrix} = \begin{bmatrix}
        \vect{F}_{ext} - \vect{F}_{int} \\
        -\vect{F}^m_{int} \\
        -\vect{F}^b_{int}
    \end{bmatrix} \,,
\end{align}
where
{\allowdisplaybreaks
\begin{align}
    & \mat{K}_{11}^{\text{geom}} = \int_{\bar{\Omega}} \mat{K}_{11}^{\text{m,geom}} \,+\, \mat{K}_{11}^{\text{b,geom}} \; \mathrm{d} \bar{\Omega} \,, \label{eq:geometric-stiffness}\\
    & \mat{K}_{12} = d \, \int_{\bar{\Omega}} \left(\mat{B}^m\right)^T \mat{C} \, \mat{N} \; \mathrm{d} {\bar{\Omega}} \,, \\
    & \mat{K}_{21} = d \, \int_{\bar{\Omega}} \mat{N}^T \mat{C} \, \mat{B}^m \; \mathrm{d} \bar{\Omega} \,, \\
    & \mat{K}_{22} = - d \, \int_{\bar{\Omega}} \mat{N}^T \mat{C} \, \mat{N} \; \mathrm{d} \bar{\Omega} \,, \label{eq51}\\
    & \mat{K}_{13} = \frac{d^3}{12} \int_{\bar{\Omega}} \left(\mat{B}^b\right)^T \mat{C} \, \mat{N} \; \mathrm{d} {\bar{\Omega}} \,, \\    
    & \mat{K}_{31} = \frac{d^3}{12} \int_{\bar{\Omega}} \mat{N}^T \mat{C} \, \mat{B}^b \; \mathrm{d} \bar{\Omega} \,, \\
    & \mat{K}_{33} = - \frac{d^3}{12} \int_{\bar{\Omega}} \mat{N}^T \mat{C} \, \mat{N} \; \mathrm{d} \bar{\Omega} \,, \label{eq54}\\
    & \vect{F}_{int} = d \, \int_{\bar{\Omega}} \left(\mat{B}^m\right)^T \mat{C} \, \vect{e}^h \; \mathrm{d} \bar{\Omega} \,+\, \frac{d^3}{12} \int_{\bar{\Omega}} \left(\mat{B}^b\right)^T \mat{C} \, \vect{k}^h \; \mathrm{d} \bar{\Omega} \,, \\
    & \vect{F}^m_{int} = d \, \int_{\bar{\Omega}} \mat{N}^T \mat{C} \left(\vect{\varepsilon}^h - \vect{e}^h\right) \; \mathrm{d} \bar{\Omega} \,, \\
    & \vect{F}^b_{int} = \frac{d^3}{12} \int_{\bar{\Omega}} \mat{N}^T \mat{C} \left(\vect{\kappa}^h - \vect{k}^h\right) \; \mathrm{d} \bar{\Omega} \,, \\
    & \vect{F}_{ext} = d \, \int_{\bar{\Omega}} \mat{B}^T \vect{F}_b \; \mathrm{d} \bar{\Omega} \,+\, d \, \int_{\bar{\Gamma}_t} \mat{B}^T \vect{F}_t \; \mathrm{d} \bar{\Gamma} \,.
\end{align}}

The quantities $\mat{B}^m$ and $\mat{B}^b$ denote the strain-displacement matrices corresponding to the membrane and bending strains, respectively. The matrices $\mat{K}_{11}^{\text{m,geom}}$ and $\mat{K}_{11}^{\text{b,geom}}$ denote the contributions to the geometric stiffness matrix due to the membrane and bending deformations, respectively. The quantities $\mat{N}$ and $\mat{B}$ denote the matrices of basis functions for discretizing the strains and the displacements, respectively. 
For more details, we refer the interested reader to the Appendix \ref{sec:shell_derivation}, where we summarize the derivation of these matrices and the force vectors.

\begin{remark}\label{remark_strain_projection2}
    The matrix equation corresponding to the alternative approach discussed in 
    Remark \ref{remark_strain_projection} takes the following form:
    \begin{align}\label{eq:matrix-eq-shell2}
        \begin{bmatrix}
            \mat{K}_{11}^{\text{b,mat}} + \Tilde{\mat{K}}_{11}^{\text{geom}} & \mat{K}_{12} \\ \mat{K}_{21} & \mat{K}_{22}            
        \end{bmatrix} \, \begin{bmatrix}
            \Delta \vect{\hat{u}}^h \\ \Delta \vect{\hat{e}}^h
        \end{bmatrix} = \begin{bmatrix}
            \vect{F}_{ext} - \Tilde{\vect{F}}_{int} \\ -\vect{F}^m_{int}
        \end{bmatrix} \,,
    \end{align}
    where 
    \begin{align}
        & \mat{K}_{11}^{\text{b,mat}} = \frac{d^3}{12} \int_{\bar{\Omega}} \left(\mat{B}^b\right)^T \mat{C} \, \mat{B}^b \; \mathrm{d} {\bar{\Omega}} \,, \\
        & \Tilde{\mat{K}}_{11}^{\text{geom}} = \int_{\bar{\Omega}} \mat{K}_{11}^{\text{m,geom}} \,+\, \Tilde{\mat{K}}_{11}^{\text{b,geom}} \; \mathrm{d} \bar{\Omega} \,, \\
        & \Tilde{\vect{F}}_{int} = d \, \int_{\bar{\Omega}} \left(\mat{B}^m\right)^T \mat{C} \, \vect{e}^h \; \mathrm{d} \bar{\Omega} \,+\, \frac{d^3}{12} \int_{\bar{\Omega}} \left(\mat{B}^b\right)^T \mat{C} \, \vect{\kappa}^h \; \mathrm{d} \bar{\Omega} \,.
    \end{align}
    We derive $\Tilde{\mat{K}}_{11}^{\text{b,geom}}$ in Appendix \ref{sec:shell_derivation}. The tilde denotes the matrices that are changed due to the alternative approach. 
    Equation \eqref{eq:matrix-eq-shell2} has a smaller size than the system \eqref{eq:matrix-eq-shell} since it has one variable field fewer. 
\end{remark}

\section{Strain condensation}\label{sec:strain_condensation}

We now proceed to the key contribution of the current article. In comparison to the discrete system that results from the standard displacement-based Galerkin formulation, the system \eqref{eq:matrix-eq-shell} has several significant disadvantages. For the same mesh resolution, it is larger, has a more complicated sparsity pattern and a larger bandwidth, and therefore requires more memory to store all submatrices and much longer computing times for its fully coupled solution. One idea is to perform static condensation of the unknowns of the strain fields. We thus arrive at a final system that has the same size as the one of a standard displacement-based formulation, but the associated computational cost turns out to be prohibitively expensive if performed in a consistent manner and without special computational tricks \cite{kikis2022two}. A follow-up idea is therefore to come up with special methods that reduce the computational cost associated with condensation to such an extent that the complete computing effort lies at least within the same order of magnitude as the standard displacement-based formulation for the same mesh\cite{kikis2022two}. If this can be achieved without compromising accuracy, the mixed formulation retains its significant advantage of mitigating locking at a cost at least in the range of the standard displacement-based formulation. In the following, we show that our idea for strain condensation that is based on the combination of approximate dual functions and row-sum lumping targets at exactly that. 

\subsection{Approximate dual functions, row-sum lumping and the condensed system}

We use the approximate dual spline functions \eqref{eq:test_func} reviewed in Section \ref{sec:l2projection} to discretize the variations associated with the independent strain fields. Inspired by \cite{zou_dual_locking_2020}, instead of \eqref{eq:discrete-test-strain-galerkin} and \eqref{eq:discrete-test-bending-galerkin}, we use the following expressions for the discrete variations of the membrane and bending strains:
\begin{align}\label{scal1}
    \delta \vect{e}^h = & \frac{1}{d} \, \mat{C}^{-T} \sum_{i=1}^{m} \, \Tilde{\mat{N}}_i \, \delta \vect{\hat{e}}_i^h 
                      = \frac{1}{d} \, \mat{C}^{-T} \Tilde{\mat{N}} \, \delta \vect{\hat{e}}^h \,, \\
  \label{scal2}  \delta \vect{k}^h = & \frac{12}{d^3} \, \mat{C}^{-T} \sum_{i=1}^{m} \, \Tilde{\mat{N}}_i \, \delta \vect{\hat{k}}_i^h 
                      = \frac{12}{d^3} \, \mat{C}^{-T} \Tilde{\mat{N}} \, \delta \vect{\hat{k}}^h \,,
\end{align}
where
\begin{align}
    \tilde{N}_i (\vect{x}) := \frac{\hat{N}_i(\vect{x})}{C(\vect{x})} \,.
\end{align}
The spline basis $\hat{N}_i$, $i=1,\ldots,\m$, consists of the approximate dual functions that correspond to the B-spline trial functions $N_i$ in \eqref{eq:discrete-trial-strain} and \eqref{eq:discrete-trial-bending}. They are modified
by multiplying with the inverse of the determinant of the Jacobian matrix of the mapping, $\detJ(\hat{\mat{x}})$, as discussed in Section \ref{sec22}. The additional scaling terms introduced in \eqref{scal1} and \eqref{scal2} correspond to the values of the inverse membrane stiffnesses and the values of the inverse bending stiffnesses, respectively. We add these scaling terms with the understanding that we only multiply the inverse values, but not the units, such that the following expressions maintain their unit consistency and their interpretation as stiffness matrices and force vectors. 

The matrices $\mat{K}_{11}^{\text{geom}}$, $\mat{K}_{12}$, and $\mat{K}_{13}$, and the force vectors $\vect{F}_{int}$ and $\vect{F}_{ext}$ in the matrix formulation \eqref{eq:matrix-eq-shell} remain the same, while all other matrices and vectors change to: 
{\allowdisplaybreaks
\begin{align}
    & \mat{K}_{21} = \int_{\bar{\Omega}} \Tilde{\mat{N}}^T \, \mat{B}^m \; \mathrm{d} \, \bar{\Omega} \,, \\
    & \mat{K}_{22} = - \int_{\bar{\Omega}} \Tilde{\mat{N}}^T \, \mat{N} \; \mathrm{d} \, \bar{\Omega} \,=\, \mat{K}_{33} \,, \label{neardiag} \\
    & \mat{K}_{31} = \int_{\bar{\Omega}} \Tilde{\mat{N}}^T \, \mat{B}^b \; \mathrm{d} \, \bar{\Omega} \,, \\
    & \vect{F}_{int}^m = \int_{\bar{\Omega}} \Tilde{\mat{N}}^T \left(\vect{\varepsilon}^h - \vect{e}^h \right) \; \mathrm{d} \, \bar{\Omega} \,, \\
    & \vect{F}_{int}^b = \int_{\bar{\Omega}} \Tilde{\mat{N}}^T \left(\vect{\kappa}^h - \vect{k}^h \right) \; \mathrm{d} \, \bar{\Omega} \,.
\end{align}}

\noindent where the values of the membrane stiffnesses and the values of the inverse bending stiffnesses cancel with the scaling of the variations (but not their units). As a consequence of this choice, row-sum lumping of the matrices $\mat{K}_{22}$ and $\mat{K}_{33}$ yields negative identity matrices, $\mat{K}_{22} = \mat{K}_{33} = -\mat{I}$. We emphasize that scaling the variations by the inverse membrane and bending stiffnesses, as shown in \eqref{scal1} and \eqref{scal2}, is essential for eliminating the stiffness terms from \eqref{neardiag}. This transformation ensures that the resulting matrices are nearly diagonal and can be efficiently diagonalized using row-sum lumping without compromising higher-order accuracy (see also further discussions in Appendix \ref{sec:equivalentProj}). If the stiffness matrix remains in \eqref{neardiag}, the resulting matrix is no longer nearly diagonal. This can be easily verified by inserting the unscaled variations discretized with the actual dual spline basis, which evidently does not yield a diagonal matrix.

In turn, we obtain direct expressions of the strain-related unknowns: 
\begin{align}
    & \Delta \vect{\hat{e}}^h = \mat{K}_{21} \, \Delta \vect{\hat{u}}^h + \vect{F}_{int}^m \,, \label{eq:strain_projection_membrane} \\
    & \Delta \vect{\hat{k}}^h = \mat{K}_{31} \, \Delta \vect{\hat{u}}^h + \vect{F}_{int}^b \,. \label{eq:strain_projection_bending}
\end{align}
The unknowns related to the discrete independent membrane and bending strain fields can thus be computed directly, such that they can be a-priori eliminated from the first row of \eqref{eq:matrix-eq-shell}. This results in the following purely displacement-based matrix equations:
\begin{align}\label{eq:matrix-eq-displacement-based}
    \underbrace{\left(\mat{K}_{11}^{\text{geom}} + \mat{K}_{12}\, \mat{K}_{21} + \mat{K}_{13}\, \mat{K}_{31}\right)}_{\mat{K}} \, \Delta \vect{\hat{u}}^h = \vect{F}_{ext} - \vect{F}_{int} - \mat{K}_{12}\,\vect{F}_{int}^m - \mat{K}_{13}\, \vect{F}_{int}^b \,.
\end{align}
From a computational perspective, the condensation procedure in \eqref{eq:matrix-eq-displacement-based} constitutes a significant improvement over the condensation procedure for the standard Galerkin mixed formulation, which we repeat here for convenience:
\begin{align}\label{eq:standard_condensation}
    \underbrace{\left(\mat{K}_{11}^{\text{geom}} - \mat{K}_{12}\,  \mat{K}_{22}^{-1}\, \mat{K}_{21} - \mat{K}_{13}\, \mat{K}_{33}^{-1}\, \mat{K}_{31}\right)}_{\mat{K}} \, \Delta \vect{\hat{u}}^h = \vect{F}_{ext} - \vect{F}_{int} + \mat{K}_{12}\, \mat{K}_{22}^{-1}\, \vect{F}_{int}^m + \mat{K}_{13}\, \mat{K}_{33}^{-1}\, \vect{F}_{int}^b \,.
\end{align}
We can observe that our condensation in \eqref{eq:matrix-eq-displacement-based} relies solely on multiplications of banded matrices. In contrast, the standard condensation procedure in \eqref{eq:standard_condensation} requires the inversion of matrices, which stands at the core of the significant increase in computational cost of standard mixed formulations. Moreover, we will show in the following that the additional cost of multiplying banded matrices is of the same order as the solution of the system of the displacement-based formulation.

\subsection{Theoretical complexity estimates} \label{estimates}

We analyze the computational complexity of the condensation procedure in terms of floating point operations, where we focus on the left-hand matrix side, as it dominates the cost. We start with the consistent Galerkin mixed formulation, where the matrices in \eqref{eq:matrix-eq-shell} have the following properties: (a) all matrices $\mat{K}_{ij}$ in \eqref{eq:matrix-eq-shell} are banded; (b) due to the structure of our discretization, we can assume that they have approximately the same size $n \times n$ and approximately the same bandwidth $b$; (c) $-\mat{K}_{22}$ and $-\mat{K}_{33}$ are symmetric and positive definite. We can therefore avoid direct matrix inversion and use Cholesky factorization instead. 
The condensation procedure for the consistent mixed formulation then consists of the following four steps:

\begin{enumerate}

\item Factorization of $\mat{K}_{22}$ and $\mat{K}_{33}$ with a complexity of $\mathcal{O}(n b^2)$.

\item Forward and back substitution to solve $\mat{K}_{22} \, \vect{x} = \mat{K}_{21}$ and $\mat{K}_{33} \, \vect{x} = \mat{K}_{31}$ with a total complexity of $\mathcal{O}(n^2 b)$. One must expect that in general, the resulting matrices $\mat{K}_{22}^{-1}\, \mat{K}_{21}$ and $\mat{K}_{33}^{-1}\, \mat{K}_{31}$ are dense.

\item Matrix-matrix multiplications $\mat{K}_{12}\, (\mat{K}_{22}^{-1}\, \mat{K}_{21})$ and $\mat{K}_{13}\, (\mat{K}_{33}^{-1}\, \mat{K}_{31})$ with a complexity of $\mathcal{O}(n^2 b)$, as one matrix in each multiplication is potentially dense.

\item Final subtraction of a dense matrix with a complexity of $\mathcal{O}(n^2)$.

\end{enumerate}

We then turn to our formulation, where $\mat{K}_{22}$ and $\mat{K}_{33}$ are negative identity matrices. For all other matrices, the above assumptions and properties hold. The condensation procedure then simplifies to the following two steps:

\begin{enumerate}

\item Matrix-matrix multiplications $\mat{K}_{12}\, \mat{K}_{21}$ and $\mat{K}_{13}\, \mat{K}_{31}$: As both matrices in each multiplication are banded, one obtains a complexity of $\mathcal{O}(n b^2)$.

\item Final subtraction: As the matrix to be subtracted is banded (although with increased bandwidth), the complexity is at most $\mathcal{O}(n \; 2b)$.

\end{enumerate}

Once we have chosen a mixed formulation and a polynomial degree $p$, the bandwidth $b$ remains fixed during $h$-refinement. Hence, when we refine the mesh, the complexity depends only on $n$. When we compare the theoretical complexity estimates, we observe that the leading term for the standard Galerkin mixed formulation is $\mathcal{O}(n^2 b)$, while for our formulation the leading term is $\mathcal{O}(n b^2)$. We observe that our formulation is linear in $n$, while the Galerkin mixed formulation is quadratic in $n$. 

It is reasonable and not unfair to assume that for the standard displacement-based formulation, the system is solved via Cholesky factorization with a complexity of $\mathcal{O}(n b^2)$. We can then conclude that in our mixed formulation, the leading cost factor is of the same order with respect to $n$ as in the standard displacement-based formulation. 

In the Galerkin mixed formulation, in contrast, the cost of the projection becomes dominating with increasing $n$. The reason is the matrix inversion, which leads to potentially dense system matrices, for which factorization in the final solution step has a complexity of $\mathcal{O}(n^3)$. These observations show that the computational expense of the Galerkin mixed formulation in this form is prohibitive and the method is practically not viable in this form.

\begin{remark}
Our analysis is heavily based on the assumption of banded matrices. When extending the analysis to general sparse matrices, the complexity becomes highly dependent on the sparsity pattern of the matrix and the amount of fill-in during factorization, where the complexity of the factorization could grow up to $\mathcal{O}(n \log{n})$\cite{trefethen2022numerical}.
\end{remark}

\subsection{Sparsity pattern, bandwidth and symmetry of the condensed system matrix}

\begin{figure}[t!]
	\centering 
    \captionsetup[subfloat]{labelfont=scriptsize,textfont=scriptsize}
	    \subfloat[Our approach (strain variables condensed)]{{\includegraphics[width=0.386\textwidth]{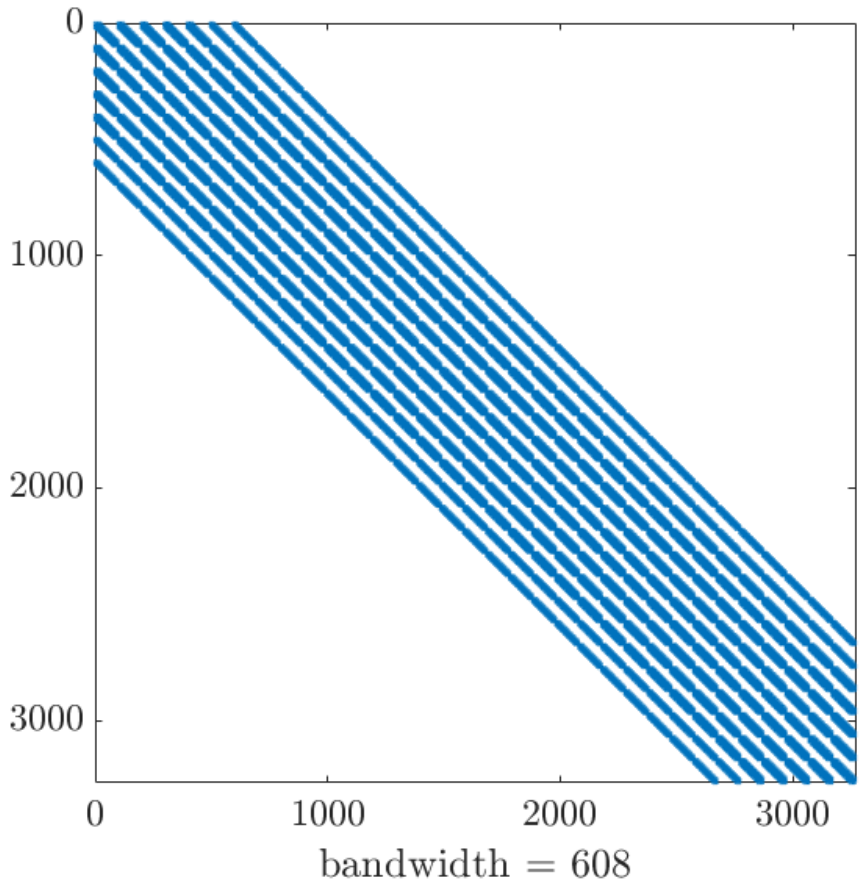} }} \hspace{0.9cm}
        \subfloat[Standard displacement-based Galerkin formulation]{{\includegraphics[width=0.386\textwidth]{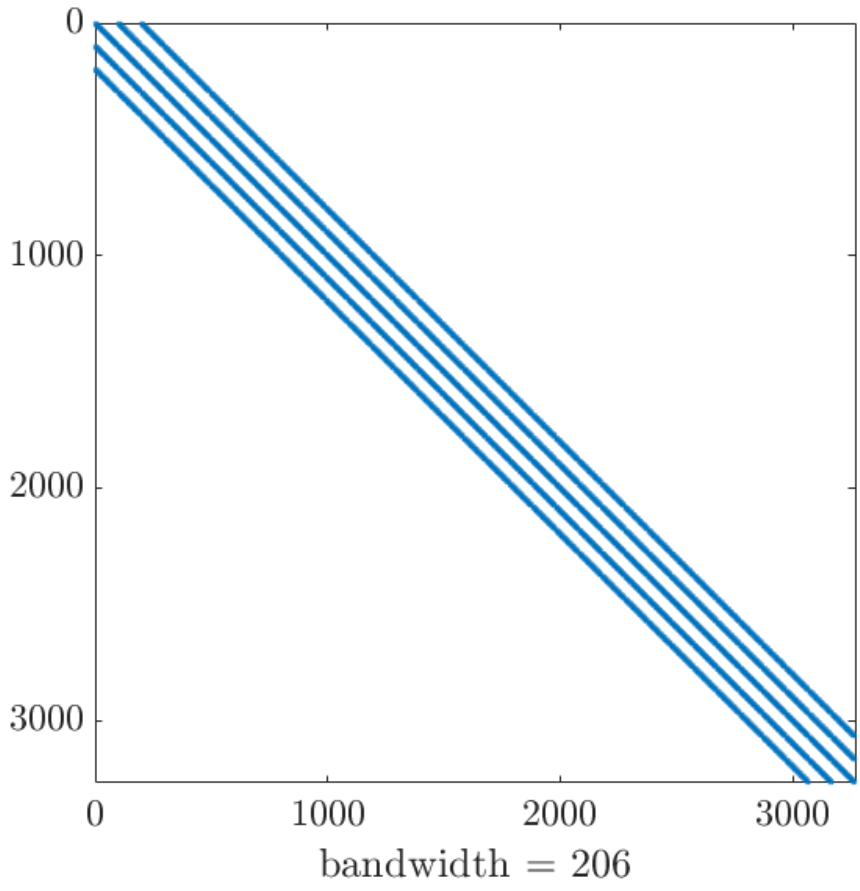} }} \hspace{.4cm}
    \vspace{0.2cm}
    \caption{Sparsity patterns of the displacement-based system matrix for a patch of quadratic B-splines on a mesh of $32 \times 32$ elements.} 
    \label{fig:K_bandwidth}
\end{figure}

In the next step, we therefore focus on comparing our mixed formulation to the standard displacement-based formulation. We have seen in our complexity analysis that in terms of $n$, they share the same complexity. We expect, however, a difference in the solution cost for the (condensed) system matrices of the two formulations due to (a) the difference in the bandwidth $b$ and (b) the (missing) symmetry of the system matrix. 
The following considerations put the additional computing effort in our formulation in perspective:
\begin{itemize}

\item The condensed system matrix in our approach remains sparse and banded because its calculation does not involve matrix inversion, unlike for the consistent Galerkin mixed formulation.  
Figure \ref{fig:K_bandwidth}a illustrates the sparsity pattern for a discretization of a square domain with quadratic B-splines on a mesh of $32 \times 32$ B\'ezier elements.

\item Figure \ref{fig:K_bandwidth}b shows the sparsity pattern for the corresponding system matrix obtained with the standard displacement-based formulation. We observe that the maximum bandwidth of the condensed matrix in our approach is larger. There are two reasons for that: the first one is that the upper bandwidth of the matrix contributions due to strain condensation, $\mat{K}_{12}\mat{K}_{21}$ and $\mat{K}_{13}\mat{K}_{31}$, results from the sum of the corresponding upper bandwidths of the multiplicands \cite{trefethen2022numerical}. The second reason is that the support of an approximate dual function is up to $3p+1$ B\'ezier elements in each parametric direction, as compared to only $p+1$ for the corresponding B-spline \cite{nguyen_mass_lumping2023}. We deduce from Figure \ref{fig:K_bandwidth} that the bandwidth increases by about a factor two due to the first reason and by another factor of 1.5 due to the increased support of the approximate dual functions.  

\item Due to the fact that we discretize the independent strain variables and their variations with different basis functions,
the off-diagonal matrices $\mat{K}_{12}$ and $\mat{K}_{13}$ are not the transpose of their counterparts $\mat{K}_{21}$ and $\mat{K}_{31}$, and the matrix contributions due to strain condensation are therefore unsymmetric. Solving a system whose coefficient matrix is unsymmetric requires a larger computational effort than solving a symmetric system of the same size, see e.g.\ \cite{trefethen2022numerical}. 

\end{itemize}

\begin{remark}\label{remark_strain_projection4} 
Due to the increased support of the approximate dual functions, also the formation of the submatrices $\mat{K}_{21}$ and $\mat{K}_{31}$ requires more operations than matrices in standard Galerkin formulations. For two-dimensional plate and shell elements, we have to expect approximately five to nine times as many basis function related operations to form the two submatrices. In practical scenarios, computationally costly routines to take into account nonlinear material behavior, such as radial return algorithms in plasticity, do not depend on the cost or number of the basis functions of the discrete test space, and hence the net increase in computational cost per quadrature point will be much lower. In addition, only two submatrices of the system matrix in \eqref{eq:matrix-eq-displacement-based} are affected. The (symmetric) submatrices $\mat{K}_{11}$, $\mat{K}_{12}$ and $\mat{K}_{13}$ are equivalent to Galerkin formulations and therefore incur no additional cost. 
Just as standard B-splines (or NURBS), the modified approximate dual basis functions can be cast into a B\'ezier or Lagrange extraction format per element \cite{borden2011isogeometric,schillinger2016lagrange}, whose derivation from relation \eqref{eq:approx_inv} and the approximate inverse of the Gramian matrix is straightforward. In comparison to the computational costs associated with condensation and solving the system, the formation cost is relatively not significant in the scenarios considered here.
\end{remark}

\begin{remark}\label{remark_strain_projection3}
    When we assume the strong satisfaction of the strain-displacement relation for the bending strains, we discretize only the remaining variation of the independent membrane strain field, see Remarks \ref{remark_strain_projection} and \ref{remark_strain_projection2}.
    The displacement-based matrix equations \eqref{eq:matrix-eq-displacement-based} then become:
    \begin{align}
        \left( \mat{K}_{11}^{\text{b,mat}} + \Tilde{\mat{K}}_{11}^{\text{geom}} + \mat{K}_{12} \mat{K}_{21} \right) \Delta \vect{\hat{u}}^h = \vect{F}_{ext} - \Tilde{\vect{F}}_{int} - \mat{K}_{12} \vect{F}_{int}^m \,.
    \end{align}
    We note that compared to \eqref{eq:matrix-eq-displacement-based}, assembling the right-hand side and the stiffness matrix requires     
    one vector-matrix multiplication fewer and 
    one matrix-matrix multiplication fewer, respectively. 
    Instead, one needs to assemble an additional matrix, $\mat{K}_{11}^{\text{b,mat}}$, and the stiffness matrix remains unsymmetric due to the product $\mat{K}_{12} \mat{K}_{21}$. 
    Although it depends on the implementation, in particular the efficiency of local matrix-matrix products versus the formation of the additional stiffness matrix, the alternative approach of considering only the membrane strains as unknown strain variables does not seem to significantly reduce the computational cost in assembly and solving the matrix equations, compared to our approach of considering both membrane and bending strains as unknown variable fields.
\end{remark}

\begin{figure}[t!]
    \centering
    \captionsetup[subfloat]{labelfont=scriptsize,textfont=scriptsize}
    \subfloat[$p=2$]{{
        \def\svgwidth{0.48\textwidth}
        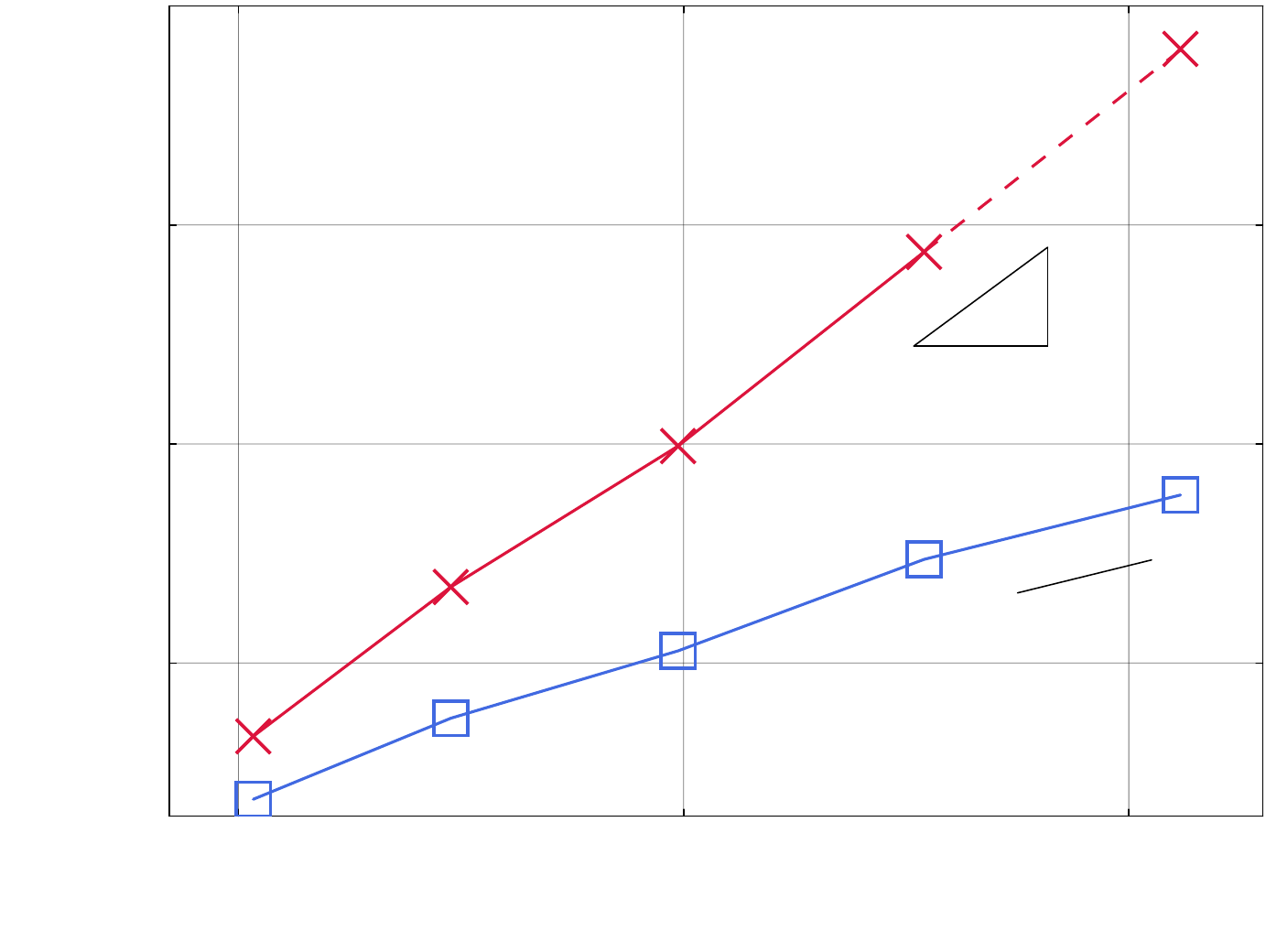 }} \hspace{0.1cm}
    \subfloat[$p=4$]{{
        \def\svgwidth{0.48\textwidth}
        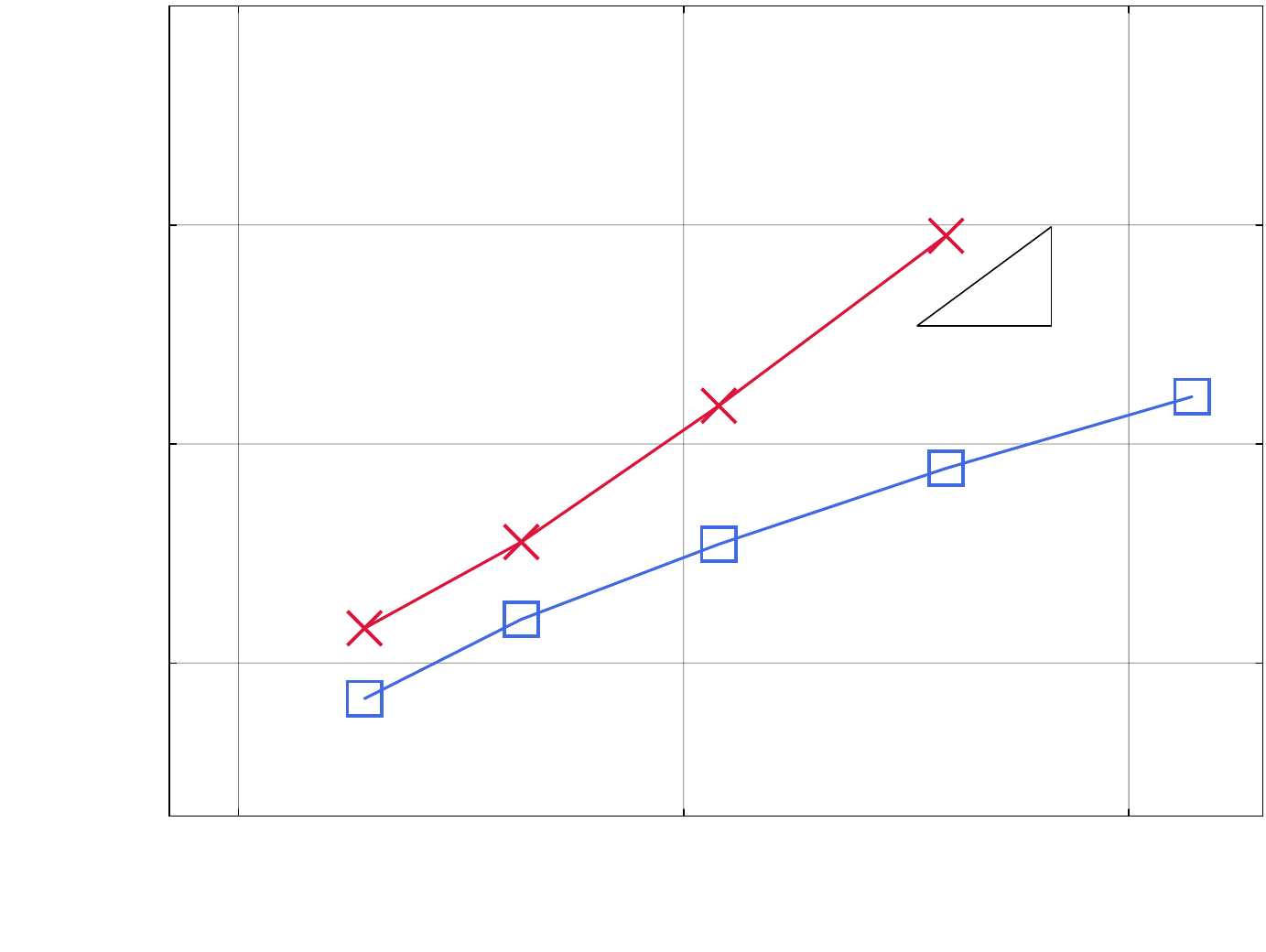 }}
    
    \vspace{0.2cm}
    {\begin{tikzpicture}
    \filldraw[red1,line width=1pt, solid] (0.0,0) -- (0.6,0);
    \filldraw[red1,line width=1pt] (0.0,0) node[right]{\scriptsize $\boldsymbol{\bigtimes}$};
    \filldraw[red1,line width=1pt] (0.7,0) node[right]{\scriptsize Mixed formulation, consistent strain projection};
    \filldraw[blue1,line width=1pt, solid] (7.8,0.05) -- (8.4,0.05);
    \filldraw[blue1,line width=1pt] (8.0,-0.08) [fill=none] rectangle ++(0.25,0.25);
    \filldraw[blue1,line width=1pt] (8.5,0) node[right]{\scriptsize Mixed formulation, lumped strain projection, approximate duals};
\end{tikzpicture}}
    \caption{Computing times for the static condensation only, measured for the hemispherical shell benchmark. The dashed portions of the curves are extrapolated.}\label{fig:tiime_static_condensation}
\end{figure}

\begin{figure}[t!]
    \centering
    \captionsetup[subfloat]{labelfont=scriptsize,textfont=scriptsize}
    \subfloat[$p=2$]{{
        \def\svgwidth{0.48\textwidth}
        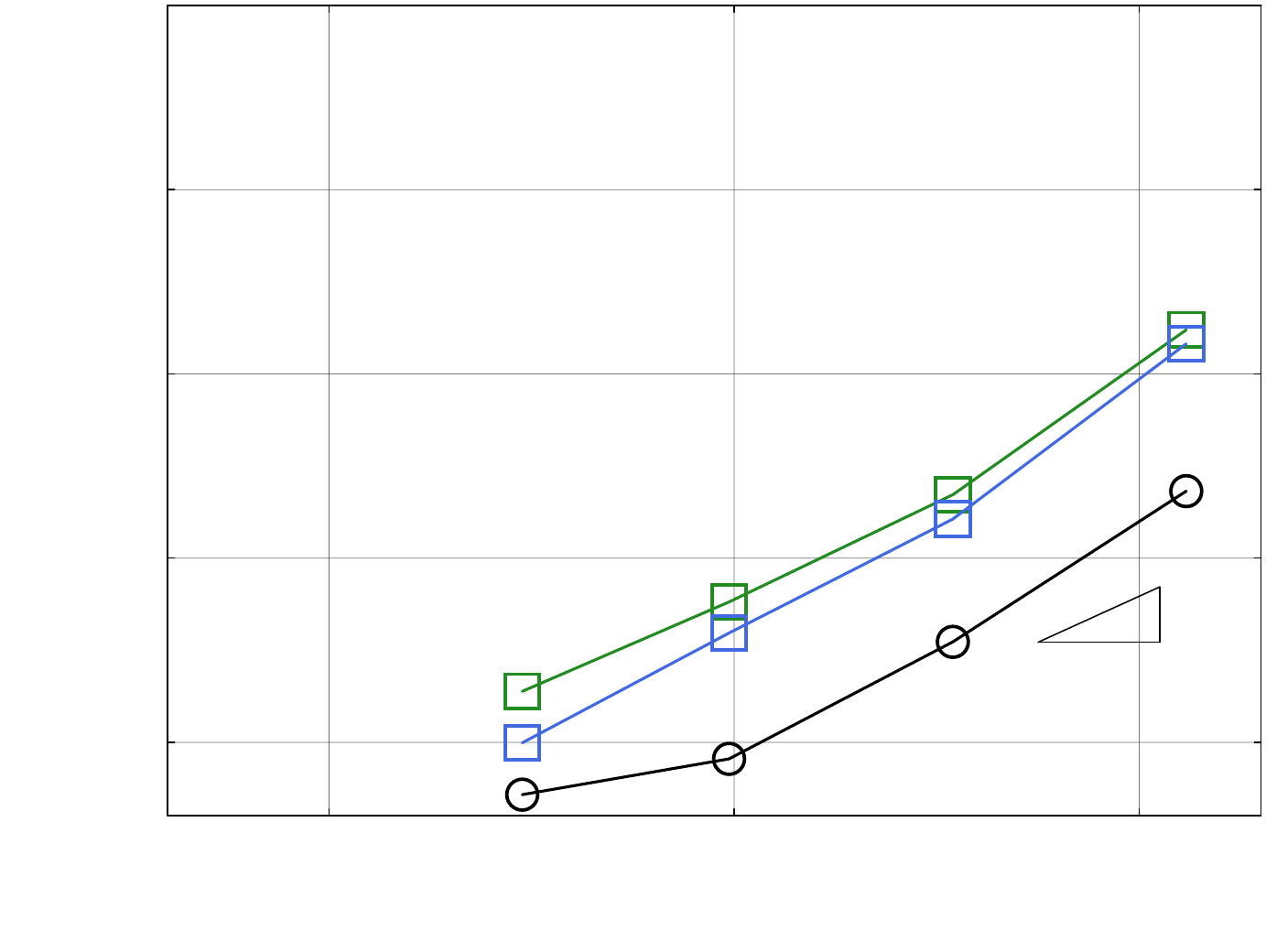 }} \hspace{0.1cm}
    \subfloat[$p=4$]{{
        \def\svgwidth{0.48\textwidth}
        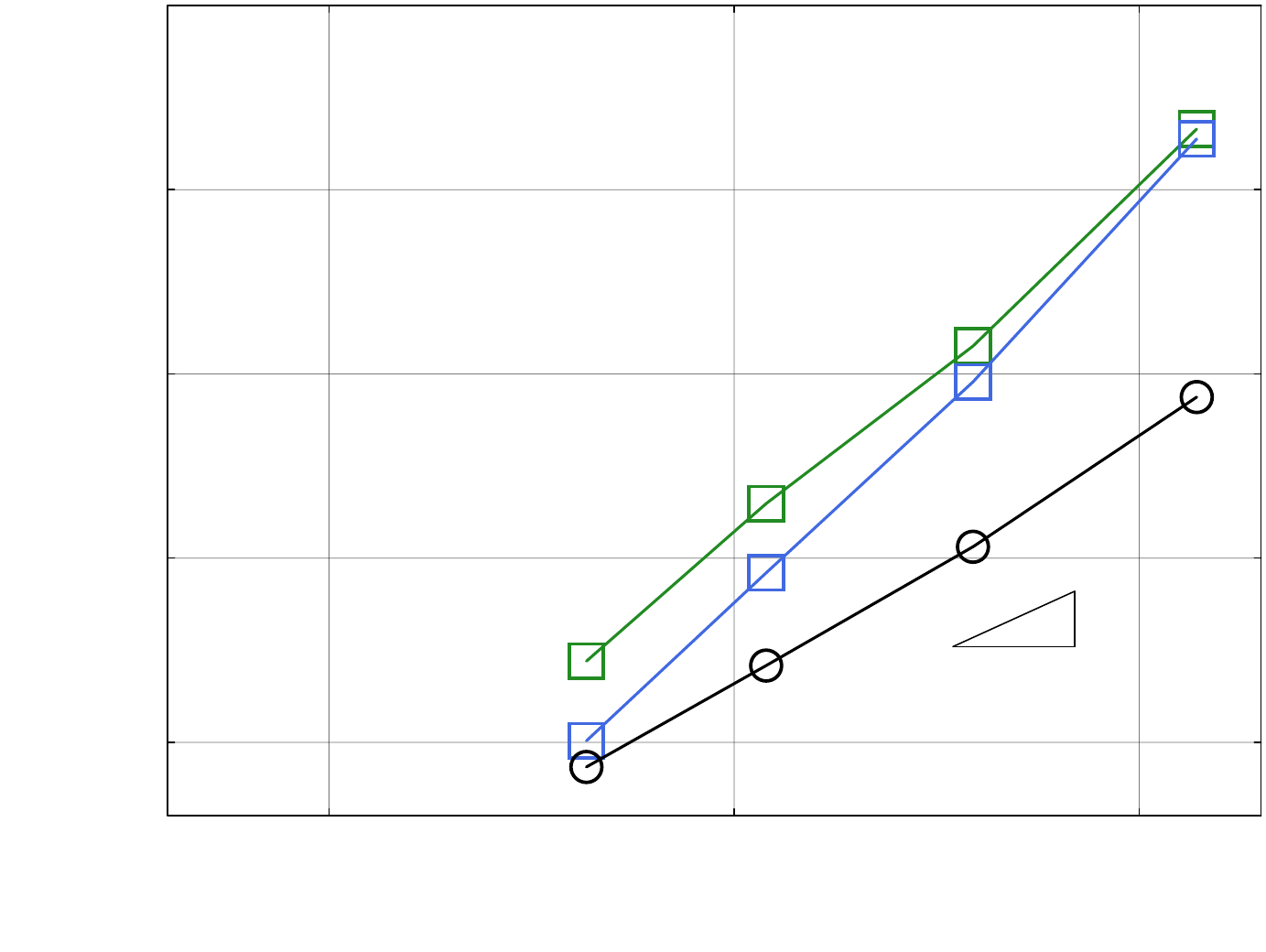 }}
    
    \vspace{0.2cm}
    {\begin{tikzpicture}
    \filldraw[black,line width=1pt, solid] (0.0,0) -- (0.6,0);
    \filldraw[black,line width=1pt] (0.3,0) [fill=none] circle (3pt);
    \filldraw[black,line width=1pt] (0.7,0) node[right]{\scriptsize Displacement-based formulation};
\end{tikzpicture}

\begin{tikzpicture}
    \filldraw[blue1,line width=1pt, solid] (2.5,0.05) -- (3.1,0.05);
    \filldraw[blue1,line width=1pt] (2.7,-0.08) [fill=none] rectangle ++(0.25,0.25);
    \filldraw[blue1,line width=1pt] (3.2,0) node[right]{\scriptsize Mixed formulation, lumped strain projection, approximate duals, without time for static condensation};
\end{tikzpicture}

\begin{tikzpicture}
    \filldraw[green1,line width=1pt, solid] (3.5,0.05) -- (4.1,0.05);
    \filldraw[green1,line width=1pt] (3.7,-0.08) [fill=none] rectangle ++(0.25,0.25);
    \filldraw[green1,line width=1pt] (4.2,0) node[right]{\scriptsize Mixed formulation, lumped strain projection, approximate duals, with time for static condensation};
\end{tikzpicture}}
    \caption{Computing times for the solution of the system of equations in our approach (with and without strain condensation) vs.\ in the standard displacement-based Galerkin formulation, measured for the hemispherical shell benchmark.}\label{fig:time_back_slash}
\end{figure}

\subsection{Computing times}

We would like to illustrate the validity of the main results of our theoretical considerations via practical computing times. The reported computing times are obtained with our Julia implementation for the hemispherical shell benchmark, discussed in more detail later in Section \ref{sec:lin_sphere}. We note that our code was not specifically optimized for any of the formulations considered. All formulations were executed on the same machine under identical conditions. The reported computing times represent the average of 100 repeated runs to minimize the impact of incidental system fluctuations.

We first focus on the strain condensation only, where we only consider the linear algebra operations discussed in Section \ref{estimates}. We entirely rely on linear algebra routines as provided by Julia: \texttt{Base.$*$} and \texttt{Base.$+$} for matrix-matrix multiplication and addition, \texttt{LinearAlgebra.Cholesky} for Cholesky factorization, and \texttt{Base.$\backslash$} for forward and back substitution. For all routines, we enforce serial execution on a single core. In Figure \ref{fig:tiime_static_condensation}, we compare the measured computing times required for the static condensation in the consistent Galerkin mixed formulation and in our approach under mesh refinement, where $n$ refers to the size of the condensed system matrix, and for polynomial degrees $p=2$ and $4$. 

The plots confirm our key observation based on our theoretical estimates: In our formulation, the computing time increases only linearly with increasing $n$, while the computational cost of the consistent Galerkin mixed formulation is prohibitively expensive. For the latter, our computational experiments indicate even a complexity of $\mathcal{O}(n^3)$, as the \texttt{Base.$*$} routine does not take advantage of the bandedness of one of the matrices in the multiplication step 3. For a (potentially small) mesh of $64\times 64$ elements and $p=2$, which corresponds to $n=13,068$ in the condensed system, the condensation process in our approach is about 10,000 times faster than the condensation in the consistent Galerkin mixed formulation. When we increase the polynomial degree, we observe that for the small problem sizes considered here, the difference becomes smaller due to the increased bandwidth of the submatrices in our formulation.

We then focus on the direct solution of the condensed system of equations with an unsymmetric matrix in our formulation versus the direct solution of the original system with a symmetric stiffness matrix in the standard displacement-based formulation. For both types of systems, we rely on Julia's \texttt{Base.$\backslash$} operator, which is expected to employ an $LU$ decomposition for the unsymmetric system and a Cholesky factorization for the symmetric system. For all routines, we enforce serial execution on a single core. In Figure \ref{fig:time_back_slash}, we compare the measured computing times required for the solution of the condensed system in our approach with and without the cost for static strain condensation and for the solution of the system of the standard displacement-based formulation under mesh refinement, where $n$ refers to the size of the condensed system matrix, and for polynomial degrees $p=2$ and $4$. 

The plots confirm the following key observations: In our new formulation, the cost for static condensation becomes insignificant relative to the cost for the solution of the unsymmetric system. For $p=2$, our formulation shows the same linear complexity as the standard displacement-based formulation. Our formulation consistently remains less than 10 times as expensive as the standard displacement-based formulation for the same number of elements, while offering locking-free solution fields. We observe, however, that for $p=4$, the difference between the computing times of the two different methods seems to increase with $n$. Consolidation of this observation 
requires further testing and analysis, which is outside the scope of this work.

Apart from these observations, a precise evaluation of our approach's computational performance is significantly influenced by the application context and the quality of the individual computer implementation. In particular, the computational aspects discussed above play a different role depending on whether a direct or iterative solver is used, or if explicit dynamics calculations are performed where the condensed system matrix is never assembled, stored, or inverted.

\section{Numerical examples}\label{sec:results}
	
In the following, we demonstrate the favorable numerical properties of higher-order accurate diagonalized strain projection based on approximate dual spline functions. We first consider a curved Euler-Bernoulli beam to assess the accuracy in terms of the preservation of higher-order convergence rates and the mitigation of membrane locking as compared to a variety of established methods. We then consider the classical shell obstacle course to showcase the performance of our approach in comparison to other mixed formulations. 
Finally, we consider two geometrically nonlinear cylindrical shell benchmarks to assess the performance of our formulation under large deformations.

\begin{figure}[t!]
	\begin{center}
		\def\svgwidth{0.7\textwidth}
		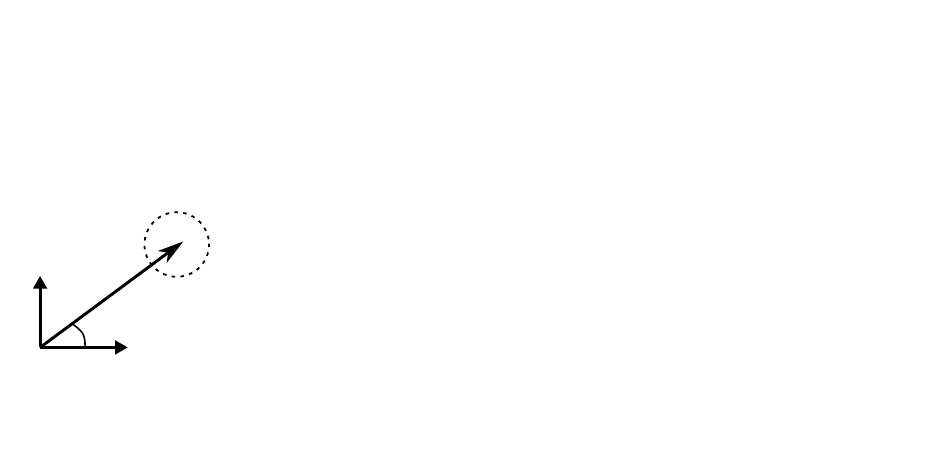
    \end{center}
    \caption{A quarter circle steel cantilever represented as a curved Euler-Bernoulli beam, along with boundary and loading conditions.}
    \label{fig:beam_geometry}
\end{figure}

\begin{figure}[ht!]
	\centering
    \captionsetup[subfloat]{labelfont=scriptsize,textfont=scriptsize}
    \subfloat[$p=2$]{{ \def\svgwidth{0.48\textwidth}
    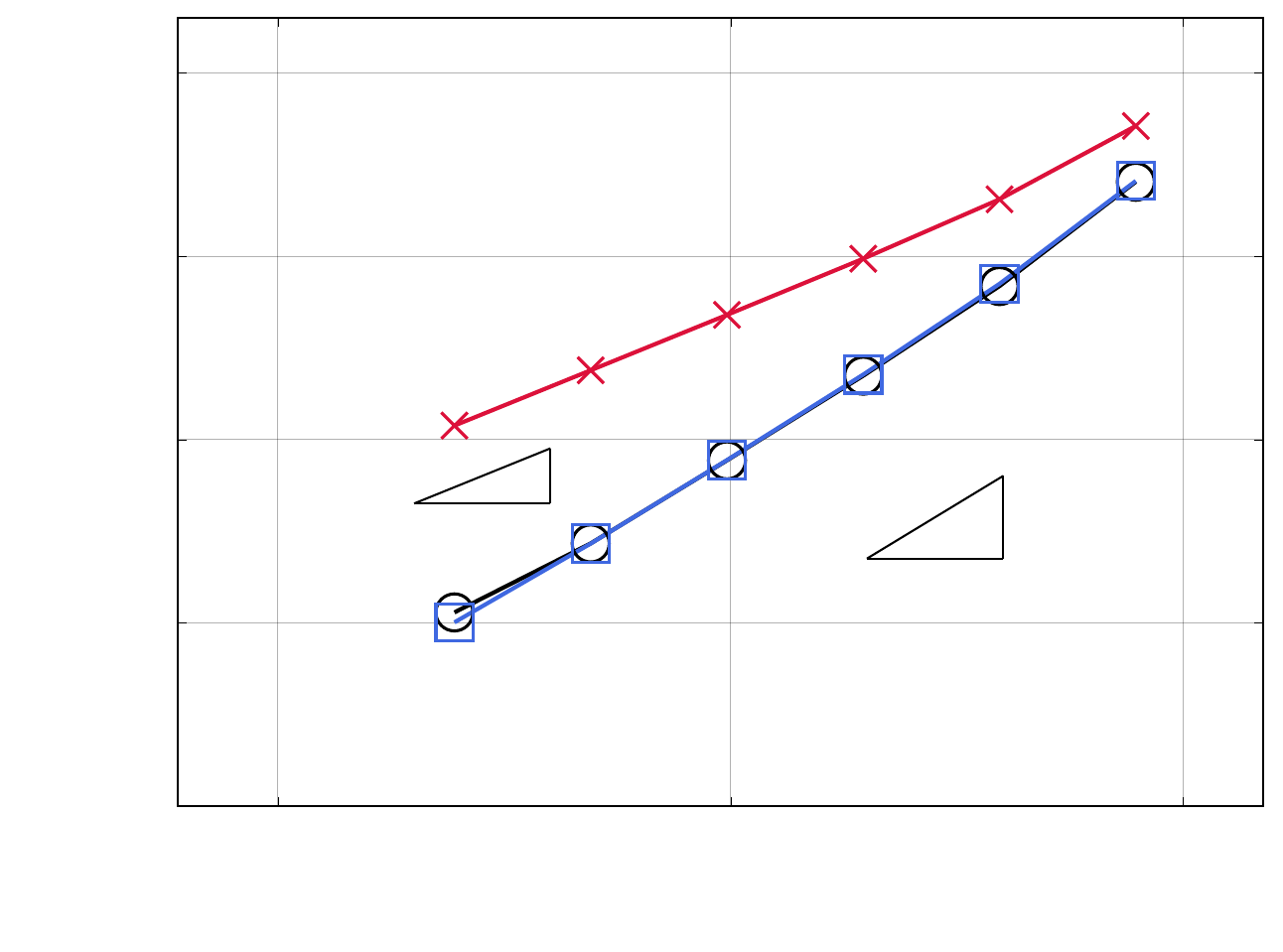 }}
    \subfloat[$p=3$]{{ \def\svgwidth{0.48\textwidth}
    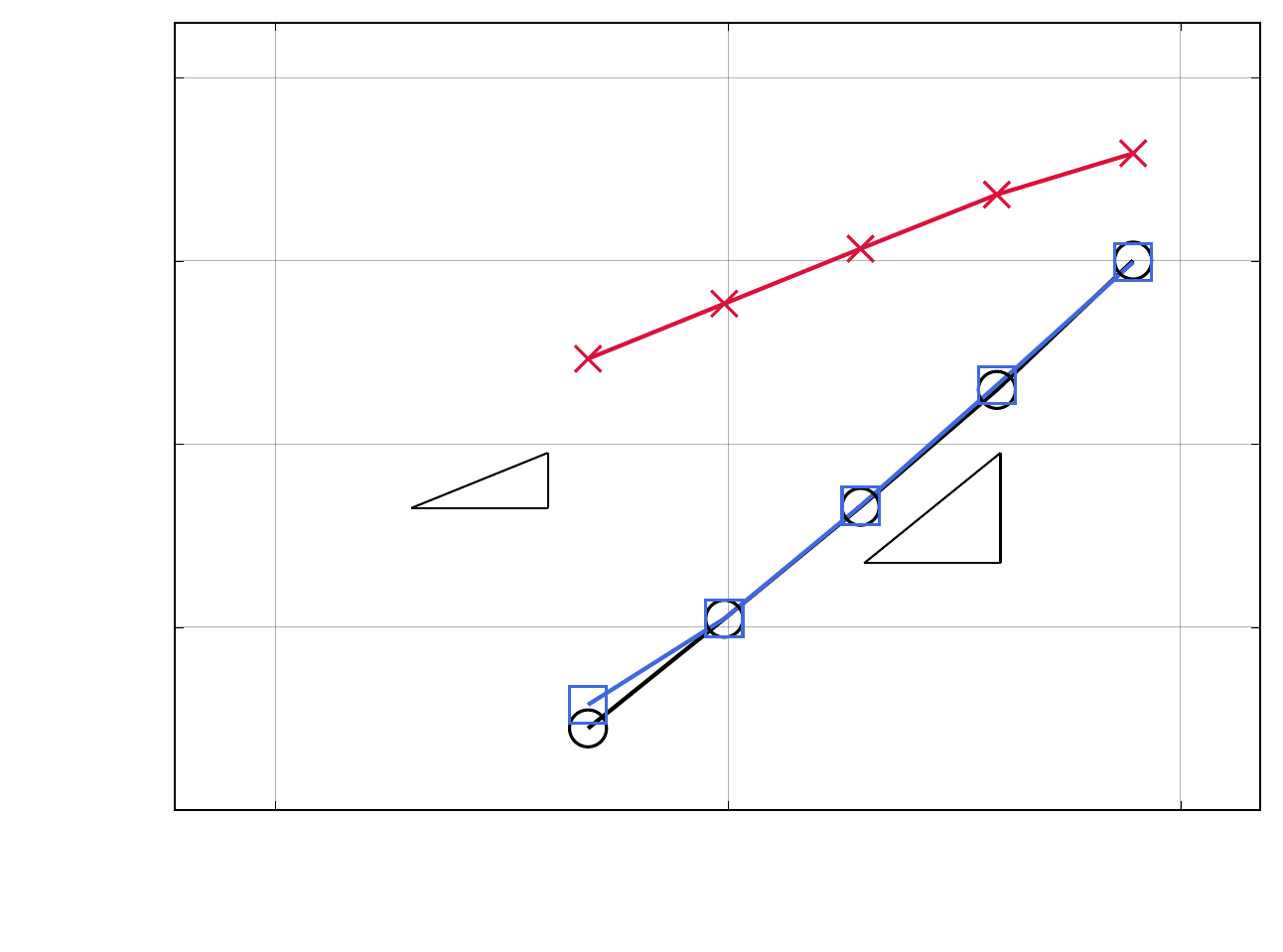 }}

    \subfloat[$p=4$]{{ \def\svgwidth{0.48\textwidth}
    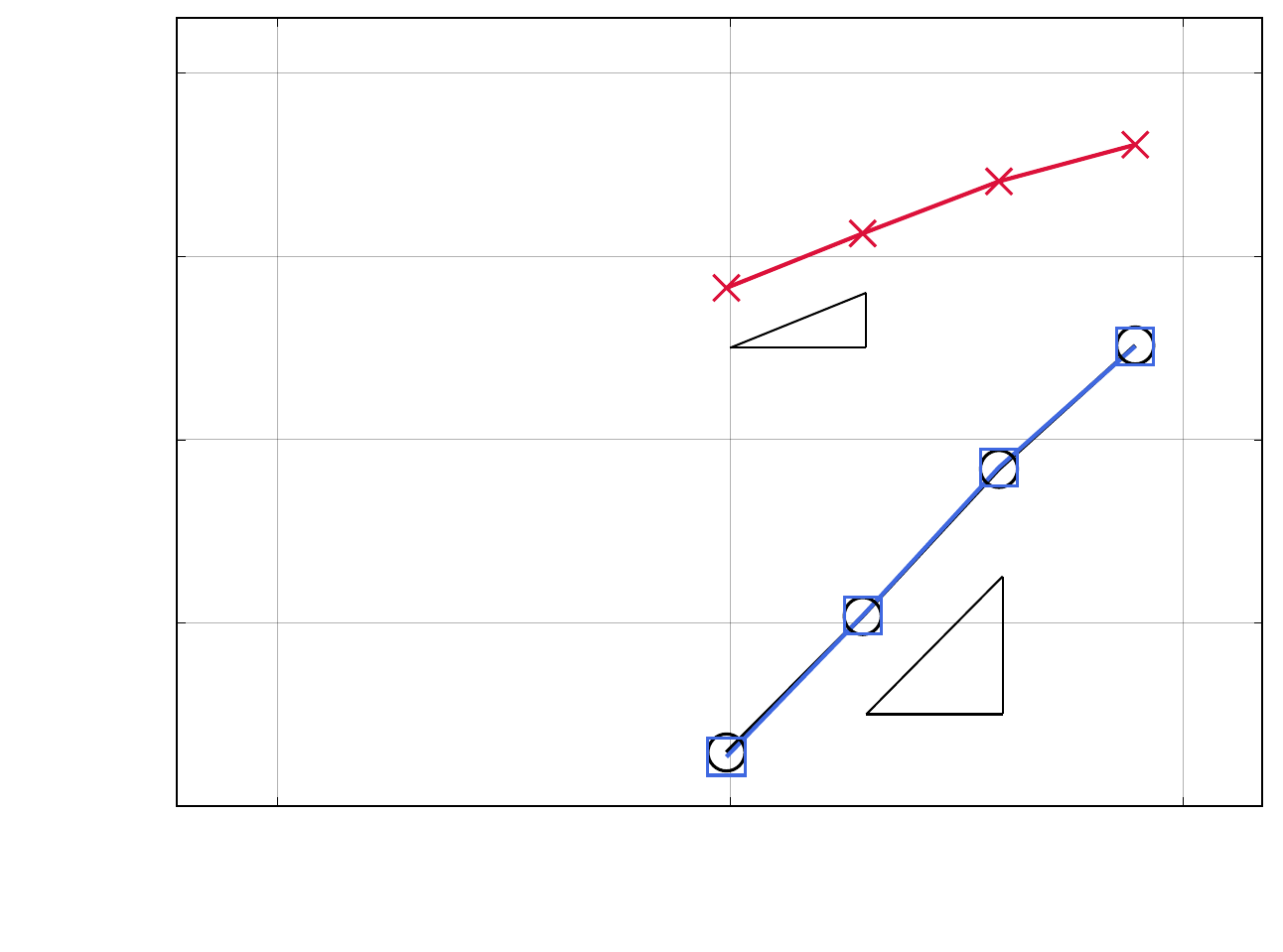 }}
    \subfloat[$p=5$]{{ \def\svgwidth{0.48\textwidth}
    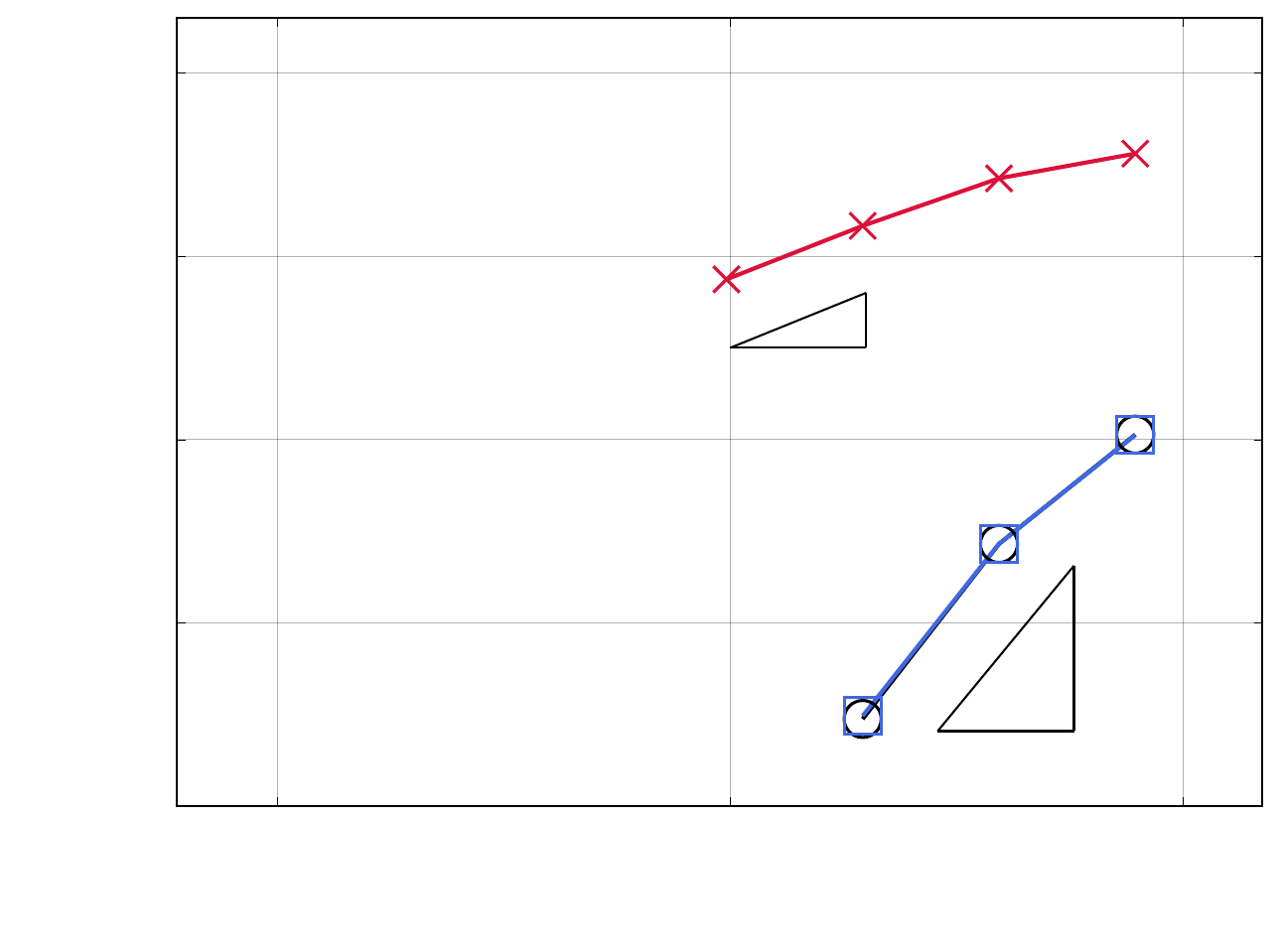 }}
    
    \vspace{0.2cm}
    \begin{tikzpicture}
    \filldraw[black,line width=1pt, solid] (0.0,0) -- (0.6,0);
    \filldraw[black,line width=1pt] (0.3,0) [fill=none] circle (3pt);
    \filldraw[black,line width=1pt] (0.7,0) node[right]{\scriptsize Galerkin mixed formulation with consistent strain condensation};
    \filldraw[blue1,line width=1pt, solid] (10.0,0.05) -- (10.7,0.05);
    \filldraw[blue1,line width=1pt] (10.2,-0.08) [fill=none] rectangle ++(0.25,0.25);
    \filldraw[blue1,line width=1pt] (10.8,0) node[right]{\scriptsize Our approach};
\end{tikzpicture}

\begin{tikzpicture}
    \filldraw[red1,line width=1pt, solid] (2.0,0) -- (2.6,0);
    \filldraw[red1,line width=1pt] (2.0,0) node[right]{\scriptsize $\boldsymbol{\bigtimes}$};
    \filldraw[red1,line width=1pt] (2.7,0) node[right]{\scriptsize Galerkin mixed formulation with lumping of the strain projection matrix};
\end{tikzpicture}
    \caption{Curved Euler-Bernoulli beam: relative error in the $L^2$-norm of the displacement field, computed with mixed formulations and different methods of strain projection.}\label{fig:beam_convergence_projection}
\end{figure}

\begin{figure}[ht!]
	\centering
    \captionsetup[subfloat]{labelfont=scriptsize,textfont=scriptsize}
    \subfloat[$p=2$]{{ \def\svgwidth{0.48\textwidth}
    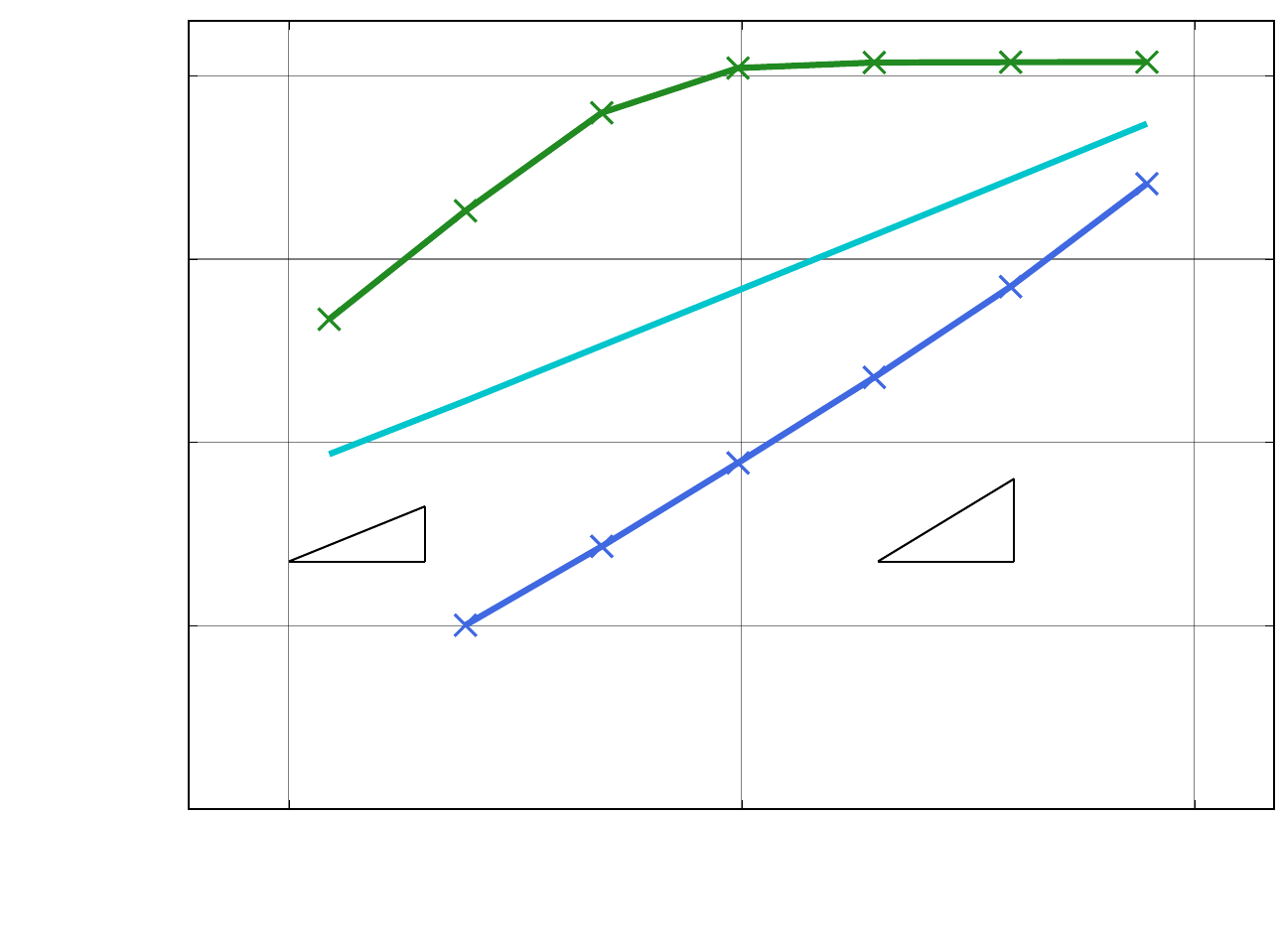 }}
    \subfloat[$p=3$]{{ \def\svgwidth{0.48\textwidth}
    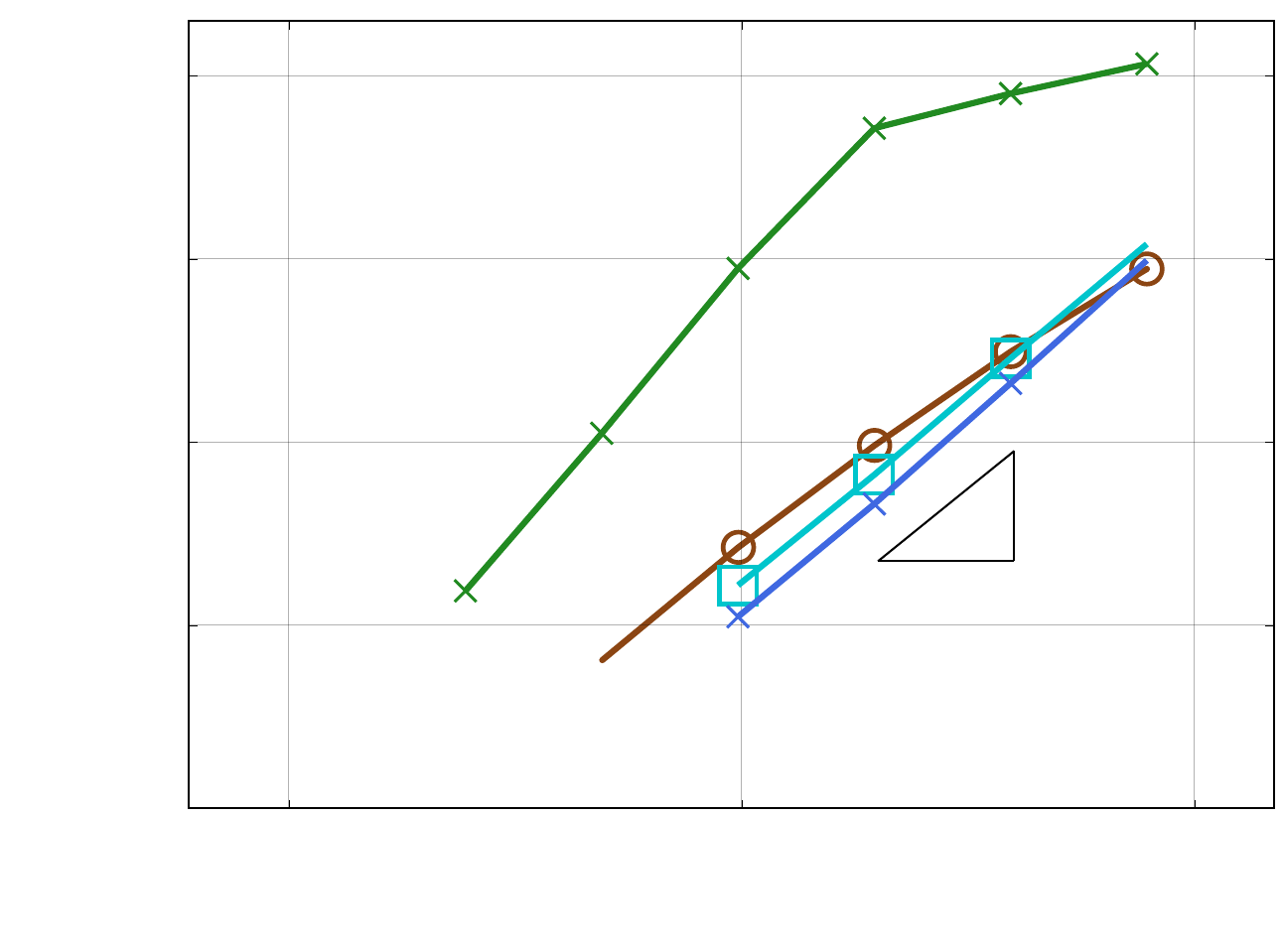 }}

    \subfloat[$p=4$]{{ \def\svgwidth{0.48\textwidth}
    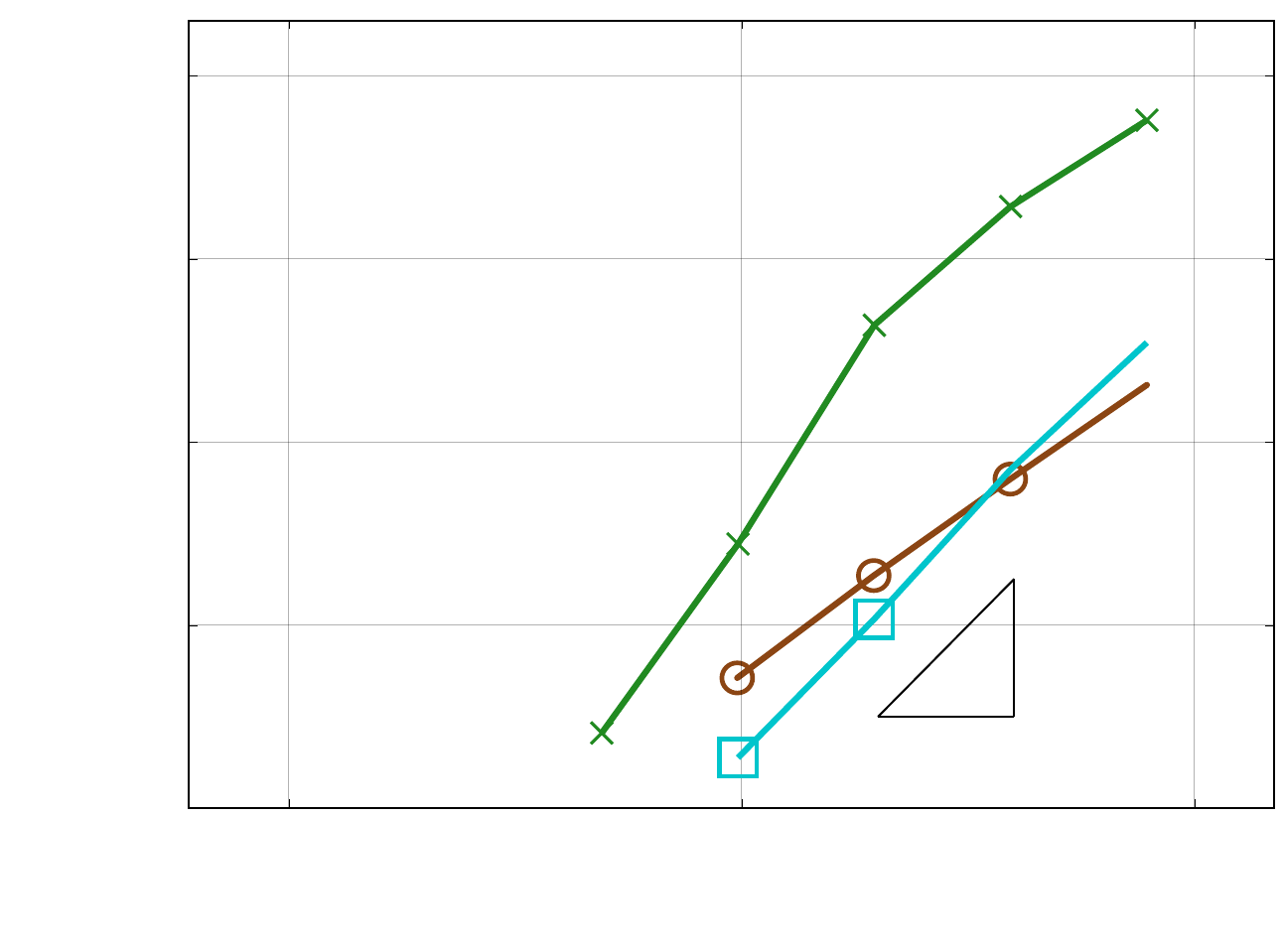 }}
    \subfloat[$p=5$]{{ \def\svgwidth{0.48\textwidth}
    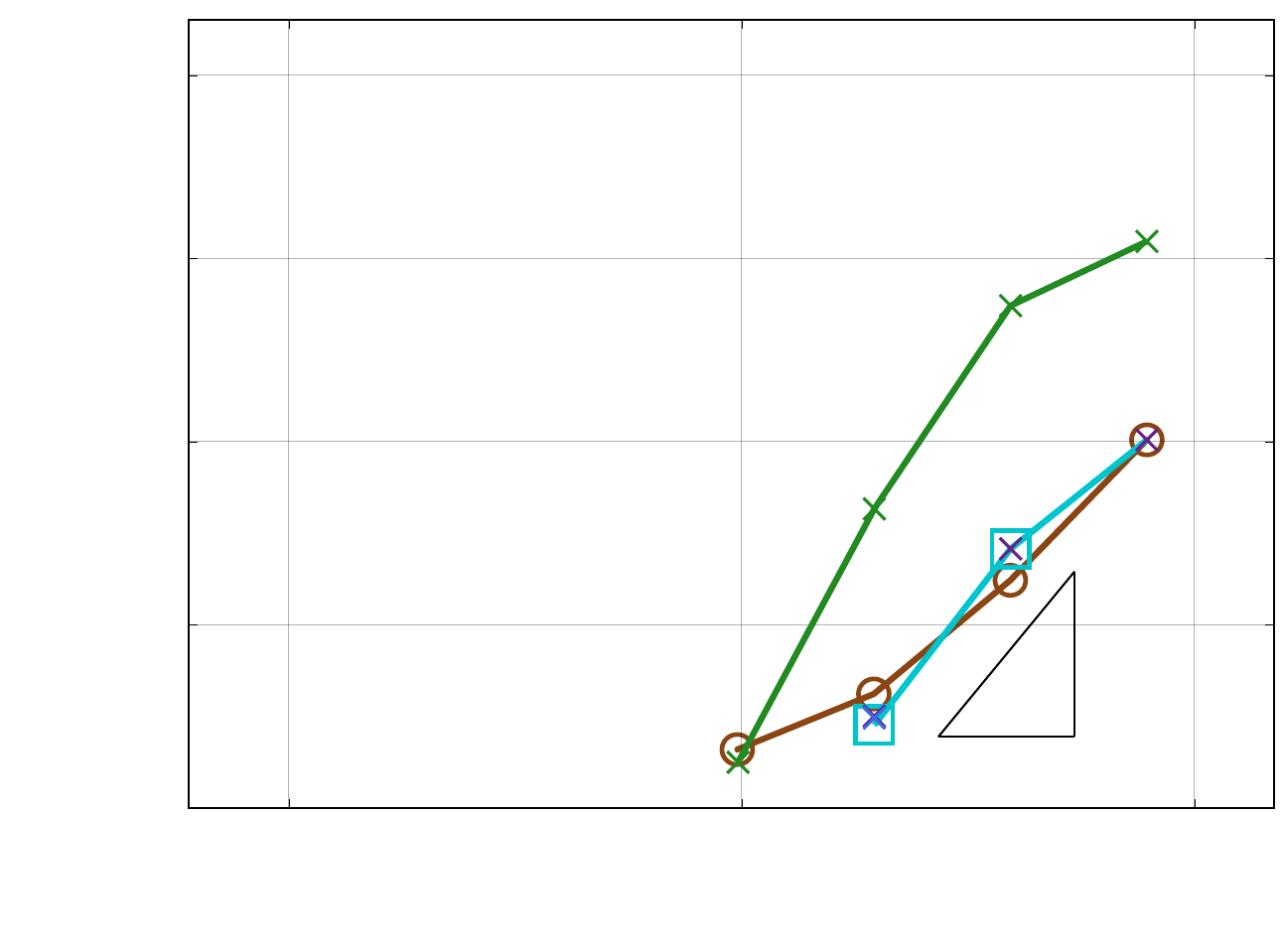 }}
    
    \vspace{0.2cm}
    \begin{tikzpicture}
    \filldraw[green1,line width=1pt, solid] (0.0,0) -- (0.6,0);
    \filldraw[green1,line width=1pt] (0.0,0) node[right]{\scriptsize $\boldsymbol{\bigtimes}$};
    \filldraw[green1,line width=1pt] (0.7,0) node[right]{\scriptsize Standard Galerkin, full integration};
    \filldraw[brown1,line width=1pt, solid] (6.0,0.05) -- (6.6,0.05);
    \filldraw[brown1,line width=1pt] (6.3,0.05) [fill=none] circle (3pt);
    \filldraw[brown1,line width=1pt] (6.7,0) node[right]{\scriptsize Standard Galerkin, {reduced integration\cite{leonetti2019simplified} ($\mathcal{S}_0^2,\mathcal{S}_1^3,\mathcal{S}_2^4,\mathcal{S}_2^5$ for $p=2,3,4,5$, respectively)}};
\end{tikzpicture}

\begin{tikzpicture}
    \filldraw[lightblue1,line width=1pt, solid] (0.5,0.05) -- (1.1,0.05);
    \filldraw[lightblue1,line width=1pt] (0.7,-0.08) [fill=none] rectangle ++(0.25,0.25);;
    \filldraw[lightblue1,line width=1pt] (1.2,0) node[right]{\scriptsize Standard Galerkin, B-bar method};
    \filldraw[purple1,line width=1pt, solid] (6.8,0) -- (7.4,0);
    \filldraw[purple1,line width=1pt] (6.8,0) node[right]{\scriptsize $\boldsymbol{\bigtimes}$};
    \filldraw[purple1,line width=1pt] (7.5,0) node[right]{\scriptsize Galerkin mixed formulation, EAS method};
    \filldraw[blue1,line width=1pt, solid] (14.0,0) -- (14.6,0);
    \filldraw[blue1,line width=1pt] (14.0,0) node[right]{\scriptsize $\boldsymbol{\bigtimes}$};
    \filldraw[blue1,line width=1pt] (14.7,0) node[right]{\scriptsize Our approach};
\end{tikzpicture}
    \caption{Curved Euler-Bernoulli beam: relative error in the $L^2$-norm of the displacement field, computed with our approach and different well-established locking-preventing mechanisms.}\label{fig:beam_convergence_locking_method}
\end{figure}

\begin{figure}[ht!]
	\centering
    \captionsetup[subfloat]{labelfont=scriptsize,textfont=scriptsize}
    \subfloat[$p=2$]{{ \def\svgwidth{0.45\textwidth}
    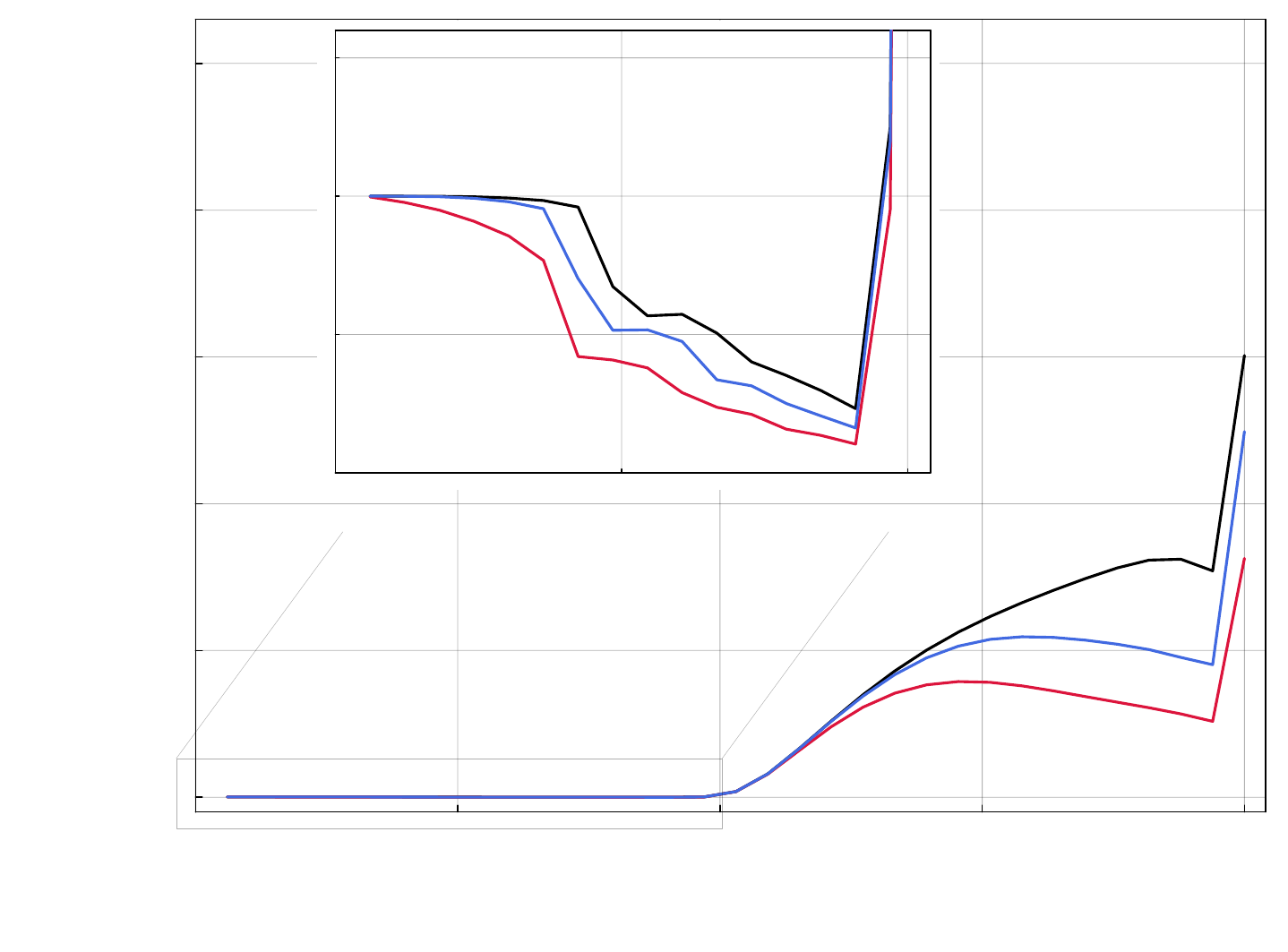 }}
    \subfloat[$p=3$]{{ \def\svgwidth{0.45\textwidth}
    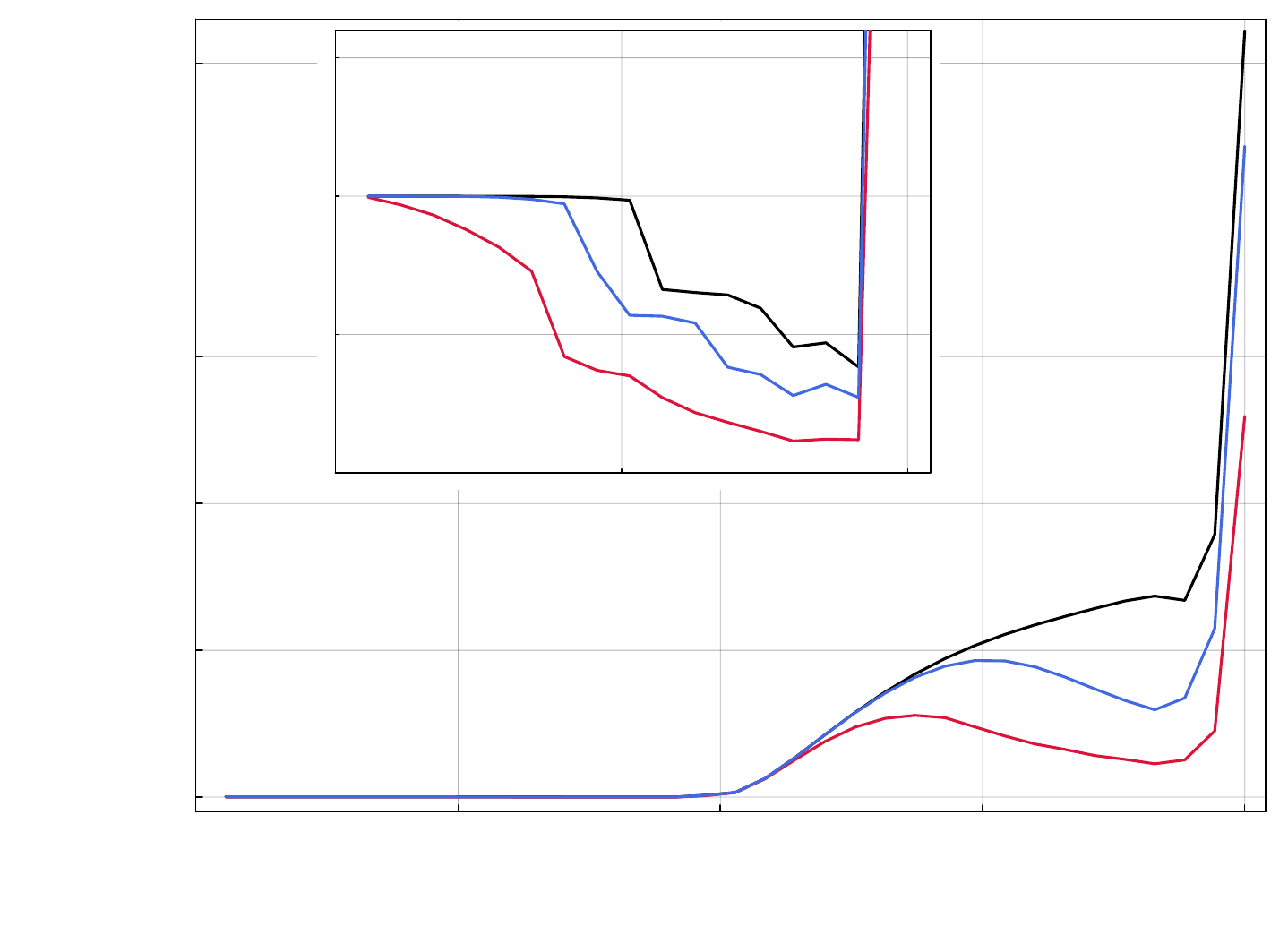 }}

    \subfloat[$p=4$]{{ \def\svgwidth{0.45\textwidth}
    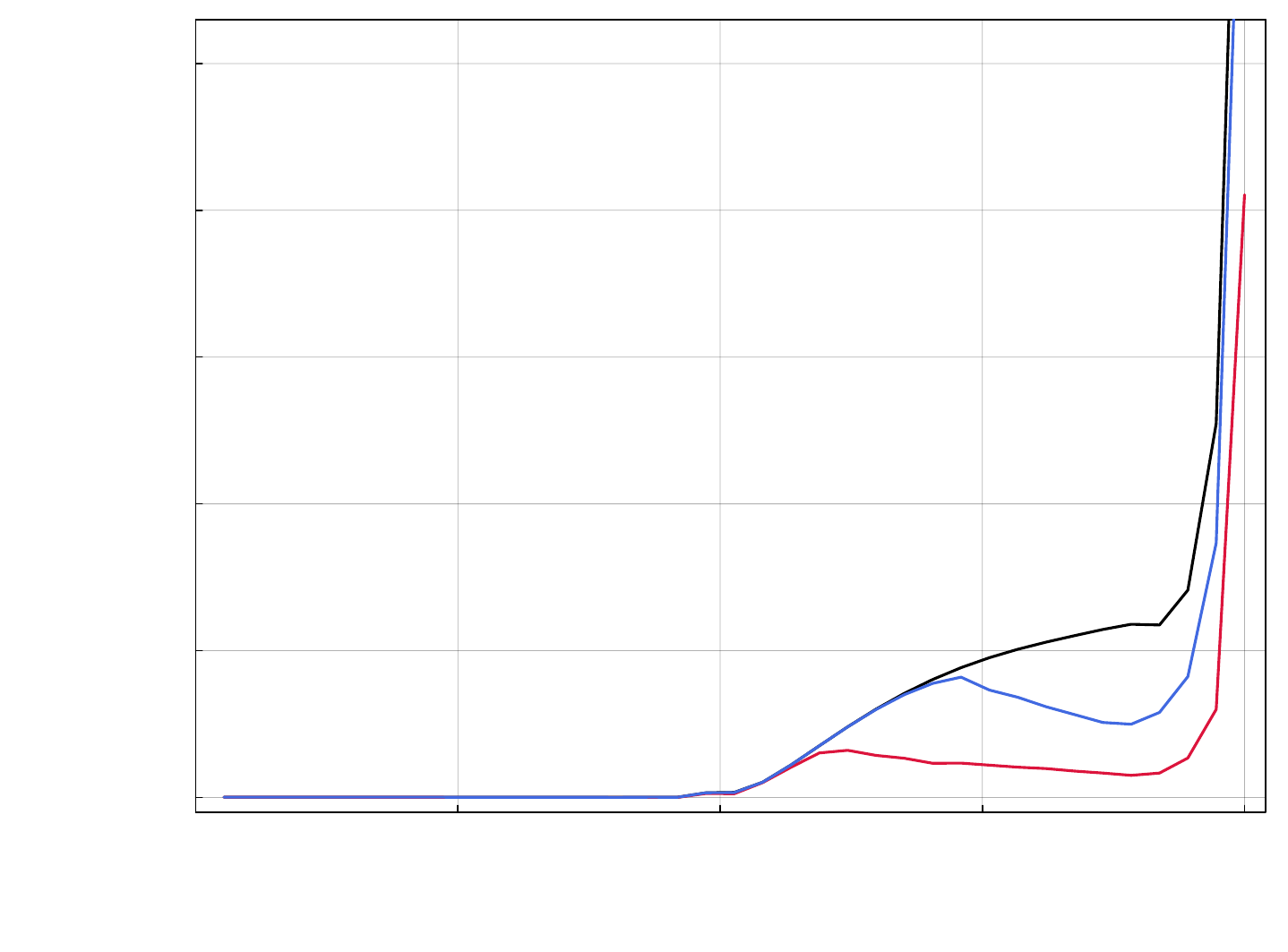 }}
    \subfloat[$p=5$]{{ \def\svgwidth{0.45\textwidth}
    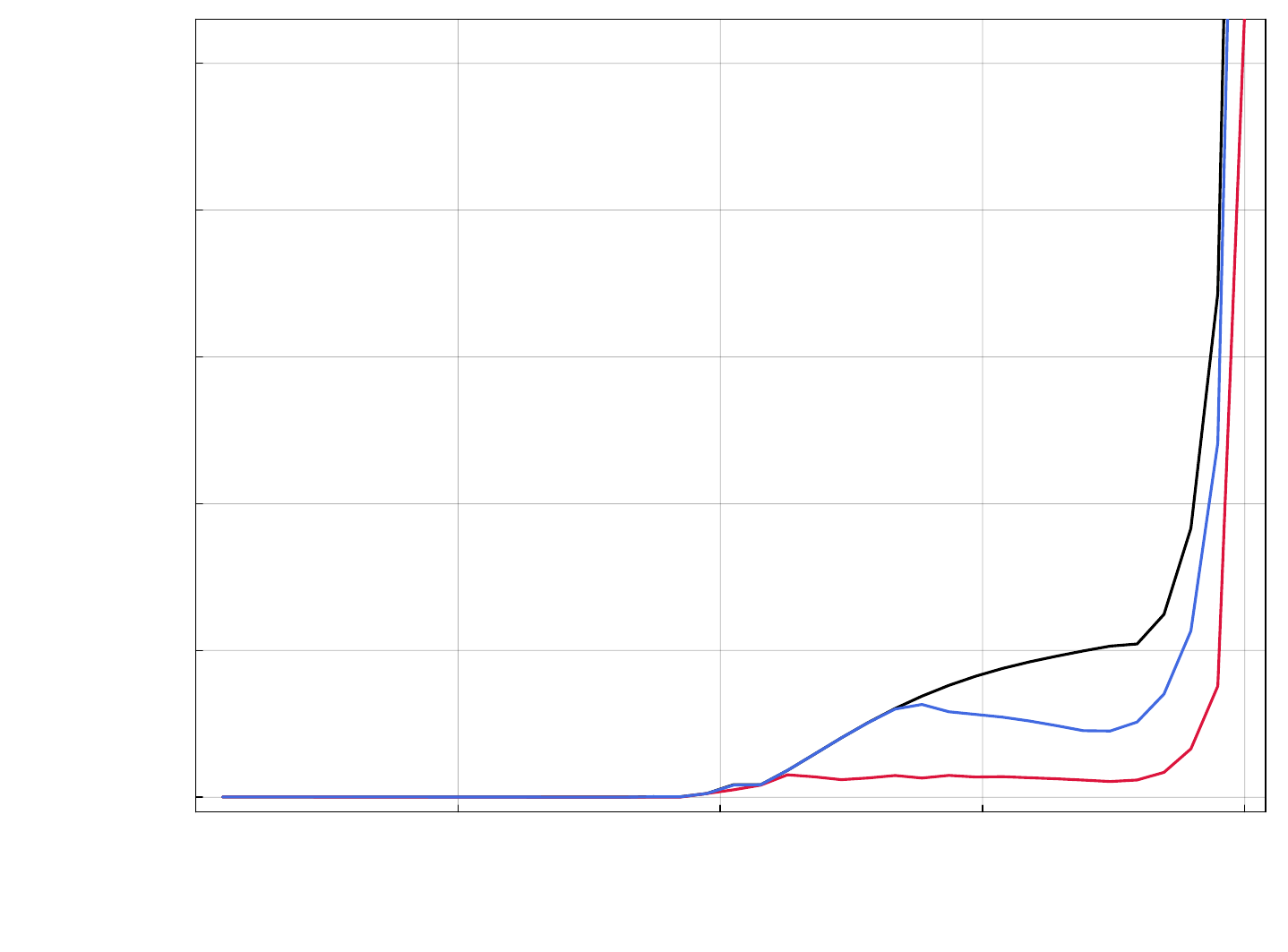 }}
    
    \vspace{0.2cm}
    \begin{tikzpicture}
    \filldraw[black,line width=1pt, solid] (0.0,0) -- (0.6,0);
    \filldraw[black,line width=1pt] (0.7,0) node[right]{\scriptsize Galerkin mixed formulation with consistent strain condensation};
    \filldraw[blue1,line width=1pt, solid] (10.0,0.0) -- (10.6,0.0);
    \filldraw[blue1,line width=1pt] (10.7,0) node[right]{\scriptsize Our approach};    
\end{tikzpicture}

\begin{tikzpicture}
    \filldraw[red1,line width=1pt, solid] (8.0,0) -- (8.6,0);
    \filldraw[red1,line width=1pt] (8.7,0) node[right]{\scriptsize Galerkin mixed formulation with lumping of the strain projection matrix};
\end{tikzpicture}
    \caption{Curved Euler-Bernoulli beam: spectrum of normalized eigenvalues, computed with mixed formulations and different methods of strain projection and a mesh of 16 B\'ezier elements.}\label{fig:beam_spectrum_projection}
\end{figure}

\begin{figure}[ht!]
	\centering
    \captionsetup[subfloat]{labelfont=scriptsize,textfont=scriptsize}
    \subfloat[$p=2$]{{ \def\svgwidth{0.45\textwidth}
    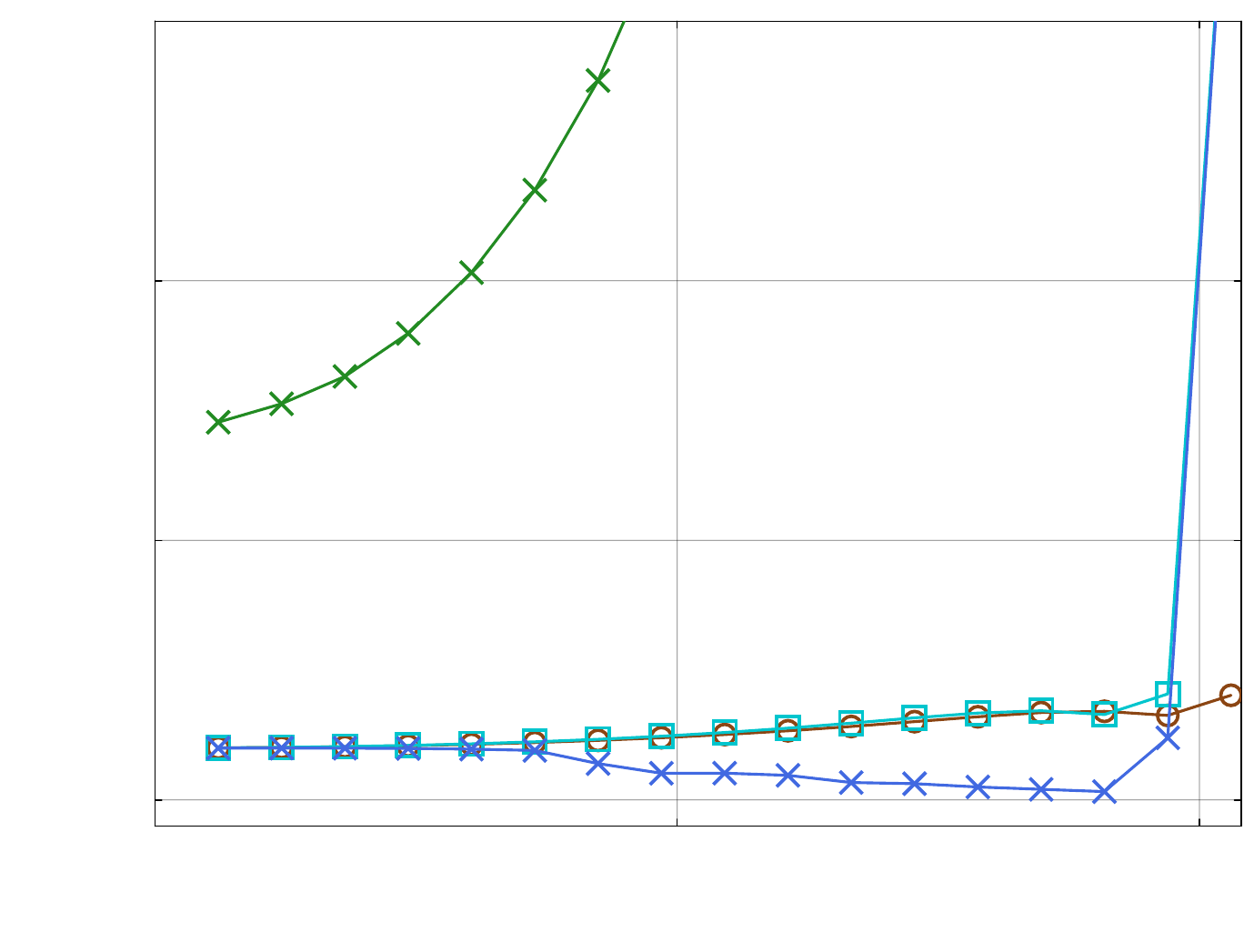 }}
    \hspace{0.2cm}
    \subfloat[$p=3$]{{ \def\svgwidth{0.45\textwidth}
    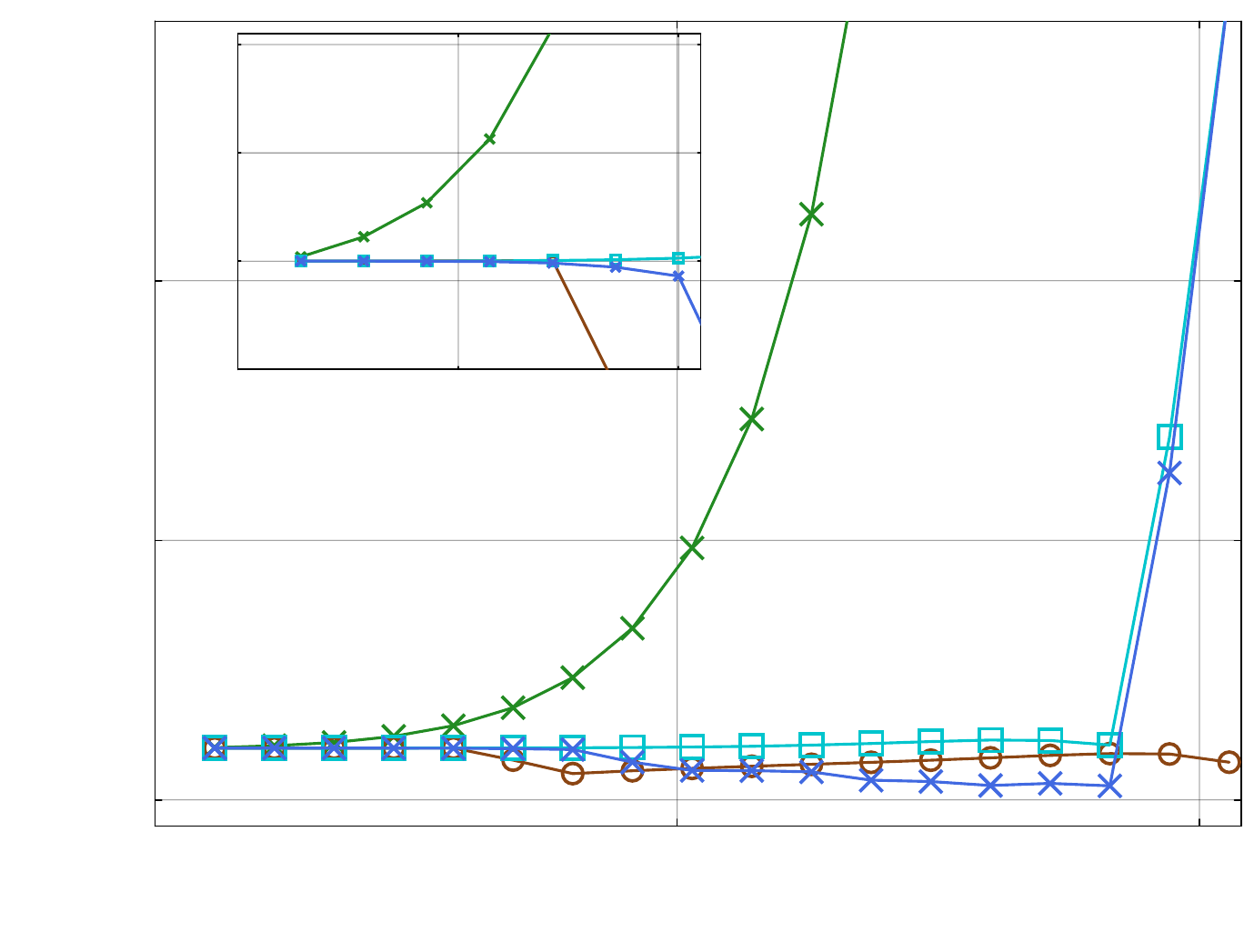 }}

    \vspace{0.2cm}
    \begin{tikzpicture}
    \filldraw[green1,line width=1pt, solid] (0.0,0) -- (0.6,0);
    \filldraw[green1,line width=1pt] (0.0,0) node[right]{\scriptsize $\boldsymbol{\bigtimes}$};
    \filldraw[green1,line width=1pt] (0.7,0) node[right]{\scriptsize Standard Galerkin, full integration};
    \filldraw[brown1,line width=1pt, solid] (6,0.05) -- (6.6,0.05);
    \filldraw[brown1,line width=1pt] (6.3,0.05) [fill=none] circle (3pt);
    \filldraw[brown1,line width=1pt] (6.7,0) node[right]{\scriptsize Standard Galerkin, {reduced integration\cite{leonetti2019simplified} ($\mathcal{S}_0^2,\mathcal{S}_1^3$ for $p=2,3$, respectively)}};
\end{tikzpicture}

\begin{tikzpicture}
    \filldraw[lightblue1,line width=1pt, solid] (0.5,0.05) -- (1.1,0.05);
    \filldraw[lightblue1,line width=1pt] (0.7,-0.08) [fill=none] rectangle ++(0.25,0.25);;
    \filldraw[lightblue1,line width=1pt] (1.2,0) node[right]{\scriptsize Standard Galerkin, B-bar method};
    \filldraw[blue1,line width=1pt, solid] (7.0,0) -- (7.6,0);
    \filldraw[blue1,line width=1pt] (7.0,0) node[right]{\scriptsize $\boldsymbol{\bigtimes}$};
    \filldraw[blue1,line width=1pt] (7.7,0) node[right]{\scriptsize Our approach};
\end{tikzpicture}
    \caption{Curved Euler-Bernoulli beam: lower half of the spectrum of the first 50\% of normalized eigenvalues, computed with our approach and different well-established locking-preventing mechanisms and a mesh of \textbf{16 elements}.}\label{fig:beam_spectrum_locking_methods_16ele}

    \vspace{0.3cm}

    \subfloat[$p=2$]{{ \def\svgwidth{0.45\textwidth}
    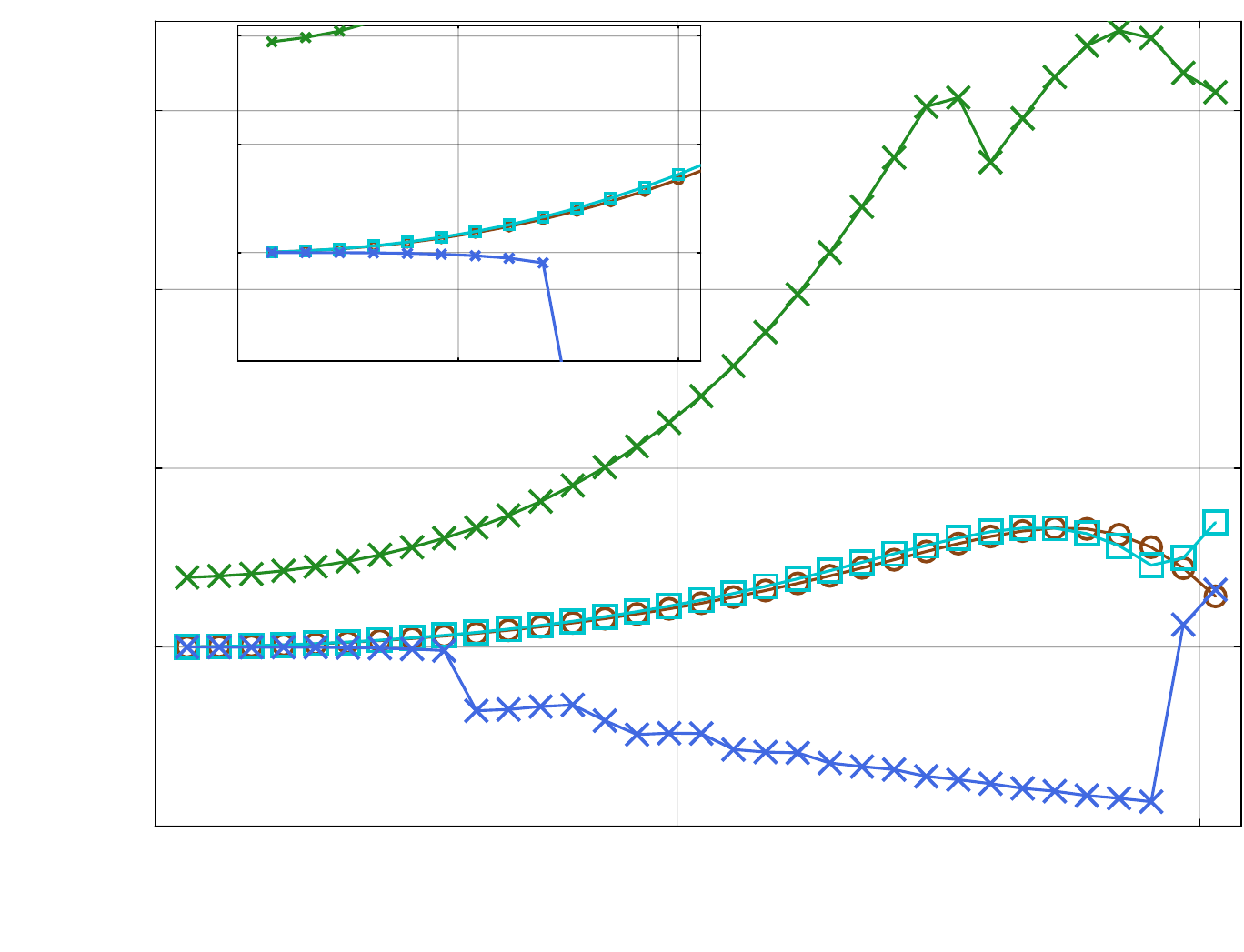 }}
    \hspace{0.2cm}
    \subfloat[$p=3$]{{ \def\svgwidth{0.45\textwidth}
    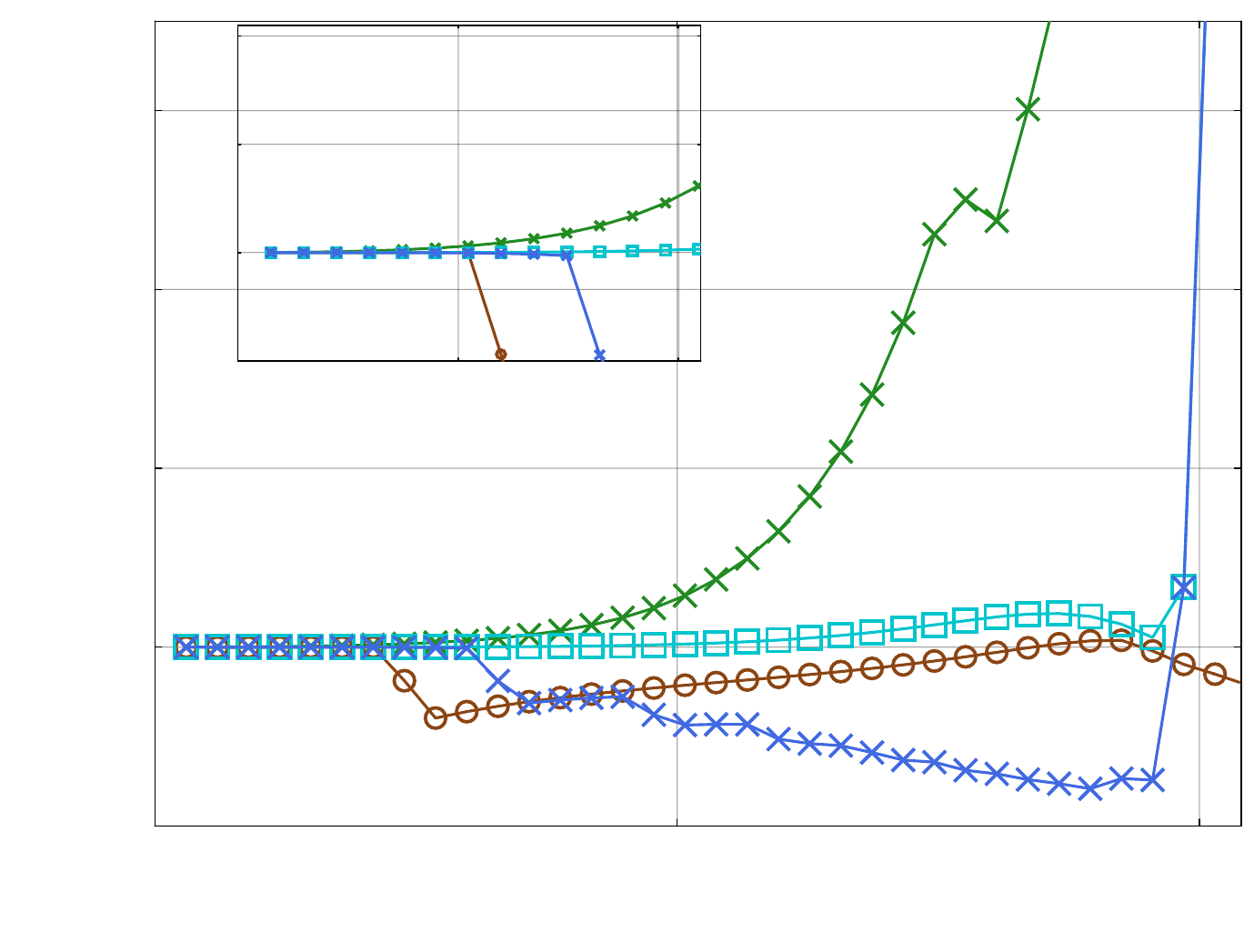 }}

    \vspace{0.2cm}
    \begin{tikzpicture}
    \filldraw[green1,line width=1pt, solid] (0.0,0) -- (0.6,0);
    \filldraw[green1,line width=1pt] (0.0,0) node[right]{\scriptsize $\boldsymbol{\bigtimes}$};
    \filldraw[green1,line width=1pt] (0.7,0) node[right]{\scriptsize Standard Galerkin, full integration};
    \filldraw[brown1,line width=1pt, solid] (6,0.05) -- (6.6,0.05);
    \filldraw[brown1,line width=1pt] (6.3,0.05) [fill=none] circle (3pt);
    \filldraw[brown1,line width=1pt] (6.7,0) node[right]{\scriptsize Standard Galerkin, {reduced integration\cite{leonetti2019simplified} ($\mathcal{S}_0^2,\mathcal{S}_1^3$ for $p=2,3$, respectively)}};
\end{tikzpicture}

\begin{tikzpicture}
    \filldraw[lightblue1,line width=1pt, solid] (0.5,0.05) -- (1.1,0.05);
    \filldraw[lightblue1,line width=1pt] (0.7,-0.08) [fill=none] rectangle ++(0.25,0.25);;
    \filldraw[lightblue1,line width=1pt] (1.2,0) node[right]{\scriptsize Standard Galerkin, B-bar method};
    \filldraw[blue1,line width=1pt, solid] (7.0,0) -- (7.6,0);
    \filldraw[blue1,line width=1pt] (7.0,0) node[right]{\scriptsize $\boldsymbol{\bigtimes}$};
    \filldraw[blue1,line width=1pt] (7.7,0) node[right]{\scriptsize Our approach};
\end{tikzpicture}
    \caption{Curved Euler-Bernoulli beam: lower half of the spectrum of the first 50\% of normalized eigenvalues, computed with our approach and different well-established locking-preventing mechanisms and a mesh of \textbf{32 elements}.}\label{fig:beam_spectrum_locking_methods_32ele}
\end{figure}

\subsection{Curved Euler-Bernoulli beam}

We consider the boundary value problem of a cantilevered curved Euler-Bernoulli beam illustrated in 
Figure \ref{fig:beam_geometry}. Our mixed formulation developed in Sections \ref{sec:mixed_form}, \ref{sec:discretization} and \ref{sec:strain_condensation} for the Kirchhoff-Love shell simplifies in a straightforward way to the case of an Euler-Bernoulli beam, and we therefore refrain from presenting details on the formulation. For a detailed exposition of the mixed formulation of the Euler-Bernoulli beam, we refer the interested reader to \cite{nguyen2022leveraging}.

\subsubsection{Convergence study}\label{sec:beam_convergence}
\label{subsub1}

To conduct a convergence study, we discretize the curved beam with spline basis functions of degree $p=2$ through $p=5$ and conduct uniform $h$-refinement.
Our reference is the analytical solution discussed in \cite{Cazzani2016} for the quarter circular Euler-Bernoulli cantilever subjected to shear load at the free end. 
In Figure \ref{fig:beam_convergence_projection}, we plot the convergence of the relative $L^2$ error of the displacement field computed with mixed formulations that evaluate the strain projection part in a different way. 
To assess our approach, we compare our formulation with independent strain variations discretized by approximate dual functions and row-sum lumping of the strain projection matrix (blue) to the Galerkin mixed formulation with consistent strain projection (black) and the Galerkin mixed formulation with row-sum lumping of the strain projection matrix (red). We note that the consistent Galerkin mixed formulation is prohibitively inefficient, as consistent strain condensation leads to a dense system matrix that is fully populated, but provides the reference accuracy. 

We observe that our approach delivers the same accuracy and optimal rates of convergence as the consistent Galerkin mixed formulation for all polynomial degrees.  
It does not suffer from the significant drop of accuracy that is characteristic for the Galerkin mixed formulation with row-sum lumping, where lumping limits the convergence rate to second-order irrespective of the polynomial degree of the basis functions. 
We note that the results observed for this example are consistent with the ones observed for the $L^2$ projection example in Section \ref{sec:l2projection}. For a comparison of the relative $L^2$ error of the displacement field, of the membrane strain field and of the bending strain field computed with our formulation and the standard displacement-based formulation for two different slenderness ratios $R/t=100$ and $1,000$, we refer the interested reader to Appendix \ref{sec:beam1}.

To assess the performance of our approach to mitigate membrane locking, we conduct a similar convergence study with different locking-preventing methods for the same curved cantilever. 
In Figure \ref{fig:beam_convergence_locking_method}, we plot the convergence of the relative $L^2$ error of the displacement field, computed with our approach (blue), a Galerkin mixed formulation with the EAS method \cite{Cardoso2012} (purple), and standard displacement-based Galerkin formulations that mitigate membrane locking via selective reduced integration via patchwise spline quadrature rules \cite{Adam2015,leonetti2019simplified} (red) and the B-bar method \cite{Bouclier2012} (light blue). For comparison, we also plot the convergence of the standard Galerkin formulation without any locking-preventing approach (green), fully integrated with $p+1$ Gauss points per B\'ezier element, which contains the full extent of membrane locking. For a full comparison of the relative $L^2$ error of the displacement field, of the membrane strain field and of the bending strain field, we refer the interested reader to Appendix \ref{sec:beam1}.

Focusing on the results obtained with the standard Galerkin formulation and full integration, we observe the typical pre-asymptotic plateau in the error curves on coarse meshes. In this region, accuracy does not improve with mesh refinement, indicating the severe impact of membrane locking on accuracy and convergence. Increasing the polynomial degree reduces this effect, thereby decreasing the size of the pre-asymptotic plateau. Using the B-bar or the EAS method completely eliminates the effect of locking, leading to optimal convergence rates of $\mathcal{O}(p)$ for $p=2$ and $\mathcal{O}(p+1)$ for $p \geq 3$ \cite{engel2002continuous,Tagliabue2014}. 

For mixed formulations, the optimal convergence rate obtained for fourth-order problems is $\mathcal{O}(p+1)$ for all $p\ge1$ \cite{engel2002continuous,Tagliabue2014}. Therefore, our approach achieves better accuracy for quadratic discretizations than any of the other methods, and the same accuracy as the B-bar and EAS methods for cubic discretizations and beyond. We also see that our approach eliminates the effect of locking, as it results in error curves that converge directly on coarse meshes with the optimal rate of $\mathcal{O}(p+1)$. We conclude that for fourth-order problems, mixed formulations such as our approach seem to be particularly attractive for quadratic discretizations, for which they achieve better accuracy and a higher rate of convergence than purely displacement-based methods.

\begin{remark}
A variety of spline quadrature rules exist that exploit the smoothness of basis functions within each patch \cite{hughes2010efficient,schillinger2014reduced,hiemstra2017optimal}. 
When applied in a reduced quadrature setting, such rules can -- when chosen appropriately -- effectively mitigate locking while preserving the accuracy benefits of smooth spline spaces \cite{Adam2015,leonetti2019simplified}.
We employed selective reduced integration in the sense that the matrix component related to bending stiffness was integrated exactly, while the matrix component related to membrane stiffness was integrated in a reduced sense.
To make sure we chose the most appropriate for our case, we tested all rules for all relevant combinations of polynomial degree $m$ and smoothness $r$ to be integrated exactly in the target space $\mathcal{S}_r^m$, where we generated the quadrature points following the procedure given in \cite{johannessen2017optimal}. We then selected the quadrature rule that produced the best accuracy.
\end{remark}

\subsubsection{Spectral analysis}

Based on our prior study in \cite{nguyen2022leveraging}, we leverage spectral analysis as an alternative tool to assess membrane locking. 
To this end, we recall the discrete eigenvalue problem of the Euler-Bernoulli beam in matrix form (see e.g. \cite{nguyen2022leveraging}):
\begin{align}\label{eq:gen_eigenval}
    \mat{K} \, \vect{U}^h_n = \lambda^h_n\,\mat{M} \, \vect{U}^h_n \,,
\end{align}
where $\vect{U}^h_n$ denotes the vector of unknown displacement coefficients corresponding to the $n^{\text{th}}$ discrete eigenmode $U^h_n$, and $\omega_n^h$ is the $n^{\text{th}}$ discrete eigenfrequency. We emphasize that in the case of a mixed formulation, the independent strain fields are already condensed out. 
For the free vibration of the cantilever illustrated in 
Figure \ref{fig:beam_geometry}, we compute the discrete eigenvalue $\lambda^h$, normalized by the reference value, $\lambda^h_{ref}$, which is computed with an ``overkill'' mesh of 1,024 B\'ezier elements. 
We remove the zero eigenvalues associated with rigid body modes, order the remaining non-zero eigenvalues in increasing order and plot the normalized values, $\lambda^h / \lambda^h_{ref}$, as a function of the normalized mode number, $n/N$.

We first consider, on a mesh of 16 B\'ezier elements, the same three strain projection variants in the mixed formulation, for which we conducted the convergence study in Figure \ref{fig:beam_convergence_projection}. In Figure \ref{fig:beam_spectrum_projection}, 
we compare the spectra obtained with the Galerkin mixed formulation with consistent strain projection (black), which we consider as the reference, the Galerkin mixed formulation with row-sum lumping of the strain projection matrix (red), and our approach based on approximate dual functions for the discretization of the independent strain variations and row-sum lumping of the strain projection matrix (blue). We also provide inset figures that zoom in on the first 50\% of the eigenvalues. 
We observe that compared to the Galerkin mixed formulation with row-sum lumping, our formulation leads to considerably better accuracy over the complete spectrum, in the sense that the spectrum is much closer to the one computed with the consistent Galerkin mixed formulation. 
We recall that the lowest modes are particularly important for the accuracy of the solution, see e.g.\ \cite{schillinger2014reduced}.
In the inset figures, we see that with increasing polynomial degree, the Galerkin mixed formulation with row-sum lumping increasingly deviates from the reference even for the lowest 5\% of the modes, while the spectrum curves of the consistent  Galerkin mixed formulation and our formulation are practically equivalent within the 20\% of the lowest modes. 
We note that the jump in the spectrum curves around $n/N = 0.5$ is due to the chosen ordering and can be eliminated by sorting out different mode types, see \cite{nguyen2022leveraging}.

We then consider again five of the locking-preventing mechanisms, for which we conducted the convergence study in Figure \ref{fig:beam_convergence_locking_method}. 
In Figures \ref{fig:beam_spectrum_locking_methods_16ele} and \ref{fig:beam_spectrum_locking_methods_32ele}, we plot the spectral analysis results obtained with a mesh of 16 quadratic and cubic B\'ezier elements and 32 quadratic and cubic B\'ezier elements, respectively. The main plots show the first 50\% of the eigenvalues, while the inset figures zoom in on the first 20\% of the eigenvalues. We compare the spectra obtained with our approach (blue) to the spectra obtained with the standard displacement-based Galerkin formulations that mitigate membrane locking via reduced integration \cite{Adam2015,leonetti2019simplified} (red) and the B-bar method \cite{Bouclier2012} (light blue). We note that we exclude the Galerkin mixed formulation with the EAS method \cite{Cardoso2012}, since the results are practically indistinguishable from the ones obtained with the B-bar method, as already observed in the convergence study above. For comparison, we also plot the spectrum of the standard Galerkin formulation without any locking-preventing approach (green), fully integrated with $p+1$ Gauss points per B\'ezier element, which includes the full extent of membrane locking. We note that jumps in the spectrum curves occur due to the chosen ordering and can be eliminated by sorting out different mode types.

The results shown here for the standard displacement-based Galerkin formulation with different locking-preventing mechanisms confirm what we reported in our earlier study \cite{nguyen2022leveraging}. We observe that the B-bar method (and the EAS method) are particularly successful in completely removing any effect of membrane locking. In addition, we observe that our approach delivers excellent spectral accuracy for the lowest modes. It exhibits a distinct decay of spectral accuracy for the medium and high modes, which is typical for mixed formulations with condensed strain variables \cite{nguyen2022leveraging}. The spectrum results in Figure \ref{fig:beam_spectrum_locking_methods_16ele}a and \ref{fig:beam_spectrum_locking_methods_32ele}a also confirm the above observation that our approach achieves better accuracy for quadratic discretizations than any of the other locking-preventing methods. 
For the case of quadratic basis functions, we observe that in the critical lower modes, the spectrum curve of our mixed formulation is much closer to the optimum value of one than the spectrum curves of all the other methods, including the B-bar method.

We emphasize that the spectral analysis presented in Figures \ref{fig:beam_spectrum_projection}--\ref{fig:beam_spectrum_locking_methods_32ele} is intended to assess membrane locking and the accuracy of the discrete eigenvalues for the curved Euler--Bernoulli beam. In this context, we distinguish between zero-energy and nonzero-energy spurious modes. While the former are associated with an additional rank deficiency of the stiffness matrix, the latter may occur even in rank-sufficient systems. Their identification requires a more detailed examination of the discrete eigenvalues and their corresponding mode shapes, see e.g.\cite{Adam2015}. Consequently, the results presented here do not constitute a systematic assessment of possible nonzero-energy spurious modes in the proposed shell formulation.

\begin{figure}[ht!]
	\centering
    \captionsetup[subfloat]{labelfont=scriptsize,textfont=scriptsize}
    \subfloat[$p=2$]{{ \def\svgwidth{0.49\textwidth}
    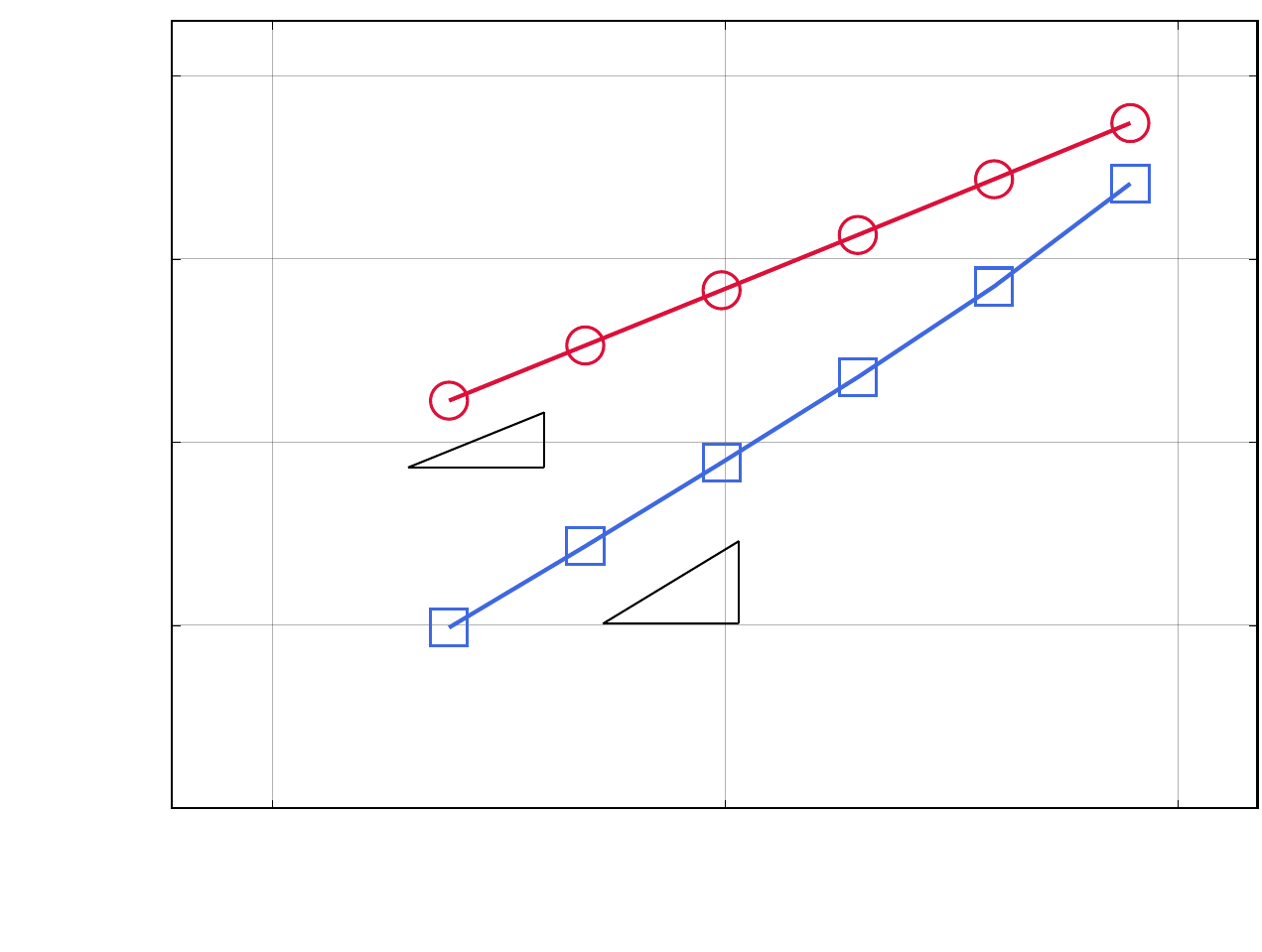 }}
    \subfloat[$p=3$]{{ \def\svgwidth{0.49\textwidth}
    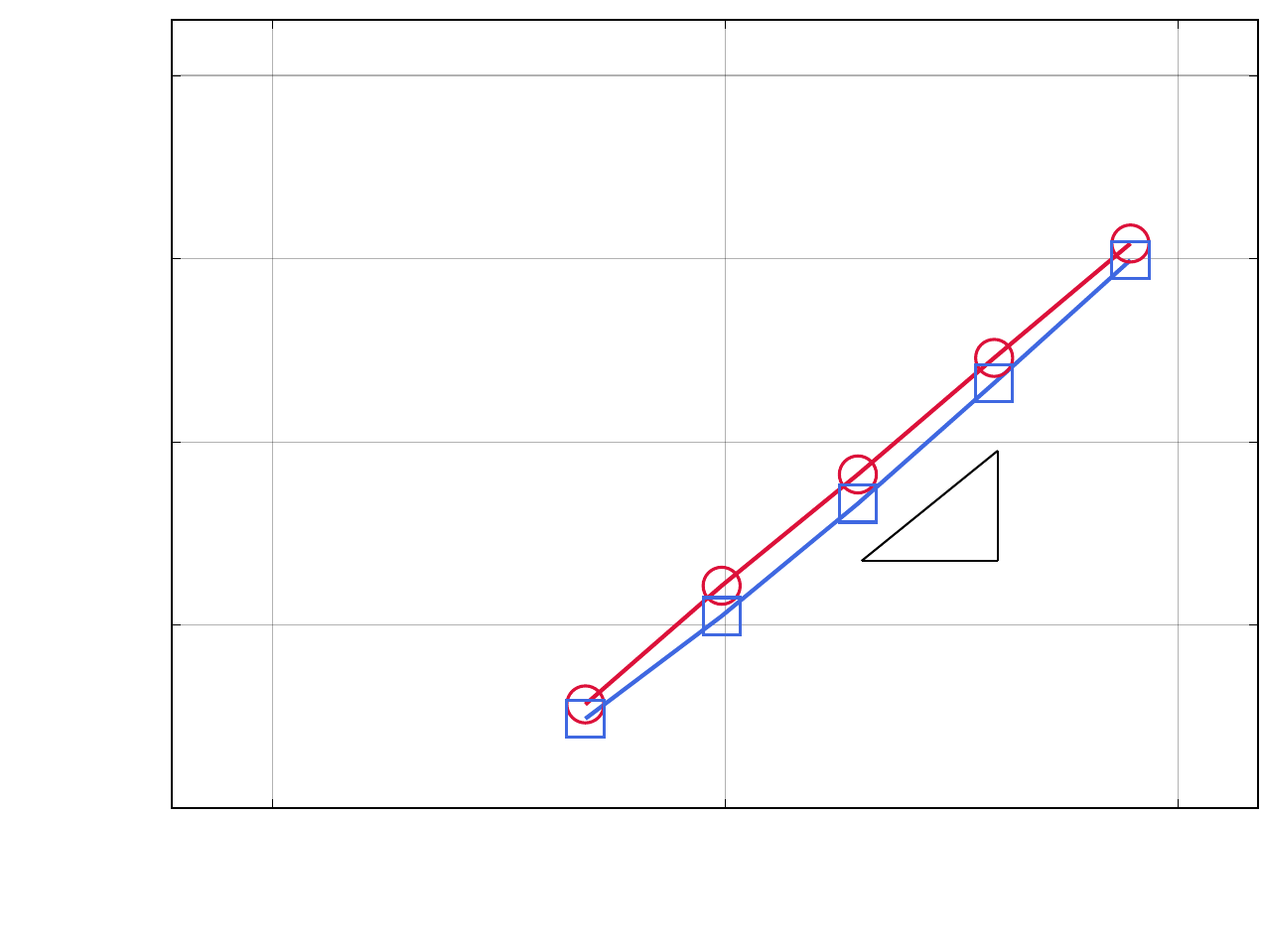 }}

    \subfloat[$p=4$]{{ \def\svgwidth{0.49\textwidth}
    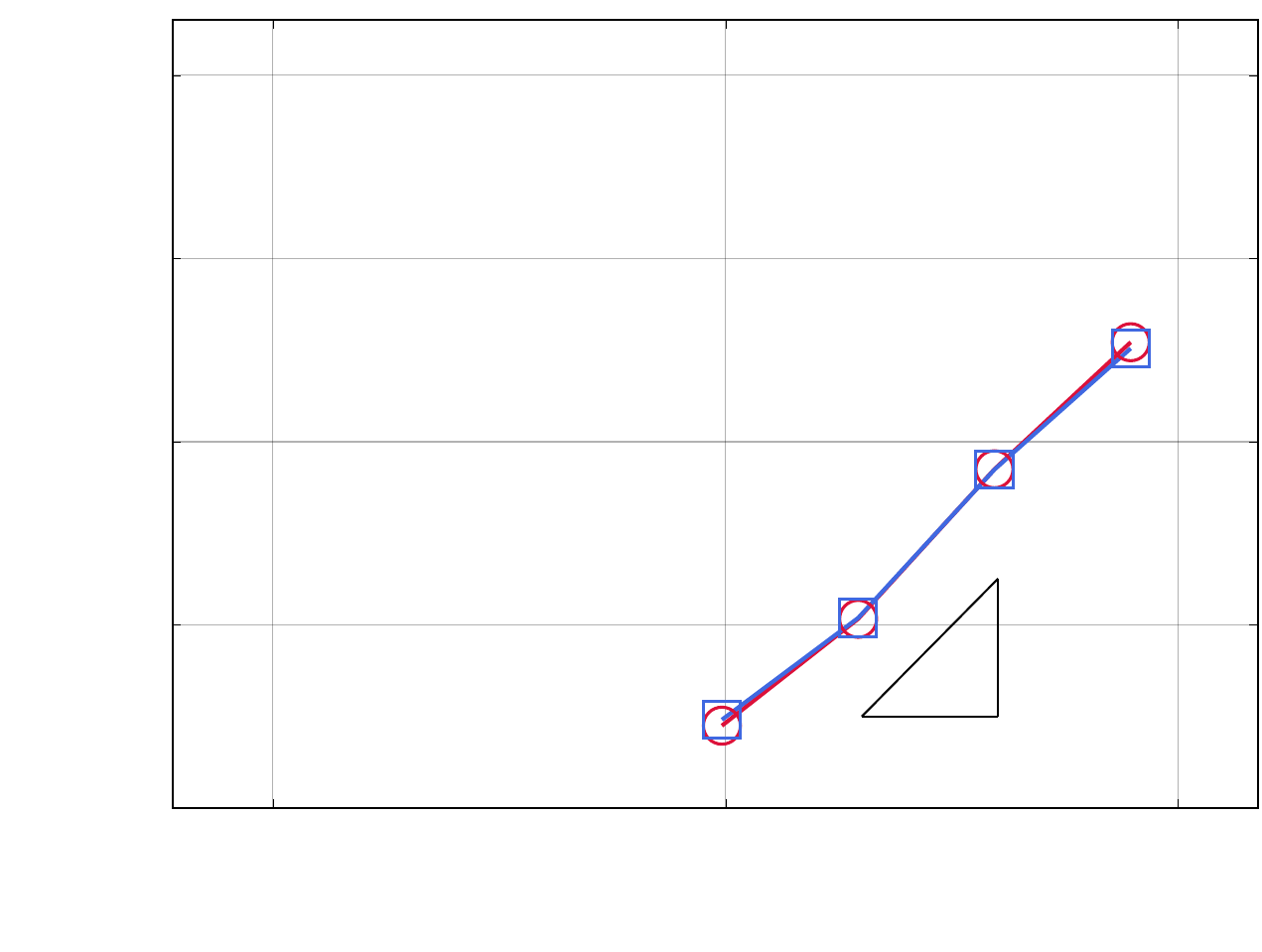 }}
    \subfloat[$p=5$]{{ \def\svgwidth{0.49\textwidth}
    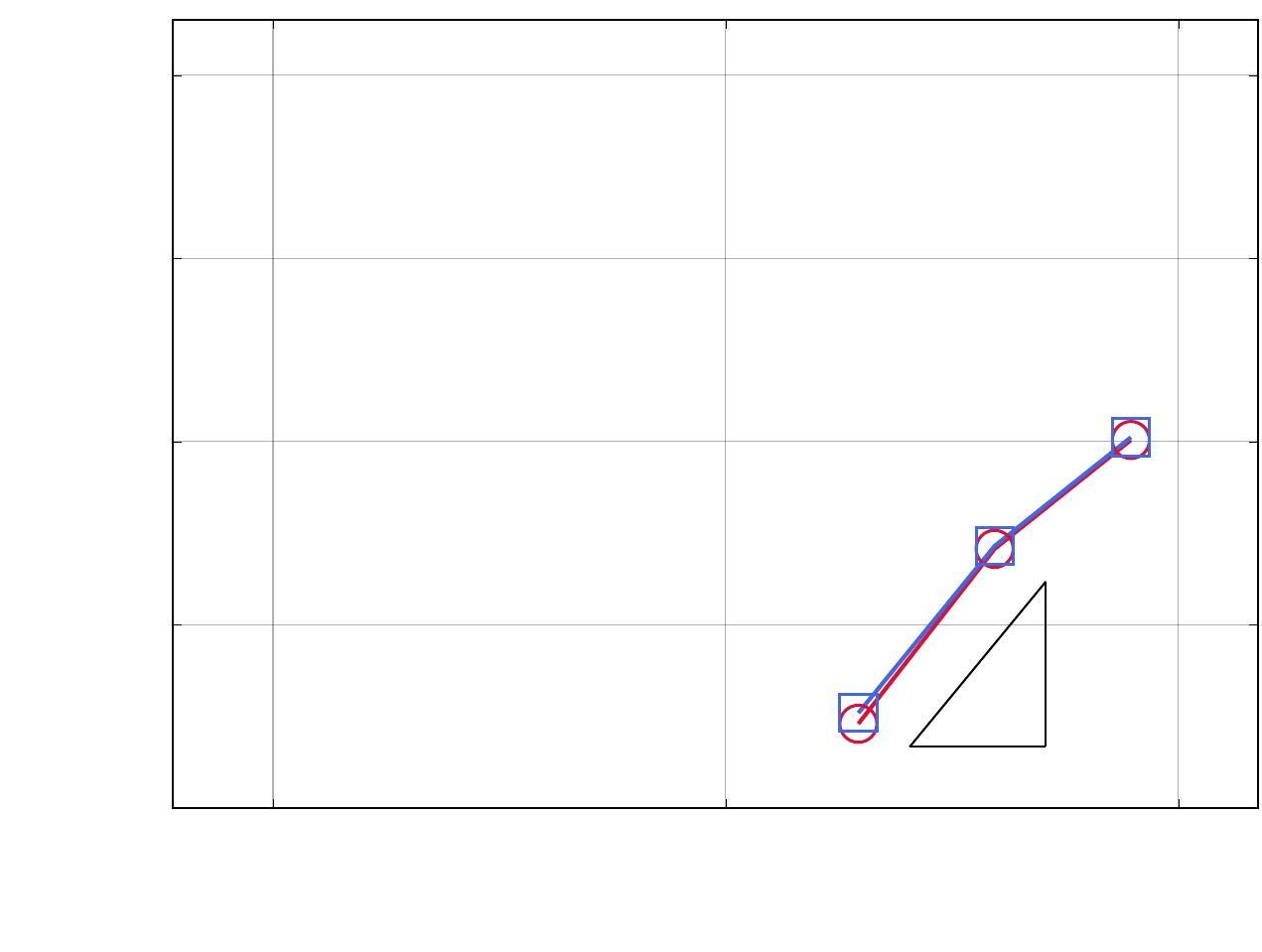 }}
    
    \vspace{0.2cm}
    \begin{tikzpicture}
    \filldraw[blue1,line width=1pt, solid] (0.0,0.05) -- (0.6,0.05);
    \filldraw[blue1,line width=1pt] (0.2,-0.08) [fill=none] rectangle ++(0.25,0.25);
    \filldraw[blue1,line width=1pt] (0.8,0) node[right]{\scriptsize Our approach with projection of both membrane and bending strains; $(\vect{u},e,k)\in \sspace{p,p-1}{n} \times \sspace{p-1,p-2}{m} \times \sspace{p-1,p-2}{m}$};  
\end{tikzpicture}

\begin{tikzpicture}
    \filldraw[red1,line width=1pt, solid] (1.0,0) -- (1.6,0);
    \filldraw[red1,line width=1pt] (1.3,0) [fill=none] circle (3pt);
    \filldraw[red1,line width=1pt] (1.7,0) node[right]{\scriptsize Our approach with projection of only membrane strains; $(\vect{u},e) \in \sspace{p,p-1}{n} \times \sspace{p-1,p-2}{m}$};
\end{tikzpicture}


    \caption{Curved Euler-Bernoulli beam: relative error in the $L^2$-norm of the displacement field, computed with our approach and two different projection variants. $\sspace{p,r}{n}$ denotes a space spanned by $n$ B-splines basis functions of polynomial degree $p$ with continuity $C^r$.}\label{fig:beam_convergence_mixed_forms}
\end{figure}

\begin{figure}[ht!]
	\centering
    \captionsetup[subfloat]{labelfont=scriptsize,textfont=scriptsize}
    \subfloat[$\varepsilon^h$]{{ \def\svgwidth{0.48\textwidth}
    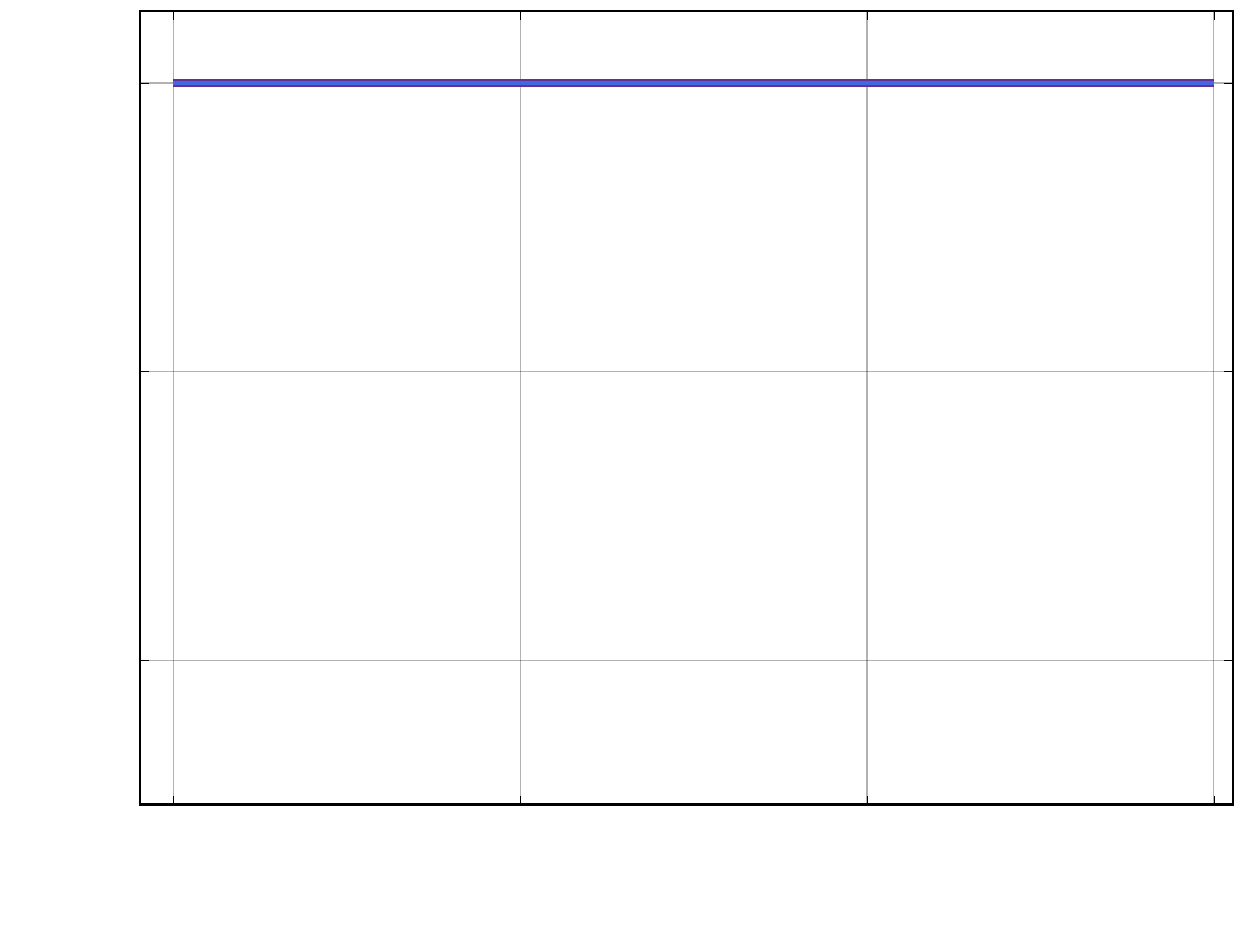 }}
    \subfloat[$\kappa^h$]{{ \def\svgwidth{0.48\textwidth}
    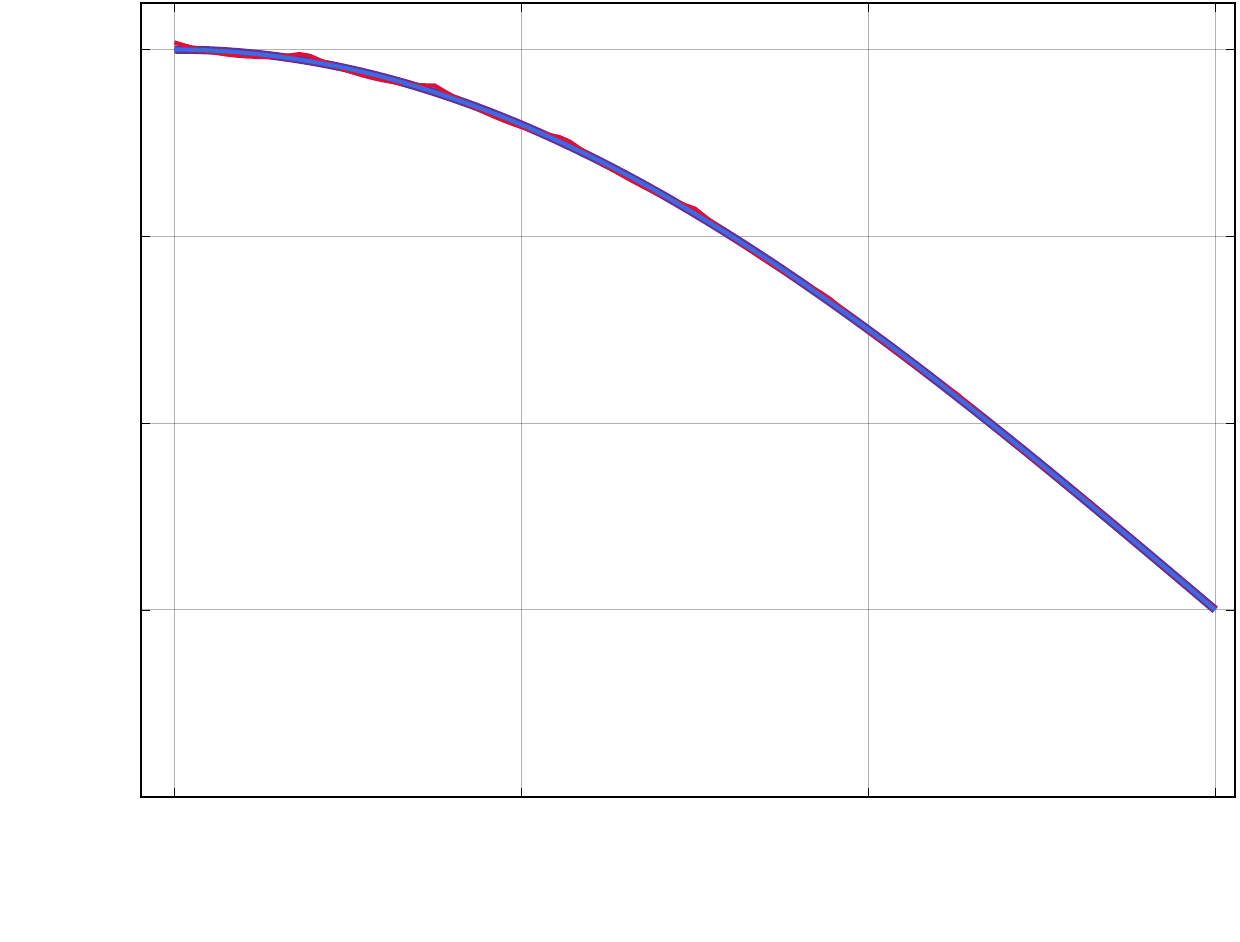 }}
    
    \vspace{0.2cm}
    \begin{tikzpicture}
    \filldraw[purple1,line width=1pt, solid] (0.0,0) -- (0.6,0);
    \filldraw[purple1,line width=1pt] (0.7,0) node[right]{\scriptsize Analytical solution};
    \filldraw[red1,line width=1pt, solid] (5.0,0) -- (5.6,0);
    \filldraw[red1,line width=1pt] (5.7,0) node[right]{\scriptsize Our approach with projection of only membrane strains; $(\vect{u},e) \in \sspace{p,p-1}{n} \times \sspace{p-1,p-2}{m}$};
\end{tikzpicture}

\begin{tikzpicture}
    \filldraw[blue1,line width=1pt, solid] (0.0,0) -- (0.6,0);
    \filldraw[blue1,line width=1pt] (0.7,0) node[right]{\scriptsize Our approach with projection of membrane and bending strains; $(\vect{u},e,k)\in \sspace{p,p-1}{n} \times \sspace{p-1,p-2}{m} \times \sspace{p-1,p-2}{m}$};
\end{tikzpicture}


    \caption{Curved Euler-Bernoulli beam: membrane strain and the change of curvature, computed with our approach and two different projection variants on a mesh of 8 cubic elements. $\sspace{p,r}{n}$ denotes a space spanned by $n$ B-splines basis functions of polynomial degree $p$ with continuity $C^r$.}\label{fig:beam_strains_mixed_forms}
\end{figure}

\subsubsection{Assessment of different strain projection variants}

We use the same benchmark of the curved cantilever beam to carry out a similar convergence study to assess the performance of our approach with the two different strain projection variants: (a) strain projection of the membrane and bending strains, where both strains are discretized with basis functions of one polynomial order lower than the displacements (plotted in blue), and (b) strain projection of only membrane strains, discretized with basis functions of one order lower than the displacements (plotted in red), along the Remarks \ref{remark_strain_projection}, \ref{remark_strain_projection2}, and \ref{remark_strain_projection3}.

In Figure \ref{fig:beam_convergence_mixed_forms}, we plot the convergence of the relative $L^2$ error of the displacement field, computed with our approach and the above strain projection variants. 
Focusing on the results obtained with quadratic basis functions, we observe that strain projection of only the membrane strains leads to a reduction of the optimal convergence rate to \ $\mathcal{O}(p)$ for $p=2$ \cite{engel2002continuous,Tagliabue2014}, since we deal with a fourth-order problem due to the strong enforcement of the bending kinematics, see also Remark \ref{remark_strain_projection}. 
In case of the projection of both the membrane and bending strains, we deal with a second-order problem, and thus are able to achieve the optimal rate of $\mathcal{O}(p+1)$ for all $p \geq 1$\cite{engel2002continuous,Tagliabue2014}. This constitutes a distinct advantage of the latter for quadratic spline discretizations. 
Focusing on the results obtained with cubic, quartic, and quintic splines, we observe that the two projection variants achieve a comparable accuracy, both in terms of convergence rate and preasymptotic error level, irrespective of the number of the projected strain fields and the degree of basis functions employed for the bending strains $\kappa$.

In Figure \ref{fig:beam_strains_mixed_forms}, we plot the discrete membrane and bending strains, obtained with the two projection variants, as a function of the angular coordinate $\theta$. For comparison, we also plot the analytical solution (see \cite{Cazzani2016} for its derivation). 
We include inset figures which focus on the results close to the boundary with clamped support. 
We observe that the projection of only the membrane strains results in bending strains that exhibit distinct oscillations. In the case of projecting both membrane and bending strains, these oscillations are absent. 
These observations confirm that our choice of projecting both membrane and bending strains has some advantages over the alternative of projecting only the membrane strains, and we therefore adopt this variant in the remainder of this work.

\begin{figure}[t!]
	\centering    
    \def\svgwidth{1\textwidth}
    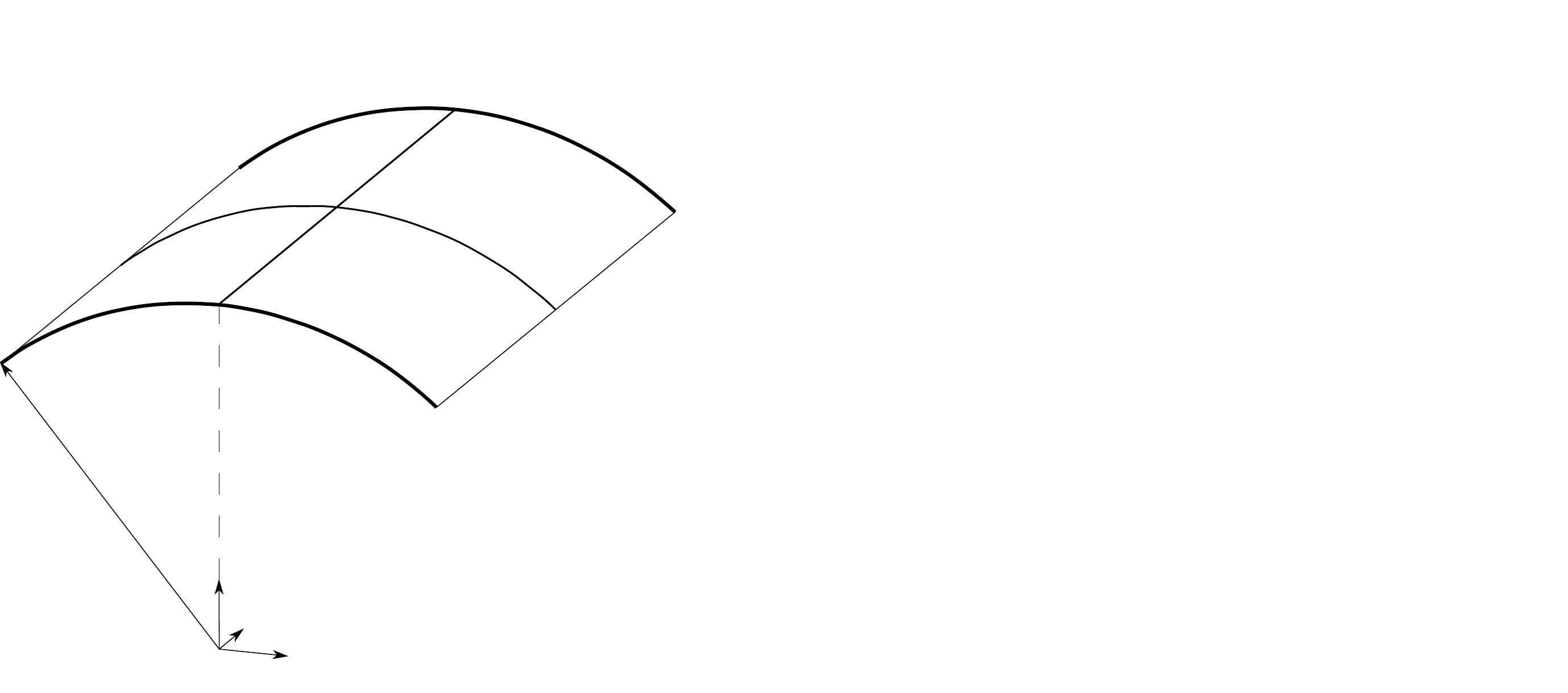

    \vspace{0.3cm}    
    \caption{Scordelis-Lo roof constrained by rigid diaphragms and subjected to its self-weight of $90.0$ per unit area: problem sketch along with a plot of the total displacement field, computed with our approach and cubic splines.} \label{fig:roof_geometry}
\end{figure}

\begin{figure}[t!]
    \centering
    \captionsetup[subfloat]{labelfont=scriptsize,textfont=scriptsize}
    \subfloat[Quadratic splines - $p=2$]{{
        \def\svgwidth{0.43\textwidth}
\begingroup%
  \makeatletter%
  \providecommand\color[2][]{%
    \errmessage{(Inkscape) Color is used for the text in Inkscape, but the package 'color.sty' is not loaded}%
    \renewcommand\color[2][]{}%
  }%
  \providecommand\transparent[1]{%
    \errmessage{(Inkscape) Transparency is used (non-zero) for the text in Inkscape, but the package 'transparent.sty' is not loaded}%
    \renewcommand\transparent[1]{}%
  }%
  \providecommand\rotatebox[2]{#2}%
  \newcommand*\fsize{\dimexpr\f@size pt\relax}%
  \newcommand*\lineheight[1]{\fontsize{\fsize}{#1\fsize}\selectfont}%
  \ifx\svgwidth\undefined%
    \setlength{\unitlength}{705bp}%
    \ifx\svgscale\undefined%
      \relax%
    \else%
      \setlength{\unitlength}{\unitlength * \real{\svgscale}}%
    \fi%
  \else%
    \setlength{\unitlength}{\svgwidth}%
  \fi%
  \global\let\svgwidth\undefined%
  \global\let\svgscale\undefined%
  \makeatother%
  \begin{picture}(1,0.60106383)%
    \lineheight{1}%
    \setlength\tabcolsep{0pt}%
    \put(0,0){\includegraphics[width=\unitlength,page=1]{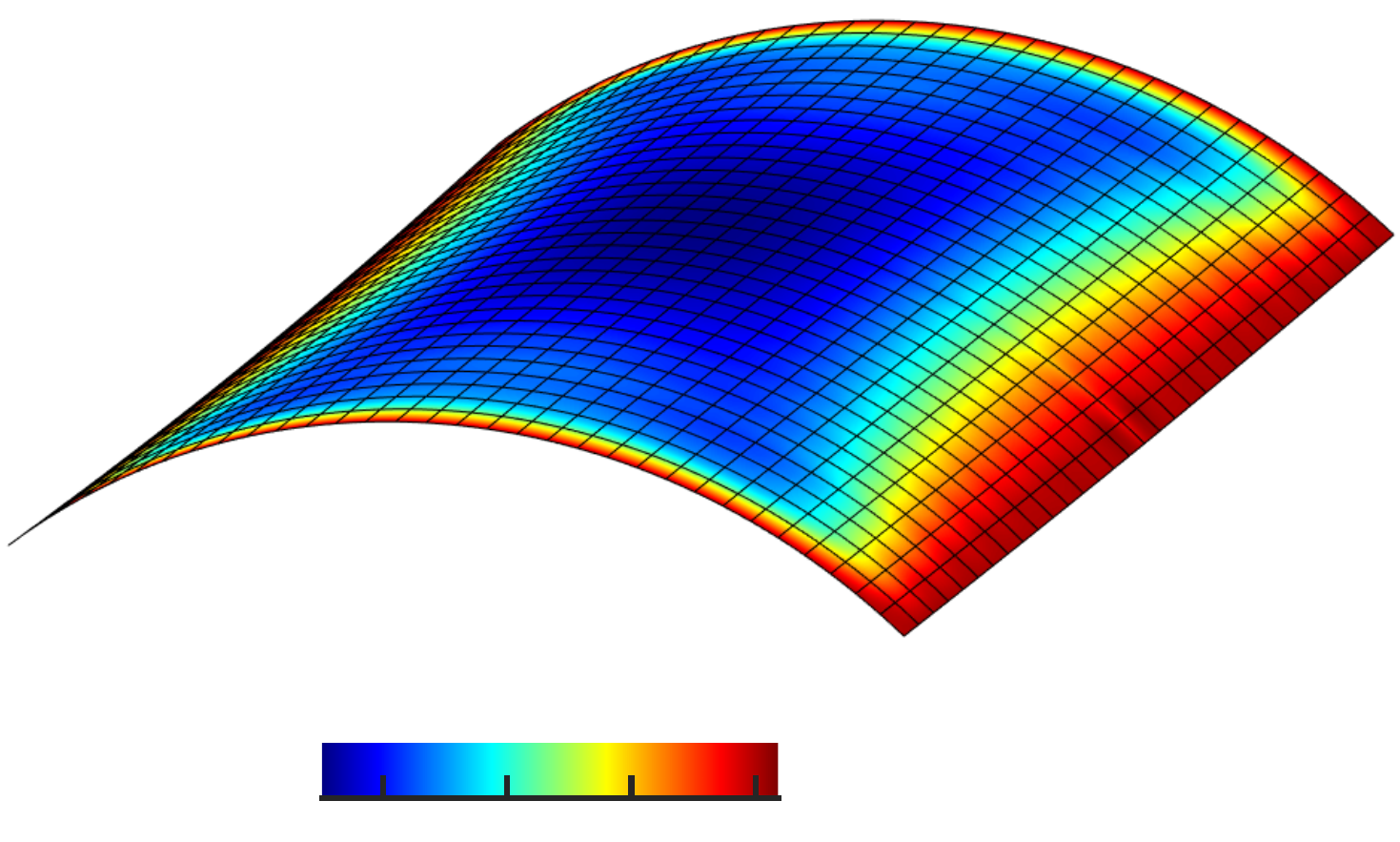}}%
    \put(0.56184079,0.06966052){\color[rgb]{0,0,0}\makebox(0,0)[lt]{\lineheight{1.25}\smash{\begin{tabular}[t]{l}$n^{11}$\end{tabular}}}}%
    \put(0.5318018,0.00015757){\color[rgb]{0,0,0}\makebox(0,0)[lt]{\lineheight{1.25}\smash{\begin{tabular}[t]{l}0\end{tabular}}}}%
    \put(0.43955475,0.00023142){\color[rgb]{0,0,0}\makebox(0,0)[lt]{\lineheight{1.25}\smash{\begin{tabular}[t]{l}-1\end{tabular}}}}%
    \put(0.35081964,0.00023142){\color[rgb]{0,0,0}\makebox(0,0)[lt]{\lineheight{1.25}\smash{\begin{tabular}[t]{l}-2\end{tabular}}}}%
    \put(0.26117207,0.00030527){\color[rgb]{0,0,0}\makebox(0,0)[lt]{\lineheight{1.25}\smash{\begin{tabular}[t]{l}-3\end{tabular}}}}%
    \put(0.56339524,0.03127054){\color[rgb]{0,0,0}\makebox(0,0)[lt]{\lineheight{1.25}\smash{\begin{tabular}[t]{l}$\times 10^{3}$\end{tabular}}}}%
  \end{picture}%
\endgroup%

        }} \hspace{1cm}
    \subfloat[Cubic splines - $p=3$]{{
        \def\svgwidth{0.43\textwidth}
\begingroup%
  \makeatletter%
  \providecommand\color[2][]{%
    \errmessage{(Inkscape) Color is used for the text in Inkscape, but the package 'color.sty' is not loaded}%
    \renewcommand\color[2][]{}%
  }%
  \providecommand\transparent[1]{%
    \errmessage{(Inkscape) Transparency is used (non-zero) for the text in Inkscape, but the package 'transparent.sty' is not loaded}%
    \renewcommand\transparent[1]{}%
  }%
  \providecommand\rotatebox[2]{#2}%
  \newcommand*\fsize{\dimexpr\f@size pt\relax}%
  \newcommand*\lineheight[1]{\fontsize{\fsize}{#1\fsize}\selectfont}%
  \ifx\svgwidth\undefined%
    \setlength{\unitlength}{705bp}%
    \ifx\svgscale\undefined%
      \relax%
    \else%
      \setlength{\unitlength}{\unitlength * \real{\svgscale}}%
    \fi%
  \else%
    \setlength{\unitlength}{\svgwidth}%
  \fi%
  \global\let\svgwidth\undefined%
  \global\let\svgscale\undefined%
  \makeatother%
  \begin{picture}(1,0.60106383)%
    \lineheight{1}%
    \setlength\tabcolsep{0pt}%
    \put(0,0){\includegraphics[width=\unitlength,page=1]{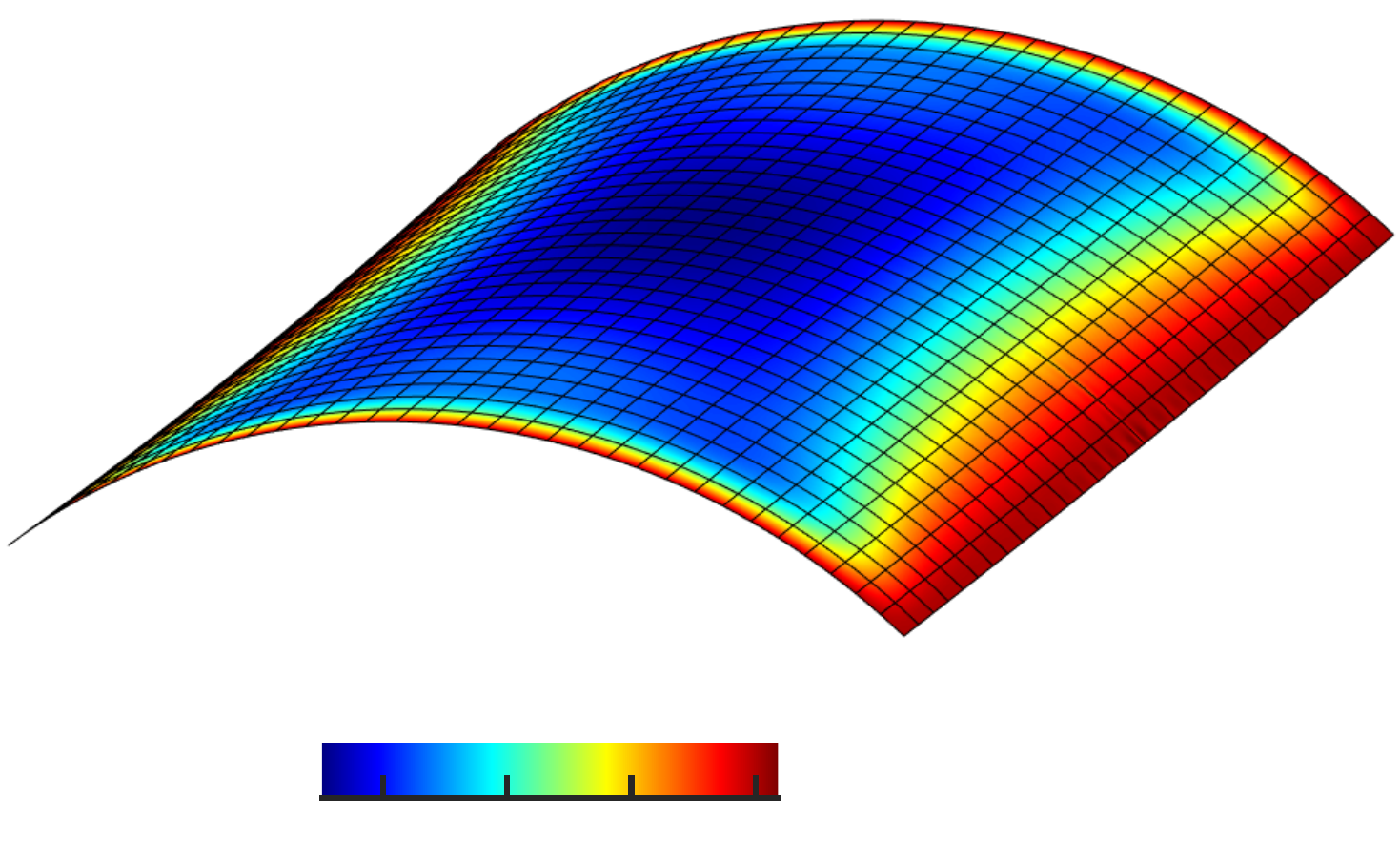}}%
    \put(0.56184079,0.06966059){\color[rgb]{0,0,0}\makebox(0,0)[lt]{\lineheight{1.25}\smash{\begin{tabular}[t]{l}$n^{11}$\end{tabular}}}}%
    \put(0.53180179,0.00015757){\color[rgb]{0,0,0}\makebox(0,0)[lt]{\lineheight{1.25}\smash{\begin{tabular}[t]{l}0\end{tabular}}}}%
    \put(0.43955475,0.00023142){\color[rgb]{0,0,0}\makebox(0,0)[lt]{\lineheight{1.25}\smash{\begin{tabular}[t]{l}-1\end{tabular}}}}%
    \put(0.35081964,0.00023142){\color[rgb]{0,0,0}\makebox(0,0)[lt]{\lineheight{1.25}\smash{\begin{tabular}[t]{l}-2\end{tabular}}}}%
    \put(0.26117205,0.00030527){\color[rgb]{0,0,0}\makebox(0,0)[lt]{\lineheight{1.25}\smash{\begin{tabular}[t]{l}-3\end{tabular}}}}%
    \put(0.56339524,0.03127054){\color[rgb]{0,0,0}\makebox(0,0)[lt]{\lineheight{1.25}\smash{\begin{tabular}[t]{l}$\times 10^{3}$\end{tabular}}}}%
  \end{picture}%
\endgroup%

        }}
\caption{Scordelis-Lo roof: membrane stress resultant $n^{11}$ plotted for the complete structure and slenderness ratio $R/t = 100$, computed with our approach on a mesh of $16 \times 16$ elements on the quarter domain.}\label{fig:shell_stress_slenderness_roof1}
\end{figure}

\begin{figure}[t!]
    \centering
\captionsetup[subfloat]{labelfont=scriptsize,textfont=scriptsize}
    \subfloat[Quadratic splines - $p=2$]{{
        \def\svgwidth{0.43\textwidth}
\begingroup%
  \makeatletter%
  \providecommand\color[2][]{%
    \errmessage{(Inkscape) Color is used for the text in Inkscape, but the package 'color.sty' is not loaded}%
    \renewcommand\color[2][]{}%
  }%
  \providecommand\transparent[1]{%
    \errmessage{(Inkscape) Transparency is used (non-zero) for the text in Inkscape, but the package 'transparent.sty' is not loaded}%
    \renewcommand\transparent[1]{}%
  }%
  \providecommand\rotatebox[2]{#2}%
  \newcommand*\fsize{\dimexpr\f@size pt\relax}%
  \newcommand*\lineheight[1]{\fontsize{\fsize}{#1\fsize}\selectfont}%
  \ifx\svgwidth\undefined%
    \setlength{\unitlength}{705bp}%
    \ifx\svgscale\undefined%
      \relax%
    \else%
      \setlength{\unitlength}{\unitlength * \real{\svgscale}}%
    \fi%
  \else%
    \setlength{\unitlength}{\svgwidth}%
  \fi%
  \global\let\svgwidth\undefined%
  \global\let\svgscale\undefined%
  \makeatother%
  \begin{picture}(1,0.60106383)%
    \lineheight{1}%
    \setlength\tabcolsep{0pt}%
    \put(0,0){\includegraphics[width=\unitlength,page=1]{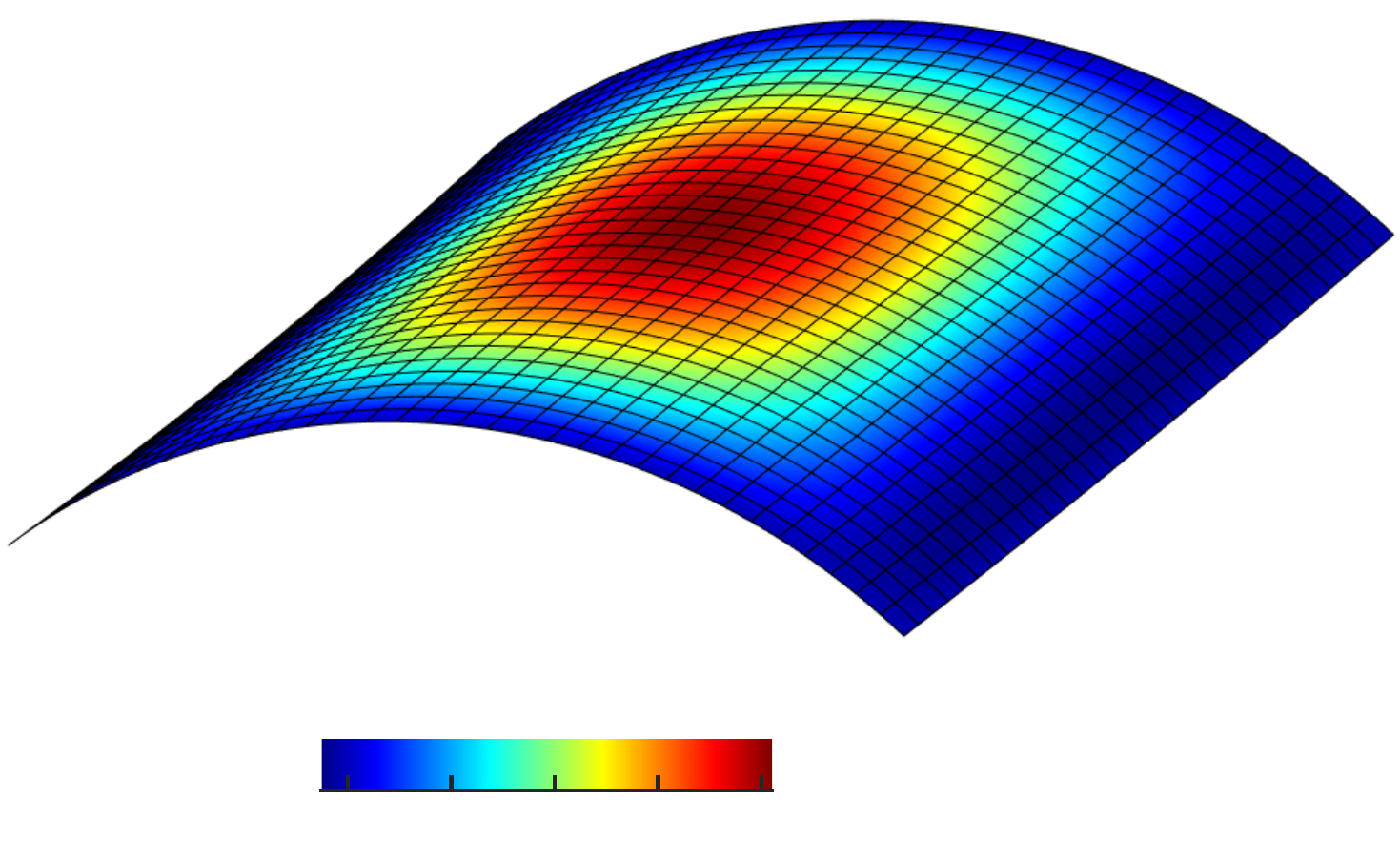}}%
    \put(0.23519842,0.00325889){\color[rgb]{0,0,0}\makebox(0,0)[lt]{\lineheight{1.25}\smash{\begin{tabular}[t]{l}$0.0$\end{tabular}}}}%
    \put(0.31058307,0.00335479){\color[rgb]{0,0,0}\makebox(0,0)[lt]{\lineheight{1.25}\smash{\begin{tabular}[t]{l}$0.5$\end{tabular}}}}%
    \put(0.38060663,0.00302747){\color[rgb]{0,0,0}\makebox(0,0)[lt]{\lineheight{1.25}\smash{\begin{tabular}[t]{l}$1.0$\end{tabular}}}}%
    \put(0.45551222,0.00228473){\color[rgb]{0,0,0}\makebox(0,0)[lt]{\lineheight{1.25}\smash{\begin{tabular}[t]{l}$1.5$\end{tabular}}}}%
    \put(0.52920058,0.0031224){\color[rgb]{0,0,0}\makebox(0,0)[lt]{\lineheight{1.25}\smash{\begin{tabular}[t]{l}$2.0$\end{tabular}}}}%
    \put(0.56015021,0.07362505){\color[rgb]{0,0,0}\makebox(0,0)[lt]{\lineheight{1.25}\smash{\begin{tabular}[t]{l}$m^{11}$\end{tabular}}}}%
    \put(0.55507767,0.03422462){\color[rgb]{0,0,0}\makebox(0,0)[lt]{\lineheight{1.25}\smash{\begin{tabular}[t]{l}$\times10^3$\end{tabular}}}}%
  \end{picture}%
\endgroup%

        }} \hspace{1cm}
    \subfloat[Cubic splines - $p=3$]{{
        \def\svgwidth{0.43\textwidth}
\begingroup%
  \makeatletter%
  \providecommand\color[2][]{%
    \errmessage{(Inkscape) Color is used for the text in Inkscape, but the package 'color.sty' is not loaded}%
    \renewcommand\color[2][]{}%
  }%
  \providecommand\transparent[1]{%
    \errmessage{(Inkscape) Transparency is used (non-zero) for the text in Inkscape, but the package 'transparent.sty' is not loaded}%
    \renewcommand\transparent[1]{}%
  }%
  \providecommand\rotatebox[2]{#2}%
  \newcommand*\fsize{\dimexpr\f@size pt\relax}%
  \newcommand*\lineheight[1]{\fontsize{\fsize}{#1\fsize}\selectfont}%
  \ifx\svgwidth\undefined%
    \setlength{\unitlength}{705bp}%
    \ifx\svgscale\undefined%
      \relax%
    \else%
      \setlength{\unitlength}{\unitlength * \real{\svgscale}}%
    \fi%
  \else%
    \setlength{\unitlength}{\svgwidth}%
  \fi%
  \global\let\svgwidth\undefined%
  \global\let\svgscale\undefined%
  \makeatother%
  \begin{picture}(1,0.60106383)%
    \lineheight{1}%
    \setlength\tabcolsep{0pt}%
    \put(0,0){\includegraphics[width=\unitlength,page=1]{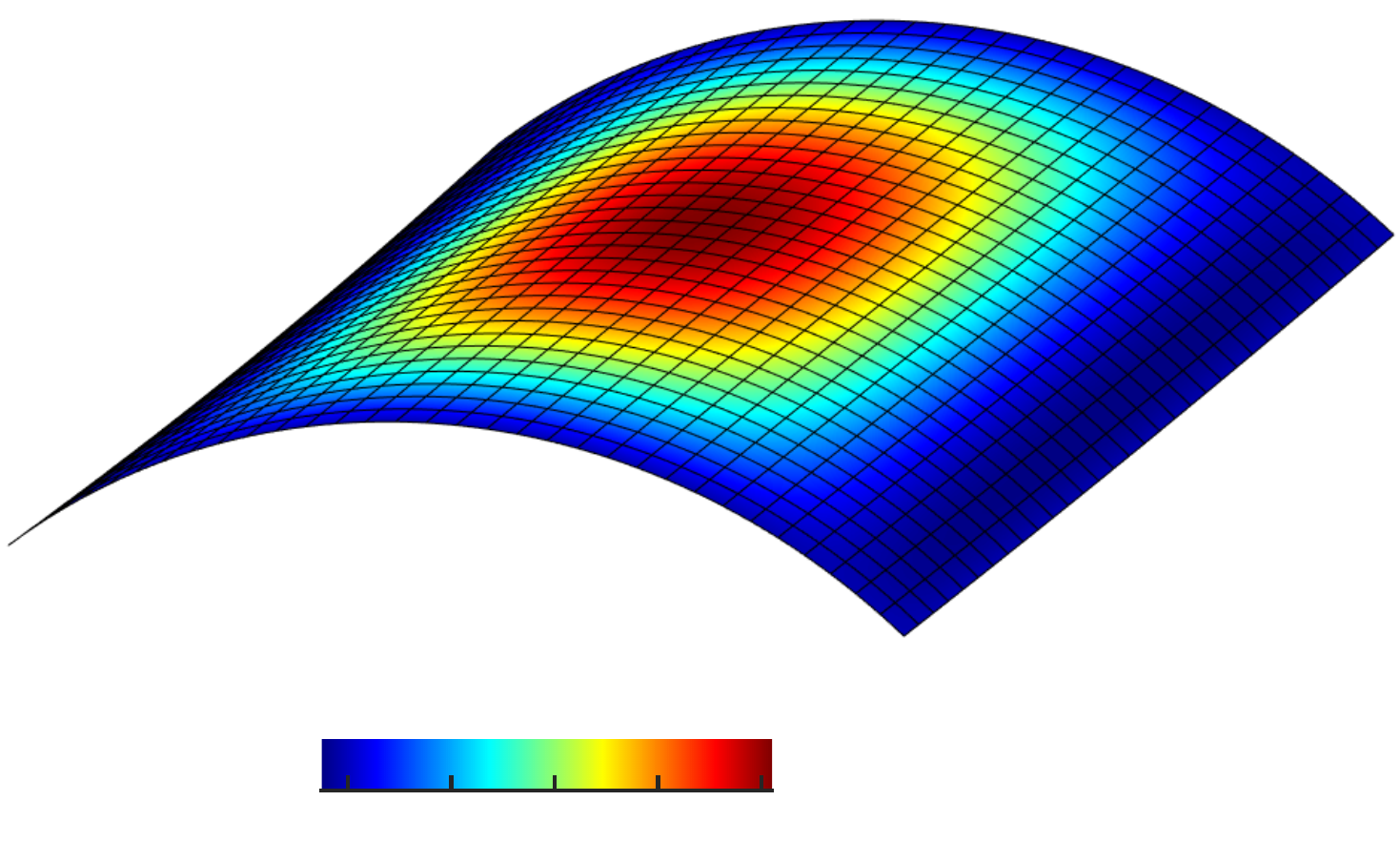}}%
    \put(0.2351982,0.00325889){\color[rgb]{0,0,0}\makebox(0,0)[lt]{\lineheight{1.25}\smash{\begin{tabular}[t]{l}$0.0$\end{tabular}}}}%
    \put(0.31058285,0.00335479){\color[rgb]{0,0,0}\makebox(0,0)[lt]{\lineheight{1.25}\smash{\begin{tabular}[t]{l}$0.5$\end{tabular}}}}%
    \put(0.3806064,0.00302747){\color[rgb]{0,0,0}\makebox(0,0)[lt]{\lineheight{1.25}\smash{\begin{tabular}[t]{l}$1.0$\end{tabular}}}}%
    \put(0.45551199,0.00228473){\color[rgb]{0,0,0}\makebox(0,0)[lt]{\lineheight{1.25}\smash{\begin{tabular}[t]{l}$1.5$\end{tabular}}}}%
    \put(0.52920039,0.0031224){\color[rgb]{0,0,0}\makebox(0,0)[lt]{\lineheight{1.25}\smash{\begin{tabular}[t]{l}$2.0$\end{tabular}}}}%
    \put(0.56015002,0.07362499){\color[rgb]{0,0,0}\makebox(0,0)[lt]{\lineheight{1.25}\smash{\begin{tabular}[t]{l}$m^{11}$\end{tabular}}}}%
    \put(0.55507748,0.03422462){\color[rgb]{0,0,0}\makebox(0,0)[lt]{\lineheight{1.25}\smash{\begin{tabular}[t]{l}$\times10^3$\end{tabular}}}}%
  \end{picture}%
\endgroup%

        }}
    \caption{Scordelis-Lo roof: bending stress resultant $m^{11}$ plotted for the complete structure and slenderness ratio $R/t = 100$, computed with our approach on a mesh of $16 \times 16$ elements on the quarter domain.}\label{fig:shell_stress_slenderness_roof2}
\end{figure}

\begin{figure}[t!]
    \centering
    \captionsetup[subfloat]{labelfont=scriptsize,textfont=scriptsize}
    \subfloat[$p=2$, $R/t=100$]{{
        \def\svgwidth{0.47\textwidth}
        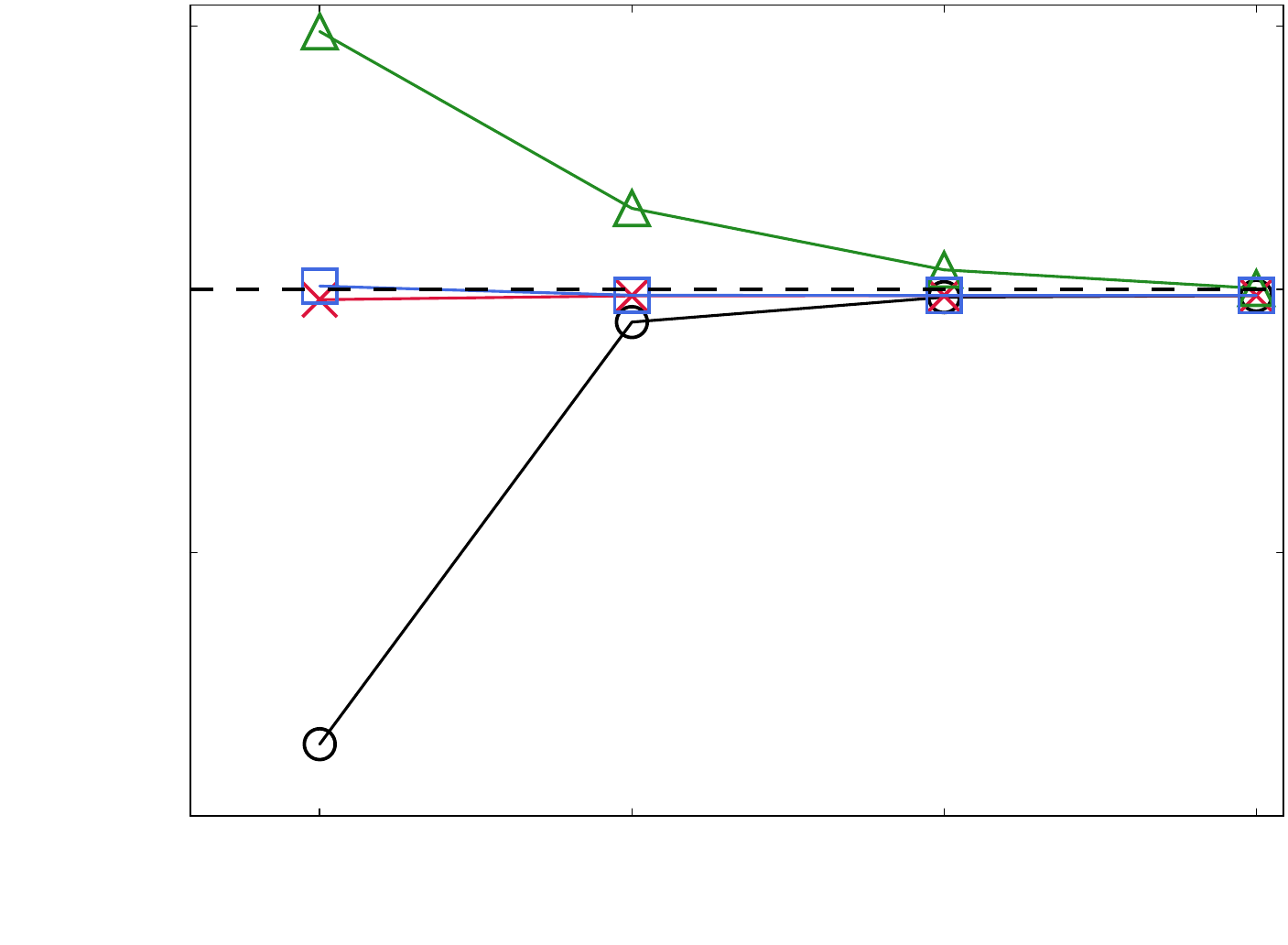
        }} \hspace{0.2cm}
    \subfloat[$p=3$, $R/t=100$]{{
        \def\svgwidth{0.47\textwidth}
        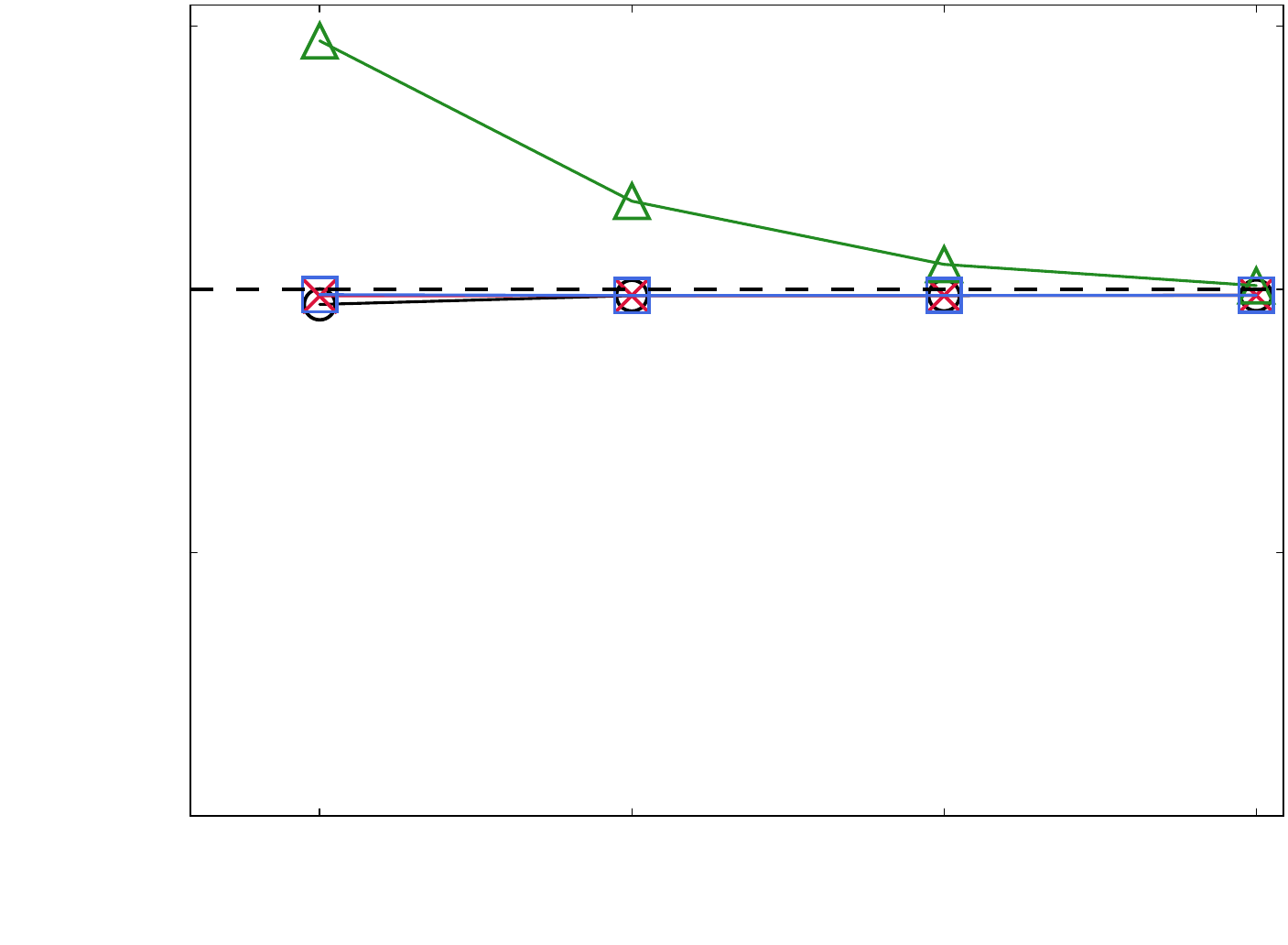
        }}

    \subfloat[$p=2$, $R/t=1,000$]{{
        \def\svgwidth{0.47\textwidth}
        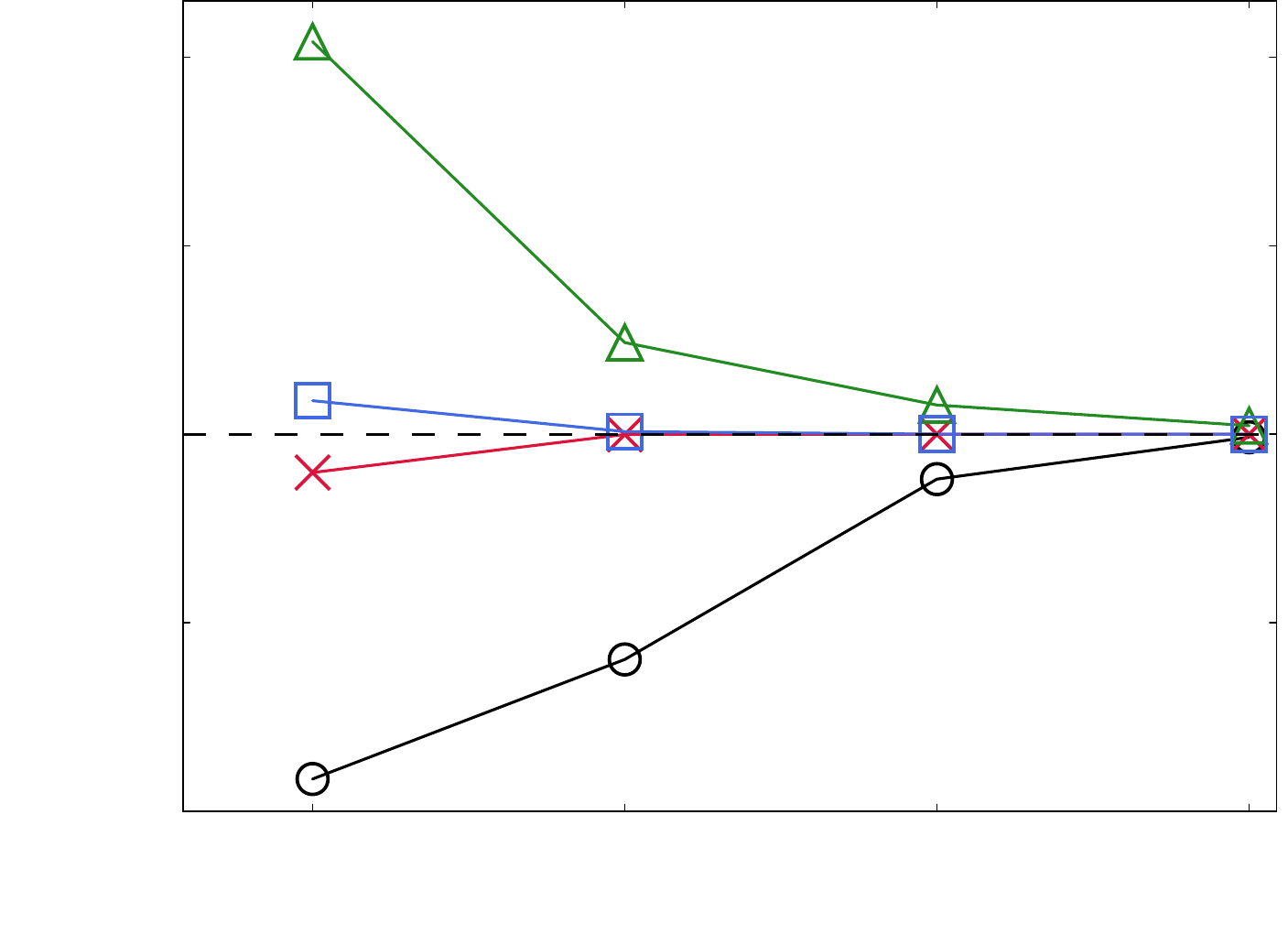
        }} \hspace{0.2cm}
    \subfloat[$p=3$, $R/t=1,000$]{{
        \def\svgwidth{0.47\textwidth}
        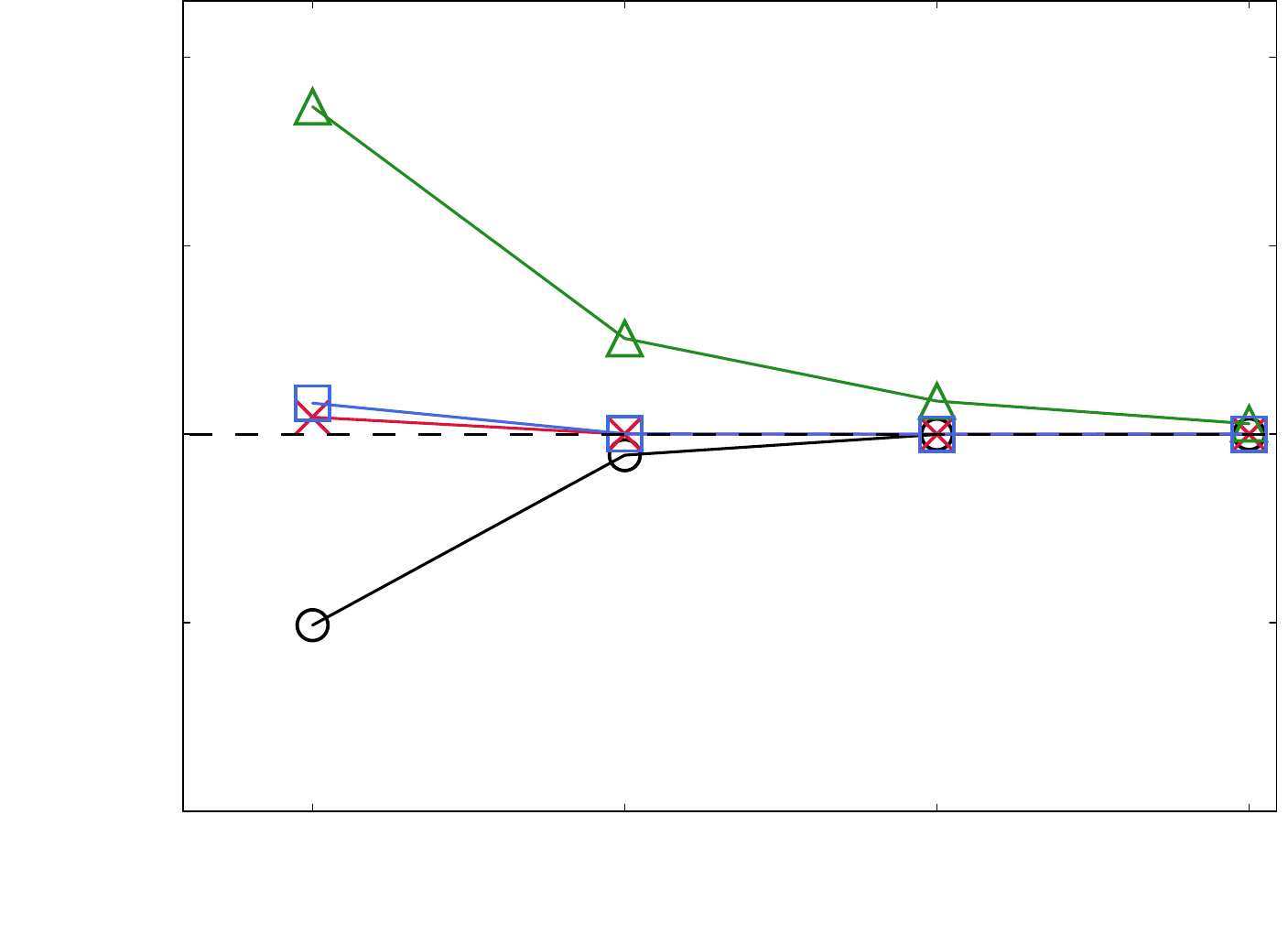
        }}

    \vspace{0.2cm}
    {\begin{tikzpicture}
    \filldraw[black,line width=1pt, solid] (0.0,0) -- (0.6,0);
    \filldraw[black,line width=1pt] (0.3,0) [fill=none] circle (3pt);
    \filldraw[black,line width=1pt] (0.7,0) node[right]{\scriptsize Displacement-based formulation};
    \filldraw[red1,line width=1pt, solid] (7.5,0) -- (8.1,0);
    \filldraw[red1,line width=1pt] (7.5,0) node[right]{\scriptsize $\boldsymbol{\bigtimes}$};
    \filldraw[red1,line width=1pt] (8.2,0) node[right]{\scriptsize Mixed formulation, consistent strain projection};
\end{tikzpicture}

\begin{tikzpicture}
    \filldraw[green1,line width=1pt, solid] (0.0,0) -- (0.6,0);
    \filldraw[green1,line width=1pt] (0.0,0) node[right]{\scriptsize $\boldsymbol{\Delta}$};
    \filldraw[green1,line width=1pt] (0.7,0) node[right]{\scriptsize Mixed formulation, lumped strain projection, B-splines};
    \filldraw[blue1,line width=1pt, solid] (8.5,0.05) -- (9.1,0.05);
    \filldraw[blue1,line width=1pt] (8.7,-0.08) [fill=none] rectangle ++(0.25,0.25);
    \filldraw[blue1,line width=1pt] (9.2,0) node[right]{\scriptsize Mixed formulation, lumped strain projection, approximate duals};
\end{tikzpicture}}
    \caption{{Scordelis-Lo roof}: convergence of the normalized vertical displacement at the midpoint of the free edge, computed with different formulations for two different slenderness ratios.}\label{fig:shell_tip_roof}
\end{figure}

\begin{figure}[t!]
    \centering
    \captionsetup[subfloat]{labelfont=scriptsize,textfont=scriptsize}
    \subfloat[$p=2$]{{\def\svgwidth{0.47\textwidth}
    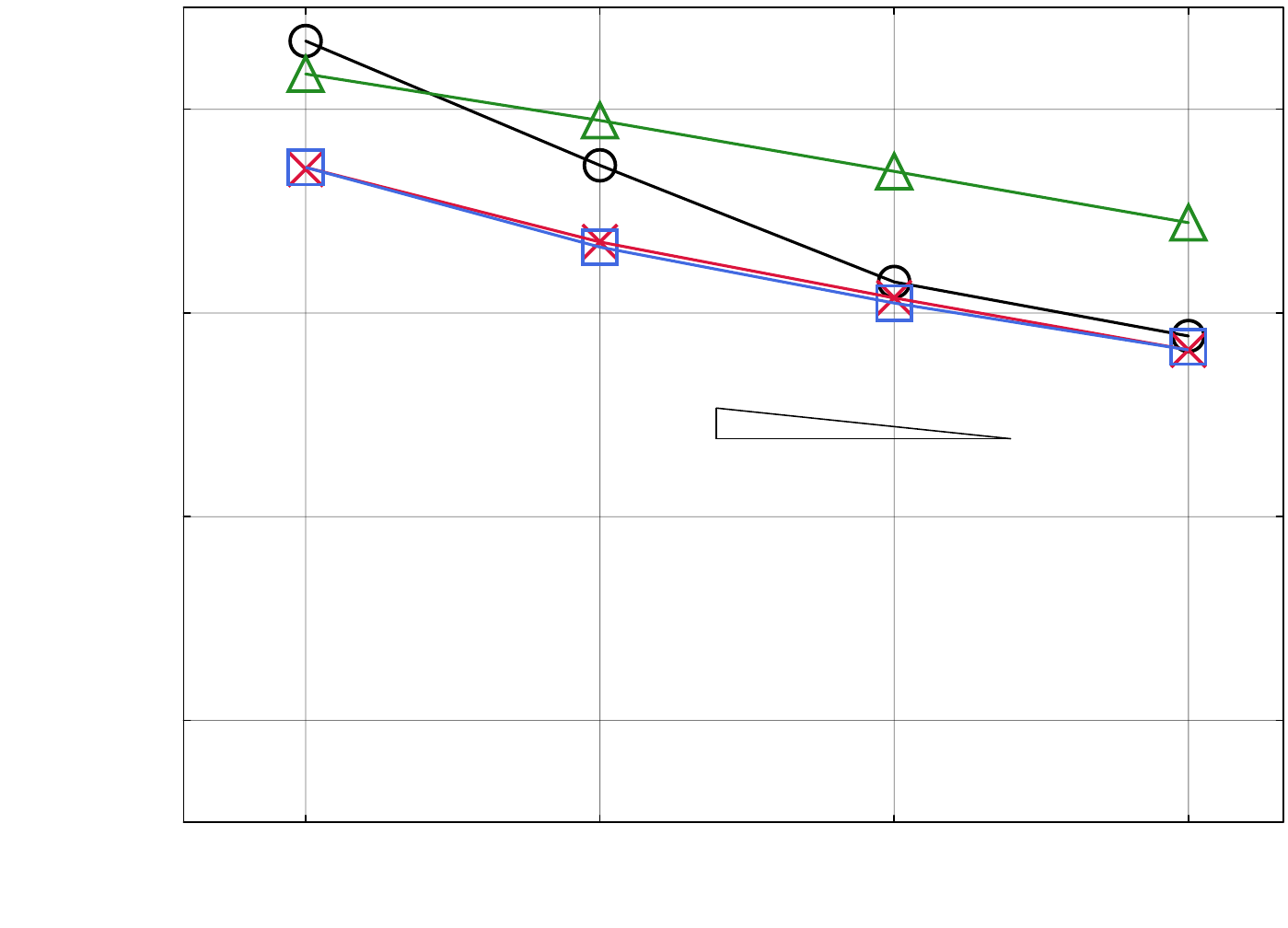 }} \hspace{0.2cm}
    \subfloat[$p=3$]{{\def\svgwidth{0.47\textwidth}
    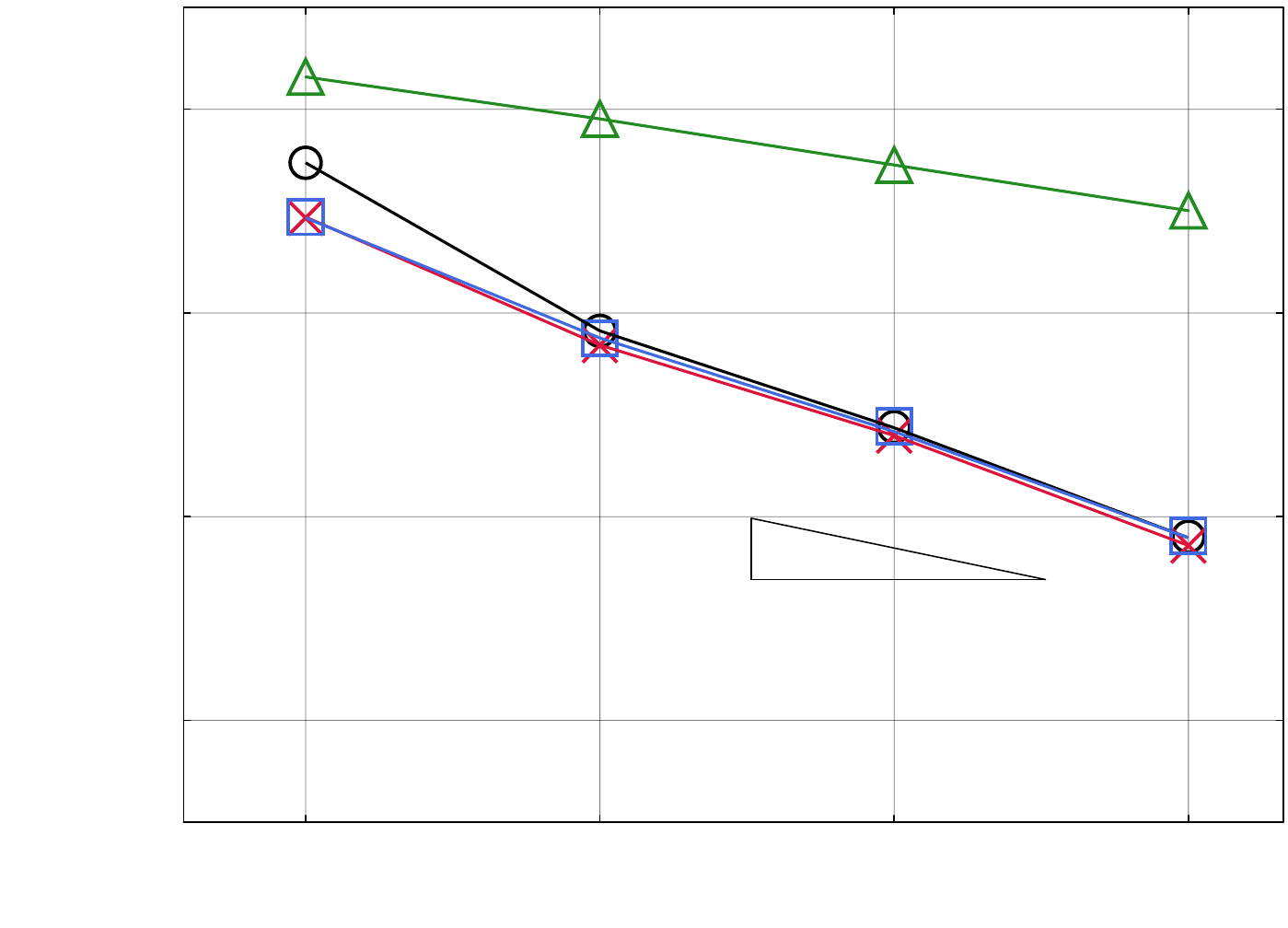 }}

    \subfloat[$p=4$]{{\def\svgwidth{0.47\textwidth}
    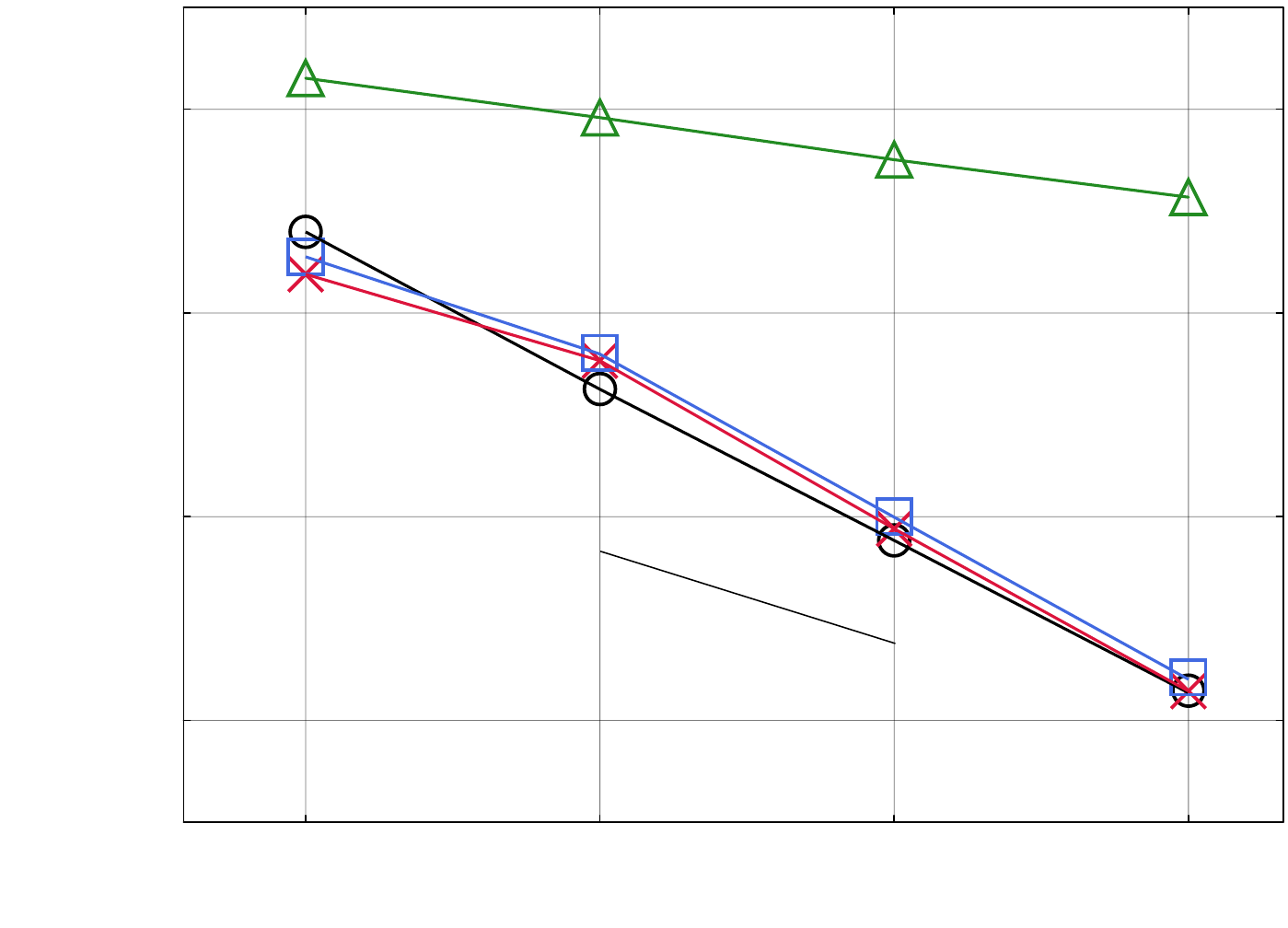 }} \hspace{0.2cm}
    \subfloat[$p=5$]{{\def\svgwidth{0.47\textwidth}
    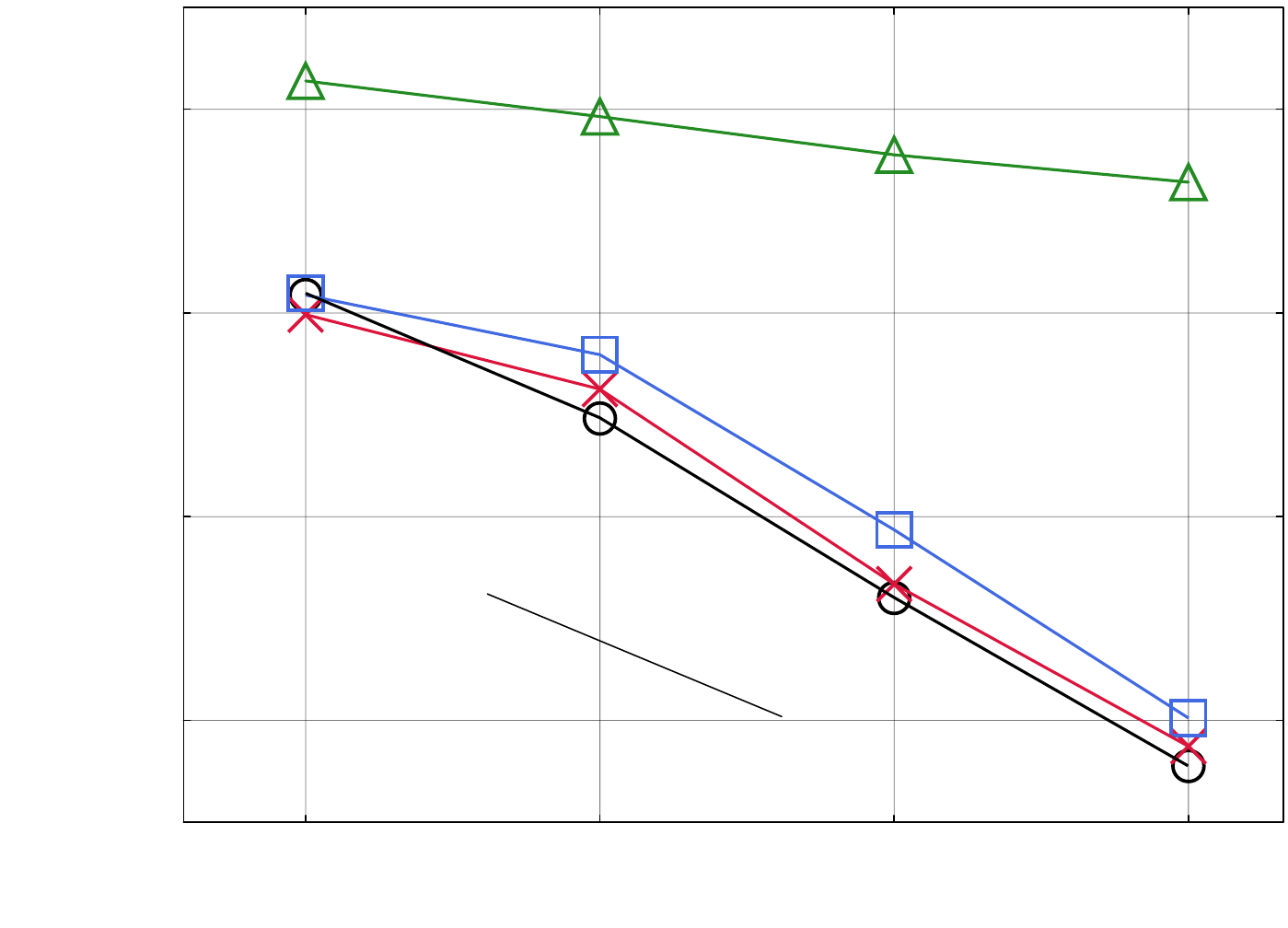 }}

    \vspace{0.2cm}
    {\begin{tikzpicture}
    \filldraw[black,line width=1pt, solid] (0.0,0) -- (0.6,0);
    \filldraw[black,line width=1pt] (0.3,0) [fill=none] circle (3pt);
    \filldraw[black,line width=1pt] (0.7,0) node[right]{\scriptsize Displacement-based formulation};
    \filldraw[red1,line width=1pt, solid] (7.5,0) -- (8.1,0);
    \filldraw[red1,line width=1pt] (7.5,0) node[right]{\scriptsize $\boldsymbol{\bigtimes}$};
    \filldraw[red1,line width=1pt] (8.2,0) node[right]{\scriptsize Mixed formulation, consistent strain projection};
\end{tikzpicture}

\begin{tikzpicture}
    \filldraw[green1,line width=1pt, solid] (0.0,0) -- (0.6,0);
    \filldraw[green1,line width=1pt] (0.0,0) node[right]{\scriptsize $\boldsymbol{\Delta}$};
    \filldraw[green1,line width=1pt] (0.7,0) node[right]{\scriptsize Mixed formulation, lumped strain projection, B-splines};
    \filldraw[blue1,line width=1pt, solid] (8.5,0.05) -- (9.1,0.05);
    \filldraw[blue1,line width=1pt] (8.7,-0.08) [fill=none] rectangle ++(0.25,0.25);
    \filldraw[blue1,line width=1pt] (9.2,0) node[right]{\scriptsize Mixed formulation, lumped strain projection, approximate duals};
\end{tikzpicture}}
    \caption{{Scordelis-Lo roof}: convergence of the relative error in the $L^2$ norm of the bending moment $m^{12}$, computed with different formulations.}\label{fig:shell_h2converge_roof}
\end{figure}


\begin{figure}[ht!]
	\centering
    \def\svgwidth{1\textwidth}
    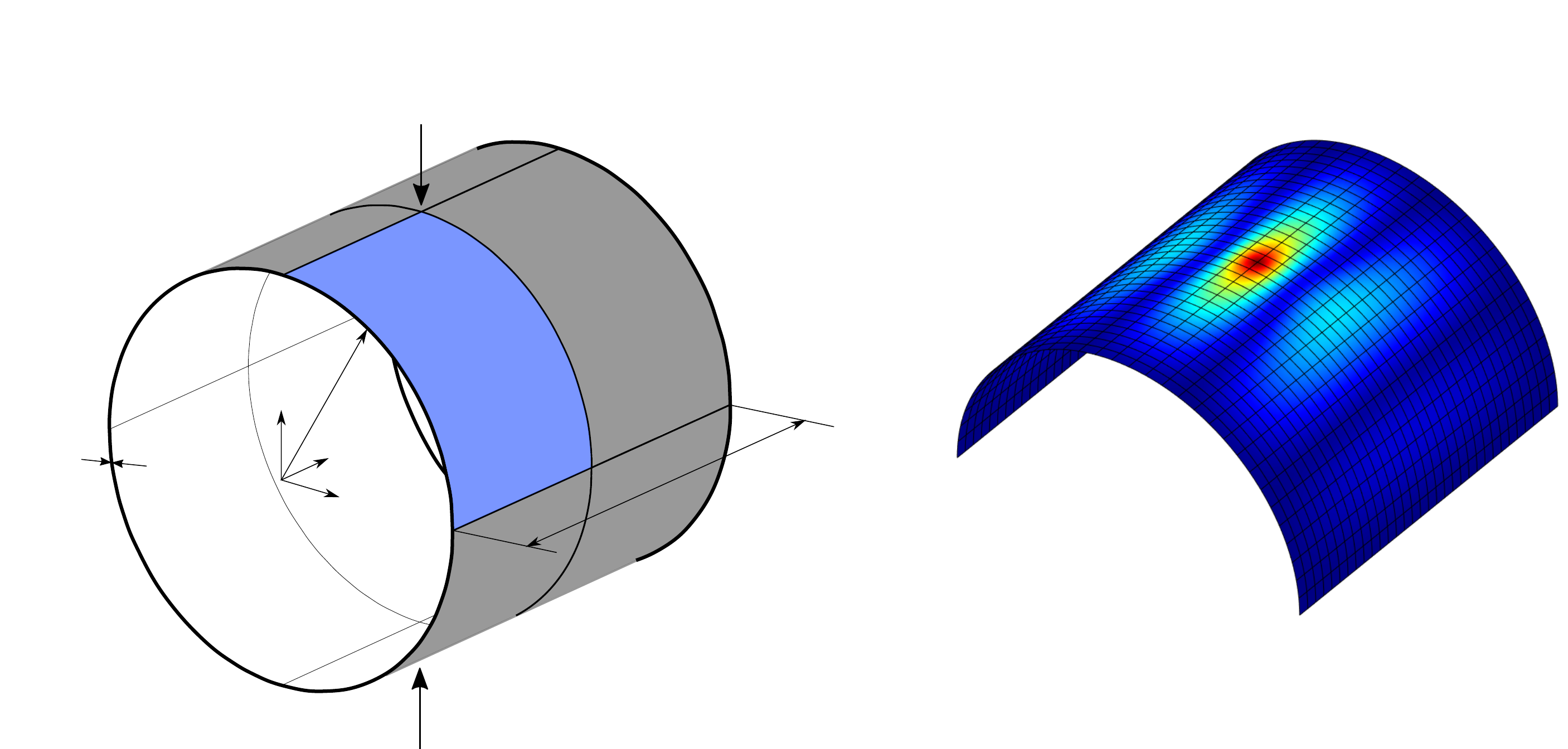

    \caption{Pinched cylinder constrained by rigid diaphragms: problem sketch along with a plot of the total displacements, computed with our approach and cubic splines.} \label{fig:cylinder_geometry}
\end{figure}

\begin{figure}[t!]
    \centering
    \captionsetup[subfloat]{labelfont=scriptsize,textfont=scriptsize}    
    \subfloat[$n^{12}$]{{
        \def\svgwidth{0.43\textwidth}
\begingroup%
  \makeatletter%
  \providecommand\color[2][]{%
    \errmessage{(Inkscape) Color is used for the text in Inkscape, but the package 'color.sty' is not loaded}%
    \renewcommand\color[2][]{}%
  }%
  \providecommand\transparent[1]{%
    \errmessage{(Inkscape) Transparency is used (non-zero) for the text in Inkscape, but the package 'transparent.sty' is not loaded}%
    \renewcommand\transparent[1]{}%
  }%
  \providecommand\rotatebox[2]{#2}%
  \newcommand*\fsize{\dimexpr\f@size pt\relax}%
  \newcommand*\lineheight[1]{\fontsize{\fsize}{#1\fsize}\selectfont}%
  \ifx\svgwidth\undefined%
    \setlength{\unitlength}{622.5bp}%
    \ifx\svgscale\undefined%
      \relax%
    \else%
      \setlength{\unitlength}{\unitlength * \real{\svgscale}}%
    \fi%
  \else%
    \setlength{\unitlength}{\svgwidth}%
  \fi%
  \global\let\svgwidth\undefined%
  \global\let\svgscale\undefined%
  \makeatother%
  \begin{picture}(1,0.93975904)%
    \lineheight{1}%
    \setlength\tabcolsep{0pt}%
    \put(0,0){\includegraphics[width=\unitlength,page=1]{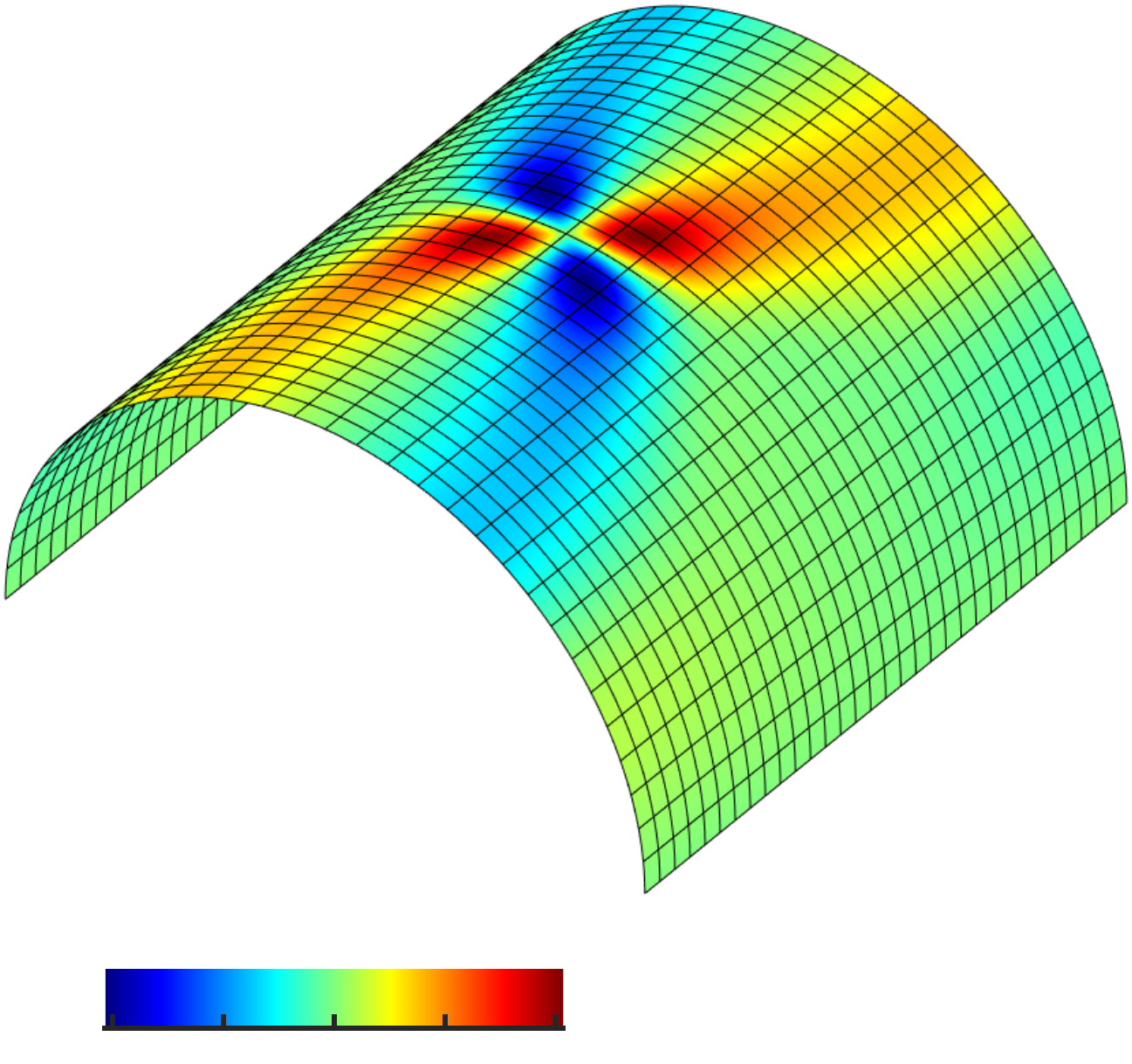}}%
    \put(0.51327187,0.08199375){\color[rgb]{0,0,0}\makebox(0,0)[lt]{\lineheight{1.25}\smash{\begin{tabular}[t]{l}$n^{12}$\end{tabular}}}}%
    \put(0.51292961,0.03480869){\color[rgb]{0,0,0}\makebox(0,0)[lt]{\lineheight{1.25}\smash{\begin{tabular}[t]{l}$\times 10^{-3}$\end{tabular}}}}%
    \put(0.38360354,0.00027459){\color[rgb]{0,0,0}\makebox(0,0)[lt]{\lineheight{1.25}\smash{\begin{tabular}[t]{l}8\end{tabular}}}}%
    \put(0.28545238,0.0003498){\color[rgb]{0,0,0}\makebox(0,0)[lt]{\lineheight{1.25}\smash{\begin{tabular}[t]{l}0\end{tabular}}}}%
    \put(0.17559168,0.0003498){\color[rgb]{0,0,0}\makebox(0,0)[lt]{\lineheight{1.25}\smash{\begin{tabular}[t]{l}-8\end{tabular}}}}%
    \put(0.06703924,0.00016008){\color[rgb]{0,0,0}\makebox(0,0)[lt]{\lineheight{1.25}\smash{\begin{tabular}[t]{l}-16\end{tabular}}}}%
    \put(0.4730011,0.00027459){\color[rgb]{0,0,0}\makebox(0,0)[lt]{\lineheight{1.25}\smash{\begin{tabular}[t]{l}16\end{tabular}}}}%
  \end{picture}%
\endgroup%

        }}\hspace{0.4cm}
    \subfloat[$m^{11}$]{{
        \def\svgwidth{0.43\textwidth}
        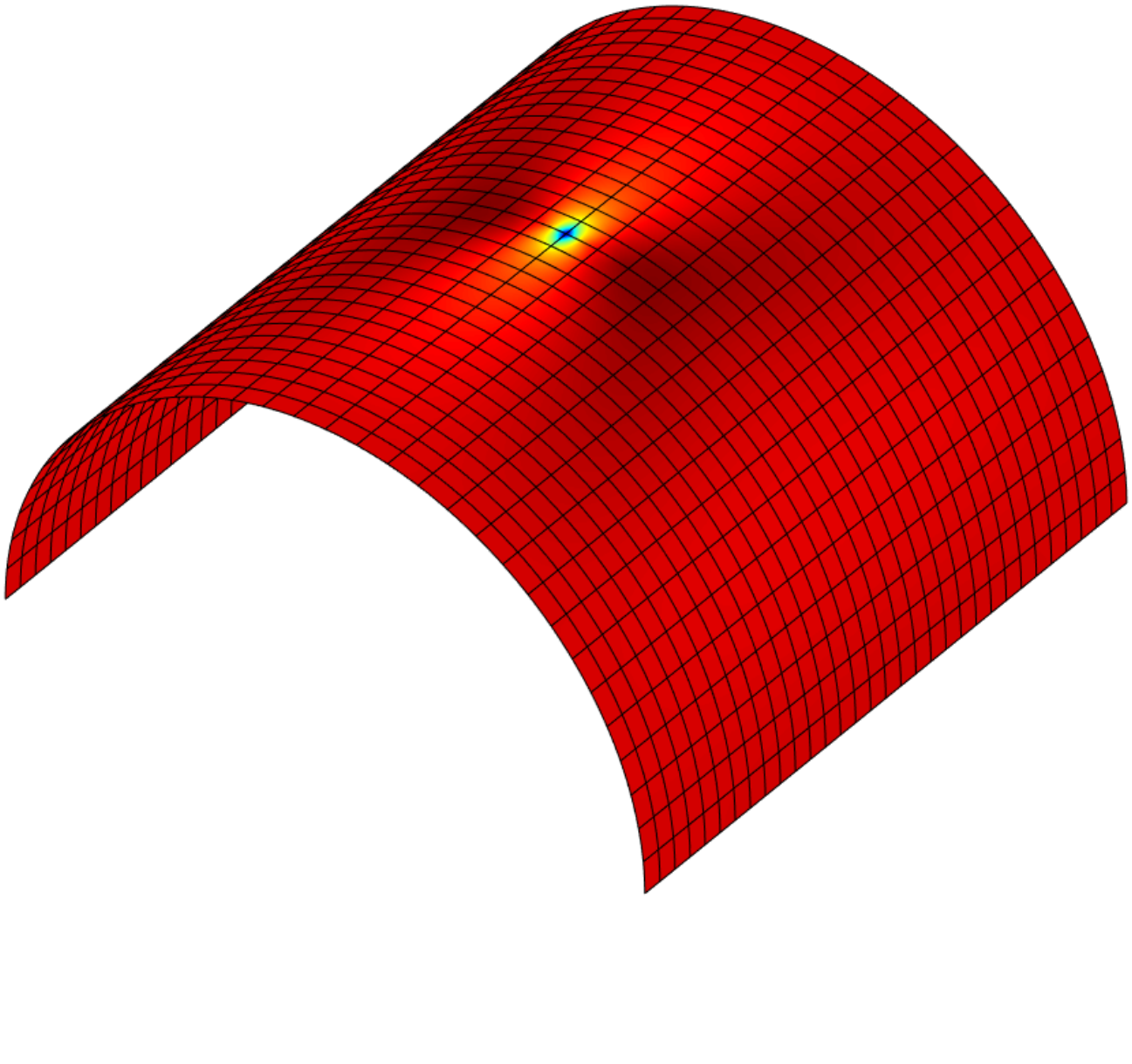
        }}

    \caption{{Pinched cylinder}: the membrane stress resultant $n^{12}$ and the bending moment $m^{11}$ for slenderness ratio $R/t=100$, computed with our approach and cubic splines on a mesh of $16 \times 16$ elements.}\label{fig:shell_stress_slenderness_cylinder}
\end{figure}

\begin{figure}[t!]
     \centering
    \captionsetup[subfloat]{labelfont=scriptsize,textfont=scriptsize}
    \subfloat[$p=2$, $R/t=100$]{{
        \def\svgwidth{0.47\textwidth}
        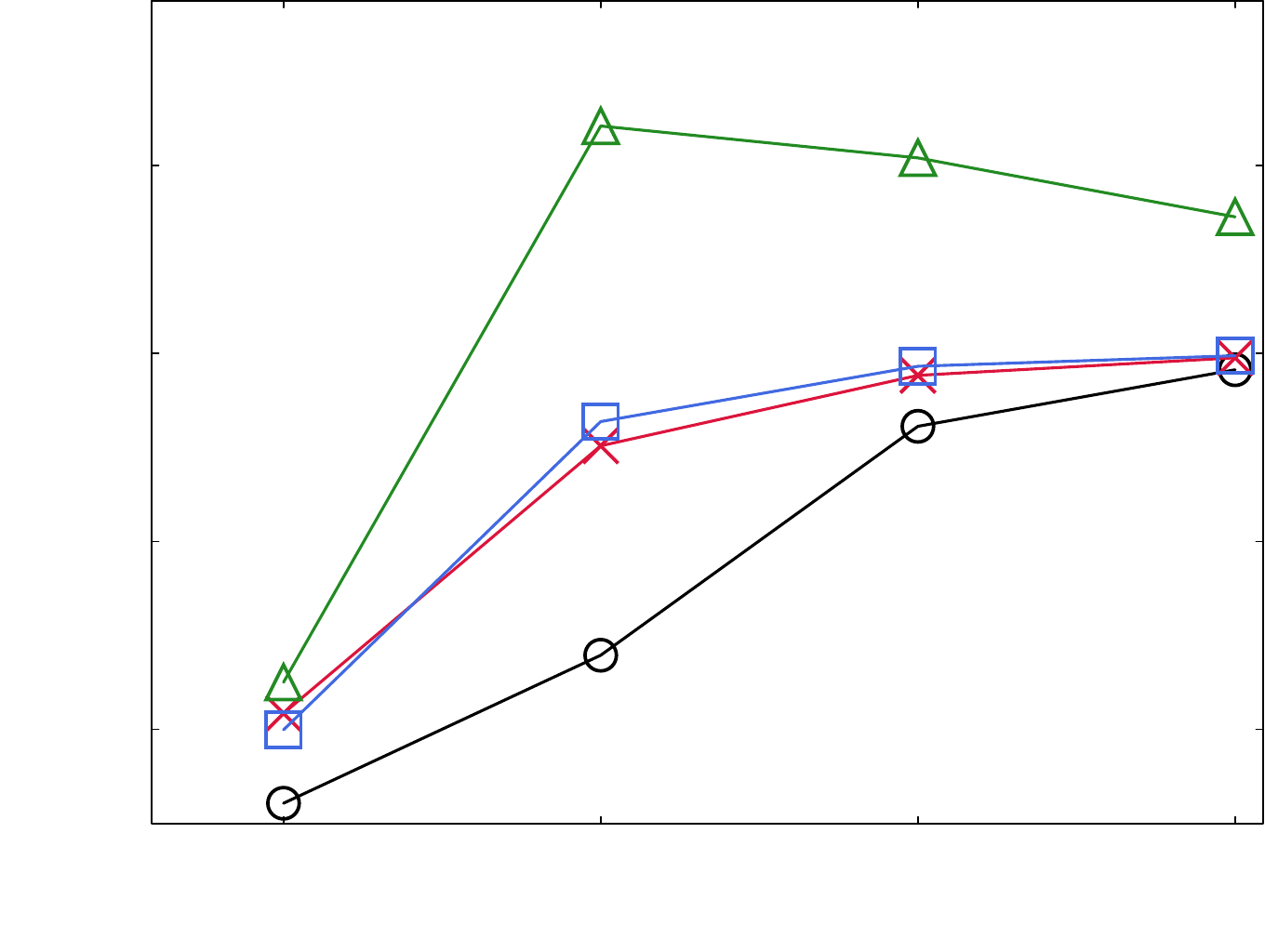
        }} \hspace{0.2cm}
    \subfloat[$p=3$, $R/t=100$]{{
        \def\svgwidth{0.47\textwidth}
        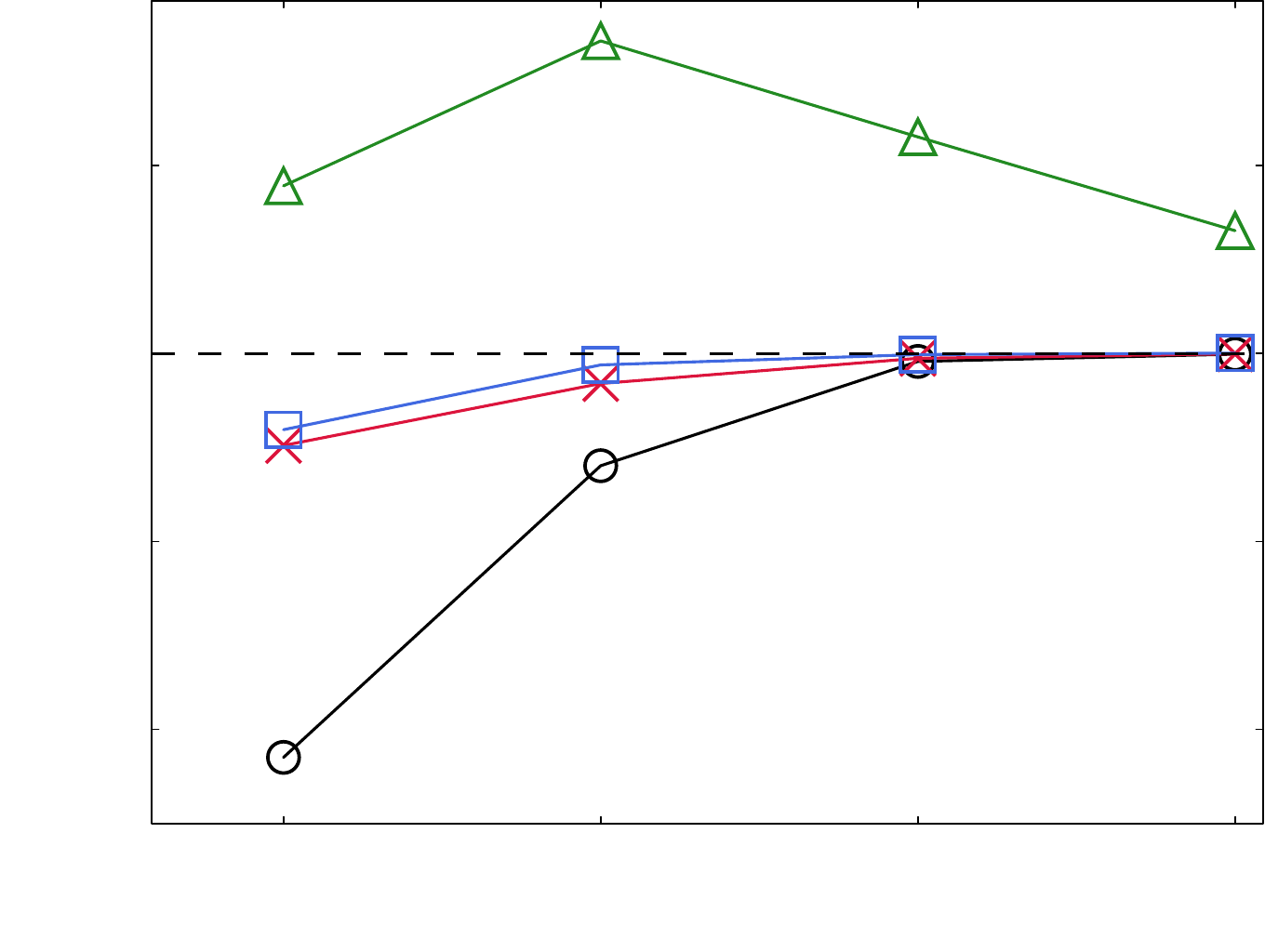
        }}

    \subfloat[$p=2$, $R/t=1,000$]{{
        \def\svgwidth{0.47\textwidth}
        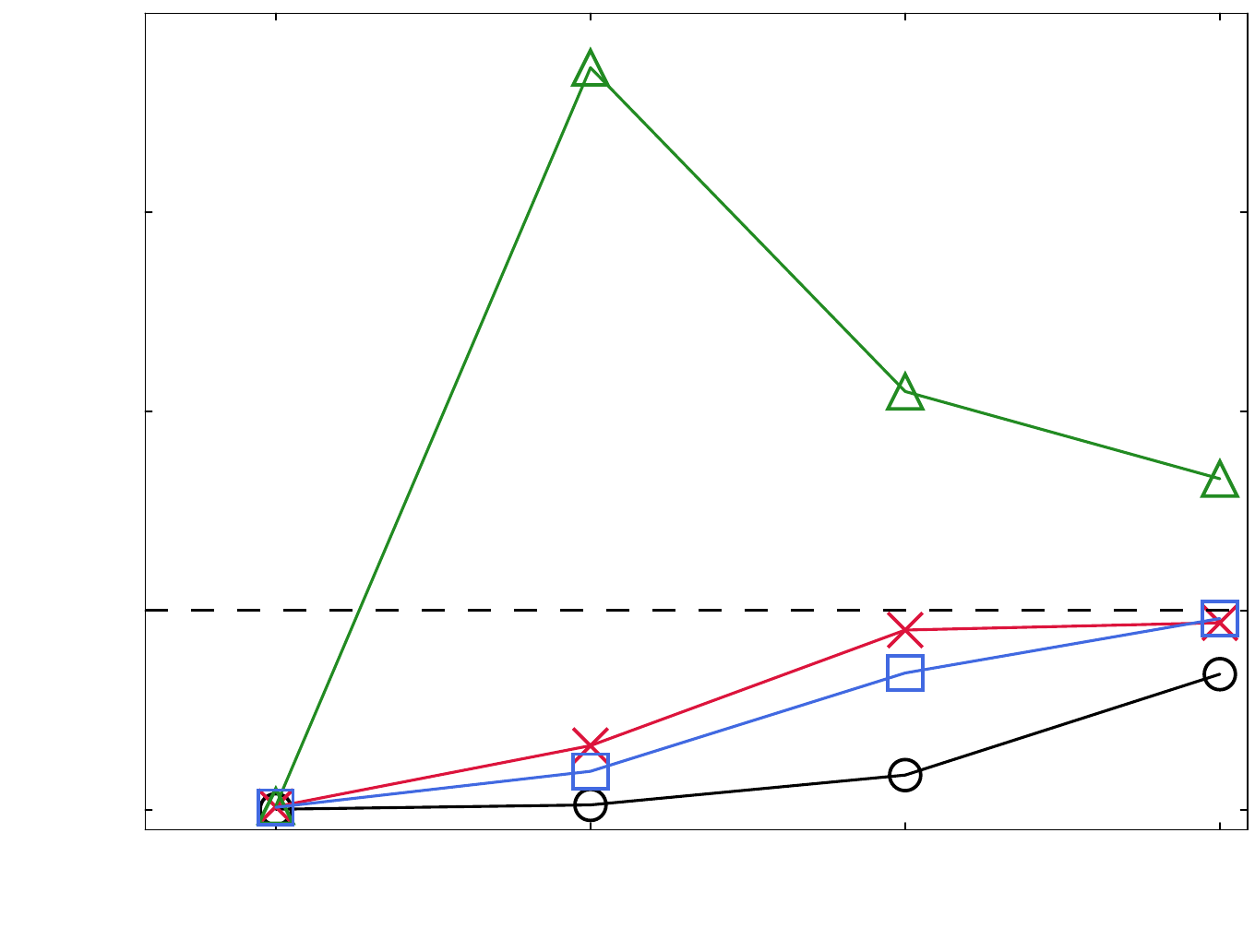
        }} \hspace{0.2cm}
    \subfloat[$p=3$, $R/t=1,000$]{{
        \def\svgwidth{0.47\textwidth}
        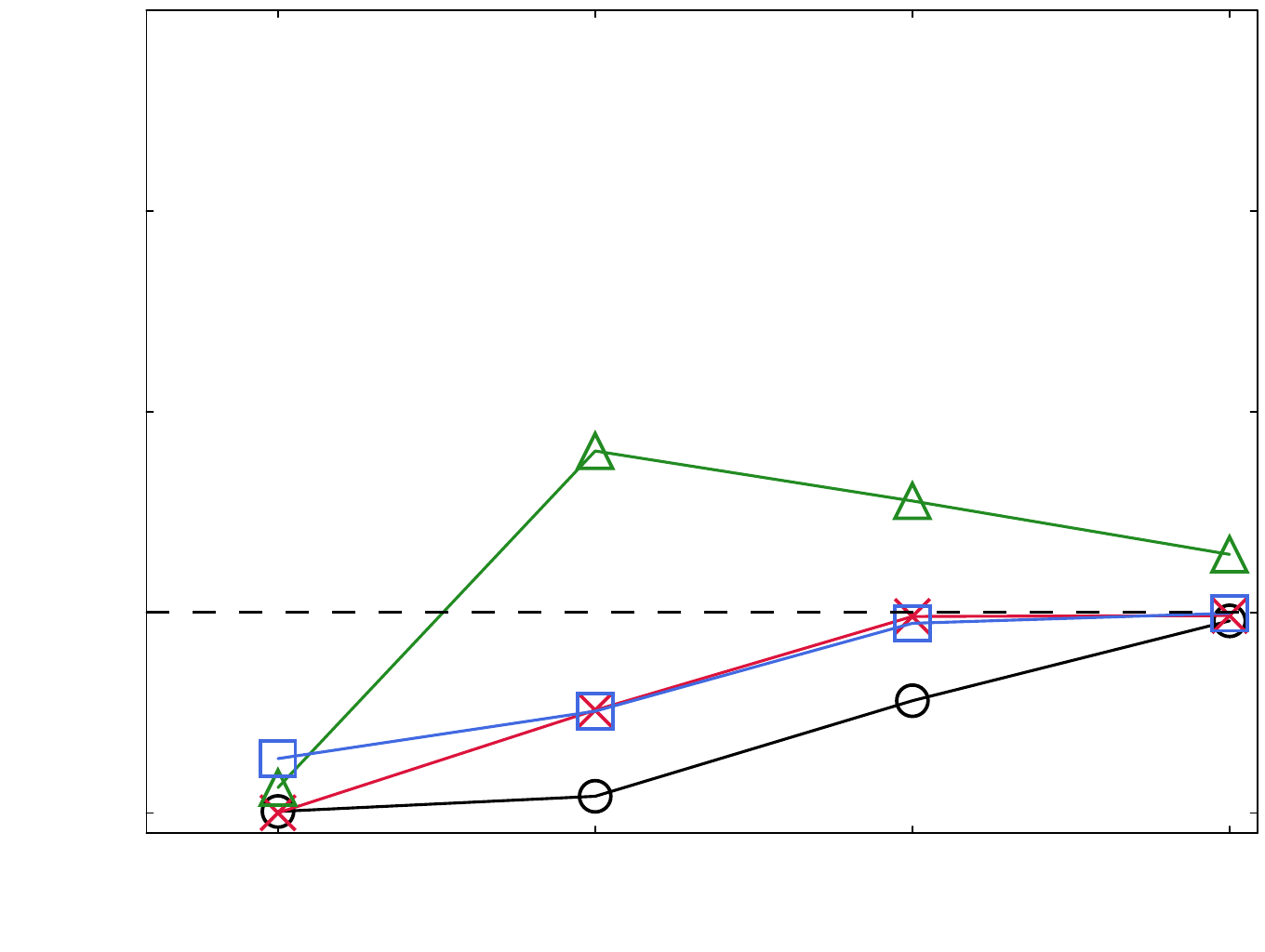
        }}

    \vspace{0.15cm}
    {\begin{tikzpicture}
    \filldraw[black,line width=1pt, solid] (0.0,0) -- (0.6,0);
    \filldraw[black,line width=1pt] (0.3,0) [fill=none] circle (3pt);
    \filldraw[black,line width=1pt] (0.7,0) node[right]{\scriptsize Displacement-based formulation};
    \filldraw[red1,line width=1pt, solid] (7.5,0) -- (8.1,0);
    \filldraw[red1,line width=1pt] (7.5,0) node[right]{\scriptsize $\boldsymbol{\bigtimes}$};
    \filldraw[red1,line width=1pt] (8.2,0) node[right]{\scriptsize Mixed formulation, consistent strain projection};
\end{tikzpicture}

\begin{tikzpicture}
    \filldraw[green1,line width=1pt, solid] (0.0,0) -- (0.6,0);
    \filldraw[green1,line width=1pt] (0.0,0) node[right]{\scriptsize $\boldsymbol{\Delta}$};
    \filldraw[green1,line width=1pt] (0.7,0) node[right]{\scriptsize Mixed formulation, lumped strain projection, B-splines};
    \filldraw[blue1,line width=1pt, solid] (8.5,0.05) -- (9.1,0.05);
    \filldraw[blue1,line width=1pt] (8.7,-0.08) [fill=none] rectangle ++(0.25,0.25);
    \filldraw[blue1,line width=1pt] (9.2,0) node[right]{\scriptsize Mixed formulation, lumped strain projection, approximate duals};
\end{tikzpicture}} 
    \caption{{Pinched cylinder}: convergence of the normalized displacement under the load, computed with different formulations for two different slenderness ratios.} \label{fig:shell_tip_cylinder}
\end{figure}


\begin{figure}[ht!]
	\centering
    \def\svgwidth{1\textwidth}
    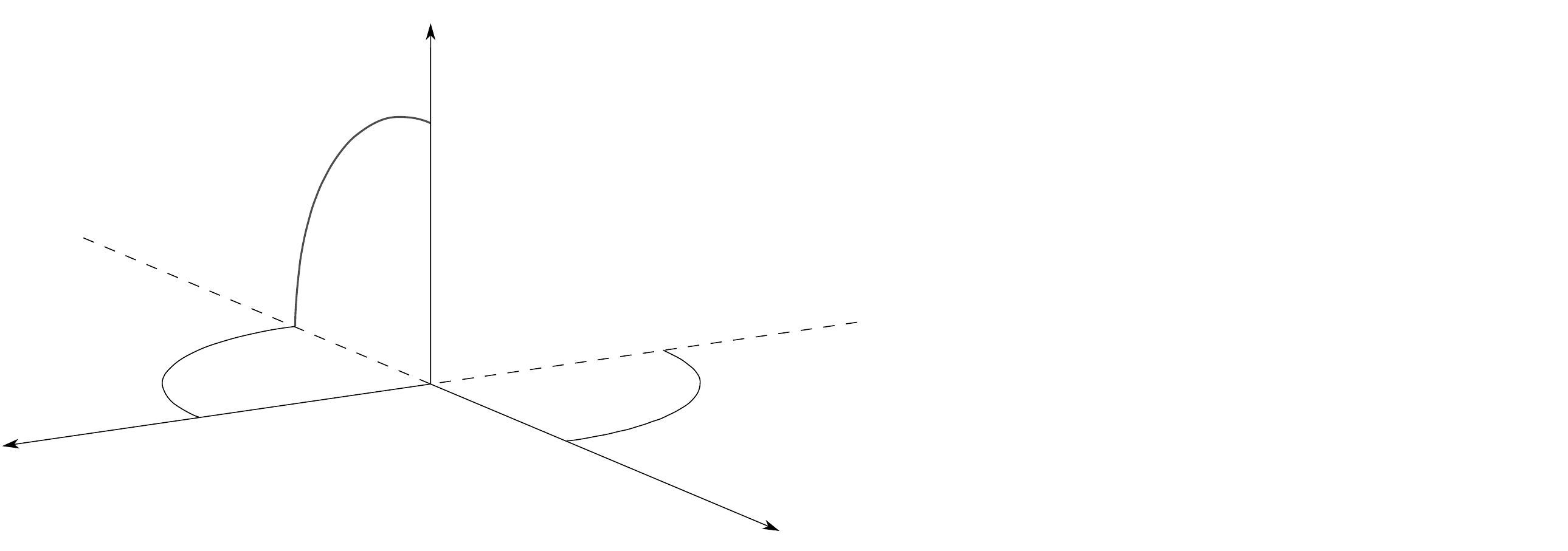

	\caption{Hemispherical shell fixed at the top: problem sketch along with a plot of the total displacements, computed with our approach and cubic splines.} \label{fig:sphere_geometry}
\end{figure}

\begin{figure}[t!]
    \centering
    \captionsetup[subfloat]{labelfont=scriptsize,textfont=scriptsize}
    \subfloat[$n^{11}$]{{
        \def\svgwidth{0.33\textwidth}
\begingroup%
  \makeatletter%
  \providecommand\color[2][]{%
    \errmessage{(Inkscape) Color is used for the text in Inkscape, but the package 'color.sty' is not loaded}%
    \renewcommand\color[2][]{}%
  }%
  \providecommand\transparent[1]{%
    \errmessage{(Inkscape) Transparency is used (non-zero) for the text in Inkscape, but the package 'transparent.sty' is not loaded}%
    \renewcommand\transparent[1]{}%
  }%
  \providecommand\rotatebox[2]{#2}%
  \newcommand*\fsize{\dimexpr\f@size pt\relax}%
  \newcommand*\lineheight[1]{\fontsize{\fsize}{#1\fsize}\selectfont}%
  \ifx\svgwidth\undefined%
    \setlength{\unitlength}{427.5bp}%
    \ifx\svgscale\undefined%
      \relax%
    \else%
      \setlength{\unitlength}{\unitlength * \real{\svgscale}}%
    \fi%
  \else%
    \setlength{\unitlength}{\svgwidth}%
  \fi%
  \global\let\svgwidth\undefined%
  \global\let\svgscale\undefined%
  \makeatother%
  \begin{picture}(1,1.16666667)%
    \lineheight{1}%
    \setlength\tabcolsep{0pt}%
    \put(0,0){\includegraphics[width=\unitlength,page=1]{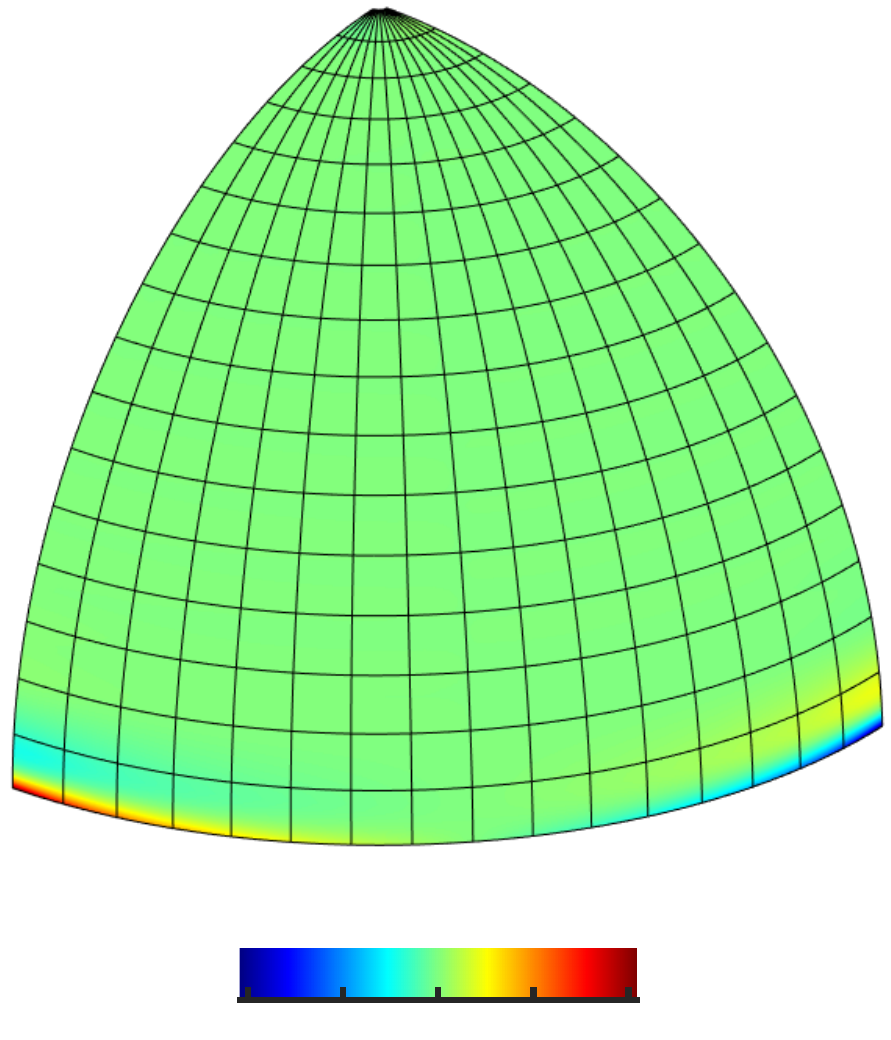}}%
    \put(0.73396098,0.08765582){\color[rgb]{0,0,0}\makebox(0,0)[lt]{\lineheight{1.25}\smash{\begin{tabular}[t]{l}$n^{11}$\end{tabular}}}}%
    \put(0.24689958,0.00017305){\color[rgb]{0,0,0}\makebox(0,0)[lt]{\lineheight{1.25}\smash{\begin{tabular}[t]{l}-90\end{tabular}}}}%
    \put(0.48297635,0.00017305){\color[rgb]{0,0,0}\makebox(0,0)[lt]{\lineheight{1.25}\smash{\begin{tabular}[t]{l}0\end{tabular}}}}%
    \put(0.57821062,0.00017305){\color[rgb]{0,0,0}\makebox(0,0)[lt]{\lineheight{1.25}\smash{\begin{tabular}[t]{l}45\end{tabular}}}}%
    \put(0.35964245,0.00017305){\color[rgb]{0,0,0}\makebox(0,0)[lt]{\lineheight{1.25}\smash{\begin{tabular}[t]{l}-45\end{tabular}}}}%
    \put(0.68698353,0.00017305){\color[rgb]{0,0,0}\makebox(0,0)[lt]{\lineheight{1.25}\smash{\begin{tabular}[t]{l}90\end{tabular}}}}%
  \end{picture}%
\endgroup%

        }}\hspace{1.2cm}
    \subfloat[$m^{11}$]{{
        \def\svgwidth{0.33\textwidth}
\begingroup%
  \makeatletter%
  \providecommand\color[2][]{%
    \errmessage{(Inkscape) Color is used for the text in Inkscape, but the package 'color.sty' is not loaded}%
    \renewcommand\color[2][]{}%
  }%
  \providecommand\transparent[1]{%
    \errmessage{(Inkscape) Transparency is used (non-zero) for the text in Inkscape, but the package 'transparent.sty' is not loaded}%
    \renewcommand\transparent[1]{}%
  }%
  \providecommand\rotatebox[2]{#2}%
  \newcommand*\fsize{\dimexpr\f@size pt\relax}%
  \newcommand*\lineheight[1]{\fontsize{\fsize}{#1\fsize}\selectfont}%
  \ifx\svgwidth\undefined%
    \setlength{\unitlength}{427.5bp}%
    \ifx\svgscale\undefined%
      \relax%
    \else%
      \setlength{\unitlength}{\unitlength * \real{\svgscale}}%
    \fi%
  \else%
    \setlength{\unitlength}{\svgwidth}%
  \fi%
  \global\let\svgwidth\undefined%
  \global\let\svgscale\undefined%
  \makeatother%
  \begin{picture}(1,1.16666667)%
    \lineheight{1}%
    \setlength\tabcolsep{0pt}%
    \put(0,0){\includegraphics[width=\unitlength,page=1]{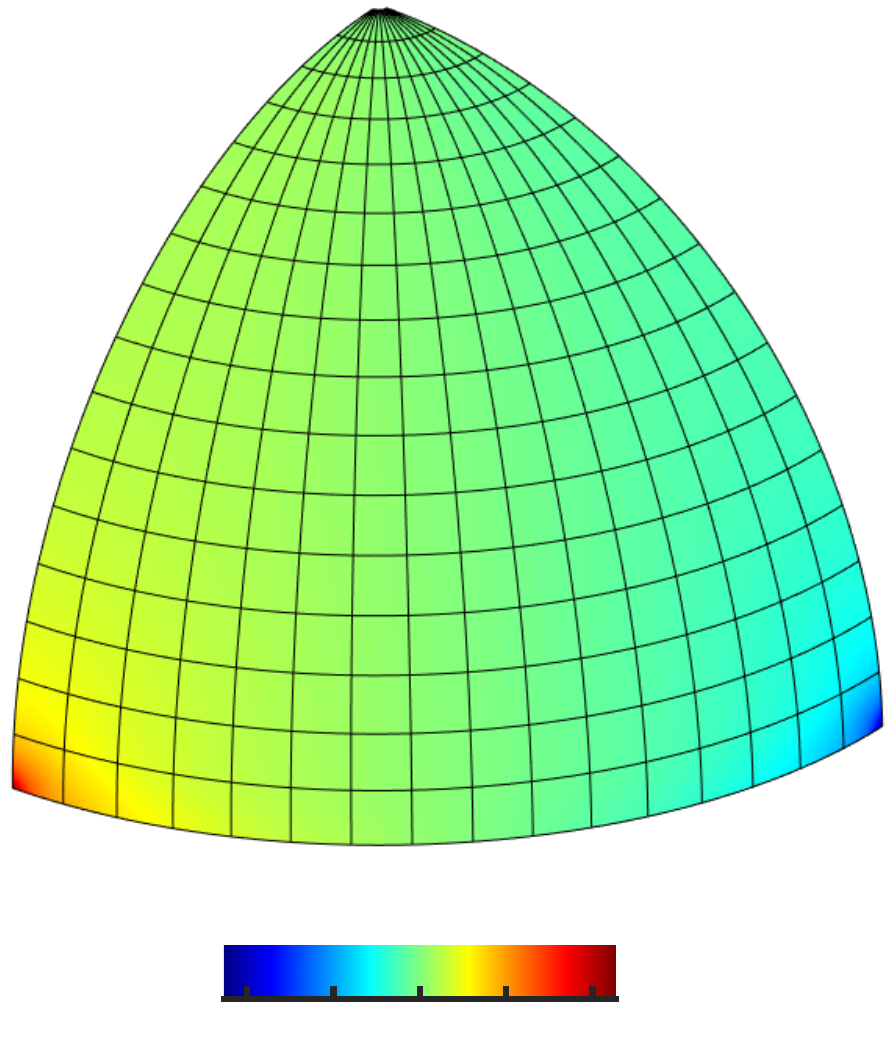}}%
    \put(0.70519355,0.09410625){\color[rgb]{0,0,0}\makebox(0,0)[lt]{\lineheight{1.25}\smash{\begin{tabular}[t]{l}$m^{11}$\end{tabular}}}}%
    \put(0.21571352,0.00025727){\color[rgb]{0,0,0}\makebox(0,0)[lt]{\lineheight{1.25}\smash{\begin{tabular}[t]{l}-2.4\end{tabular}}}}%
    \put(0.33503542,0.00030569){\color[rgb]{0,0,0}\makebox(0,0)[lt]{\lineheight{1.25}\smash{\begin{tabular}[t]{l}-1.2\end{tabular}}}}%
    \put(0.46214223,0.00017529){\color[rgb]{0,0,0}\makebox(0,0)[lt]{\lineheight{1.25}\smash{\begin{tabular}[t]{l}0\end{tabular}}}}%
    \put(0.54267362,0.00045017){\color[rgb]{0,0,0}\makebox(0,0)[lt]{\lineheight{1.25}\smash{\begin{tabular}[t]{l}1.2\end{tabular}}}}%
    \put(0.64754085,0.00024812){\color[rgb]{0,0,0}\makebox(0,0)[lt]{\lineheight{1.25}\smash{\begin{tabular}[t]{l}2.4\end{tabular}}}}%
  \end{picture}%
\endgroup%

        }}

    \caption{{Hemispherical shell}: the membrane stress resultant $n^{11}$ and the bending moment $m^{11}$ for slenderness ratio $R/t=250$, computed with our approach and cubic splines on a mesh of $16 \times 16$ elements.}\label{fig:shell_stress_slenderness_sphere}
\end{figure}

\begin{figure}[ht!]
    \centering
    \captionsetup[subfloat]{labelfont=scriptsize,textfont=scriptsize}
    \subfloat[$p=2$, $R/t=250$]{{
        \def\svgwidth{0.47\textwidth}
        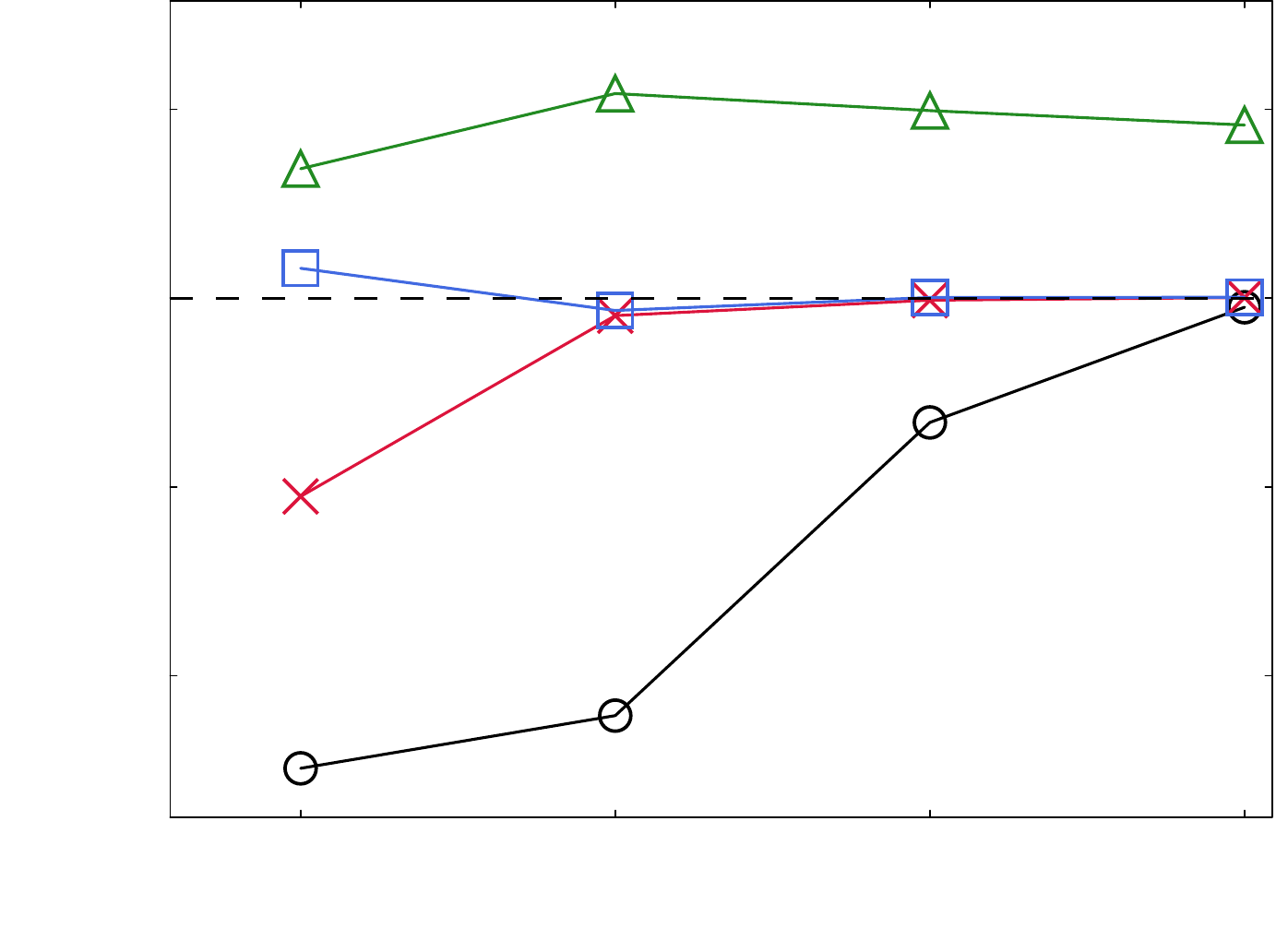
        }} \hspace{0.2cm}
    \subfloat[$p=3$, $R/t=250$]{{
        \def\svgwidth{0.47\textwidth}
        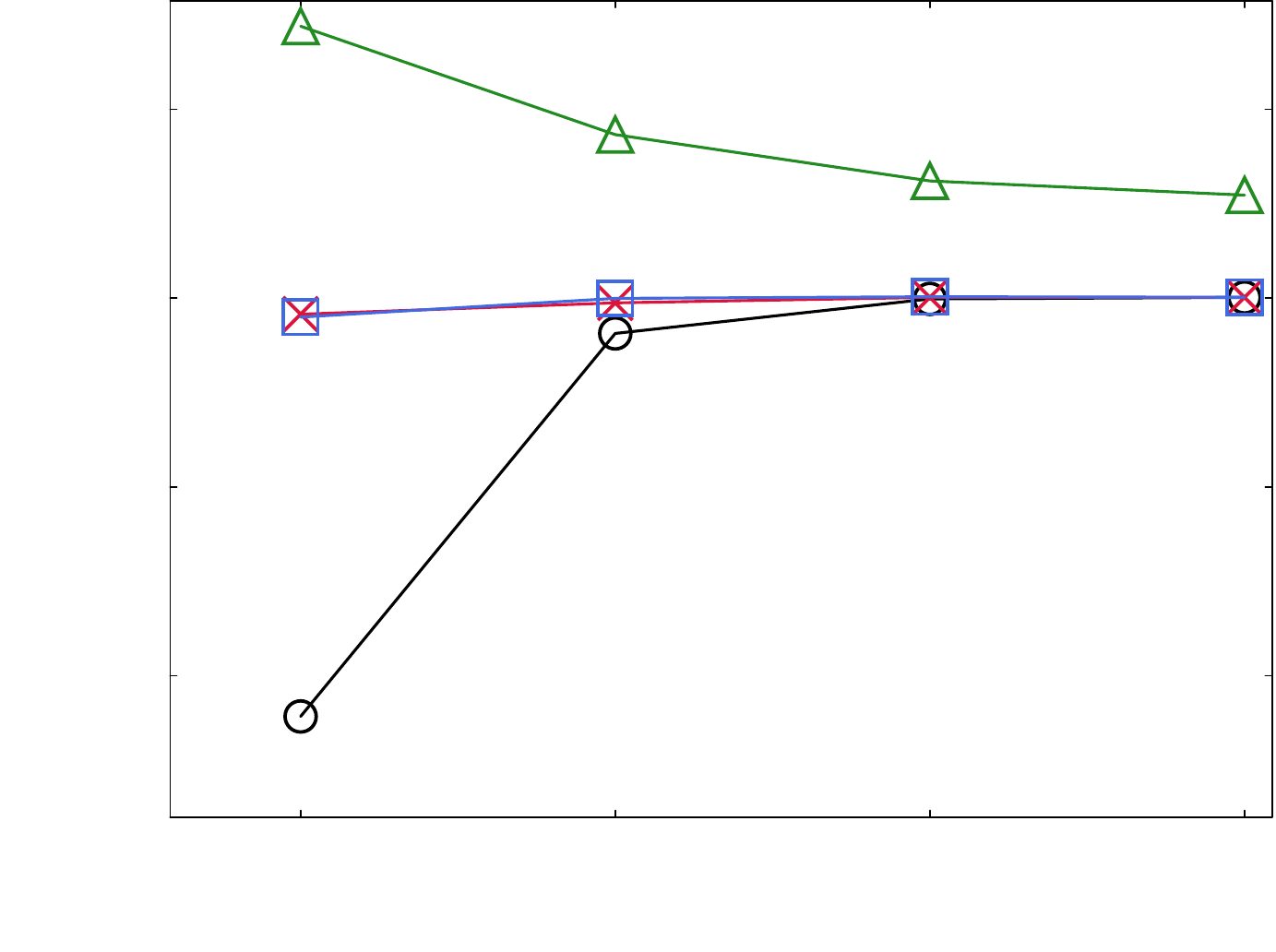
        }}

    \subfloat[$p=2$, $R/t=2,500$]{{
        \def\svgwidth{0.47\textwidth}
        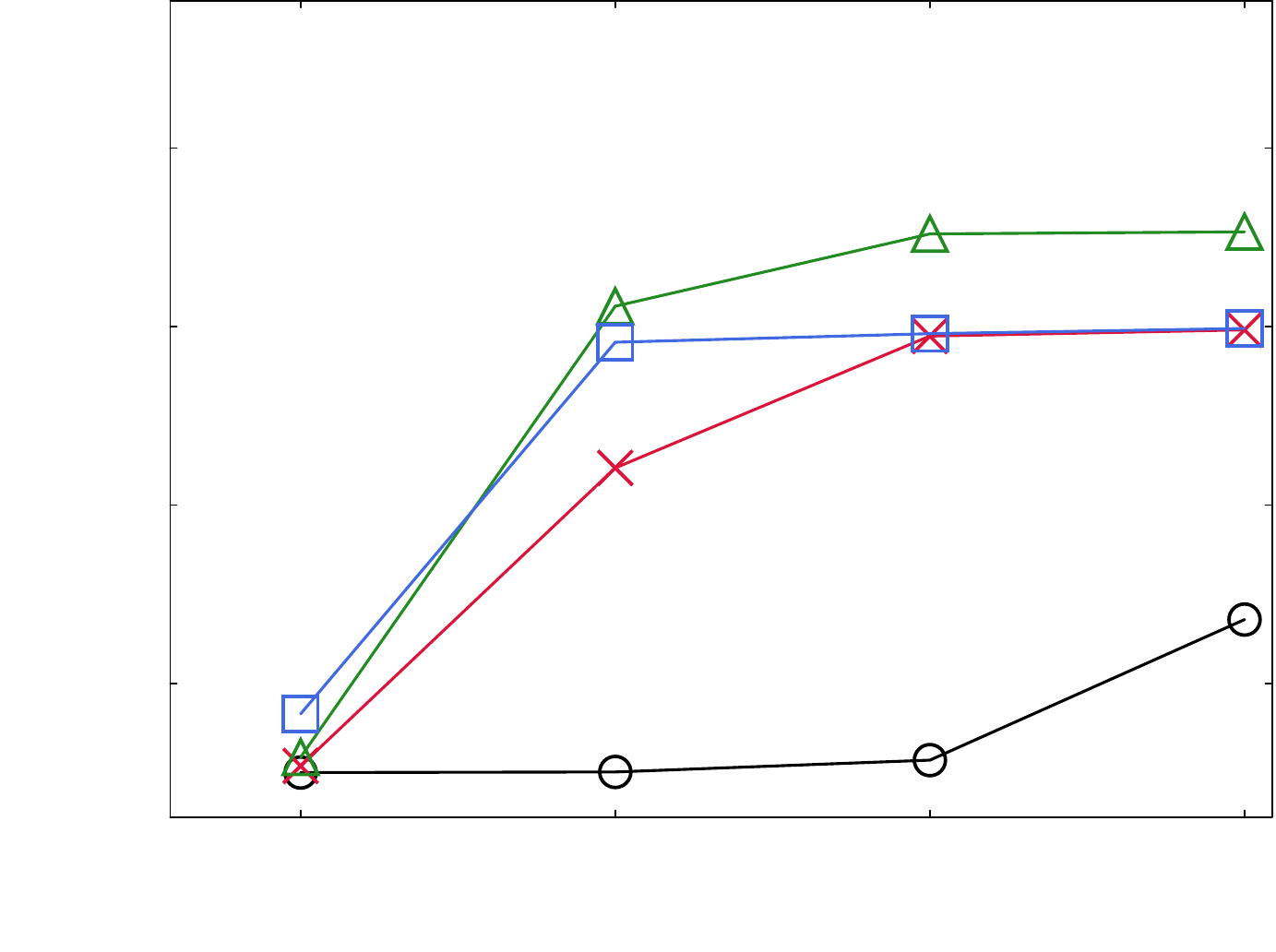
        }} \hspace{0.2cm}
    \subfloat[$p=3$, $R/t=2,500$]{{
        \def\svgwidth{0.47\textwidth}
        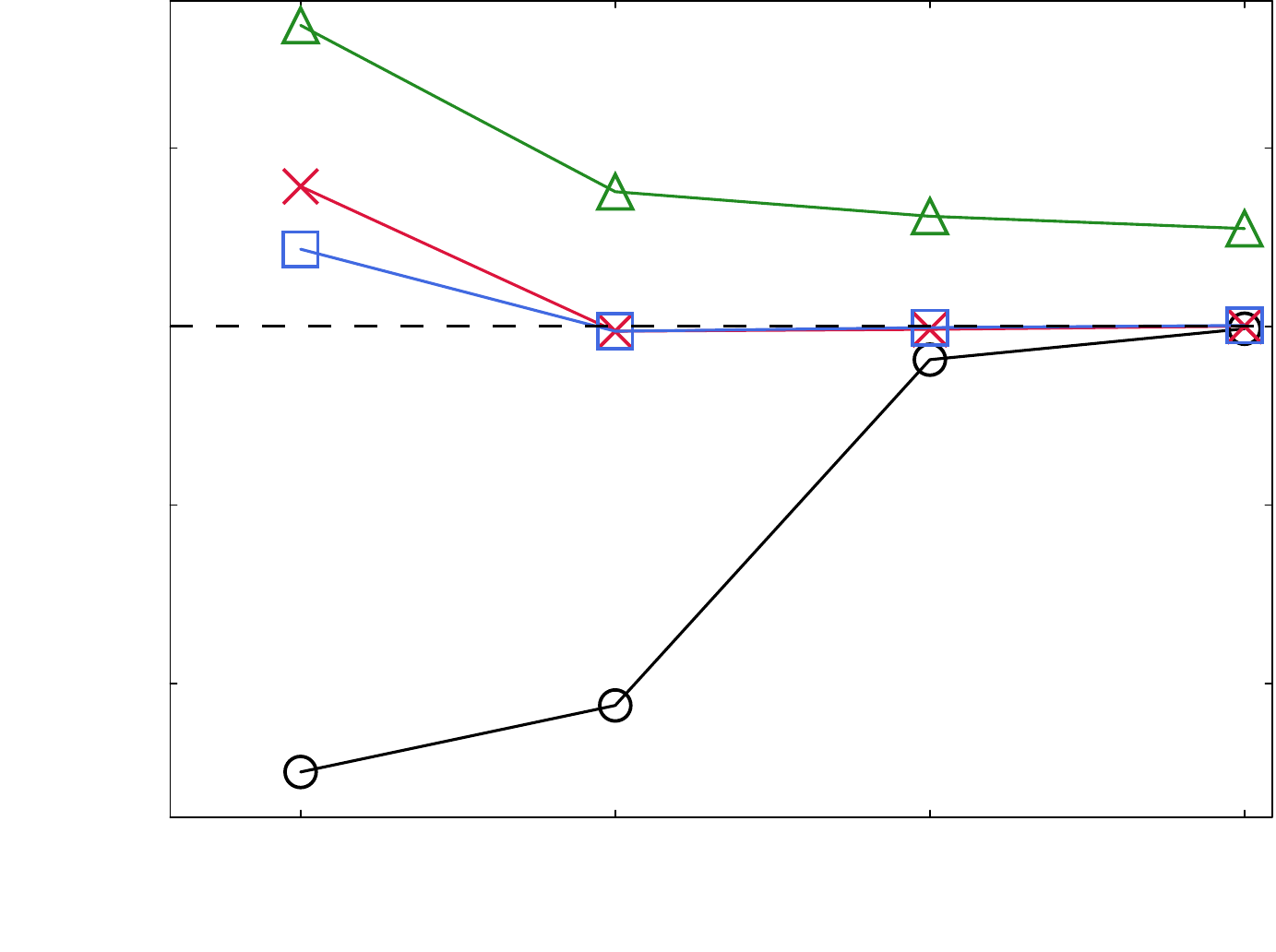
        }}

    \vspace{0.2cm}
    {\begin{tikzpicture}
    \filldraw[black,line width=1pt, solid] (0.0,0) -- (0.6,0);
    \filldraw[black,line width=1pt] (0.3,0) [fill=none] circle (3pt);
    \filldraw[black,line width=1pt] (0.7,0) node[right]{\scriptsize Displacement-based formulation};
    \filldraw[red1,line width=1pt, solid] (7.5,0) -- (8.1,0);
    \filldraw[red1,line width=1pt] (7.5,0) node[right]{\scriptsize $\boldsymbol{\bigtimes}$};
    \filldraw[red1,line width=1pt] (8.2,0) node[right]{\scriptsize Mixed formulation, consistent strain projection};
\end{tikzpicture}

\begin{tikzpicture}
    \filldraw[green1,line width=1pt, solid] (0.0,0) -- (0.6,0);
    \filldraw[green1,line width=1pt] (0.0,0) node[right]{\scriptsize $\boldsymbol{\Delta}$};
    \filldraw[green1,line width=1pt] (0.7,0) node[right]{\scriptsize Mixed formulation, lumped strain projection, B-splines};
    \filldraw[blue1,line width=1pt, solid] (8.5,0.05) -- (9.1,0.05);
    \filldraw[blue1,line width=1pt] (8.7,-0.08) [fill=none] rectangle ++(0.25,0.25);
    \filldraw[blue1,line width=1pt] (9.2,0) node[right]{\scriptsize Mixed formulation, lumped strain projection, approximate duals};
\end{tikzpicture}}    
    \caption{{Hemispherical shell}: convergence of the normalized radial displacement under the load, computed with different formulations for two different slenderness ratios.}\label{fig:shell_tip_sphere}
\end{figure}

\begin{figure}[ht!]
    \centering
    \captionsetup[subfloat]{labelfont=scriptsize,textfont=scriptsize}
    \subfloat[$p=2$]{{\def\svgwidth{0.47\textwidth}
    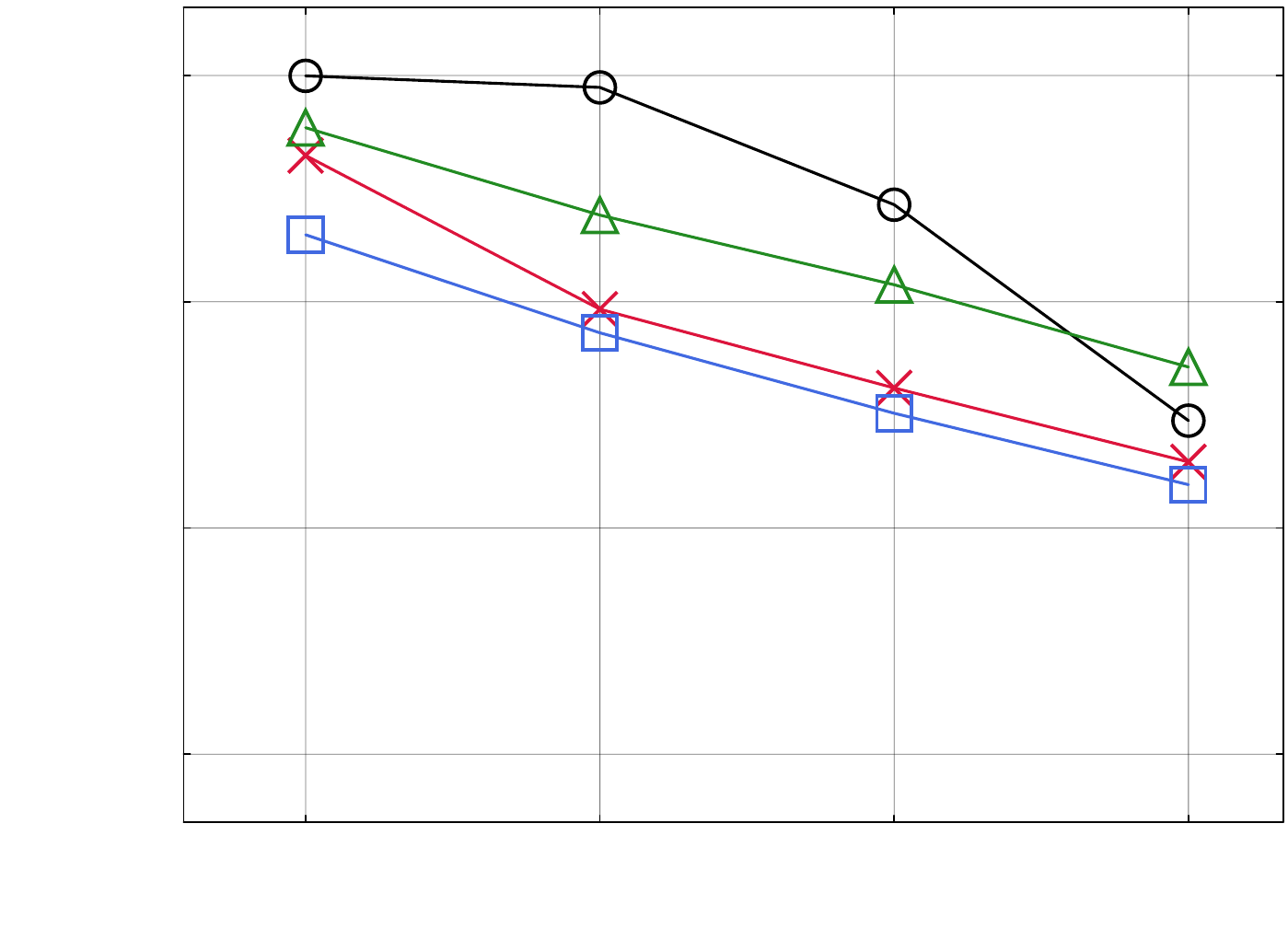 }} \hspace{0.2cm}
    \subfloat[$p=3$]{{\def\svgwidth{0.47\textwidth}
    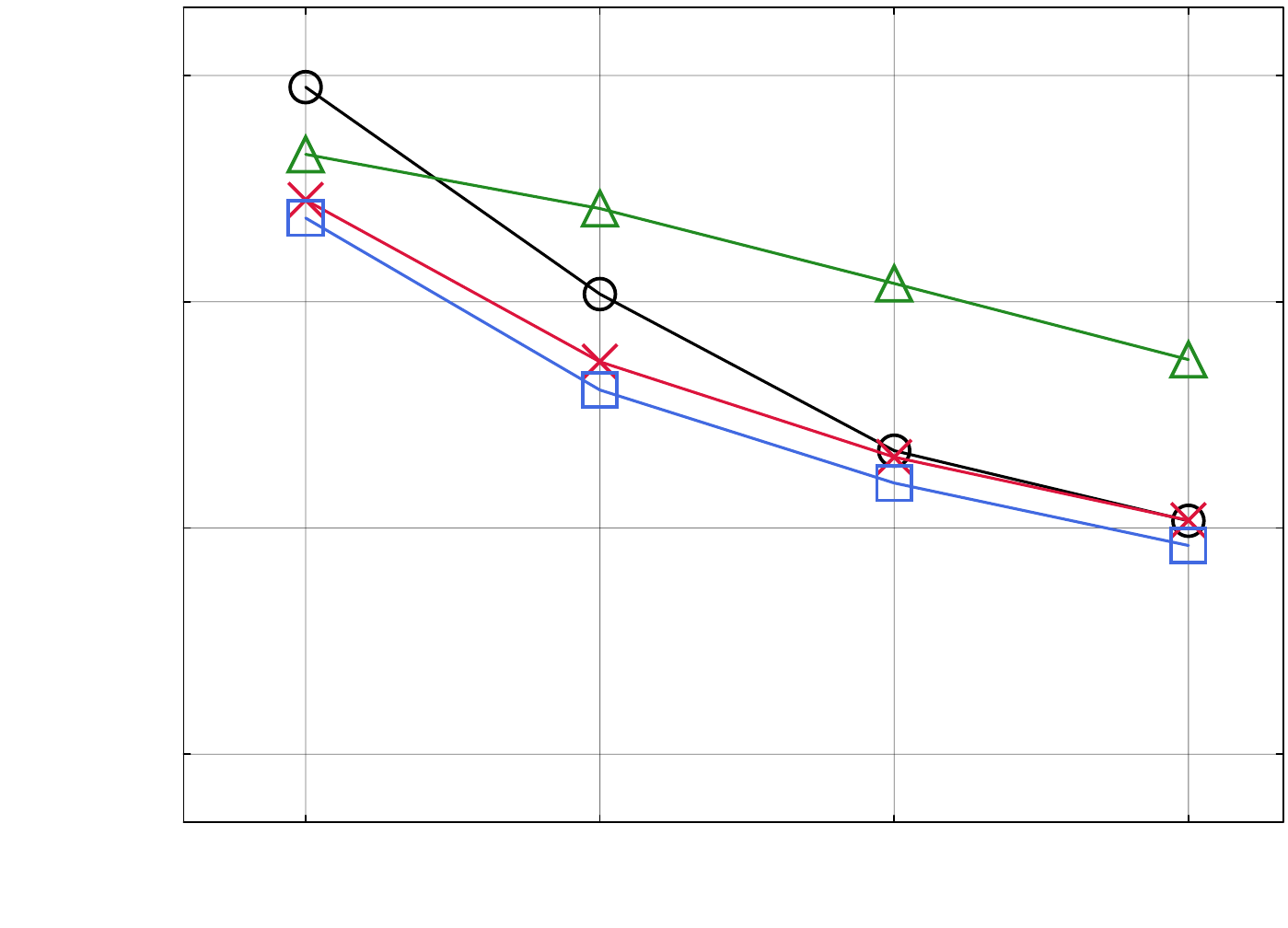 }}

    \subfloat[$p=4$]{{\def\svgwidth{0.47\textwidth}
    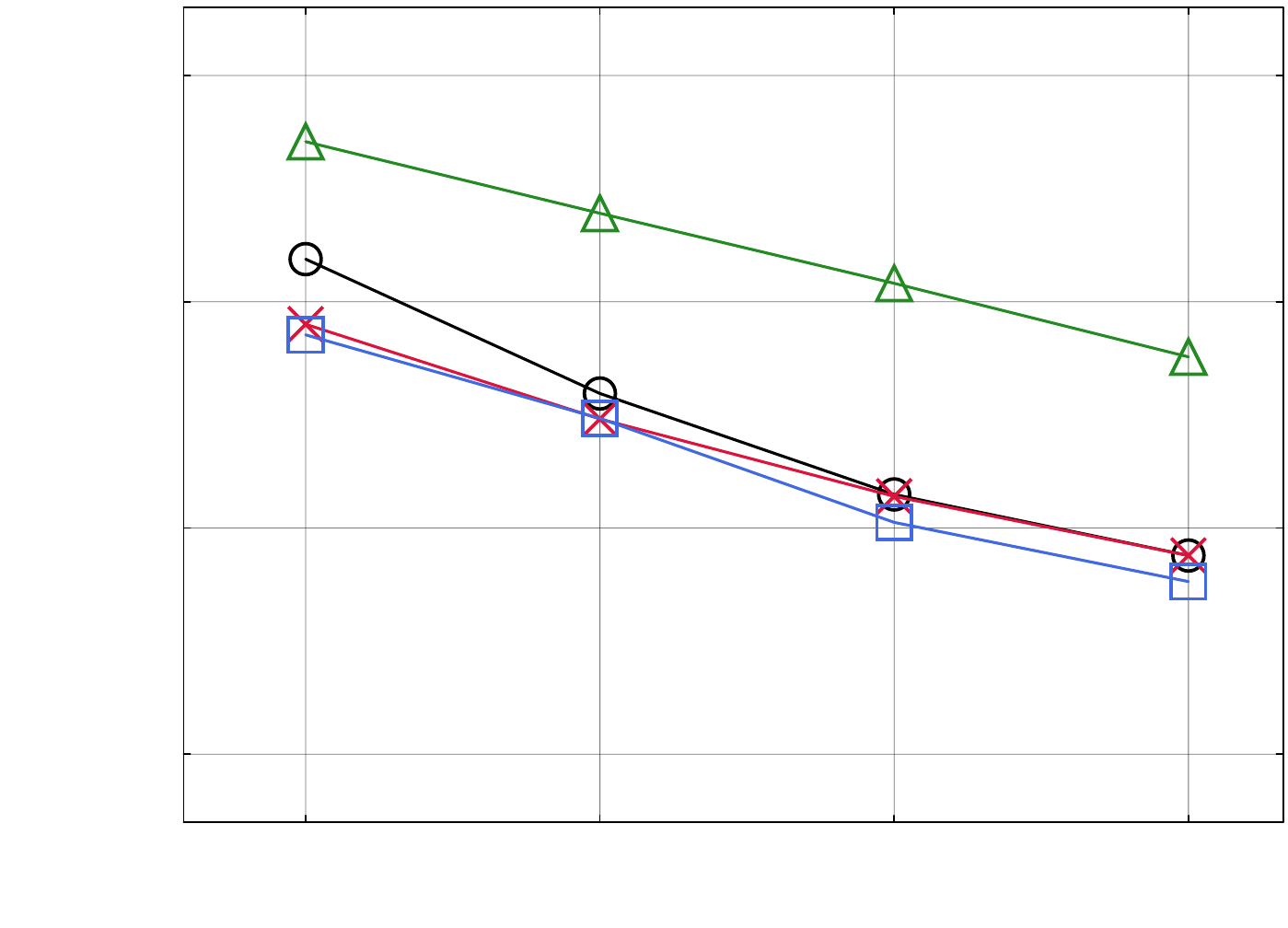 }} \hspace{0.2cm}
    \subfloat[$p=5$]{{\def\svgwidth{0.47\textwidth}
    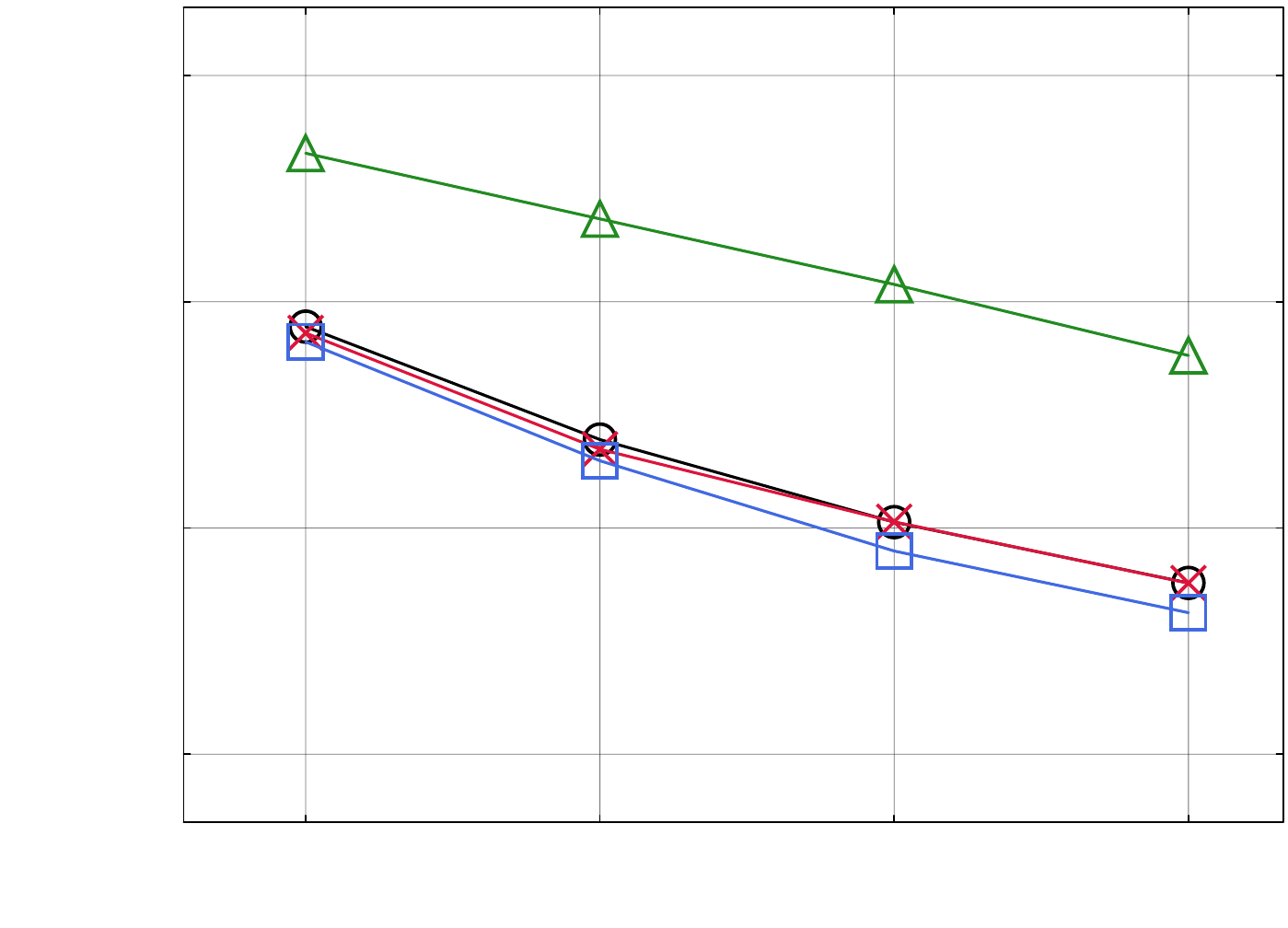 }}

    \vspace{0.2cm}
    {\begin{tikzpicture}
    \filldraw[black,line width=1pt, solid] (0.0,0) -- (0.6,0);
    \filldraw[black,line width=1pt] (0.3,0) [fill=none] circle (3pt);
    \filldraw[black,line width=1pt] (0.7,0) node[right]{\scriptsize Displacement-based formulation};
    \filldraw[red1,line width=1pt, solid] (7.5,0) -- (8.1,0);
    \filldraw[red1,line width=1pt] (7.5,0) node[right]{\scriptsize $\boldsymbol{\bigtimes}$};
    \filldraw[red1,line width=1pt] (8.2,0) node[right]{\scriptsize Mixed formulation, consistent strain projection};
\end{tikzpicture}

\begin{tikzpicture}
    \filldraw[green1,line width=1pt, solid] (0.0,0) -- (0.6,0);
    \filldraw[green1,line width=1pt] (0.0,0) node[right]{\scriptsize $\boldsymbol{\Delta}$};
    \filldraw[green1,line width=1pt] (0.7,0) node[right]{\scriptsize Mixed formulation, lumped strain projection, B-splines};
    \filldraw[blue1,line width=1pt, solid] (8.5,0.05) -- (9.1,0.05);
    \filldraw[blue1,line width=1pt] (8.7,-0.08) [fill=none] rectangle ++(0.25,0.25);
    \filldraw[blue1,line width=1pt] (9.2,0) node[right]{\scriptsize Mixed formulation, lumped strain projection, approximate duals};
\end{tikzpicture}}
    \caption{{Hemispherical shell}: convergence of the relative error in the $L^2$ norm of the bending moment $m^{12}$, computed with different formulations.}\label{fig:shell_h2converge_sphere}
\end{figure}

\subsection{The classical shell obstacle course}\label{sec:shell_obstacle_course}

We now move to the Kirchhoff-Love shell problem and demonstrate the performance of our mixed formulation for the three elastostatic examples of the classical shell obstacle course \cite{belytschko_locking_1985}. We compare our formulation based on approximate dual functions for the discretization of the independent strain variations and row-sum lumping of the resulting strain projection matrix to the standard displacement-based Galerkin formulation \cite{Kiendl_shell_2009}, the (computationally prohibitively expensive) Galerkin mixed formulation with consistent strain projection, which serves as the reference in terms of accuracy, and the (notoriously inaccurate) Galerkin mixed formulation with row-sum lumping of the strain projection matrix, which offers equivalent advantages in the strain condensation.

\subsubsection{Scordelis-Lo roof}\label{sec:lin_roof}

The first example consists of the Scordelis-Lo roof subjected to its self-weight of 90.0 per unit area, illustrated in Figure \ref{fig:roof_geometry}.
For discretization, we exploit the symmetry and model only a quarter of the structure marked in blue. We note that all our statements on element numbers and mesh sizes refer to the quarter problem, although we plot solution fields for the complete structure for better visibility. Figure~\ref{fig:roof_geometry} also illustrates the total displacement field, computed with our formulation and cubic splines on a $16 \times 16$ mesh. In Figures \ref{fig:shell_stress_slenderness_roof1} and \ref{fig:shell_stress_slenderness_roof2}, we plot one membrane stress resultant and one bending moment for slenderness ratio $R/t=100$, obtained with quadratic and cubic basis functions, respectively. We observe that the stress resultant fields are free of oscillations that typically indicate locking-prone behavior, thereby confirming that our approach is indeed locking-free. Readers interested in plots for the slenderness ratio $R/t=1,000$ are referred to Figure \ref{Scord1} and \ref{Scord2} in Appendix \ref{sec:addshells}.

In Figure \ref{fig:shell_tip_roof}, we plot the convergence of the normalized displacement at the midpoint of the free edge, for which a reference solution exists, see e.g.\ \cite{belytschko_locking_1985,Kiendl_shell_2009}, for slenderness ratios $R/t = 100$ and $1,000$. 
We observe that the standard Galerkin formulation (black) leads to inaccurate results on coarse meshes, likely a result of membrane locking. This effect is reduced when we increase the polynomial degree of the basis functions from quadratics to cubics. We also see that the Galerkin mixed formulation with row-sum lumping of the strain projection matrix (green) leads to significant errors that do not vanish with increasing polynomial degrees. Consistent with the observations for the beam example, we see that our strain projection approach (blue) leads to results that are very close to the ones of the consistent Galerkin mixed formulation (red).

In the next step, we carry out uniform $h$-refinement using the sequence of 3, 6, 12, and 24 \Bezier elements per edge and smooth splines of degree $p=2$ through $p=5$. 
In Figure \ref{fig:shell_h2converge_roof}, we plot the convergence of the relative error in the $L^2$ norm of the bending moment $m^{12}$. 
The error is computed with respect to an overkill solution, computed with the Galerkin mixed formulation on a mesh of $128\times128$ \Bezier elements and smooth B-splines of degree $p=7$. 
We observe that the standard Galerkin formulation (black) shows a slightly higher error level on coarse meshes, but converges to the same error level as the consistent Galerkin mixed formulation (red). 
Increasing the polynomial degree of the basis functions reduces this error difference also for coarser meshes.  
When we compare the results obtained with the consistent mixed Galerkin formulation (red) and the mixed Galerkin formulation with lumped strain projection matrix (green), we see that the latter leads to significantly larger errors and the same linear convergence irrespective of the polynomial degree. 
This is consistent with the observations above and illustrates the dominating impact of row-sum lumping on the accuracy and convergence of the corresponding discretization scheme. 
When we focus on our formulation (blue) and the consistent Galerkin mixed formulation (red), 
we observe that the former shows a slightly higher error level than the latter. Nevertheless, the error associated with our approach remains very close (within the same order of magnitude) and converges with the optimal rate of convergence, $\mathcal{O}(p-1)$ \cite{engel2002continuous,Tagliabue2014}, when the mesh is refined.

\begin{remark}
We also carried out test simulations with geometrically nonlinear versions of the Scordelis-Lo roof. Readers interested in the results are referred to Appendix \ref{sec:addshells}.
\end{remark}

\subsubsection{Pinched cylinder}\label{sec:lin_cyl}

The second example consists of a pinched cylinder subjected to two opposite unit loads in the middle, 
illustrated in Figure \ref{fig:cylinder_geometry}. 
We again exploit the symmetry and model only one-eighth of the structure marked in blue, to which all our statements on element numbers and mesh sizes refer, but we plot solution fields for a half cylinder for better visibility. Figure~\ref{fig:cylinder_geometry} also illustrates the total displacement field, computed with our formulation and cubic splines on a $16 \times 16$ mesh. In Figure \ref{fig:shell_stress_slenderness_cylinder}, we plot one membrane stress resultant and one bending moment for slenderness ratio $R/t=100$. We observe that the stress resultant fields are free of oscillations that typically indicate locking-prone behavior, thereby confirming that our approach is indeed locking-free. Readers interested in plots for the slenderness ratio $R/t=1,000$ are referred to Figure \ref{Pinch1} in Appendix \ref{sec:addshells}.

In Figure \ref{fig:shell_tip_cylinder}, we plot the convergence of the normalized displacement at point A under the pinching load, for which reference solutions exist, see e.g. \cite{belytschko_locking_1985,Kiendl_shell_2009}, for slenderness ratios $R/t = 100$ and $1,000$. We observe the same behavior as before. The standard Galerkin formulation (black) leads to largely inaccurate results that are too stiff, likely amplified by membrane locking, in particular for coarse meshes. The results of our formulation (blue) are considerably better and very close to the results of the consistent Galerkin mixed formulation (red). Row-sum lumping of the strain projection matrix in the Galerkin mixed formulation (green) leads to a persistent error and prevents convergence. We report that the errors in the $L^2$ norm of the membrane stress resultants and the bending moments are dominated by the stress singularity under the loading point, so that its convergence is linear for all methods. Therefore, we refrain from including the corresponding plots here.

\subsubsection{Hemispherical shell}\label{sec:lin_sphere}

The third example consists of a hemispherical shell subjected to two opposite loads in the $xy$-plane, illustrated in Figure \ref{fig:sphere_geometry}. We also exploit the symmetry and model only a quarter of the structure marked in blue. 
Figure~\ref{fig:sphere_geometry} also illustrates the total displacement field, computed with our formulation and cubic splines on a $16 \times 16$ mesh. In Figure \ref{fig:shell_stress_slenderness_sphere}, we plot one membrane stress resultant and one bending moment for slenderness ratio $R/t=250$. We observe that the stress resultant fields are free of oscillations that typically indicate locking-prone behavior, thereby confirming that our approach is indeed locking-free.

In Figure \ref{fig:shell_tip_sphere}, we plot the convergence of the normalized radial displacement at the pinched point under the load, for which reference solutions exist, see e.g.\ \cite{belytschko_locking_1985,Kiendl_shell_2009}, for slenderness ratios $R/t = 250$ and $2,500$. Similarly to the Scordelis-Lo roof and the pinched cylinder, we observe that the standard Galerkin formulation (black) exhibits results on coarse meshes that are largely too stiff, likely due to membrane locking. 
This effect is reduced when we increase the polynomial degree of the basis functions from quadratics to cubics. We also see that the Galerkin mixed formulation with row-sum lumping of the strain projection matrix (green) leads to significant errors that do not vanish with increasing polynomial degrees. Consistent with the observations for the beam example, we see that our strain projection approach (blue) effectively eliminates the locking effect on coarse meshes and leads to results that are very close to and for the most part practically indistinguishable from the ones of the consistent Galerkin mixed formulation (red).

In the next step, we carry out uniform $h$-refinement at fixed polynomial degrees $p=2$ through $p=5$. 
In Figure \ref{fig:shell_h2converge_sphere}, we plot the convergence of the relative error in the $L^2$ norm of the bending moment $m^{12}$ with respect to an overkill solution, computed again with the Galerkin mixed formulation on a mesh of $128\times128$ \Bezier elements and splines of degree $p=7$. 

We observe that the standard Galerkin formulation (black) leads to significant errors on coarse meshes. Particularly in the case of quadratic spline functions, we see a clear pre-asymptotic plateau similar to the one observed earlier for the curved Euler-Bernoulli beam, a typical sign of membrane locking. Increasing the polynomial degree of the basis functions removes this plateau, also for coarser meshes. We observe that our formulation (blue) effectively removes the locking effect, also on coarse meshes with moderate polynomial degree, and its results are practically indistinguishable from the results of the consistent Galerkin mixed formulation (red).

\subsection{Geometrically nonlinear benchmarks}

\begin{figure}[ht!]
	\centering
    \def\svgwidth{1\textwidth}
    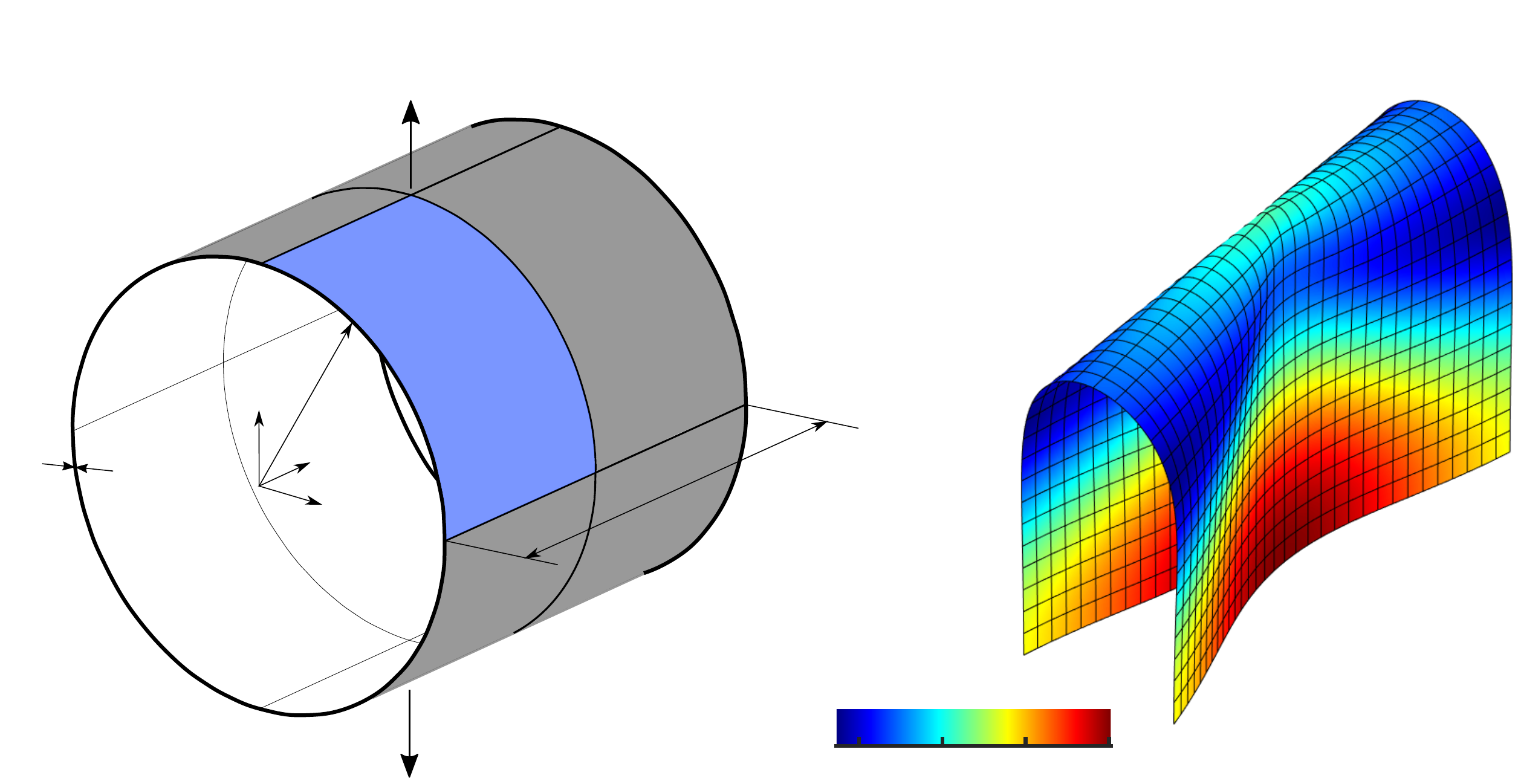

	\caption{Pullout of a cylindrical shell: problem sketch along with a plot of the total displacements, computed with our approach (octant geometry, mesh of $16 \times 16$ cubic elements).} \label{fig:cyl1}
\end{figure}

\begin{figure}[t!]
    \centering
    \captionsetup[subfloat]{labelfont=scriptsize,textfont=scriptsize}
    \subfloat[$n^{12}$]{{
        \def\svgwidth{0.4\textwidth}
        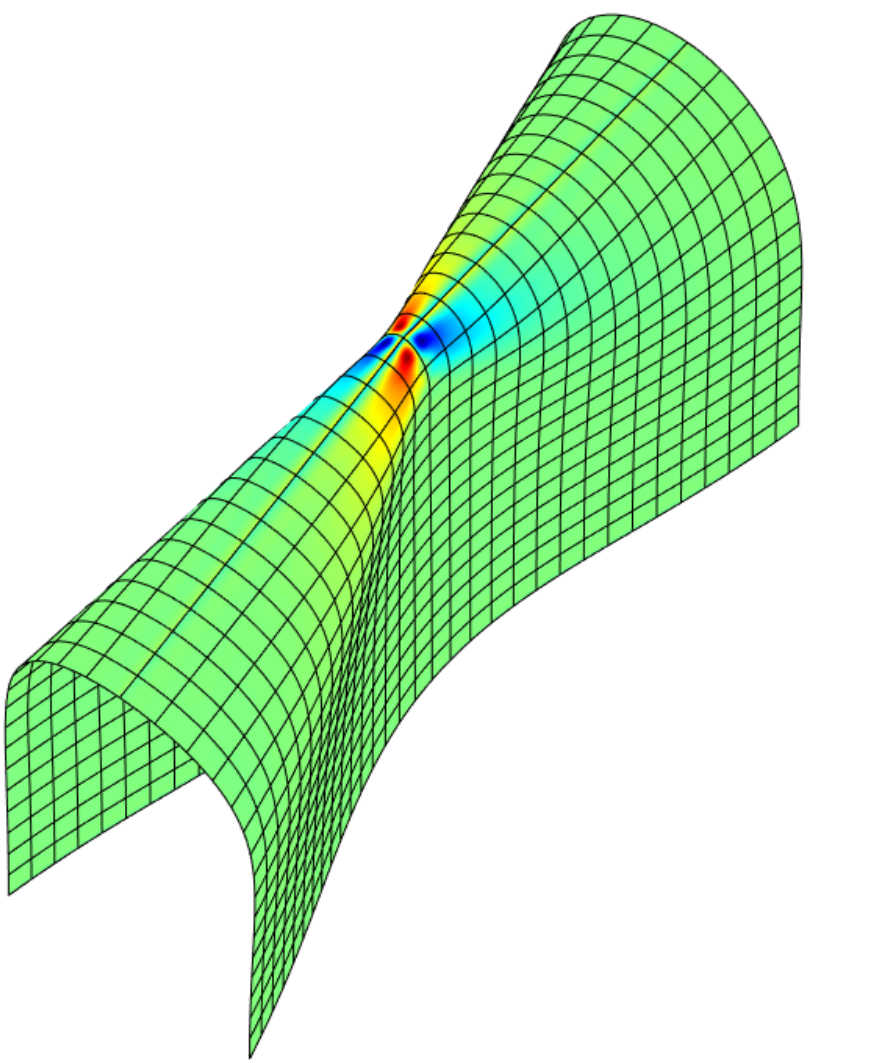
        }}\hspace{1.5cm}
    \subfloat[$m^{11}$]{{
        \def\svgwidth{0.4\textwidth}
\begingroup%
  \makeatletter%
  \providecommand\color[2][]{%
    \errmessage{(Inkscape) Color is used for the text in Inkscape, but the package 'color.sty' is not loaded}%
    \renewcommand\color[2][]{}%
  }%
  \providecommand\transparent[1]{%
    \errmessage{(Inkscape) Transparency is used (non-zero) for the text in Inkscape, but the package 'transparent.sty' is not loaded}%
    \renewcommand\transparent[1]{}%
  }%
  \providecommand\rotatebox[2]{#2}%
  \newcommand*\fsize{\dimexpr\f@size pt\relax}%
  \newcommand*\lineheight[1]{\fontsize{\fsize}{#1\fsize}\selectfont}%
  \ifx\svgwidth\undefined%
    \setlength{\unitlength}{420bp}%
    \ifx\svgscale\undefined%
      \relax%
    \else%
      \setlength{\unitlength}{\unitlength * \real{\svgscale}}%
    \fi%
  \else%
    \setlength{\unitlength}{\svgwidth}%
  \fi%
  \global\let\svgwidth\undefined%
  \global\let\svgscale\undefined%
  \makeatother%
  \begin{picture}(1,1.21428571)%
    \lineheight{1}%
    \setlength\tabcolsep{0pt}%
    \put(0,0){\includegraphics[width=\unitlength,page=1]{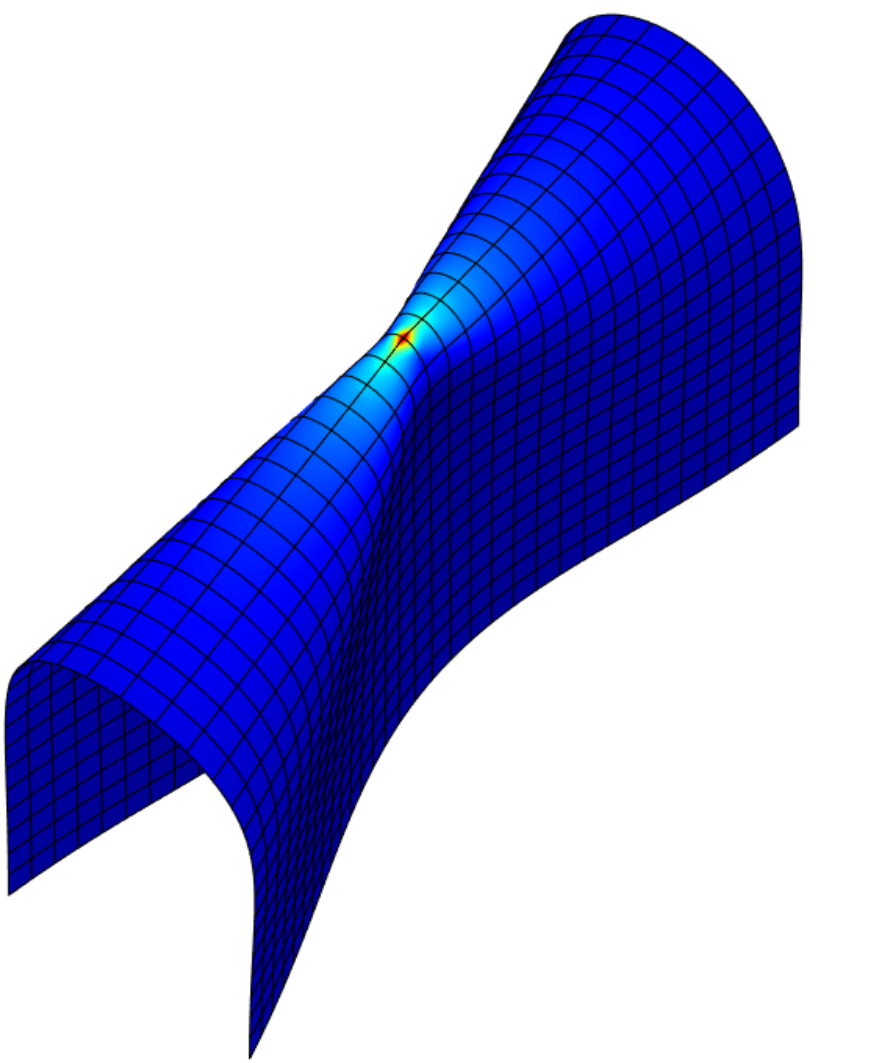}}%
    \put(0.50419677,0.04942017){\color[rgb]{0,0,0}\makebox(0,0)[lt]{\lineheight{1.25}\smash{\begin{tabular}[t]{l}0\end{tabular}}}}%
    \put(0.62716811,0.04882017){\color[rgb]{0,0,0}\makebox(0,0)[lt]{\lineheight{1.25}\smash{\begin{tabular}[t]{l}2\end{tabular}}}}%
    \put(0.75216811,0.04942017){\color[rgb]{0,0,0}\makebox(0,0)[lt]{\lineheight{1.25}\smash{\begin{tabular}[t]{l}4\end{tabular}}}}%
    \put(0.8753824,0.04942017){\color[rgb]{0,0,0}\makebox(0,0)[lt]{\lineheight{1.25}\smash{\begin{tabular}[t]{l}6\end{tabular}}}}%
    \put(0.95302591,0.15640714){\color[rgb]{0,0,0}\makebox(0,0)[lt]{\lineheight{1.25}\smash{\begin{tabular}[t]{l}$m^{11}$\end{tabular}}}}%
    \put(0.95176316,0.09290448){\color[rgb]{0,0,0}\makebox(0,0)[lt]{\lineheight{1.25}\smash{\begin{tabular}[t]{l}$\times 10^{3}$\end{tabular}}}}%
    \put(0,0){\includegraphics[width=\unitlength,page=2]{fig/nonlinear_shell/cylinder_nonlin_deformed_m11.pdf}}%
  \end{picture}%
\endgroup%

        }}

    \caption{Pullout of a cylindrical shell: the membrane stress resultant $n^{12}$ and the bending moment $m^{11}$, computed with our approach (octant geometry, mesh of $16 \times 16$ cubic elements).}\label{fig:cyl2}
\end{figure}

\begin{figure}[t!]
    \centering
    \captionsetup[subfloat]{labelfont=scriptsize,textfont=scriptsize}
    \subfloat[$p=2$]{{ \def\svgwidth{0.44\textwidth}
        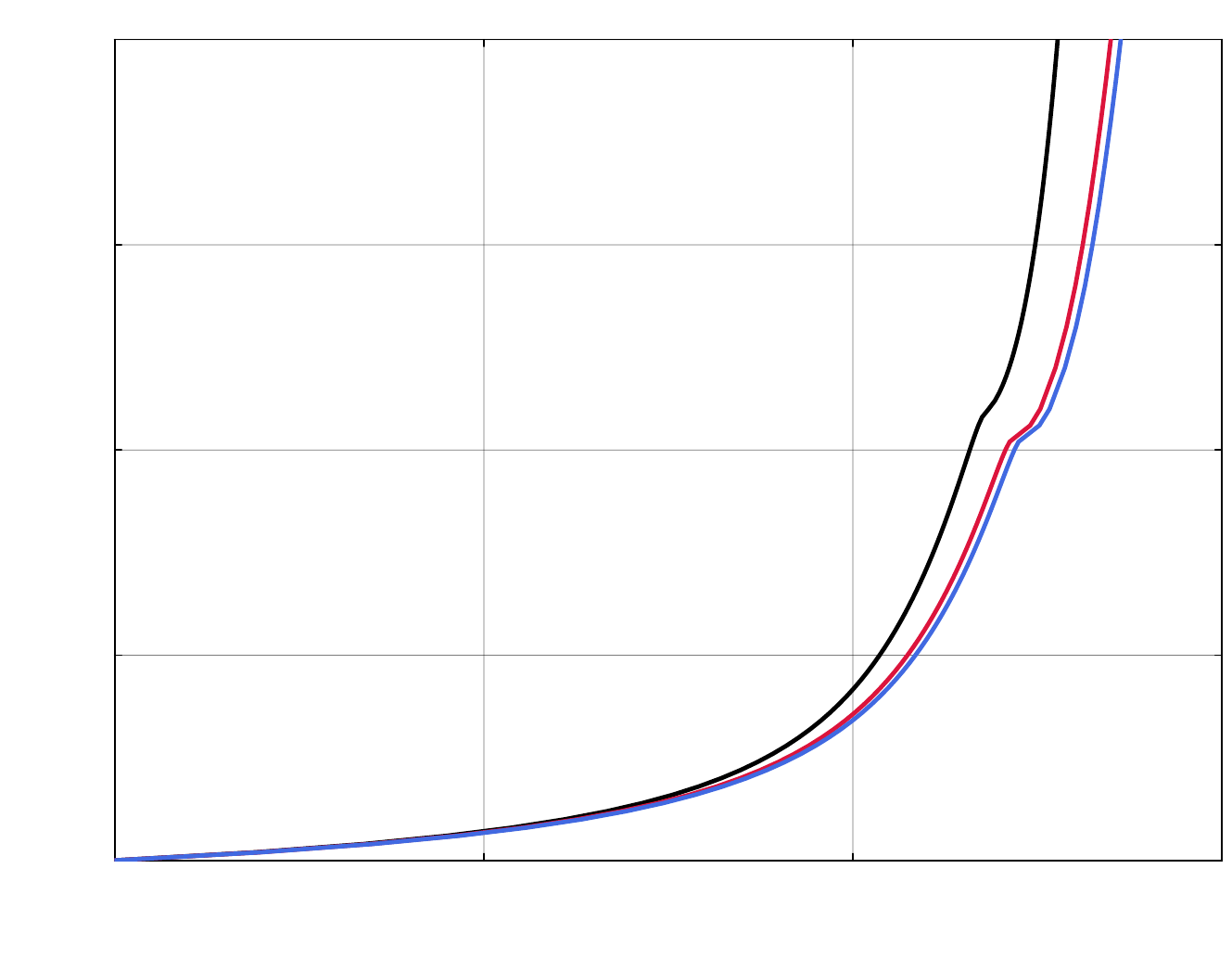 }}
    \subfloat[$p=3$]{{ \def\svgwidth{0.44\textwidth}
        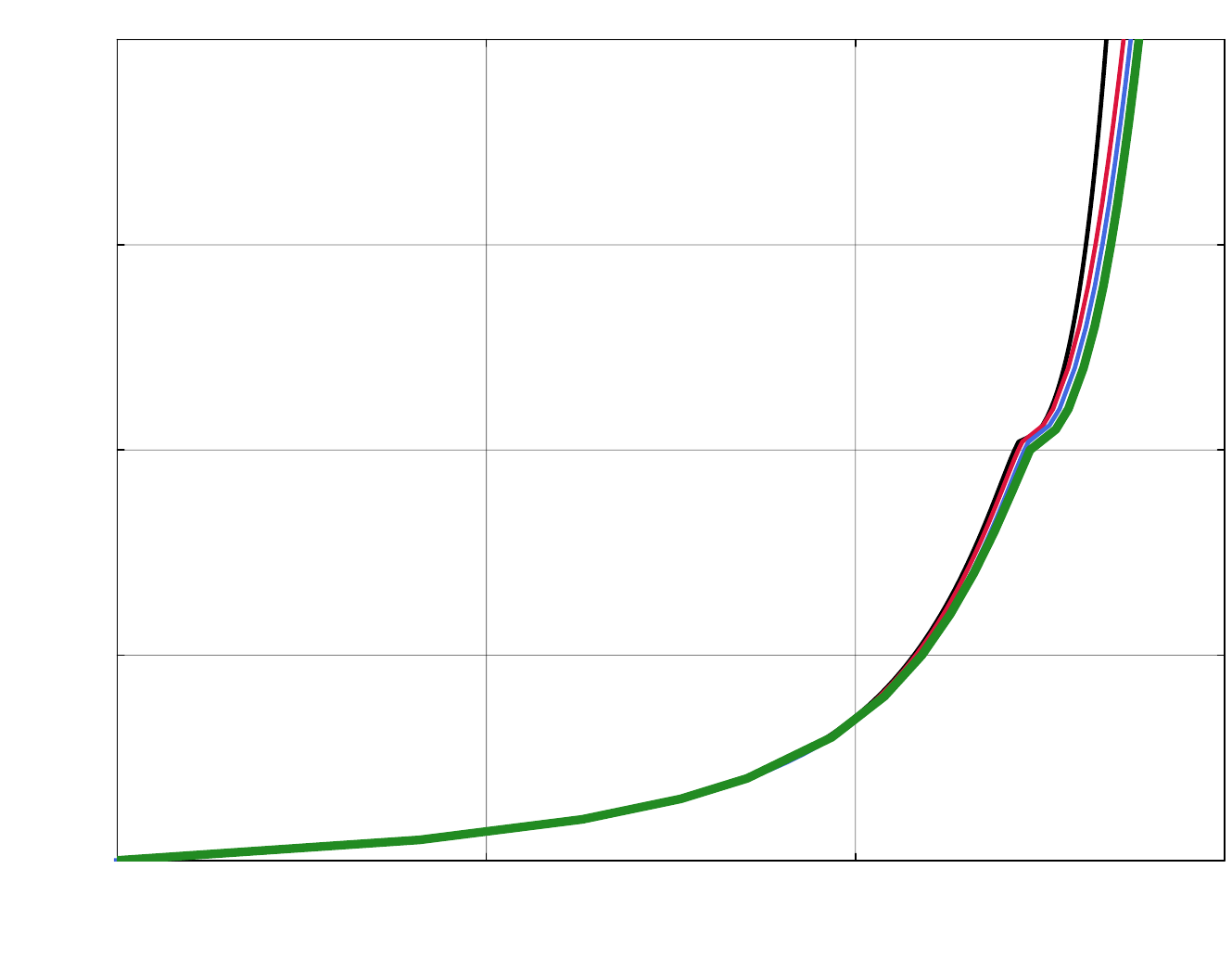 }}

    \vspace{0.2cm}
    {\begin{tikzpicture}
    \filldraw[green1,line width=1pt, solid] (0.0,0) -- (0.6,0);
    \filldraw[green1,line width=1pt] (0.7,0) node[right]{\scriptsize Reference solution\cite{Sze_shell_benchmark_2004}};
    \filldraw[black,line width=1pt, solid] (4.5,0) -- (5.1,0);
    \filldraw[black,line width=1pt] (5.2,0) node[right]{\scriptsize Displacement-based formulation};
    \filldraw[red1,line width=1pt, solid] (10.5,0) -- (11.1,0);
    \filldraw[red1,line width=1pt] (11.2,0) node[right]{\scriptsize Mixed formulation, consistent strain projection};
\end{tikzpicture}

\begin{tikzpicture}    
    \filldraw[blue1,line width=1pt, solid] (4.5,0.05) -- (5.1,0.05);
    \filldraw[blue1,line width=1pt] (5.2,0) node[right]{\scriptsize Mixed formulation, lumped strain projection, approximate duals};
\end{tikzpicture}}
    \caption{Pullout of a cylindrical shell: load-deflection curve of the vertical displacement under the load, computed with different formulations (octant geometry, mesh of $16 \times 16$ quadratic and cubic elements).}\label{fig:cyl3}
\end{figure}

\begin{figure}[t!]
    \centering
    \captionsetup[subfloat]{labelfont=scriptsize,textfont=scriptsize}
    \subfloat[$p=2$]{{ \def\svgwidth{0.44\textwidth}
        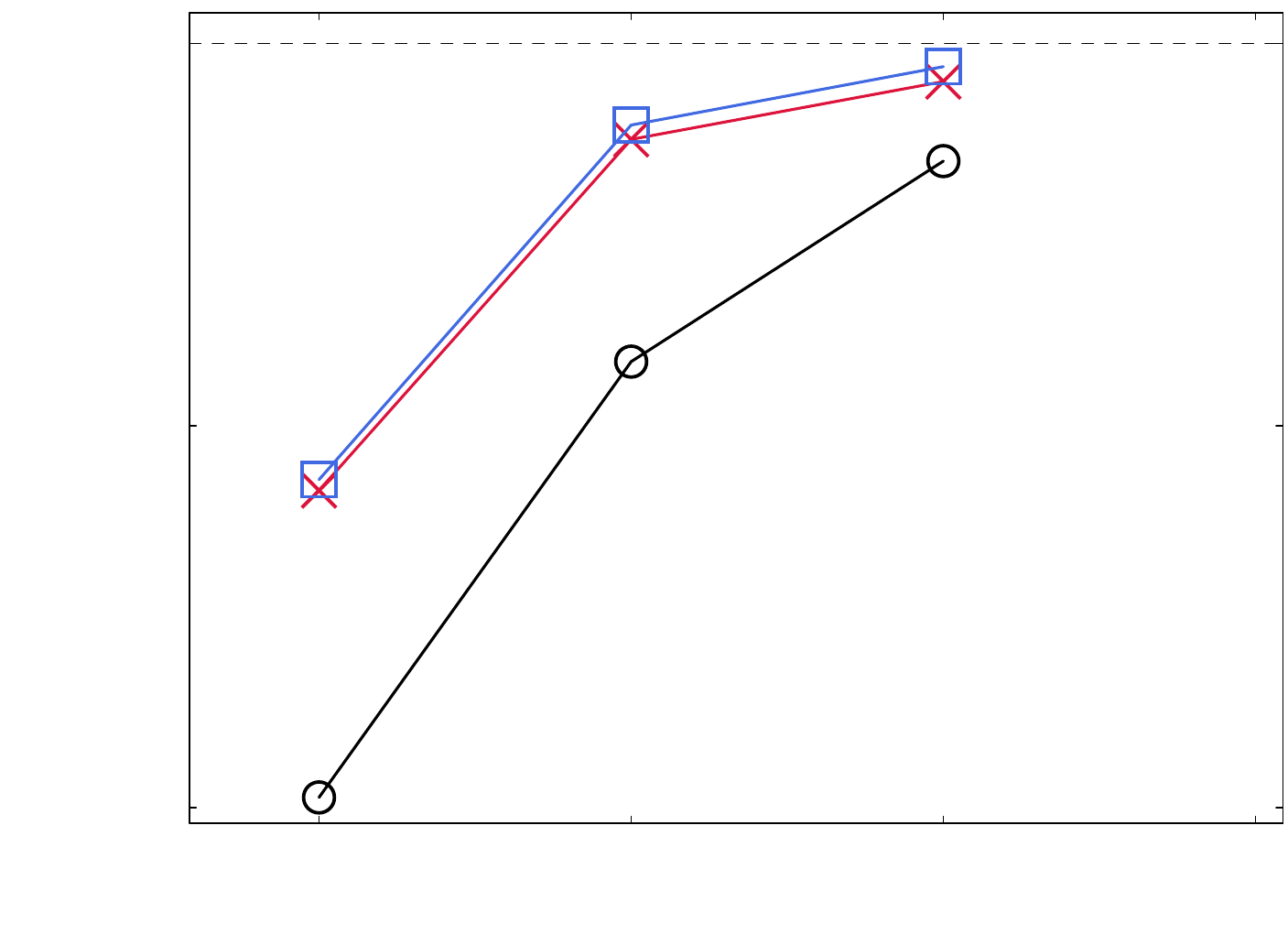 }}
    \subfloat[$p=3$]{{ \def\svgwidth{0.44\textwidth}
        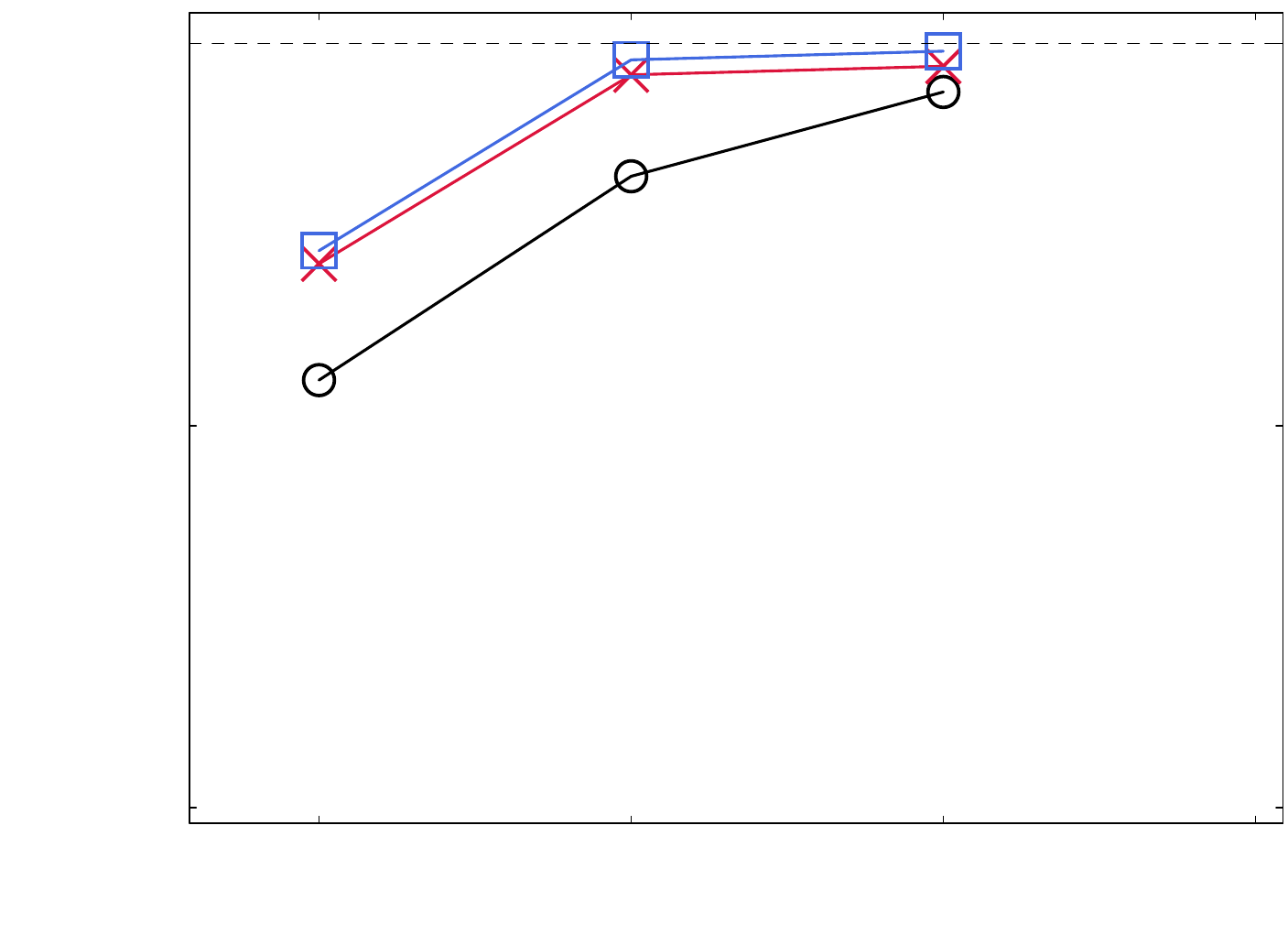 }}

    \vspace{0.2cm}
    {\begin{tikzpicture}
    \filldraw[black,line width=1pt, solid] (0.0,0) -- (0.6,0);
    \filldraw[black,line width=1pt] (0.3,0) [fill=none] circle (3pt);
    \filldraw[black,line width=1pt] (0.7,0) node[right]{\scriptsize Displacement-based formulation};
    \filldraw[red1,line width=1pt, solid] (7.5,0) -- (8.1,0);
    \filldraw[red1,line width=1pt] (7.5,0) node[right]{\scriptsize $\boldsymbol{\bigtimes}$};
    \filldraw[red1,line width=1pt] (8.2,0) node[right]{\scriptsize Mixed formulation, consistent strain projection};
\end{tikzpicture}

\begin{tikzpicture}
    \filldraw[blue1,line width=1pt, solid] (4.5,0.05) -- (5.1,0.05);
    \filldraw[blue1,line width=1pt] (4.7,-0.08) [fill=none] rectangle ++(0.25,0.25);
    \filldraw[blue1,line width=1pt] (5.2,0) node[right]{\scriptsize Mixed formulation, lumped strain projection, approximate duals};
\end{tikzpicture}}
    \caption{Pullout of a cylindrical shell: convergence of the normalized vertical displacement under the load, computed with different formulations (octant geometry, meshes of quadratic and cubic elements).}\label{fig:cyl4}
\end{figure}


\begin{figure}[t!]
    \centering
    \captionsetup[subfloat]{labelfont=scriptsize,textfont=scriptsize}
    \subfloat[$|\boldsymbol{u}|$]{{\includegraphics[width=0.4\textwidth]{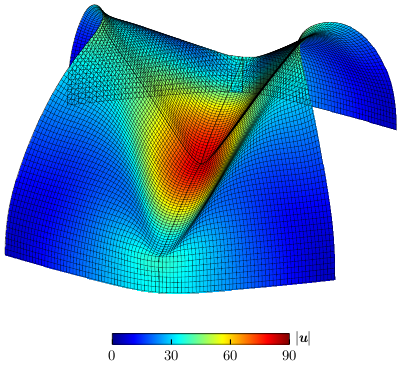}}}

    \subfloat[$n^{12}$]{{\includegraphics[width=0.4\textwidth]{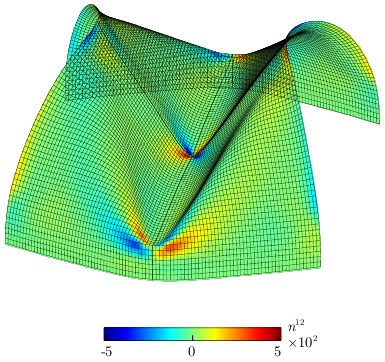}}}\hspace{1.5cm}
    \subfloat[$m^{11}$]{{\includegraphics[width=0.4\textwidth]{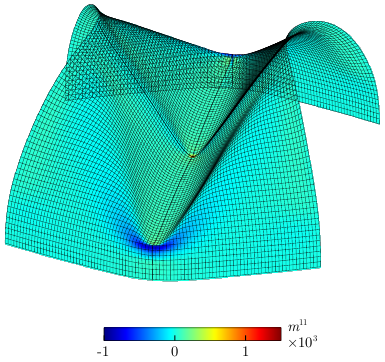}}}

    \caption{Pinched cylindrical shell: total displacements, membrane stress resultant $n^{12}$, and bending moment $m^{11}$ at the maximum load, computed with our approach (octant geometry, mesh of $48 \times 48$ cubic elements).}\label{fig:pincyl2}
\end{figure}

\begin{figure}[t!]
    \centering
    \captionsetup[subfloat]{labelfont=scriptsize,textfont=scriptsize}
    \subfloat[$p=2$]{{ \includegraphics[trim={0cm 0cm 0cm 0cm},clip,width=0.47\textwidth]{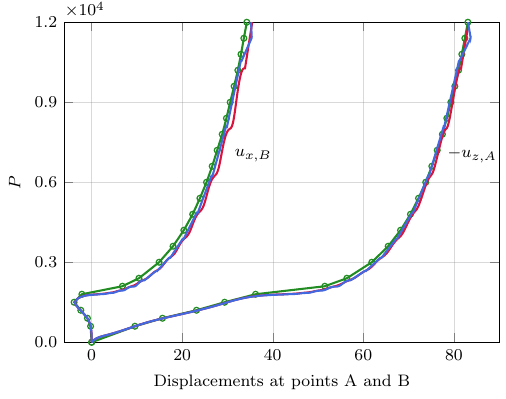} }}
    \subfloat[$p=3$]{{\includegraphics[trim={0cm 0cm 0cm 0cm},clip,width=0.47\textwidth]{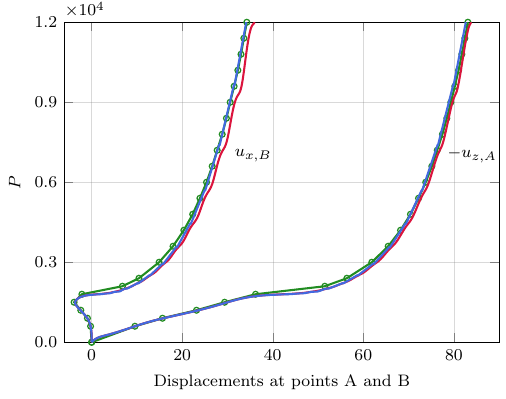} }}

    \vspace{0.2cm}
    {\begin{tikzpicture}
    \filldraw[green1,line width=1pt, solid] (0.0,0) -- (0.6,0);
    \draw[green1,line width=0.6pt,fill=white] (0.3,0) circle (1.4pt);
    \filldraw[green1,line width=1pt] (0.7,0) node[right]{\scriptsize Reference solution\cite{Sze_shell_benchmark_2004}};
    \filldraw[black,line width=1pt, solid] (4.5,0) -- (5.1,0);
    \filldraw[black,line width=1pt] (5.2,0) node[right]{\scriptsize Displacement-based formulation};
    \filldraw[red1,line width=1pt, solid] (10.5,0) -- (11.1,0);
    \filldraw[red1,line width=1pt] (11.2,0) node[right]{\scriptsize Mixed formulation, consistent strain projection};
\end{tikzpicture}

\begin{tikzpicture}
    \filldraw[blue1,line width=1pt, solid] (4.5,0.05) -- (5.1,0.05);
    \filldraw[blue1,line width=1pt] (5.2,0) node[right]{\scriptsize Mixed formulation, lumped strain projection, approximate duals};
\end{tikzpicture}}
    \caption{Pinched cylindrical shell: load-deflection curves at points A and B, computed with our formulation and with the consistent Galerkin mixed formulation, both with strain spaces of the same degree as the displacements, and with the standard displacement-based formulation (octant geometry, mesh of $48 \times 48$ quadratic and cubic elements).}\label{fig:pincyl3}
\end{figure}


We now assess the performance of our formulation in the geometrically nonlinear regime. To this end, we consider two established benchmarks from\cite{Sze_shell_benchmark_2004}: 
the pullout of an unconstrained cylindrical shell and 
the pinched cylindrical shell. 
We compare the load--displacement curves obtained with our formulation against the reference solutions in \cite{Sze_shell_benchmark_2004} and the results obtained with other formulations. 
We illustrate that our approach 
accurately captures the geometrically nonlinear response of both benchmarks, including the strongly nonlinear pinched cylinder, while producing displacement and stress resultant fields without pronounced oscillations.

\subsubsection{Pullout of a cylindrical shell}

The first example consists of an open-ended cylindrical shell subjected to two opposite pulling forces, illustrated in Figure \ref{fig:cyl1}. We exploit the symmetry and model only one-eighth of the structure, to which all our statements on element numbers and mesh sizes refer. 
Figure \ref{fig:cyl1} also illustrates the total displacement field, computed with our formulation and cubic splines on a $16 \times 16$ mesh at the maximum load of $P = 40,000$. 
In Figure \ref{fig:cyl2}, we plot the corresponding membrane stress resultant $n^{12}$ and the bending moment $m^{11}$. Apart from the localized concentrations near the point load, the stress resultant fields exhibit no pronounced oscillations.

In Figure \ref{fig:cyl3}, we compare the load--displacement curves at point A obtained with quadratic and cubic splines on a $16 \times 16$ mesh. 
We observe that the standard displacement-based formulation exhibits a stiffer response, particularly for quadratic splines, whereas our formulation follows the reference solution \cite{Sze_shell_benchmark_2004} more closely, with results comparable to those of the consistent Galerkin mixed formulation. 
This observation is supported by the convergence results in Figure \ref{fig:cyl4}, where we plot the normalized displacement at point A under the maximum load for different mesh resolutions. 
We observe that our formulation achieves practically the same accuracy as the consistent Galerkin mixed formulation, also for relatively coarse meshes, confirming that our strain projection preserves the accuracy of the nonlinear response in this example.

\subsubsection{Pinched cylindrical shell}

The second example consists of a cylindrical shell with rigid diaphragms at both ends, subjected to two opposite compressive point loads, similar to the linear benchmark in Section \ref{sec:lin_cyl}.
We follow the nonlinear benchmark in \cite{Sze_shell_benchmark_2004},
with Young's modulus $E = 3 \cdot 10^4$, Poisson's ratio $\nu = 0.3$, radius $R = 100$, length $L = 200$, thickness $d = 1$, and maximum load $P_{\text{max}} = 12,000$.
We again exploit the symmetry, model only one-eighth of the structure, and discretize it with a mesh of \(48\times48\) quadratic and cubic elements, which corresponds to the mesh resolution of the reference solution \cite{Sze_shell_benchmark_2004}. 
To trace the strongly nonlinear equilibrium path, including its limit points, we employ arc-length continuation from the unloaded state, supplemented by line search.

\begin{remark}\label{rem:pinched_fold}
    We note that this benchmark requires a sufficiently fine mesh to reproduce the response accurately up to the maximum load, particularly because the deformation concentrates in sharp folds as the pinching load increases. 
    As these folds move through the mesh, their accurate representation requires a sufficient number of elements across them. Otherwise, the computed response exhibits a periodic variation, whose amplitude strongly depends on the mesh and the discrete spaces, and the computation of the equilibrium path can become unstable (see e.g.\ \cite{samimi2011selfadaptive,radenkovic2021nonlinear}).
    In our computations, mesh refinement reduced this variation and stabilized the computation.
\end{remark}

For this benchmark, we discretize the independent strains with the same spline spaces as the displacements, that is, with splines of degree $p$ and continuity $C^{p-1}$ in both parametric directions. 
The reason for this choice is the higher accuracy of the row-sum lumped strain projection in these richer spaces, which reduces the additional flexibility of the condensed system with respect to its consistent counterpart.
In this example, even a small additional flexibility is critical, since the cylinder itself softens at a critical point, where large rotations couple the membrane and bending strains.
The two softening effects add up and can render the computation of the equilibrium path unstable.
We note that strain spaces of the same degree as the displacements do not by themselves remove membrane locking (see Section \ref{sec:discretization}).
On the fine mesh that this benchmark requires, however, the effect of membrane locking is insignificant (cf.\ Section \ref{sec:lin_cyl}, where membrane locking mainly affects coarse meshes).

Figure \ref{fig:pincyl2} illustrates the total displacement field, the membrane stress resultant $n^{12}$, and the bending moment $m^{11}$ at the maximum load, computed with our formulation and cubic splines.
The stress resultant fields exhibit localized concentrations near the pinching region and no pronounced oscillations elsewhere.
In Figure \ref{fig:pincyl3}, we compare the load--displacement curves at points A and B obtained with quadratic and cubic splines against the reference solution in \cite{Sze_shell_benchmark_2004}, 
the consistent Galerkin mixed formulation with the same strain spaces (red curves), 
and the standard displacement-based formulation (black curves).
We observe that our formulation captures the pronounced nonlinear response and closely follows the reference curves, the curves of the consistent formulation, and those of the standard displacement-based formulation up to the maximum load.
On this fine mesh, the standard displacement-based formulation also closely follows the reference solution, consistent with the insignificant effect of membrane locking on fine meshes discussed above.
Its load--displacement curves exhibit less pronounced variations than those of our formulation, indicating that, for this benchmark, our formulation is less robust than the displacement-based formulation (see also Remark~\ref{rem:pinched_fold}).
Increasing the polynomial degree from quadratic to cubic splines further improves the agreement with the reference solution for our formulation, consistent with the observations for the preceding benchmarks.
Overall, these results show that, on a sufficiently fine mesh, our formulation reproduces the strongly nonlinear equilibrium response of this benchmark.

\section{Summary and conclusions}\label{sec:conclusions}

We presented a novel discretization strategy for the mixed variational formulation of the Kirchhoff-Love shell problem, derived from the modified Hellinger-Reissner variational principle. For mitigating membrane locking, we discretized the independent strain fields with spline basis functions that are one degree lower than those used for the displacements. The key novelty of our approach is the utilization of approximate dual spline functions to discretize the variations of the independent strain fields. We demonstrated that the resulting nearly diagonal submatrices of the strain projection equations can then be diagonalized by row-sum lumping without loss of higher-order accuracy. In comparison to the consistent Galerkin mixed formulation, the diagonalized projection matrices enable a significant and decisive cost reduction for the condensation of the independent strain variables and the solution of the condensed system of equations. In comparison to the standard displacement-based (but locking-prone) Galerkin formulation, our formulation involves the condensation procedure, an increase in bandwidth and an unsymmetric condensed system matrix. We showed via theoretical estimates and computing times that for the same mesh, the overall computational cost of our (locking-free) formulation is still much larger, but remains within the same complexity.

We demonstrated the higher-order accuracy of our approach and the mitigation of membrane locking with numerical benchmarks, including a linear elastic curved Euler-Bernoulli beam and the linear elastic examples of the shell obstacle course. 
We further assessed its performance in the geometrically nonlinear regime using the Scordelis-Lo roof and the pullout of a cylindrical shell. 
We also showed for these examples that our approach achieves practically the same accuracy as the Galerkin mixed formulation with consistent strain condensation (which is prohibitively expensive from a computing viewpoint), while it eliminates membrane locking as effectively as the best out of a number of well-established locking-preventing methods, applied in the standard Galerkin formulation. 
Moreover, for the more challenging geometrically nonlinear pinched-cylinder benchmark, we demonstrated that our formulation, with strain spaces of the same degree as the displacements on the fine mesh that this benchmark requires, accurately reproduces the strongly nonlinear response up to the maximum load, in close agreement with the reference solution and with the consistent Galerkin mixed formulation.
However, our results indicated that, for this benchmark, our formulation is less robust than the standard displacement-based formulation, whose load--displacement curves exhibit less pronounced variations.

Our results confirmed that for quadratic discretizations, our approach based on the projection of both the membrane and bending strains is particularly attractive, as mixed formulations achieve better accuracy than any of the other methods in the context of the standard displacement-based Galerkin formulation. In addition, our numerical experiments indicated that for the same quadratic discretizations, the computing time for condensation and system solution is consistently smaller than 10 times the computing time for the displacement-based formulation. The potential of our formulation becomes evident when we consider that, in our benchmark examples, a two-step $h$-refinement -- resulting in a fourfold increase in the number of elements per parametric direction and an eightfold rise in computational cost -- was generally insufficient to eliminate membrane locking in the standard displacement-based formulation. For us, this result seems particularly noteworthy, as the industrial focus at this point in time is reported to be on quadratic spline discretizations.

These computational aspects play a different role depending on whether a direct or iterative solver is used, or if explicit dynamics calculations are performed where the condensed system matrix is never assembled, stored, or factorized. 
Ultimately, in each setting, the crucial point is that a competitive method achieves a given level of accuracy more quickly than its competitors, using comparable or fewer computing resources, 
and at the same time is reliable and robust. 
In this respect, the approach presented in this work holds promise.
Our results demonstrate the mitigation of membrane locking and the accuracy of the proposed formulation for the examples considered. A systematic investigation of possible nonzero-energy spurious modes, however, remains beyond the scope of the present study \cite{Adam2015}.
Another particularly interesting avenue for future work could be the synergistic combination of the properties of approximate dual spline functions for strain condensation and mass lumping in explicit structural dynamics \cite{nguyen_mass_lumping2023,hiemstra2023} as well as outlier removal methods for the control of the maximum time step \cite{hiemstra2021removal,nguyen2022variational}.

\bmsection*{Acknowledgments}

The authors gratefully acknowledge financial support from the German Research Foundation (Deutsche Forschungsgemeinschaft) through the DFG Emmy Noether Grant SCH 1249/2-1 and the standard DFG grant SCH 1249/5-1. 
The first author would like to express her gratitude to B.A.\ Roccia (University of Bergen) for insightful discussions and constructive suggestions. AI tools were utilized to enhance the 
language and readability of the manuscript.


\bmsection*{Conflict of interest}

The authors declare no potential conflict of interests and that they have no known competing financial interests or personal relationships that could have appeared to influence the work reported in this paper.

\bmsection*{Data availability statement}
No data were used to support this study.

\appendix

\section{Linearization of the membrane and bending strains}\label{sec:shell_derivation}

We obtain the strain-displacement matrix $\mat{B}^m$ corresponding to the membrane strains via 
substituting \eqref{eq:discrete-trial-displacement} into \eqref{eq:variation-epsilon}:
\begin{align}
    \delta \vect{\varepsilon}^h = \begin{bmatrix}
        \delta \varepsilon_{11}^h \\ \delta \varepsilon_{22}^h \\ 2\delta \varepsilon_{12}^h
    \end{bmatrix} = \sum_{i=1}^{n} \, \underbrace{\begin{bmatrix}
        \vect{a}_1^T \, B_{i,1}  \\
        \vect{a}_2^T \, B_{i,2}  \\ 
        \vect{a}_1^T \, B_{i,2} + \vect{a}_2^T \, B_{i,1} 
    \end{bmatrix}}_{\mat{B}^m_i} \, \delta \vect{\hat{u}}^h_i = \sum_{i=1}^{n} \, \mat{B}^m_i \, \delta \vect{\hat{u}}^h_i \,,
\end{align}
where $\delta \vect{\hat{u}}^h$ is the variation of the vector of unknown coefficients $\vect{\hat{u}}_i^h$ in \eqref{eq:discrete-trial-displacement}. 
Here, the covariant base vectors $\vect{a}_1$ and $\vect{a}_2$ are computed using the discrete position vector of the midsurface, $\vect{\bar{x}}^h$, in the current configuration:
\begin{align}
    \vect{a}_1 = \vect{\bar{x}}^h_{,1} \,, \qquad  \vect{a}_2 = \vect{\bar{x}}^h_{,2} \,.
\end{align}

To derive the strain-displacement matrix $\mat{B}^b$, we first compute the variation of the normalized normal vector $\vect{a}_3$ \eqref{eq:base-vector-current-config} as follows:
\begin{align}\label{eq:variation-a3}
    \delta \vect{a}_3 = \frac{1}{\left\| \vect{n}_3 \right\|} \, \underbrace{\left(\mat{I} - \vect{a}_3 \otimes \vect{a}_3\right)}_{\mat{P}} \, \delta \vect{n}_3 = \frac{1}{\left\| \vect{n}_3 \right\|} \, \mat{P} \, \delta \vect{n}_3 \,, \quad \text{with} \quad \vect{n}_3 = \vect{a}_1 \times \vect{a}_2 \,, 
\end{align}
and $\left\| \vect{n}_3 \right\|$ denotes the norm of the normal vector $\vect{n}_3$. 
Substituting this and \eqref{eq:discrete-trial-displacement} into \eqref{eq:variation-kappa} leads to:
\begin{align}
    \delta \vect{\kappa}^h = \begin{bmatrix}
        \delta \kappa_{11}^h \\ \delta \kappa_{22}^h \\ 2\delta \kappa_{12}^h
    \end{bmatrix} = \, \sum_{i=1}^{n} \, \underbrace{\begin{bmatrix}
        -\vect{a}_3^T B_{i,11} 
        - \frac{1}{\left\| \vect{n}_3 \right\|} \, \left[\vect{a}_2 \times \left(\mat{P}\,\vect{a}_{1,1}\right)\right]^T B_{i,1} 
        + \frac{1}{\left\| \vect{n}_3 \right\|} \, \left[\vect{a}_1 \times \left(\mat{P}\,\vect{a}_{1,1}\right)\right]^T B_{i,2}  \\
        -\vect{a}_3^T B_{i,22} 
        - \frac{1}{\left\| \vect{n}_3 \right\|} \, \left[\vect{a}_2 \times \left(\mat{P}\,\vect{a}_{2,2}\right)\right]^T B_{i,1} 
        + \frac{1}{\left\| \vect{n}_3 \right\|} \, \left[\vect{a}_1 \times \left(\mat{P}\,\vect{a}_{2,2}\right)\right]^T B_{i,2}  \\
        -2\vect{a}_3^T B_{i,12} 
        - \frac{2}{\left\| \vect{n}_3 \right\|} \, \left[\vect{a}_2 \times \left(\mat{P}\,\vect{a}_{1,2}\right)\right]^T B_{i,1} 
        + \frac{2}{\left\| \vect{n}_3 \right\|} \, \left[\vect{a}_1 \times \left(\mat{P}\,\vect{a}_{1,2}\right)\right]^T B_{i,2} 
    \end{bmatrix}}_{\mat{B}^b_i} \, \delta \vect{\hat{u}}^h_i = \sum_{i=1}^{n} \, \mat{B}^b_i \, \delta \vect{\hat{u}}^h_i
\end{align}

\noindent
To derive the geometric stiffness matrix \eqref{eq:geometric-stiffness}, we start with the second variation of the membrane strains. 
Substituting \eqref{eq:discrete-trial-displacement} into \eqref{eq:2ndvariation-epsilon}, we obtain:
\begin{align}
    \Delta \delta \varepsilon_{\alpha \beta}^h = 
    \sum_{i=1}^{n} \, \sum_{j=1}^{n} \, \frac{1}{2} \, \left(\delta \vect{\hat{u}}^h_i\right)^T \left(B_{i,\alpha} \, B_{j,\beta} + B_{i,\beta} \, B_{j,\alpha}\right) \, \mat{I} \, \Delta \vect{\hat{u}}^h_j \,.
\end{align}
The membrane part of the geometric stiffness matrix, $\mat{K}_{11}^{\text{m,geom}}$, associated to the $i$-th and $j$-th control point is:
\begin{align}
    \mat{K}_{11,ij}^{\text{m,geom}} = \sum_{\alpha=1}^{2} \, \sum_{\beta=1}^{2} \, \frac{1}{2}\,\left(B_{i,\alpha} \, B_{j,\beta} + B_{i,\beta} \, B_{j,\alpha}\right)\,\bar{n}_{\alpha\beta}^h \, \mat{I} \,,
\end{align}
where $\bar{n}_{\alpha\beta}^h$ is the stress resultant for the discrete normal force components corresponding to the independent membrane strain variables $\vect{e}$ in \eqref{eq:strain-var-KL}, i.e.:
\begin{align}
    \begin{bmatrix}
        \bar{n}^h_{11} \\ \bar{n}^h_{22} \\ \bar{n}^h_{12}
    \end{bmatrix} = d \, \mat{C} \, \vect{e}^h \,.
\end{align}
The variation and linearization of the normal vector $\vect{n}_3$ are:
\begin{align}
    \delta \vect{n}_3 & = \delta \vect{a}_1 \times \vect{a}_2 + \vect{a}_1 \times \delta \vect{a}_2
                        = \vect{a}_1 \times \delta \vect{a}_2 - \vect{a}_2 \times \delta \vect{a}_1 \nonumber \\
                      & = \sum_{i=1}^{n} \, \underbrace{\left(\left[\vect{a}_1\right]_{\times} \, B_{i,2} - \left[\vect{a}_2\right]_{\times} \, B_{i,1}\right)}_{\mat{A}_i} \, \delta \vect{\hat{u}}^h_i 
                        = \sum_{i=1}^{n} \,\mat{A}_i \, \delta \vect{\hat{u}}^h_i \,, \\
    \Delta \vect{n}_3 & = \sum_{i=1}^{n} \,\mat{A}_i \, \Delta \vect{\hat{u}}^h_i \,,                       
\end{align}
respectively, 
where $\left[\vect{a}_\alpha\right]_{\times}$ is the skew-symmetric matrix of vector $\vect{a}_\alpha$. 
The mixed second variation of the unnormalized normal vector is:
\begin{align}
    \Delta\delta\vect{n}_3
    = \delta\vect{a}_1 \times \Delta\vect{a}_2
    + \Delta\vect{a}_1 \times \delta\vect{a}_2\,.
\end{align} 
The linearization of the variation of the normalized normal vector $\vect{a}_3$ \eqref{eq:variation-a3} is:
\begin{align}
    \Delta \delta \vect{a}_3  
    = \frac{1}{\left\|\vect{n}_3 \right\|}\,\mat{P}\,\Delta\delta\vect{n}_3 
    -\frac{1}{\left\|\vect{n}_3 \right\|^2} \left(\vect{a}_3 \cdot \Delta \vect{n}_3\right) \, \mat{P} \, \delta \vect{n}_3  
    -\frac{1}{\left\|\vect{n}_3 \right\|} \left(\Delta \vect{a}_3 \otimes \vect{a}_3 + \vect{a}_3 \otimes \Delta \vect{a}_3\right) \, \delta \vect{n}_3\,.
\end{align}
Inserting this and \eqref{eq:discrete-trial-displacement} into \eqref{eq:2ndvariation-kappa} leads to:
\begin{align}
    \Delta\delta\kappa_{\alpha\beta}^h
      &= \sum_{i=1}^{n}\sum_{j=1}^{n}\left(\delta\vect{\hat{u}}_i^h\right)^T \nonumber \\
      &\quad\left\{
    \begin{aligned}
      &- \frac{1}{\left\|\vect{n}_3\right\|}\,B_{i,\alpha\beta}\mat{P}\mat{A}_j
       + \frac{1}{\left\|\vect{n}_3\right\|}\,\mat{A}_i\mat{P}\,B_{j,\alpha\beta} \, +\, \frac{B_{i,1}B_{j,2}-B_{i,2}B_{j,1}}{\left\|\vect{n}_3\right\|}
        \left[\mat{P}\vect{a}_{\alpha,\beta}\right]_{\times} \\
      &- \frac{1}{\left\|\vect{n}_3\right\|^2}\,\mat{A}_i
         \left[2\vect{a}_3\odot\vect{a}_{\alpha,\beta}
         + (\vect{a}_3\cdot\vect{a}_{\alpha,\beta})
         (\mat{I}-3\vect{a}_3\otimes\vect{a}_3)\right]\mat{A}_j 
    \end{aligned}
    \right\}\,\Delta\vect{\hat{u}}_j^h\,,
\end{align}
where $\odot$ denotes the symmetric product of two vectors, i.e.:
\begin{align}
    \vect{a}_3 \odot \vect{a}_{\alpha,\beta} = \frac{1}{2} \left(\vect{a}_3 \otimes \vect{a}_{\alpha,\beta} + \vect{a}_{\alpha,\beta} \otimes \vect{a}_3\right) \,.
\end{align}
The bending part of the geometric stiffness matrix, $\mat{K}_{11}^{\text{b,geom}}$, associated to the $i$-th and $j$-th control point is:
\begin{align}
    \mat{K}_{11,ij}^{\text{b,geom}}
      &= \sum_{\alpha=1}^{2}\sum_{\beta=1}^{2}\bar{m}_{\alpha\beta}^h\,
      \left\{
    \begin{aligned}
      &- \frac{1}{\left\|\vect{n}_3\right\|}\,B_{i,\alpha\beta}\mat{P}\mat{A}_j
       + \frac{1}{\left\|\vect{n}_3\right\|}\,\mat{A}_i\mat{P}\,B_{j,\alpha\beta} \, + \, \frac{B_{i,1}B_{j,2}-B_{i,2}B_{j,1}}{\left\|\vect{n}_3\right\|}
        \left[\mat{P}\vect{a}_{\alpha,\beta}\right]_{\times} \\
      &- \frac{1}{\left\|\vect{n}_3\right\|^2}\,\mat{A}_i
         \left[2\vect{a}_3\odot\vect{a}_{\alpha,\beta}
         + (\vect{a}_3\cdot\vect{a}_{\alpha,\beta})
         (\mat{I}-3\vect{a}_3\otimes\vect{a}_3)\right]\mat{A}_j
    \end{aligned}
    \right\}\,,
\end{align}
where $\bar{m}_{\alpha\beta}^h$ is the stress resultant for the discrete bending moment components corresponding to the independent variables $\vect{k}$ in \eqref{eq:strain-var-KL}, i.e.:
\begin{align}
    \begin{bmatrix}
        \bar{m}^h_{11} \\ \bar{m}^h_{22} \\ \bar{m}^h_{12}
    \end{bmatrix} = \frac{d^3}{12} \, \mat{C} \, \vect{k}^h\,.
\end{align}
Analogously, $\Tilde{\mat{K}}_{11,ij}^{\text{b,geom}}$, associated to the $i$-th and $j$-th control point is:
\begin{align}
    \Tilde{\mat{K}}_{11,ij}^{\text{b,geom}}
      &= \sum_{\alpha=1}^{2}\sum_{\beta=1}^{2}m_{\alpha\beta}^h\,
      \left\{
    \begin{aligned}
      &- \frac{1}{\left\|\vect{n}_3\right\|}\,B_{i,\alpha\beta}\mat{P}\mat{A}_j
       + \frac{1}{\left\|\vect{n}_3\right\|}\,\mat{A}_i\mat{P}\,B_{j,\alpha\beta} \,+\, \frac{B_{i,1}B_{j,2}-B_{i,2}B_{j,1}}{\left\|\vect{n}_3\right\|}
        \left[\mat{P}\vect{a}_{\alpha,\beta}\right]_{\times} \\
      &- \frac{1}{\left\|\vect{n}_3\right\|^2}\,\mat{A}_i
         \left[2\vect{a}_3\odot\vect{a}_{\alpha,\beta}
         + (\vect{a}_3\cdot\vect{a}_{\alpha,\beta})
         (\mat{I}-3\vect{a}_3\otimes\vect{a}_3)\right]\mat{A}_j
    \end{aligned}
    \right\}\,,
\end{align}
where $m_{\alpha\beta}^h$ is the stress resultant for the discrete bending moment components corresponding to the displacement-based variables $\vect{\kappa}$ in \eqref{eq:membrane_bending_strains}, i.e.:
\begin{align}
    \begin{bmatrix}
        m^h_{11} \\ m^h_{22} \\ m^h_{12}
    \end{bmatrix} = \frac{d^3}{12} \, \mat{C} \, \vect{\kappa}^h\,.
\end{align}

\Needspace{5\baselineskip}
\section{Additional results for the curved Euler-Bernoulli beam}\label{sec:beam1}

\begin{figure}[ht!]
	\centering
    \captionsetup[subfloat]{labelfont=scriptsize,textfont=scriptsize}
    \subfloat[$p=2$]{{ \def\svgwidth{0.42\textwidth}
    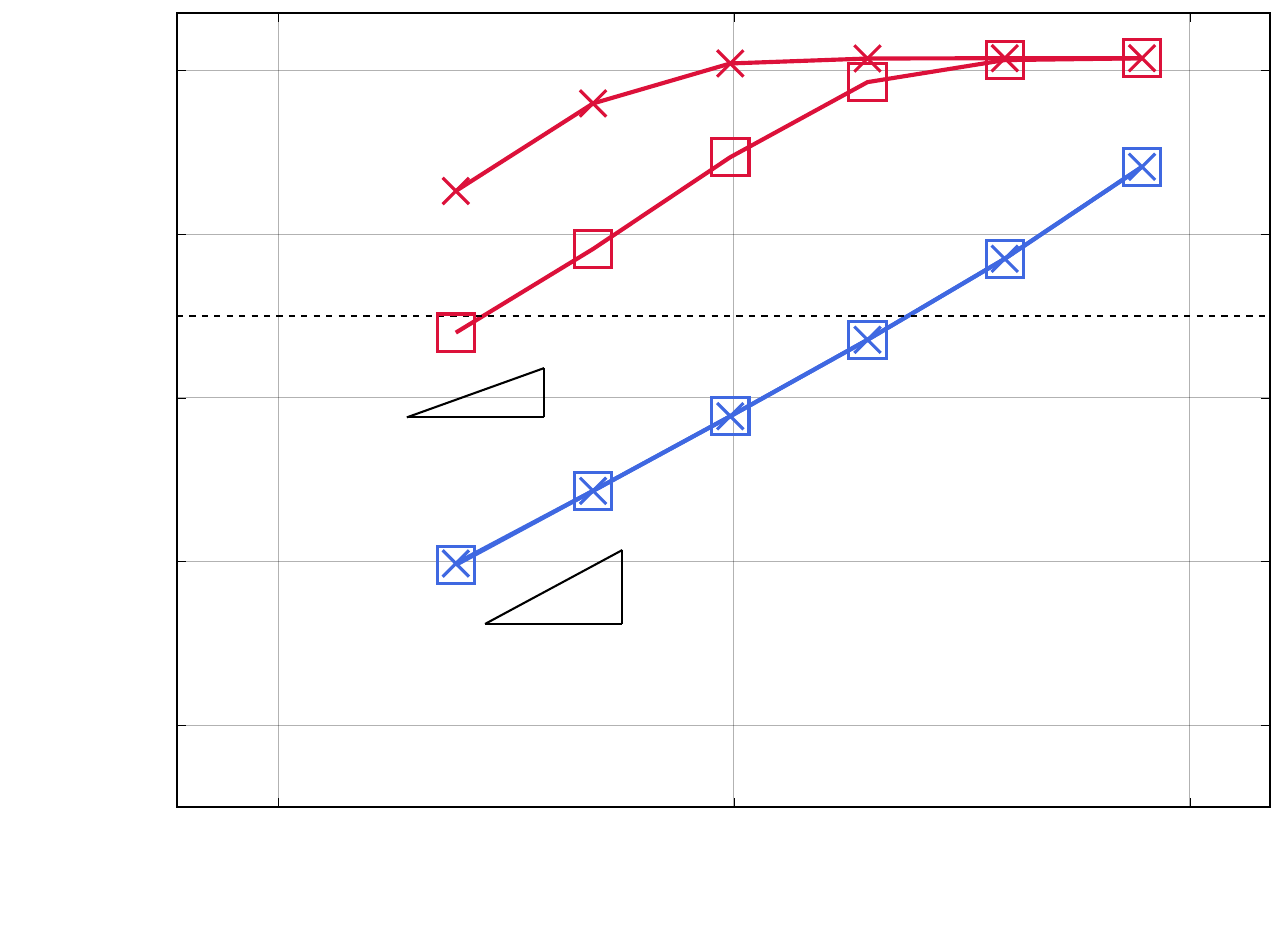 }} \hspace{0.2cm}
    \subfloat[$p=3$]{{ \def\svgwidth{0.42\textwidth}
    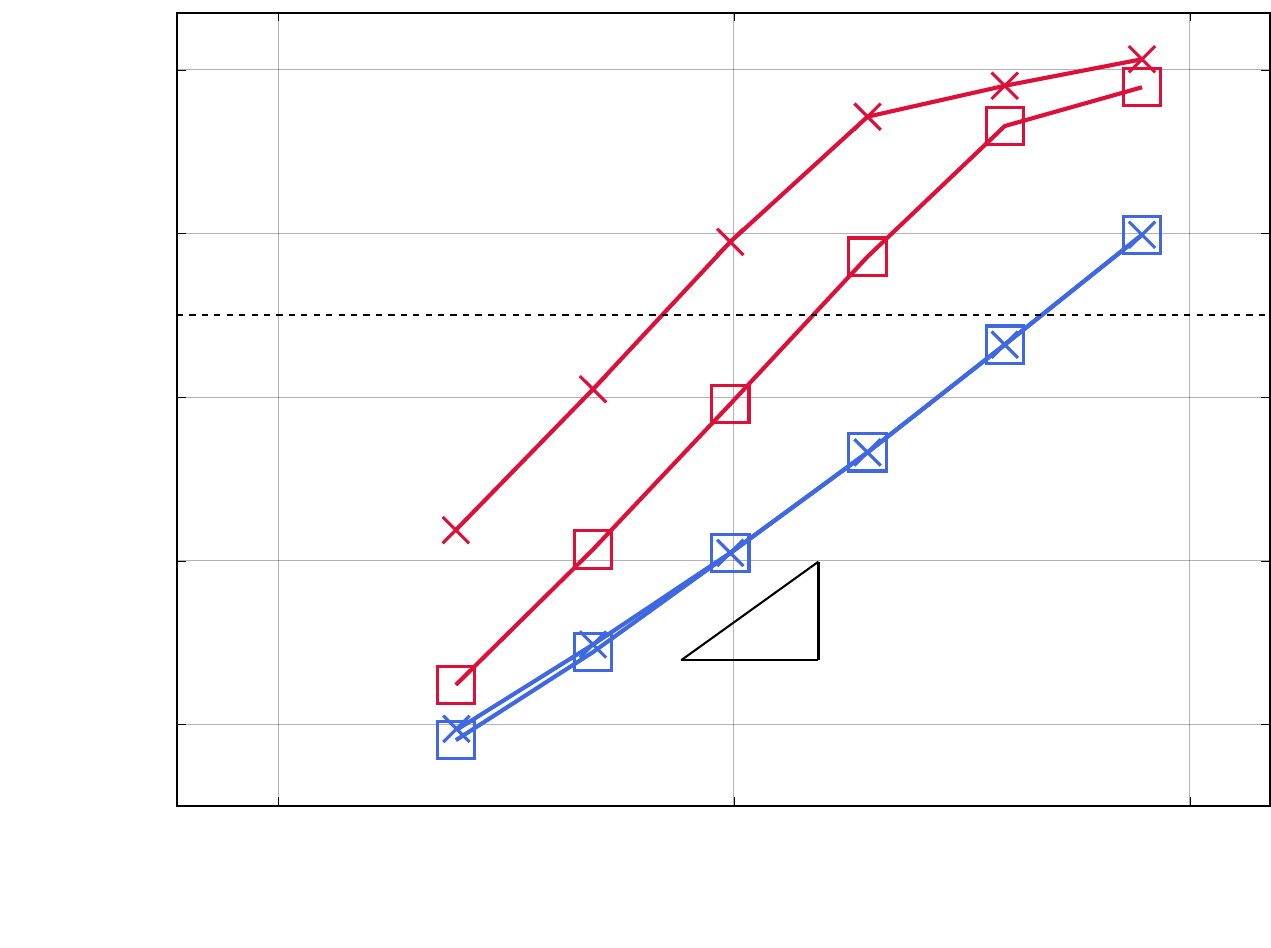 }}

    \subfloat[$p=2$]{{ \def\svgwidth{0.42\textwidth}
    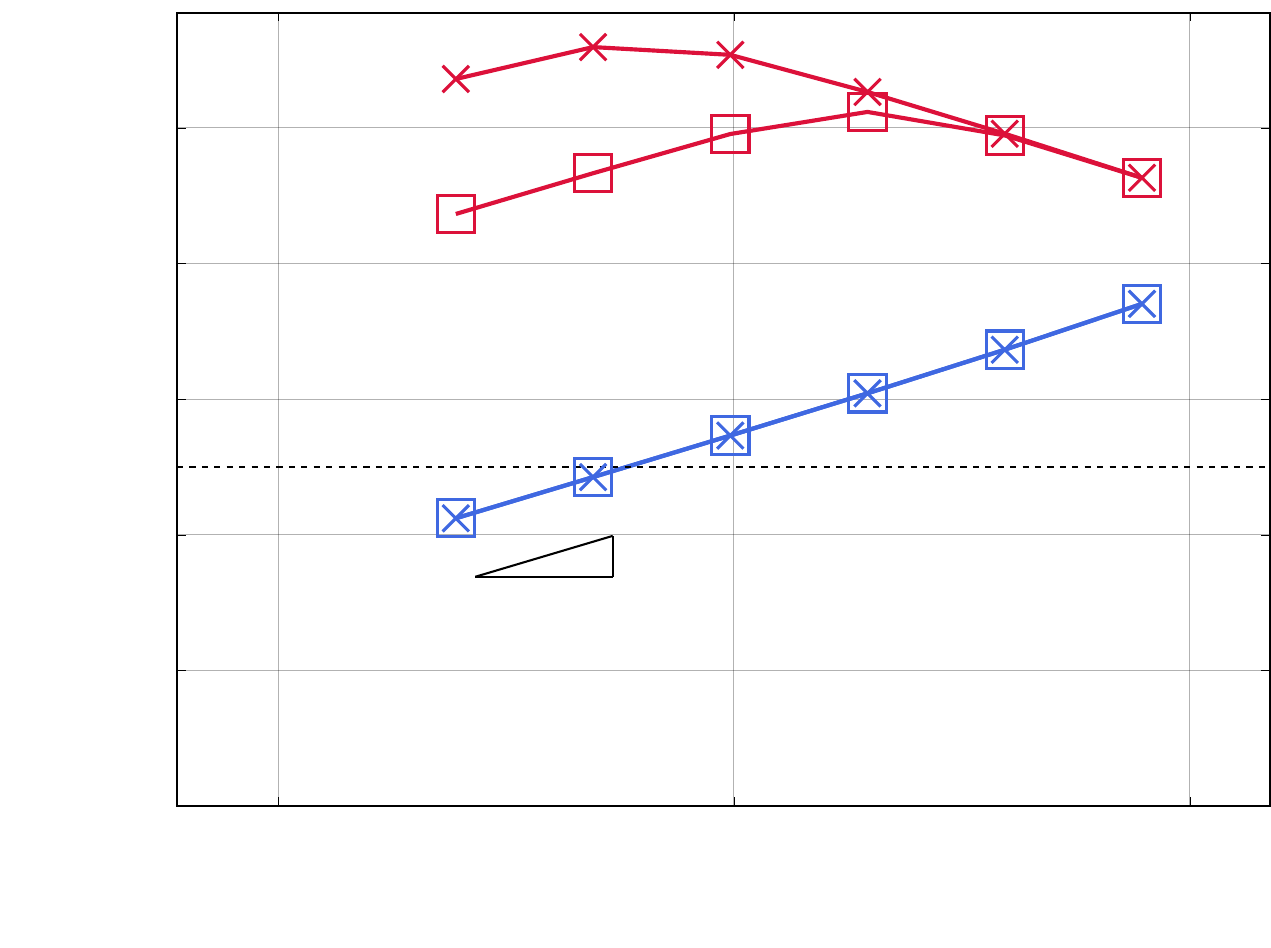 }} \hspace{0.2cm}
    \subfloat[$p=3$]{{ \def\svgwidth{0.42\textwidth}
    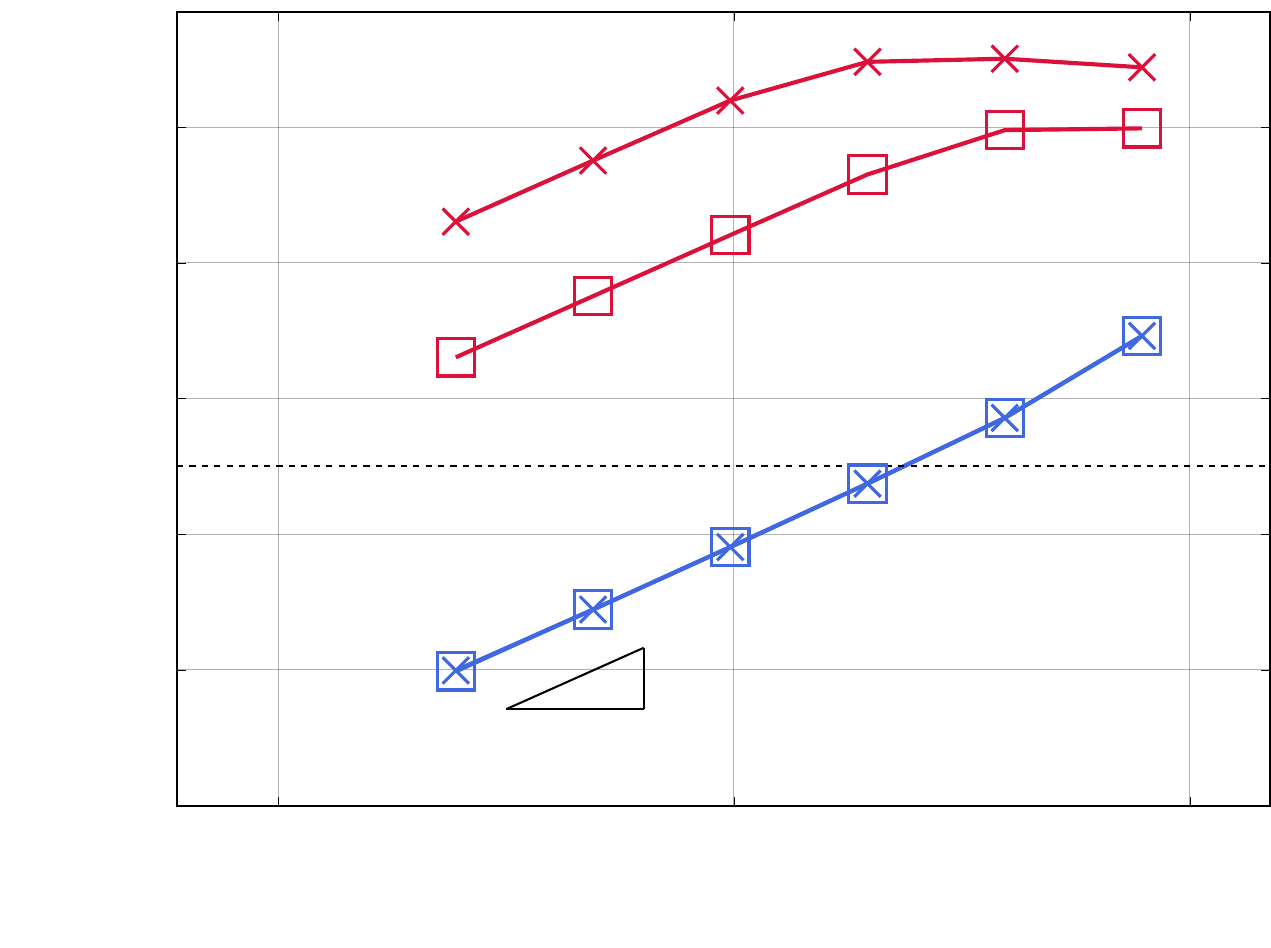 }}

    \subfloat[$p=2$]{{ \def\svgwidth{0.42\textwidth}
    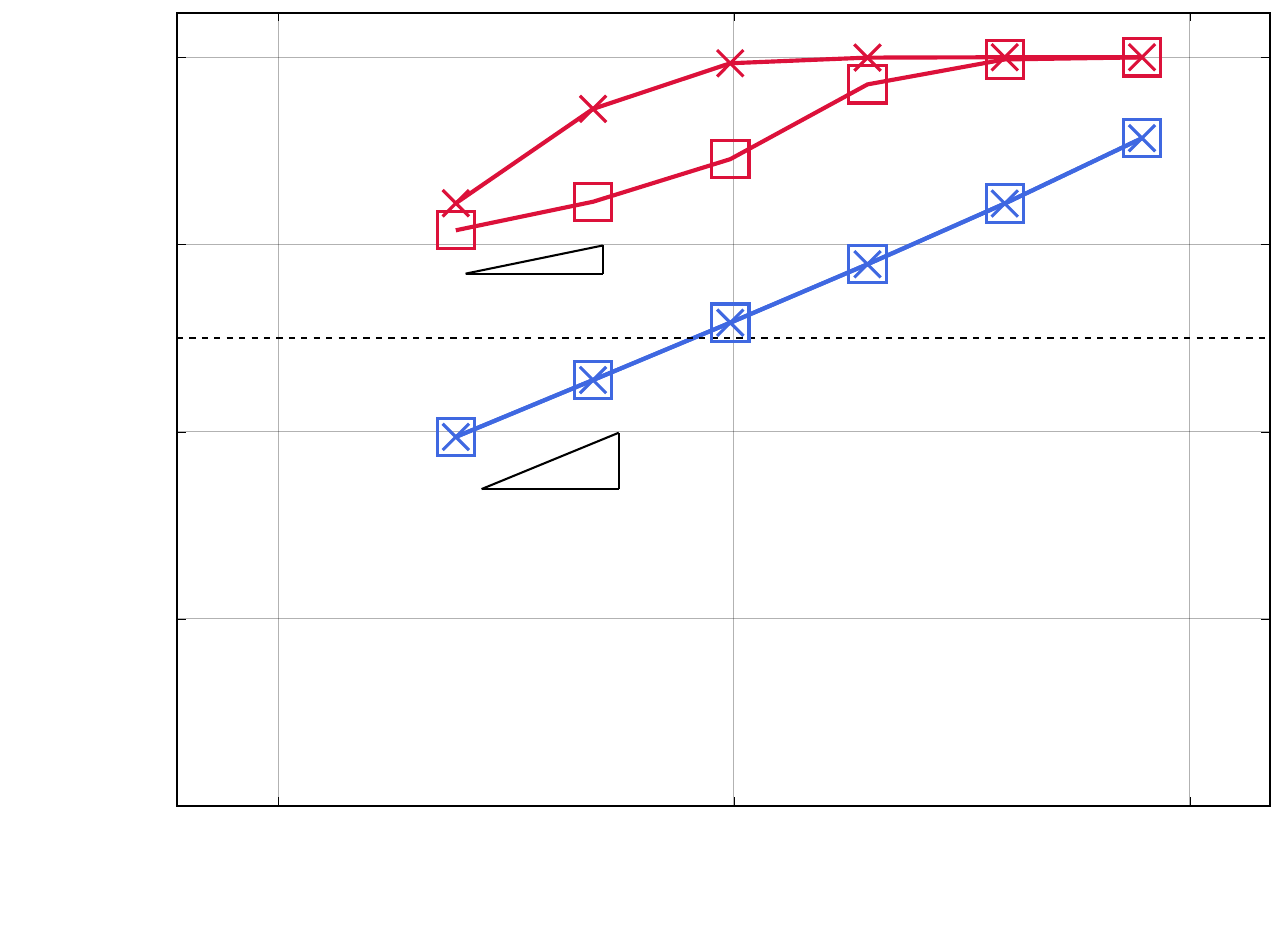 }} \hspace{0.2cm}
    \subfloat[$p=3$]{{ \def\svgwidth{0.42\textwidth}
    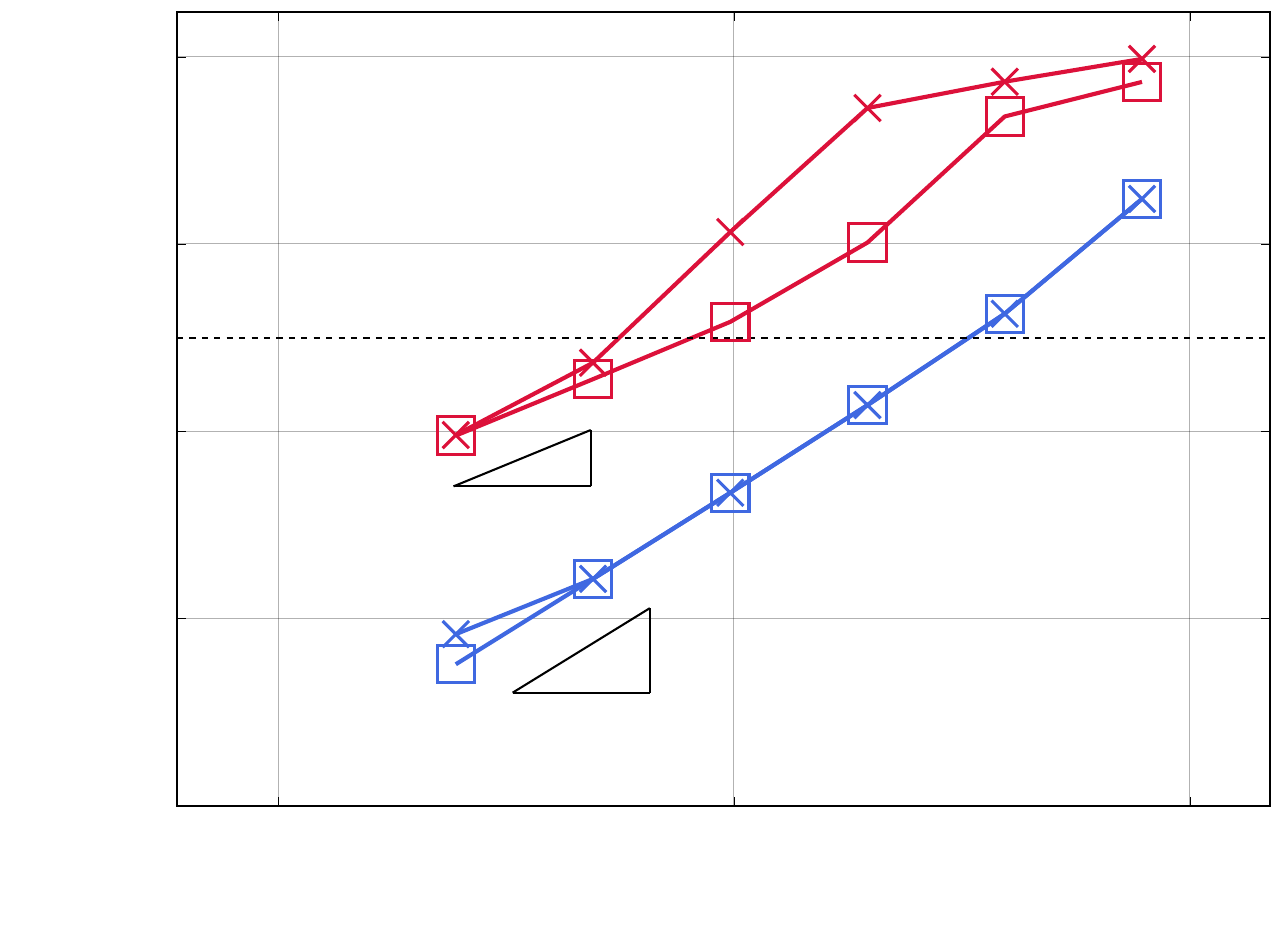 }}
    
    \vspace{0.2cm}
    \begin{tikzpicture}
    \filldraw[red1,line width=1pt, solid] (0.0,0) -- (0.6,0);
    \filldraw[red1,line width=1pt] (0.0,0) node[right]{\scriptsize $\boldsymbol{\bigtimes}$};
    \filldraw[red1,line width=1pt] (0.7,0) node[right]{\scriptsize Standard Galerkin, full integration, $R/t = 1,000$};
    \filldraw[red1,line width=1pt, solid] (8,0.05) -- (8.6,0.05);
    \filldraw[red1,line width=1pt] (8.2,-0.08) [fill=none] rectangle ++(0.25,0.25);;
    \filldraw[red1,line width=1pt] (8.7,0) node[right]{\scriptsize Standard Galerkin, full integration, $R/t = 100$};
\end{tikzpicture}

\begin{tikzpicture}
    \filldraw[blue1,line width=1pt, solid] (0.5,0) -- (1.1,0);
    \filldraw[blue1,line width=1pt] (0.5,0) node[right]{\scriptsize $\boldsymbol{\bigtimes}$};
    \filldraw[blue1,line width=1pt] (1.2,0) node[right]{\scriptsize Our approach, $R/t = 1,000$};
    \filldraw[blue1,line width=1pt, solid] (6,0.05) -- (6.6,0.05);
    \filldraw[blue1,line width=1pt] (6.2,-0.08) [fill=none] rectangle ++(0.25,0.25);;
    \filldraw[blue1,line width=1pt] (6.7,0) node[right]{\scriptsize Our approach, $R/t = 100$};
\end{tikzpicture}
    \caption{Curved Euler-Bernoulli beam: relative error in the $L^2$-norm of the displacement, membrane strain and bending strain fields, computed with our approach and the standard displacement-based formulation for slenderness ratio $R/t=100$ and $1,000$ and polynomial degrees $p=2$ and $3$.}\label{fig:beam_convergence_slenderness_ratio1}
\end{figure}

\begin{figure}[ht!]
	\centering
    \captionsetup[subfloat]{labelfont=scriptsize,textfont=scriptsize}
    \subfloat[$p=4$]{{ \def\svgwidth{0.42\textwidth}
    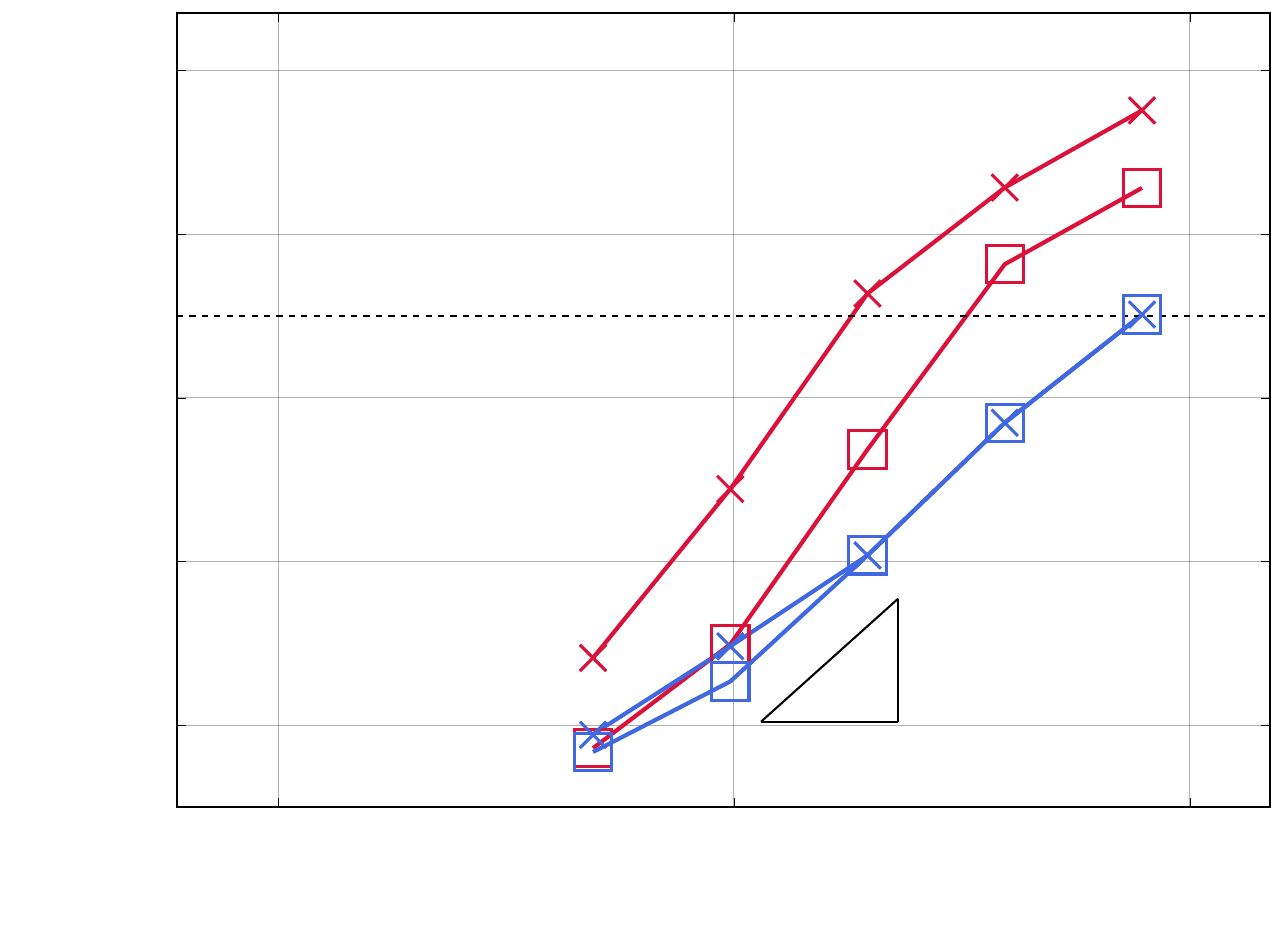 }} \hspace{0.2cm}
    \subfloat[$p=5$]{{ \def\svgwidth{0.42\textwidth}
    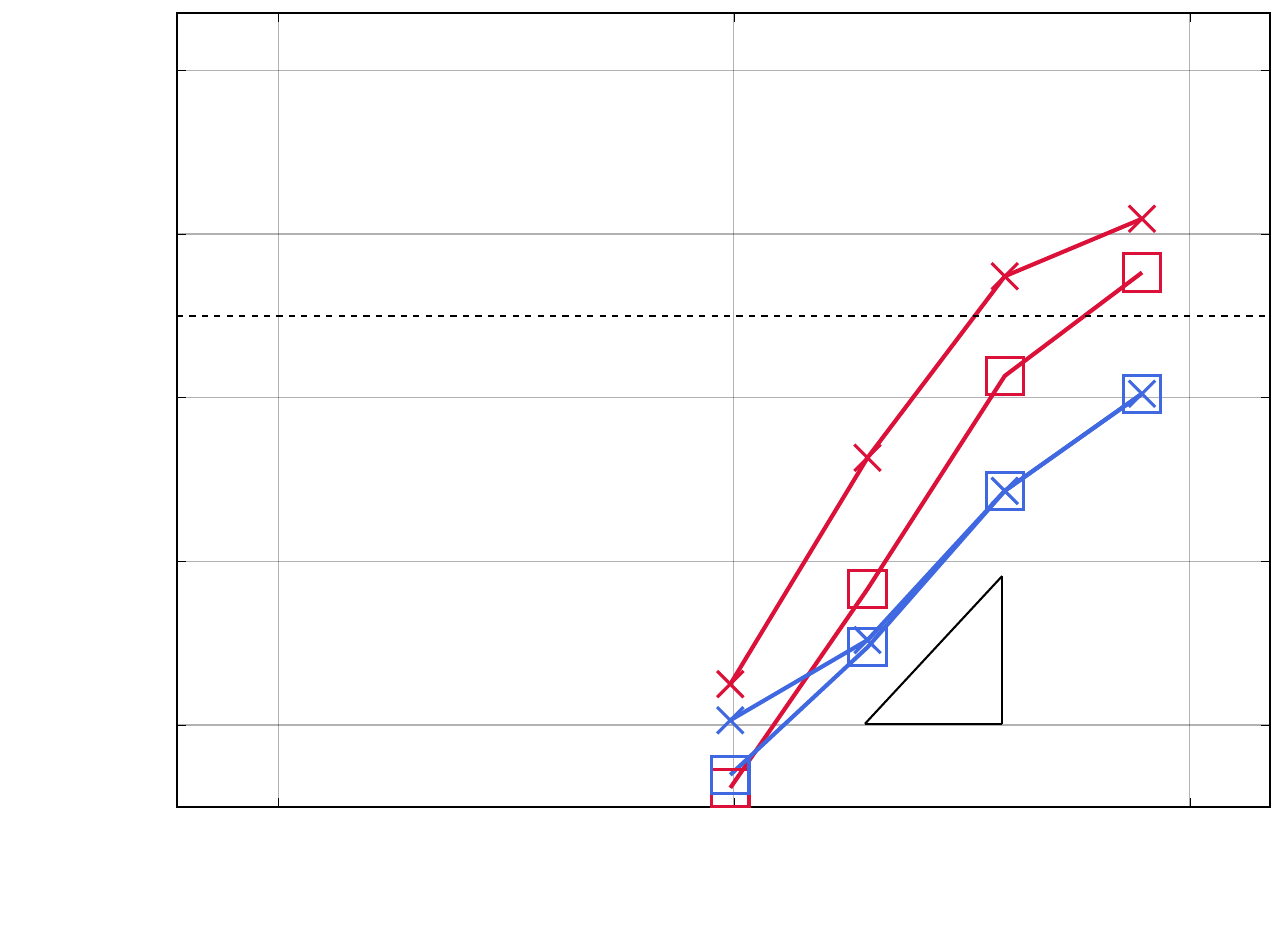 }}

    \subfloat[$p=4$]{{ \def\svgwidth{0.42\textwidth}
    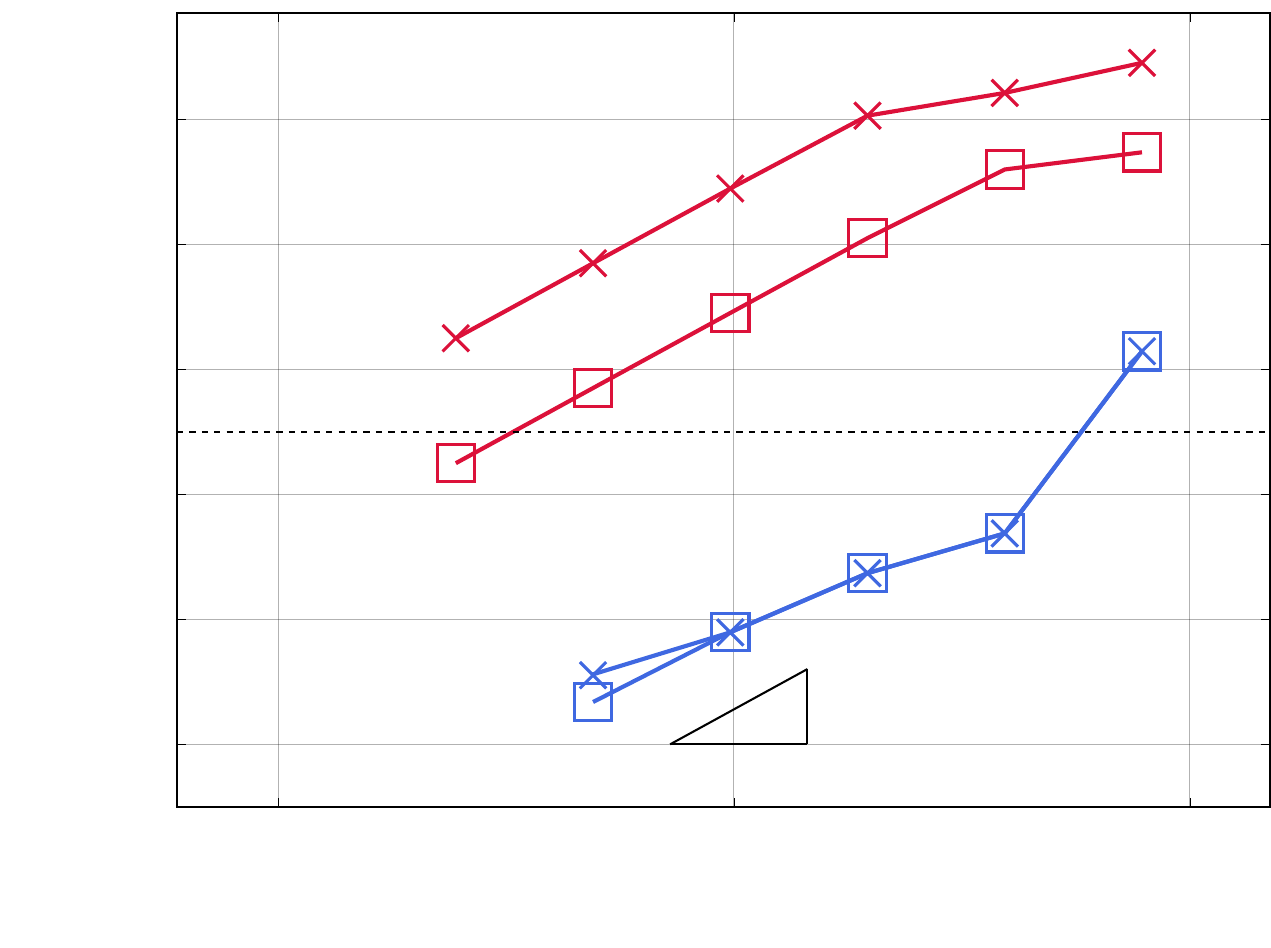 }} \hspace{0.2cm}
    \subfloat[$p=5$]{{ \def\svgwidth{0.42\textwidth}
    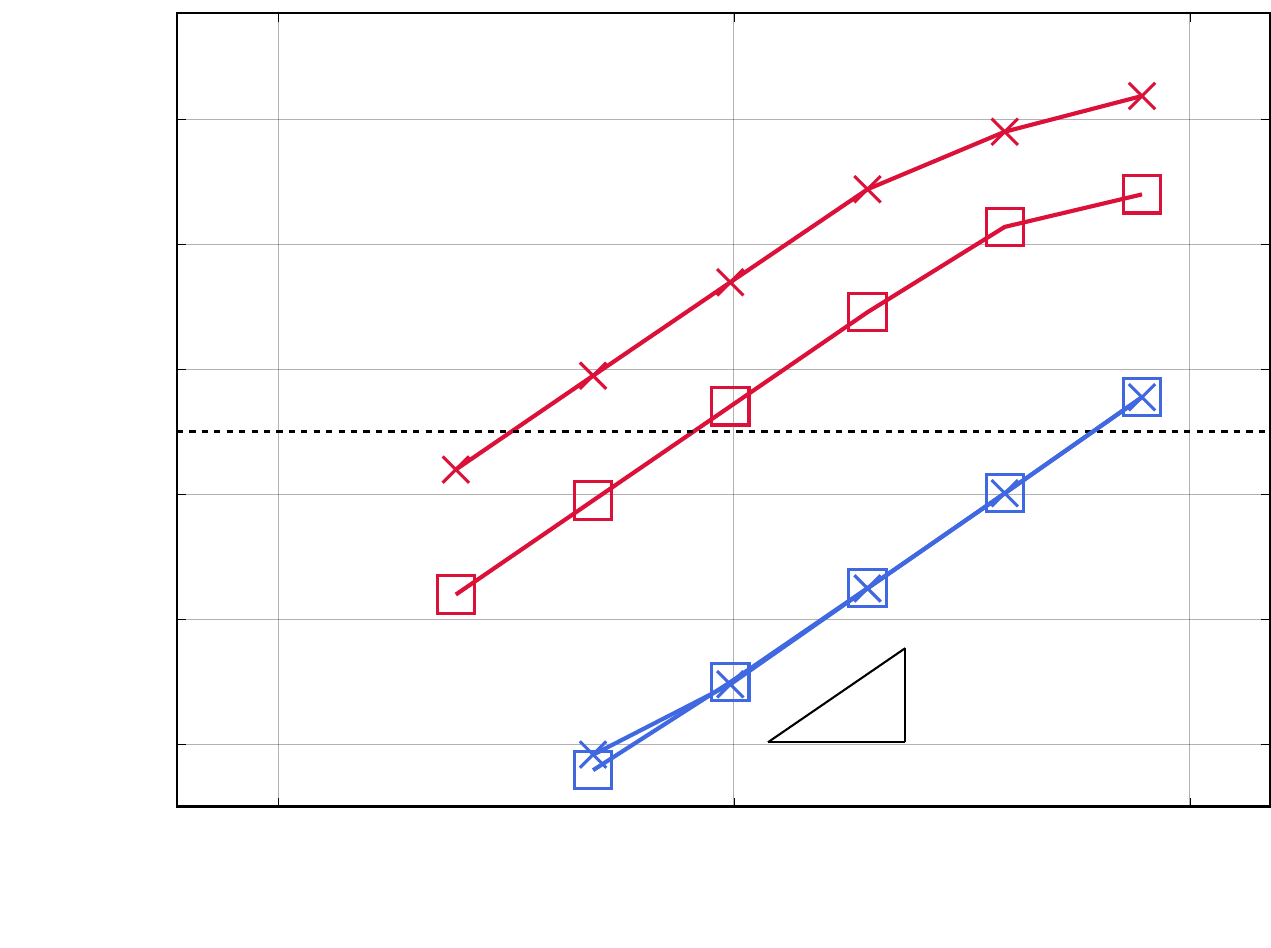 }}

    \subfloat[$p=4$]{{ \def\svgwidth{0.42\textwidth}
    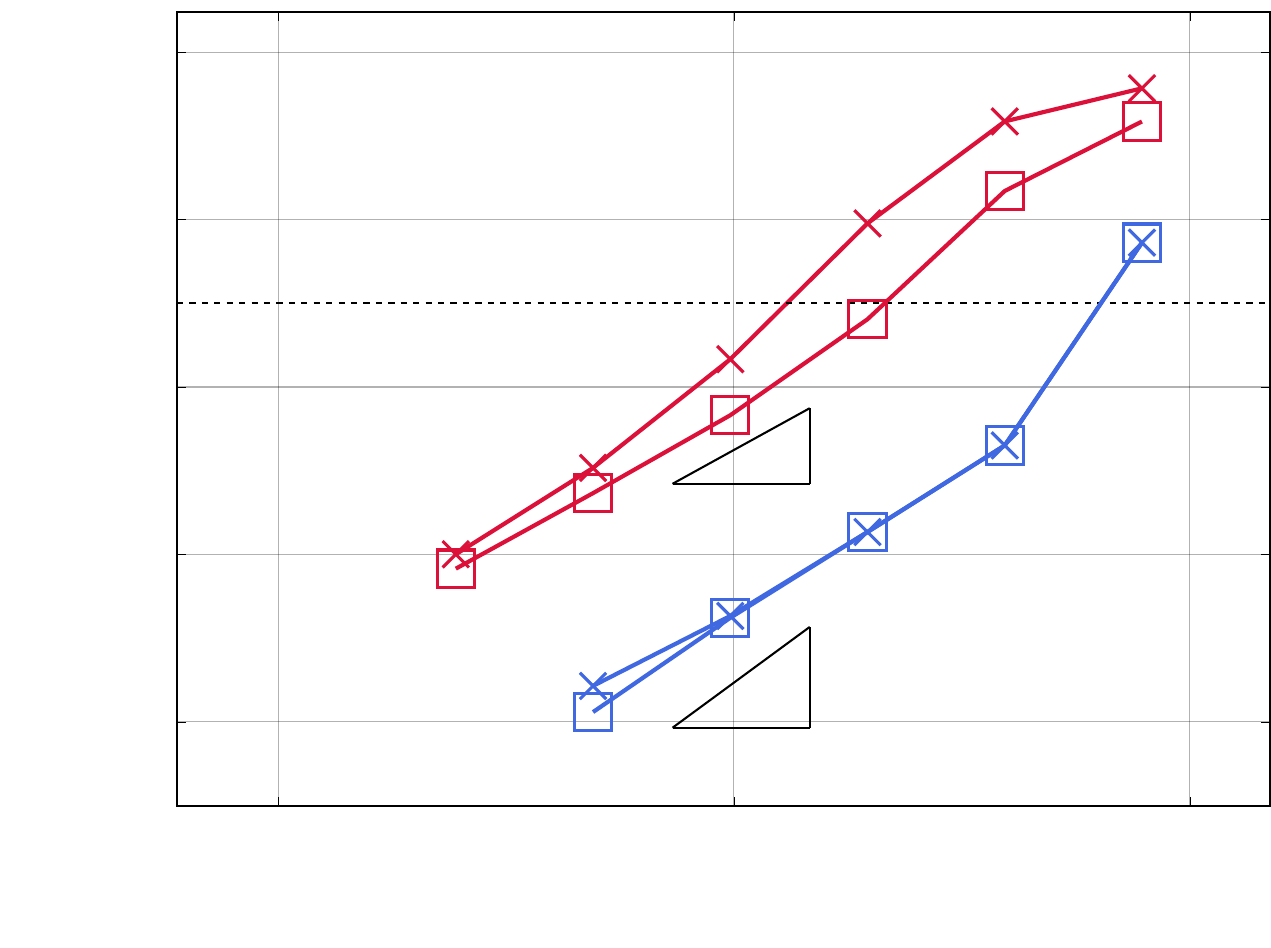 }} \hspace{0.2cm}
    \subfloat[$p=5$]{{ \def\svgwidth{0.42\textwidth}
    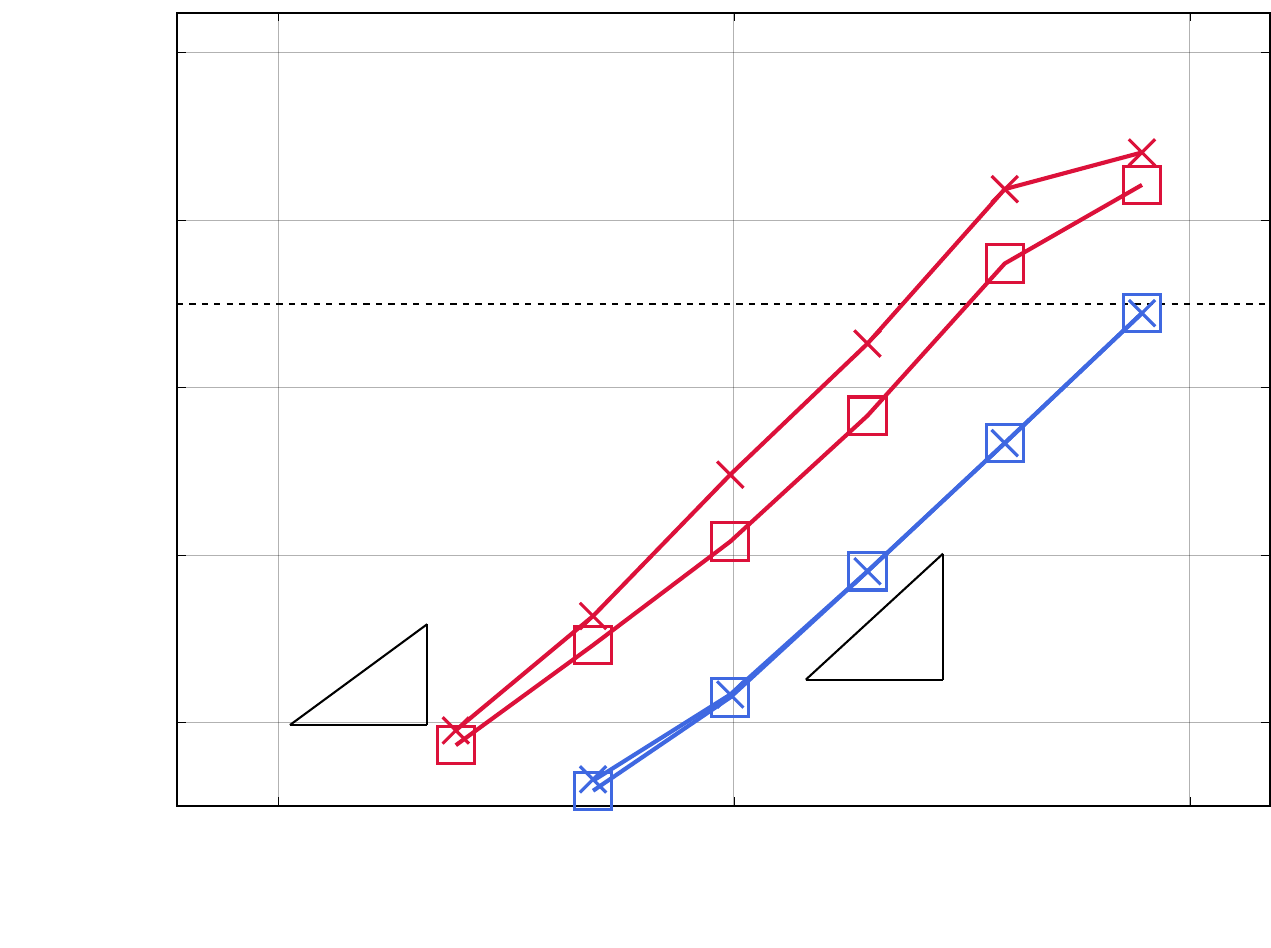 }}
    
    \vspace{0.2cm}
    \begin{tikzpicture}
    \filldraw[red1,line width=1pt, solid] (0.0,0) -- (0.6,0);
    \filldraw[red1,line width=1pt] (0.0,0) node[right]{\scriptsize $\boldsymbol{\bigtimes}$};
    \filldraw[red1,line width=1pt] (0.7,0) node[right]{\scriptsize Standard Galerkin, full integration, $R/t = 1,000$};
    \filldraw[red1,line width=1pt, solid] (8,0.05) -- (8.6,0.05);
    \filldraw[red1,line width=1pt] (8.2,-0.08) [fill=none] rectangle ++(0.25,0.25);;
    \filldraw[red1,line width=1pt] (8.7,0) node[right]{\scriptsize Standard Galerkin, full integration, $R/t = 100$};
\end{tikzpicture}

\begin{tikzpicture}
    \filldraw[blue1,line width=1pt, solid] (0.5,0) -- (1.1,0);
    \filldraw[blue1,line width=1pt] (0.5,0) node[right]{\scriptsize $\boldsymbol{\bigtimes}$};
    \filldraw[blue1,line width=1pt] (1.2,0) node[right]{\scriptsize Our approach, $R/t = 1,000$};
    \filldraw[blue1,line width=1pt, solid] (6,0.05) -- (6.6,0.05);
    \filldraw[blue1,line width=1pt] (6.2,-0.08) [fill=none] rectangle ++(0.25,0.25);;
    \filldraw[blue1,line width=1pt] (6.7,0) node[right]{\scriptsize Our approach, $R/t = 100$};
\end{tikzpicture}
    \caption{Curved Euler-Bernoulli beam: relative error in the $L^2$-norm of the displacement, membrane strain and bending strain fields, computed with our approach and the standard displacement-based formulation for slenderness ratio $R/t=100$ and $1,000$ and polynomial degrees $p=4$ and $5$.}\label{fig:beam_convergence_slenderness_ratio2}
\end{figure}

At the request of an anonymous reviewer, we provide the following additional convergence results for the curved Euler-Bernoulli beam. In Figures \ref{fig:beam_convergence_slenderness_ratio1} and \ref{fig:beam_convergence_slenderness_ratio2}, we first compare the relative $L^2$ error of the displacement field, the membrane strain field and the bending strain field computed with our formulation and the standard displacement-based formulation for slenderness ratios $R/t=100$ and $1,000$ and polynomial degrees $p=2$ to $5$. We observe that in our formulation, all fields are indeed locking-free.

\begin{figure}[ht!]
	\centering
    \captionsetup[subfloat]{labelfont=scriptsize,textfont=scriptsize}
    \subfloat[$p=2$]{{ \def\svgwidth{0.42\textwidth}
    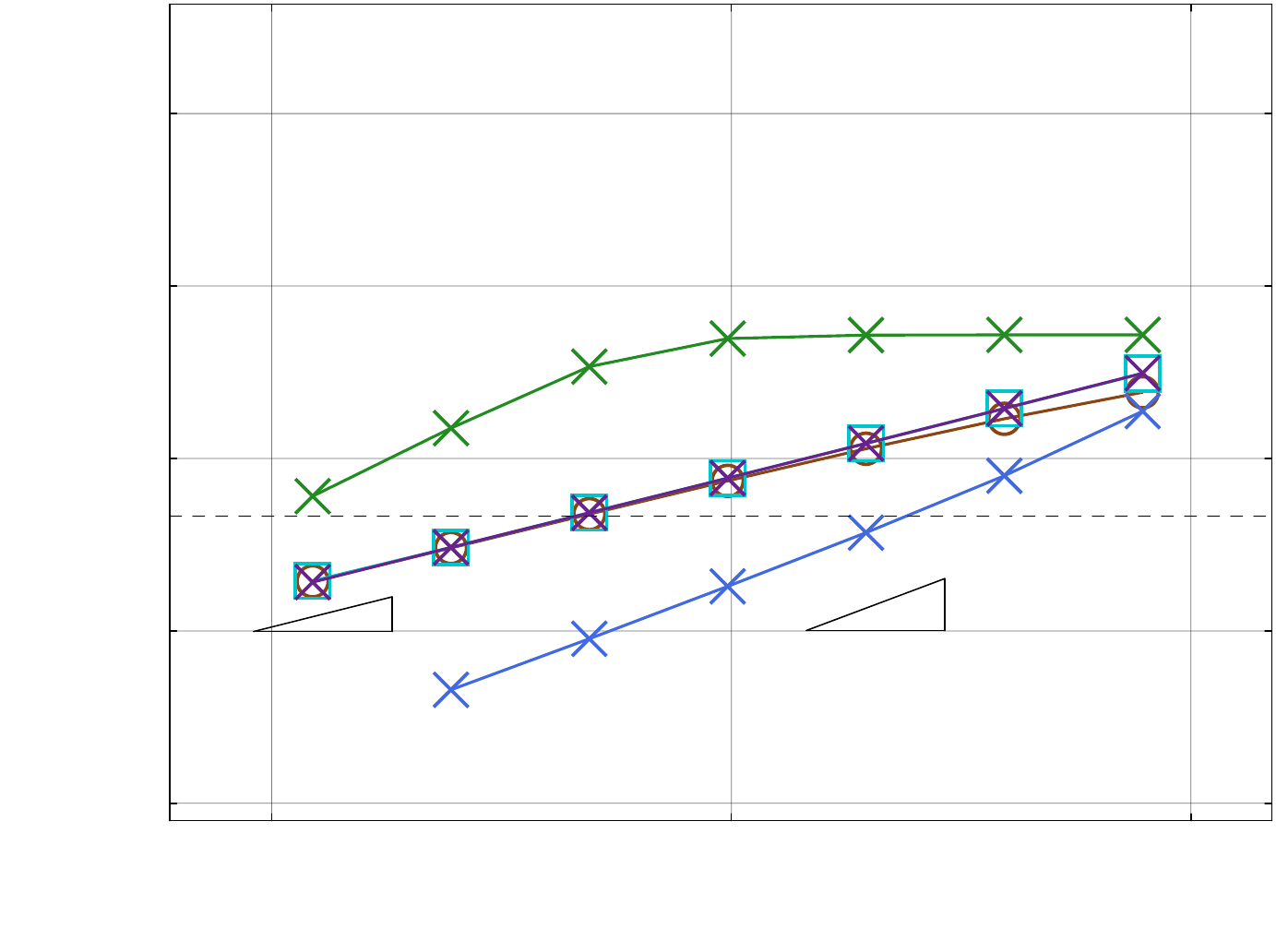 }} \hspace{0.2cm}
    \subfloat[$p=3$]{{ \def\svgwidth{0.42\textwidth}
    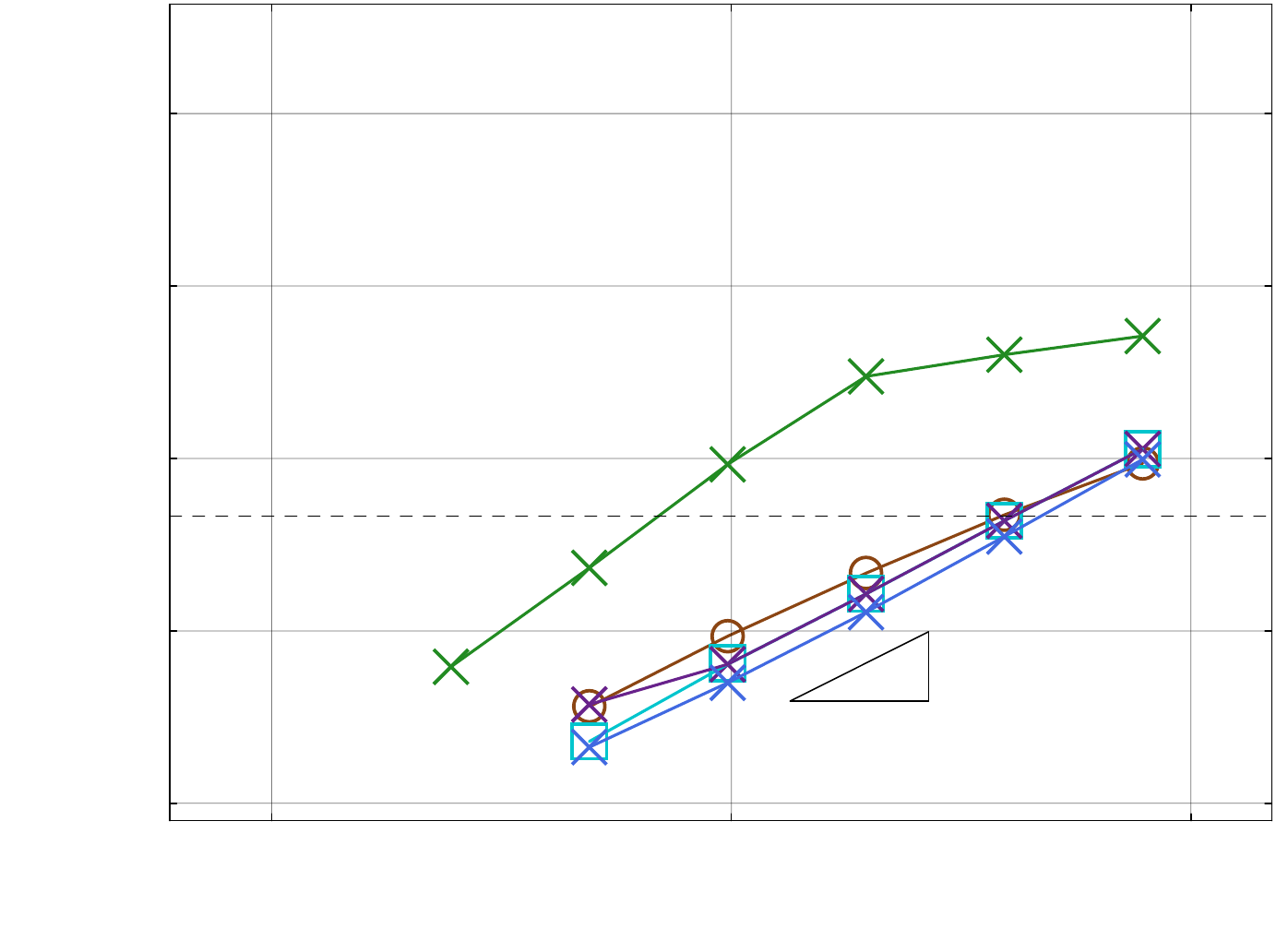 }}

    \subfloat[$p=2$]{{ \def\svgwidth{0.42\textwidth}
    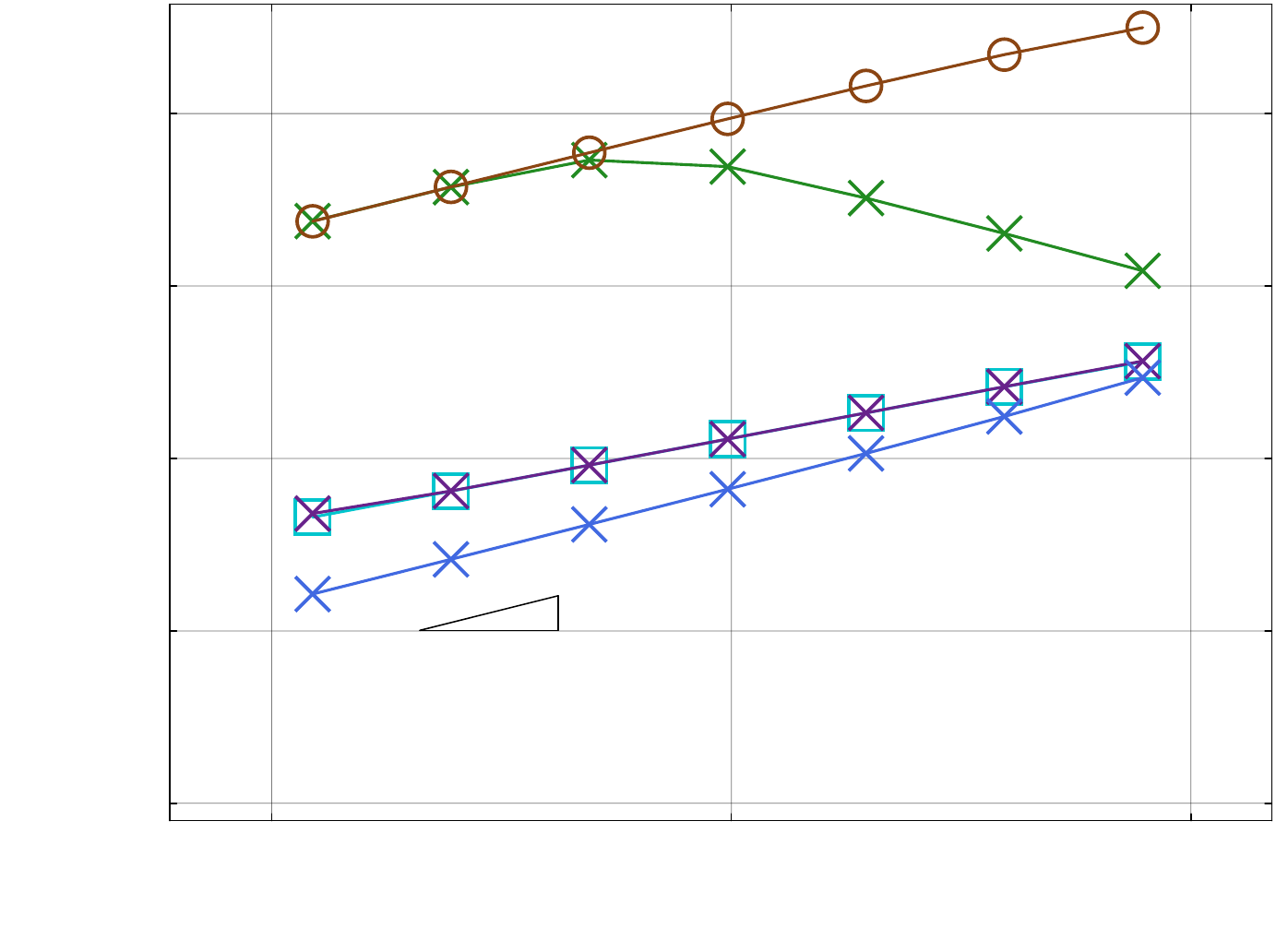 }} \hspace{0.2cm}
    \subfloat[$p=3$]{{ \def\svgwidth{0.42\textwidth}
    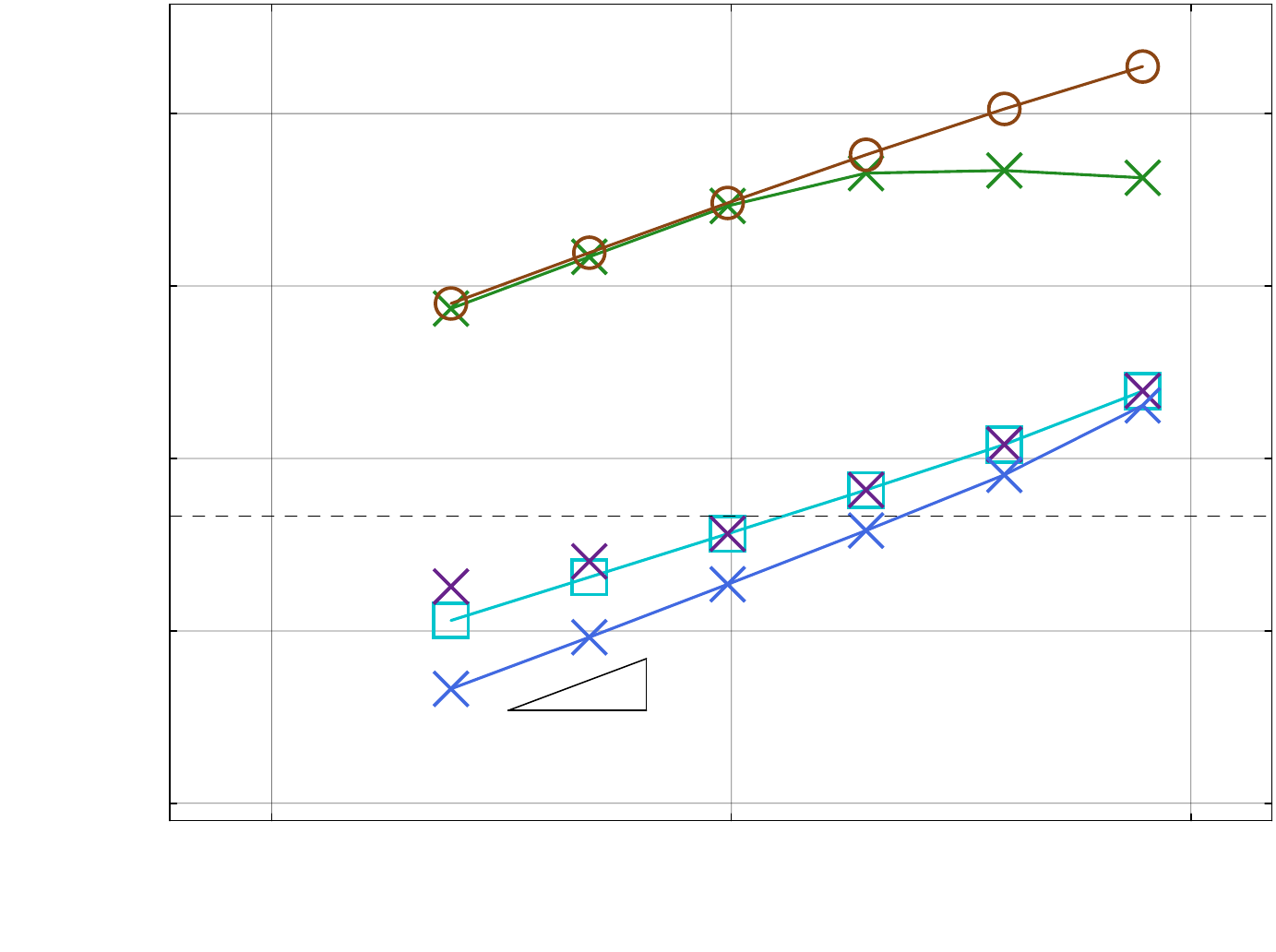 }}

    \subfloat[$p=2$]{{ \def\svgwidth{0.42\textwidth}
    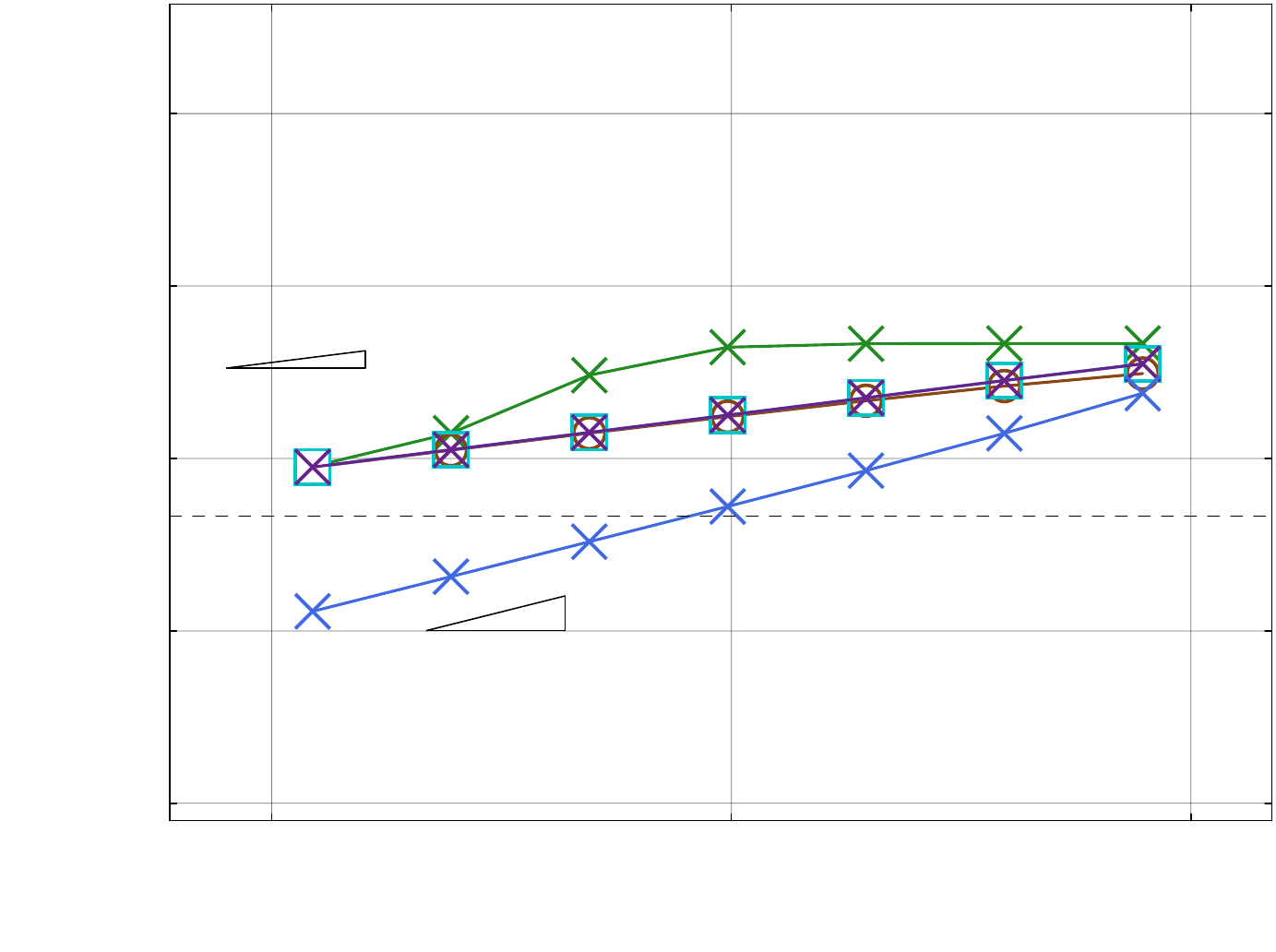 }} \hspace{0.2cm}
    \subfloat[$p=3$]{{ \def\svgwidth{0.42\textwidth}
    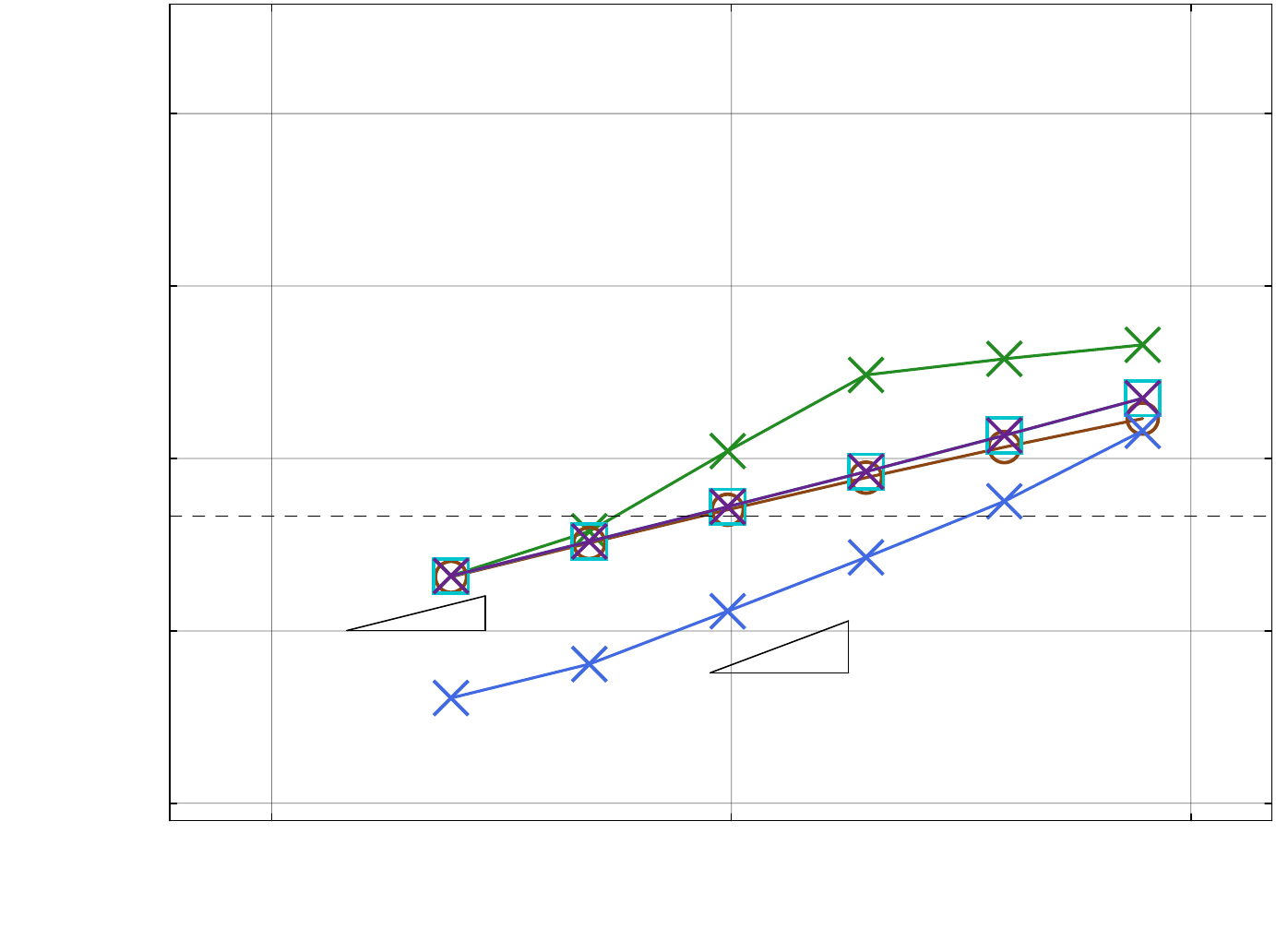 }}
    
    \vspace{0.2cm}
    \begin{tikzpicture}
    \filldraw[green1,line width=1pt, solid] (0.0,0) -- (0.6,0);
    \filldraw[green1,line width=1pt] (0.0,0) node[right]{\scriptsize $\boldsymbol{\bigtimes}$};
    \filldraw[green1,line width=1pt] (0.7,0) node[right]{\scriptsize Standard Galerkin, full integration};
    \filldraw[brown1,line width=1pt, solid] (6.0,0.05) -- (6.6,0.05);
    \filldraw[brown1,line width=1pt] (6.3,0.05) [fill=none] circle (3pt);
    \filldraw[brown1,line width=1pt] (6.7,0) node[right]{\scriptsize Standard Galerkin, {reduced integration\cite{leonetti2019simplified} ($\mathcal{S}_0^2,\mathcal{S}_1^3,\mathcal{S}_2^4,\mathcal{S}_2^5$ for $p=2,3,4,5$, respectively)}};
\end{tikzpicture}

\begin{tikzpicture}
    \filldraw[lightblue1,line width=1pt, solid] (0.5,0.05) -- (1.1,0.05);
    \filldraw[lightblue1,line width=1pt] (0.7,-0.08) [fill=none] rectangle ++(0.25,0.25);;
    \filldraw[lightblue1,line width=1pt] (1.2,0) node[right]{\scriptsize Standard Galerkin, B-bar method};
    \filldraw[purple1,line width=1pt, solid] (6.8,0) -- (7.4,0);
    \filldraw[purple1,line width=1pt] (6.8,0) node[right]{\scriptsize $\boldsymbol{\bigtimes}$};
    \filldraw[purple1,line width=1pt] (7.5,0) node[right]{\scriptsize Galerkin mixed formulation, EAS method};
    \filldraw[blue1,line width=1pt, solid] (14.0,0) -- (14.6,0);
    \filldraw[blue1,line width=1pt] (14.0,0) node[right]{\scriptsize $\boldsymbol{\bigtimes}$};
    \filldraw[blue1,line width=1pt] (14.7,0) node[right]{\scriptsize Our approach};
\end{tikzpicture}
    \caption{Curved Euler-Bernoulli beam: relative error in the $L^2$-norm of the displacement, membrane strain and bending strain fields, computed with our approach and different well-established locking-preventing mechanisms for slenderness ratio $R/t=1,000$ and polynomial degrees $p=2$ and $3$.}\label{fig:beam_convergence_methods_p23}
\end{figure}

\begin{figure}[ht!]
	\centering
    \captionsetup[subfloat]{labelfont=scriptsize,textfont=scriptsize}
    \subfloat[$p=4$]{{ \def\svgwidth{0.42\textwidth}
    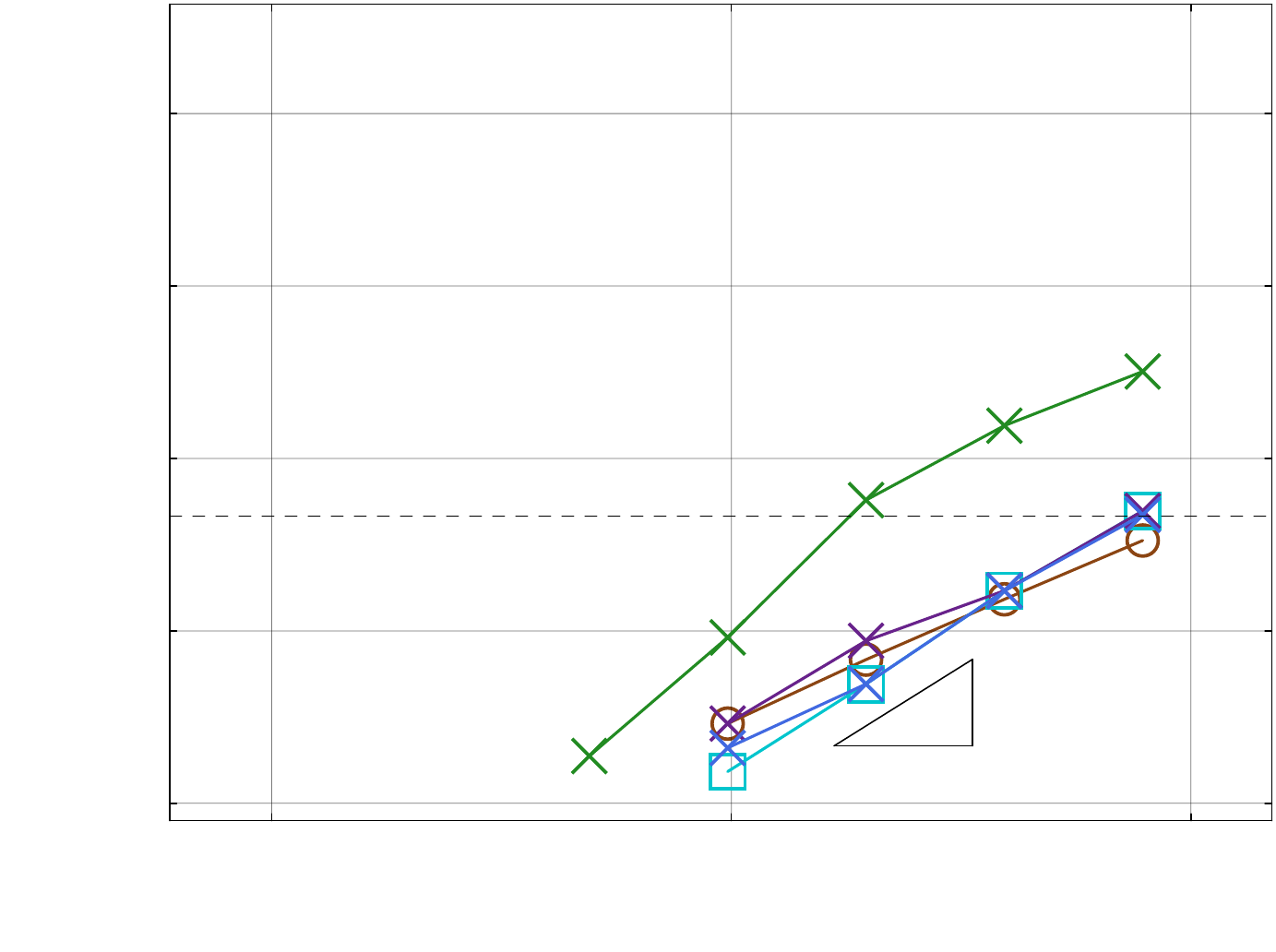 }} \hspace{0.2cm}
    \subfloat[$p=5$]{{ \def\svgwidth{0.42\textwidth}
    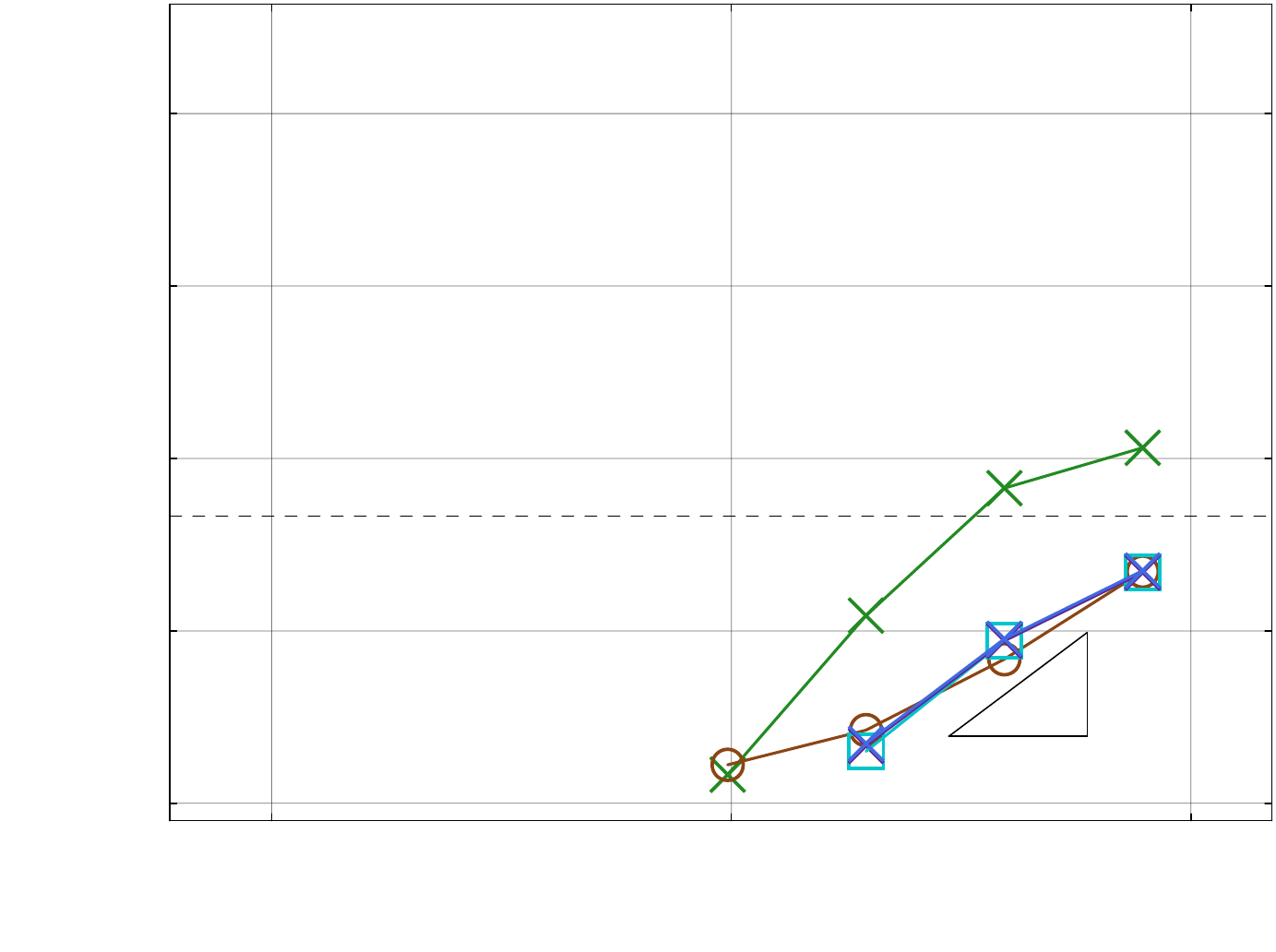 }}

    \subfloat[$p=4$]{{ \def\svgwidth{0.42\textwidth}
    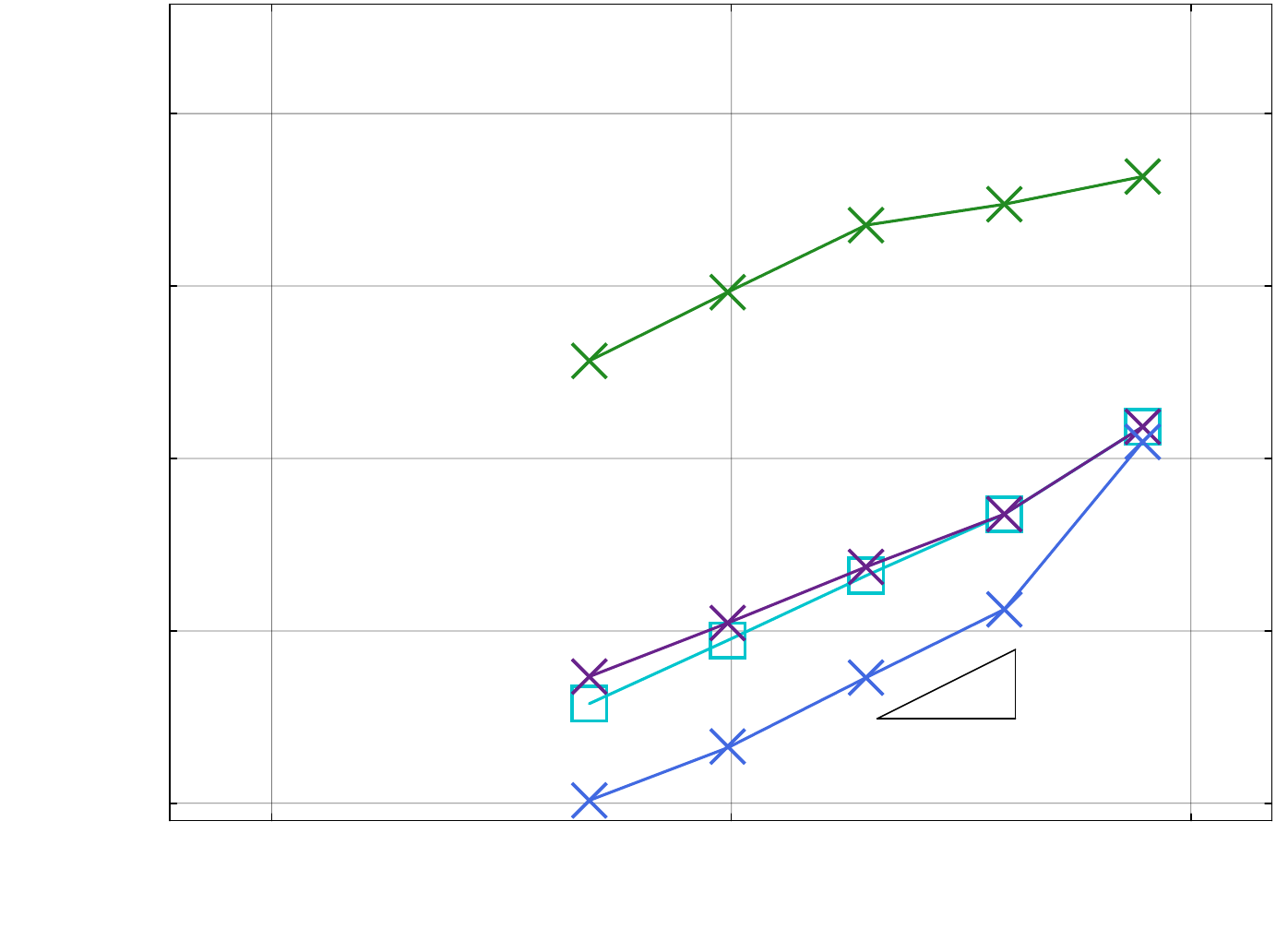 }} \hspace{0.2cm}
    \subfloat[$p=5$]{{ \def\svgwidth{0.42\textwidth}
    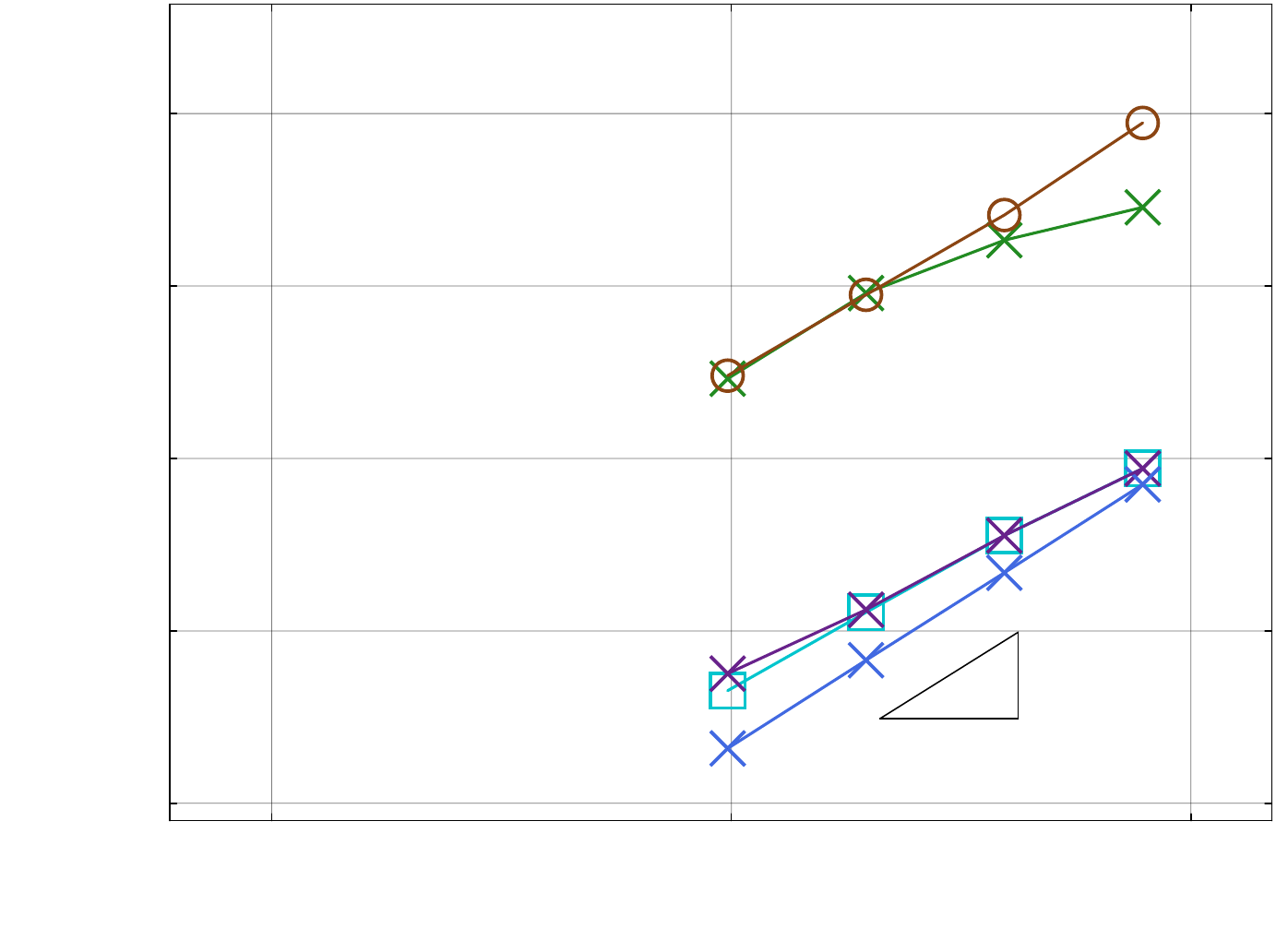 }}

    \subfloat[$p=4$]{{ \def\svgwidth{0.42\textwidth}
    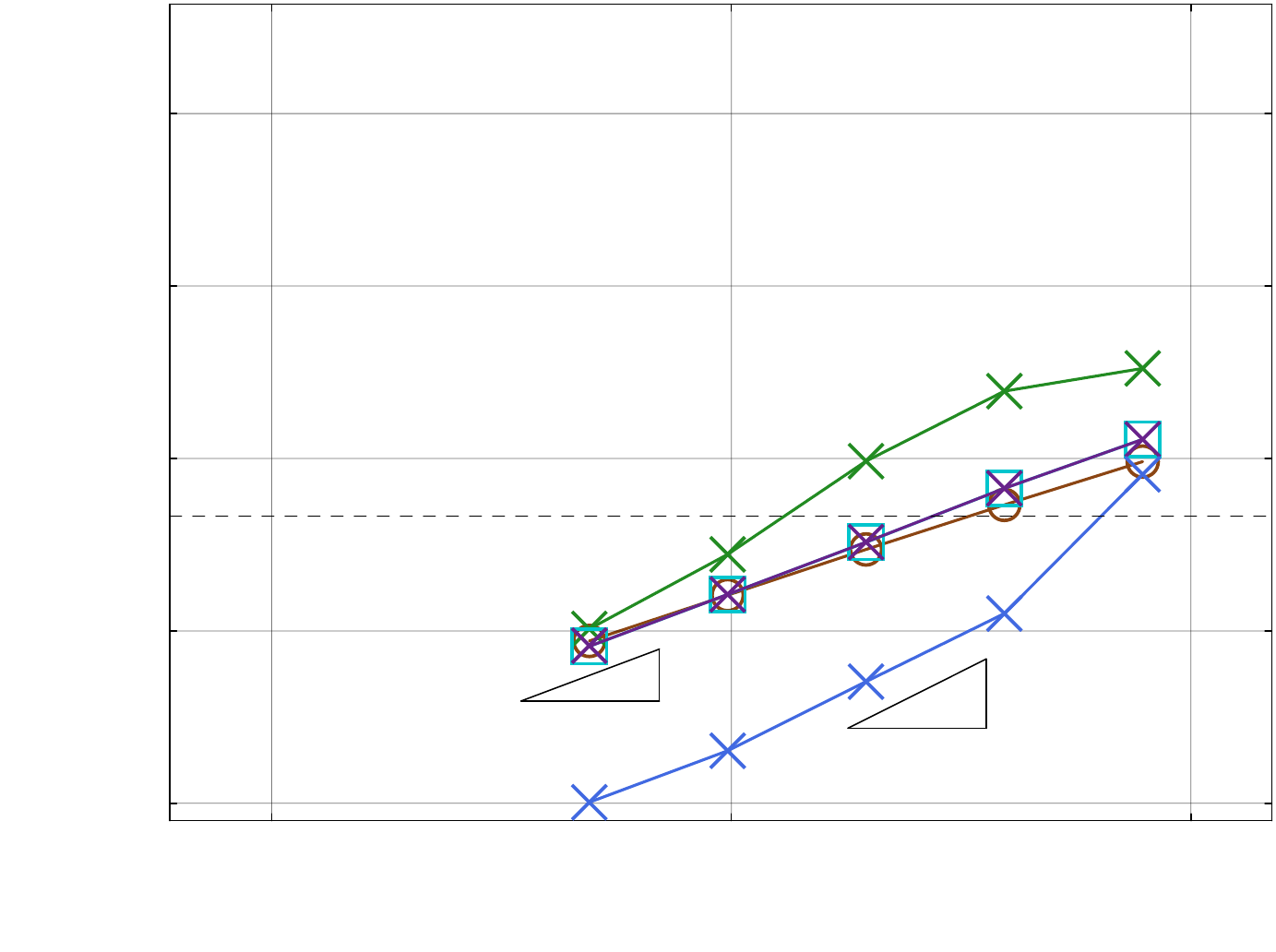 }} \hspace{0.2cm}
    \subfloat[$p=5$]{{ \def\svgwidth{0.42\textwidth}
    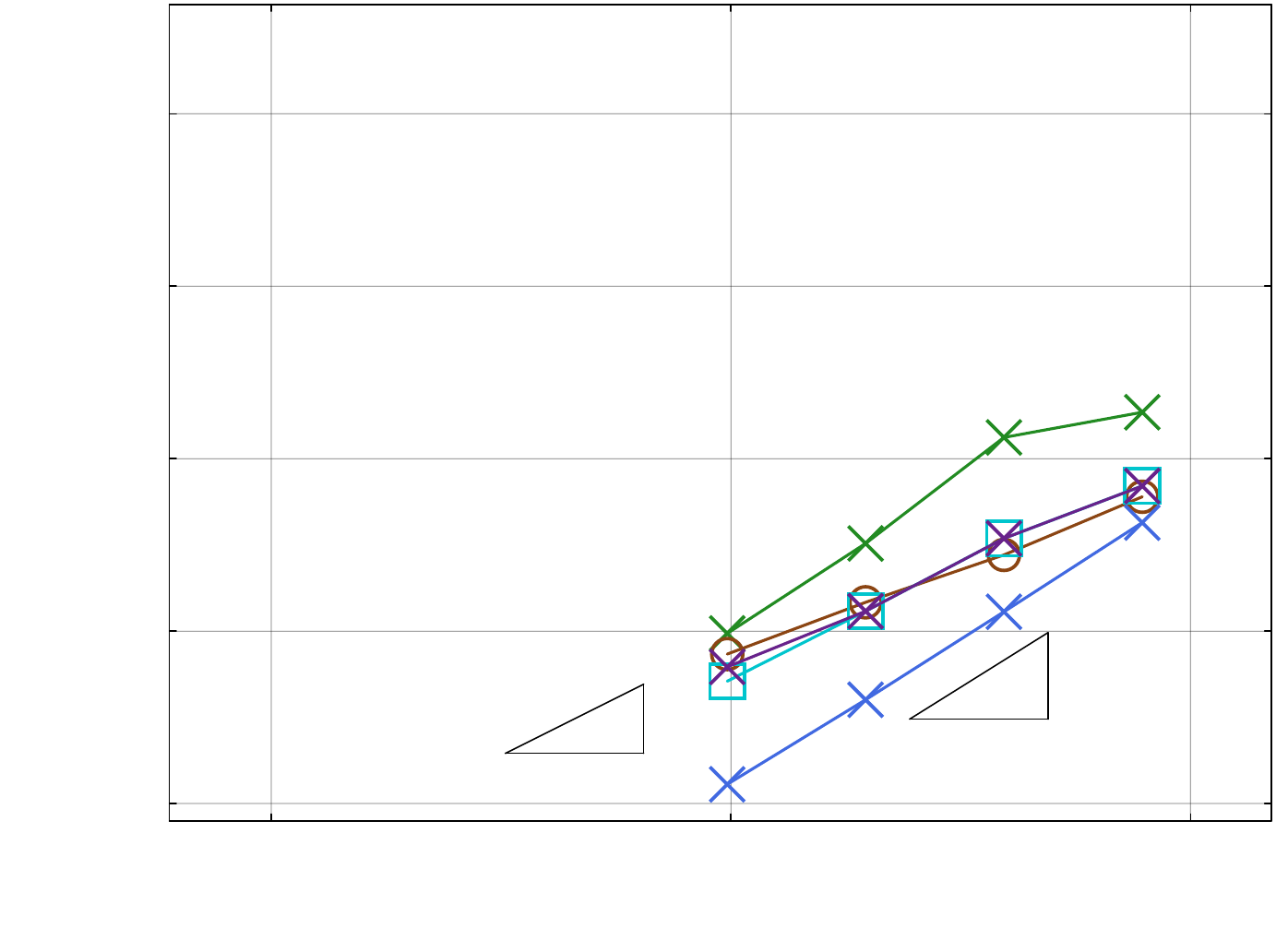 }}
    
    \vspace{0.2cm}
    \begin{tikzpicture}
    \filldraw[green1,line width=1pt, solid] (0.0,0) -- (0.6,0);
    \filldraw[green1,line width=1pt] (0.0,0) node[right]{\scriptsize $\boldsymbol{\bigtimes}$};
    \filldraw[green1,line width=1pt] (0.7,0) node[right]{\scriptsize Standard Galerkin, full integration};
    \filldraw[brown1,line width=1pt, solid] (6.0,0.05) -- (6.6,0.05);
    \filldraw[brown1,line width=1pt] (6.3,0.05) [fill=none] circle (3pt);
    \filldraw[brown1,line width=1pt] (6.7,0) node[right]{\scriptsize Standard Galerkin, {reduced integration\cite{leonetti2019simplified} ($\mathcal{S}_0^2,\mathcal{S}_1^3,\mathcal{S}_2^4,\mathcal{S}_2^5$ for $p=2,3,4,5$, respectively)}};
\end{tikzpicture}

\begin{tikzpicture}
    \filldraw[lightblue1,line width=1pt, solid] (0.5,0.05) -- (1.1,0.05);
    \filldraw[lightblue1,line width=1pt] (0.7,-0.08) [fill=none] rectangle ++(0.25,0.25);;
    \filldraw[lightblue1,line width=1pt] (1.2,0) node[right]{\scriptsize Standard Galerkin, B-bar method};
    \filldraw[purple1,line width=1pt, solid] (6.8,0) -- (7.4,0);
    \filldraw[purple1,line width=1pt] (6.8,0) node[right]{\scriptsize $\boldsymbol{\bigtimes}$};
    \filldraw[purple1,line width=1pt] (7.5,0) node[right]{\scriptsize Galerkin mixed formulation, EAS method};
    \filldraw[blue1,line width=1pt, solid] (14.0,0) -- (14.6,0);
    \filldraw[blue1,line width=1pt] (14.0,0) node[right]{\scriptsize $\boldsymbol{\bigtimes}$};
    \filldraw[blue1,line width=1pt] (14.7,0) node[right]{\scriptsize Our approach};
\end{tikzpicture}
    \caption{Curved Euler-Bernoulli beam: relative error in the $L^2$-norm of the displacement, membrane strain and bending strain fields, computed with our approach and different well-established locking-preventing mechanisms for slenderness ratio $R/t=1,000$ and polynomial degrees $p=4$ and $5$.}\label{fig:beam_convergence_methods_p45}
\end{figure}

In Figures \ref{fig:beam_convergence_methods_p23} and \ref{fig:beam_convergence_methods_p45}, we plot the convergence of the relative $L^2$ error of the displacement field, of the membrane strain field and of the bending strain field, computed with our approach (blue), a Galerkin mixed formulation with the EAS method\cite{Cardoso2012} (purple), and standard displacement-based Galerkin formulations that mitigate membrane locking via reduced integration\cite{Malkus1978} (red) and the B-bar method\cite{Bouclier2012} (light blue). We observe that our discussion in Section \ref{subsub1}, which was based on the convergence of the $L^2$ error of the displacement fields, is equivalently supported by the convergence of the $L^2$ error of the strain fields. We particularly point to the bending strain field, where mixed formulations achieve a better accuracy due to an increased convergence rate.

\section{Additional results for the shell obstacle course}\label{sec:addshells}

\begin{figure}[t!]
    \centering
    \captionsetup[subfloat]{labelfont=scriptsize,textfont=scriptsize}
    \subfloat[Quadratic splines - $p=2$]{{
        \def\svgwidth{0.43\textwidth}
\begingroup%
  \makeatletter%
  \providecommand\color[2][]{%
    \errmessage{(Inkscape) Color is used for the text in Inkscape, but the package 'color.sty' is not loaded}%
    \renewcommand\color[2][]{}%
  }%
  \providecommand\transparent[1]{%
    \errmessage{(Inkscape) Transparency is used (non-zero) for the text in Inkscape, but the package 'transparent.sty' is not loaded}%
    \renewcommand\transparent[1]{}%
  }%
  \providecommand\rotatebox[2]{#2}%
  \newcommand*\fsize{\dimexpr\f@size pt\relax}%
  \newcommand*\lineheight[1]{\fontsize{\fsize}{#1\fsize}\selectfont}%
  \ifx\svgwidth\undefined%
    \setlength{\unitlength}{705bp}%
    \ifx\svgscale\undefined%
      \relax%
    \else%
      \setlength{\unitlength}{\unitlength * \real{\svgscale}}%
    \fi%
  \else%
    \setlength{\unitlength}{\svgwidth}%
  \fi%
  \global\let\svgwidth\undefined%
  \global\let\svgscale\undefined%
  \makeatother%
  \begin{picture}(1,0.60106383)%
    \lineheight{1}%
    \setlength\tabcolsep{0pt}%
    \put(0,0){\includegraphics[width=\unitlength,page=1]{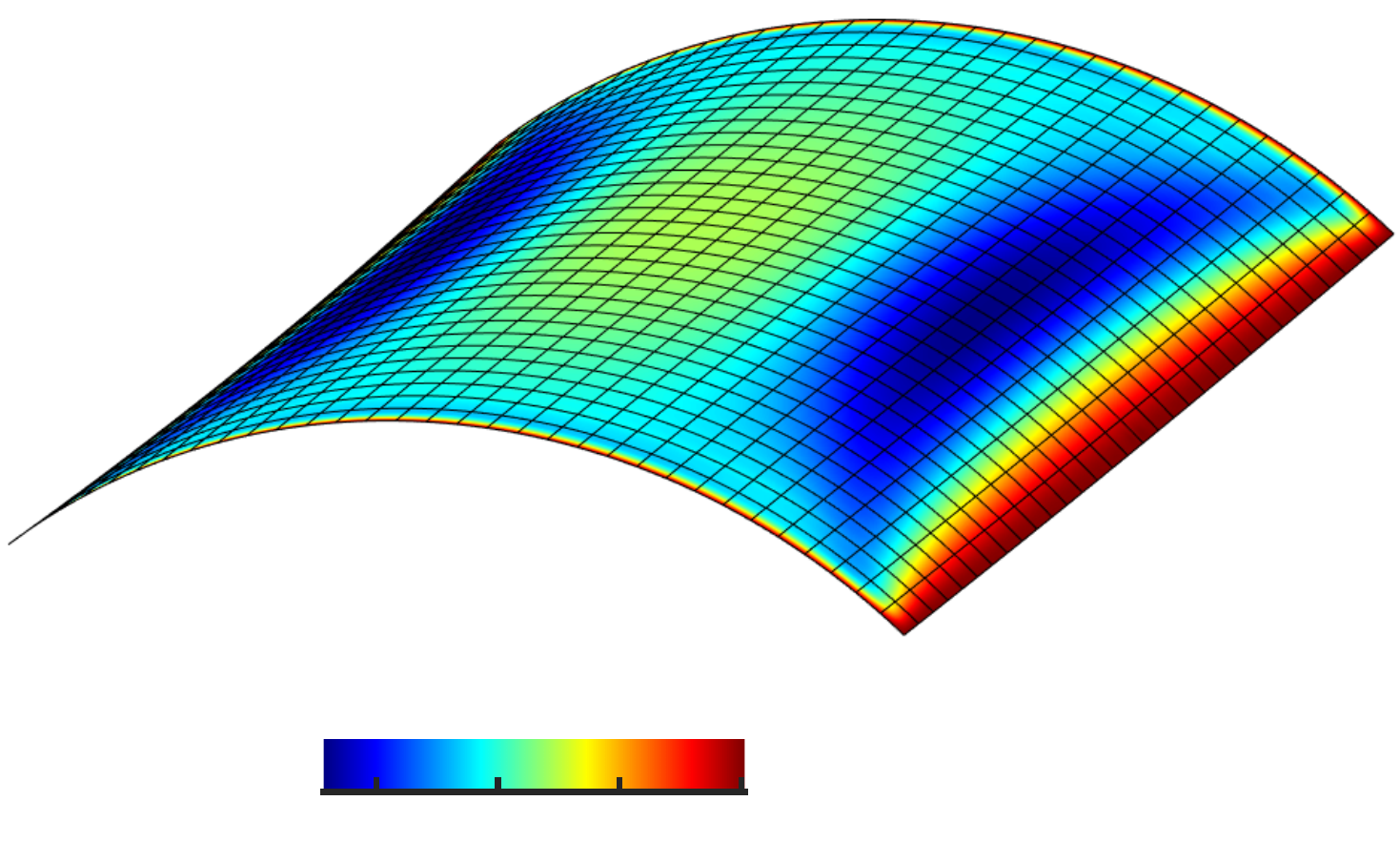}}%
    \put(0.54002038,0.05882767){\color[rgb]{0,0,0}\makebox(0,0)[lt]{\lineheight{1.25}\smash{\begin{tabular}[t]{l}$n^{11}$\end{tabular}}}}%
    \put(0.52113929,0.00013862){\color[rgb]{0,0,0}\makebox(0,0)[lt]{\lineheight{1.25}\smash{\begin{tabular}[t]{l}0\end{tabular}}}}%
    \put(0.41855161,0.0002037){\color[rgb]{0,0,0}\makebox(0,0)[lt]{\lineheight{1.25}\smash{\begin{tabular}[t]{l}-10\end{tabular}}}}%
    \put(0.33478518,0.0002037){\color[rgb]{0,0,0}\makebox(0,0)[lt]{\lineheight{1.25}\smash{\begin{tabular}[t]{l}-20\end{tabular}}}}%
    \put(0.25015732,0.00026884){\color[rgb]{0,0,0}\makebox(0,0)[lt]{\lineheight{1.25}\smash{\begin{tabular}[t]{l}-30\end{tabular}}}}%
  \end{picture}%
\endgroup%

        }} \hspace{1cm}
    \subfloat[Cubic splines - $p=3$]{{
        \def\svgwidth{0.43\textwidth}
\begingroup%
  \makeatletter%
  \providecommand\color[2][]{%
    \errmessage{(Inkscape) Color is used for the text in Inkscape, but the package 'color.sty' is not loaded}%
    \renewcommand\color[2][]{}%
  }%
  \providecommand\transparent[1]{%
    \errmessage{(Inkscape) Transparency is used (non-zero) for the text in Inkscape, but the package 'transparent.sty' is not loaded}%
    \renewcommand\transparent[1]{}%
  }%
  \providecommand\rotatebox[2]{#2}%
  \newcommand*\fsize{\dimexpr\f@size pt\relax}%
  \newcommand*\lineheight[1]{\fontsize{\fsize}{#1\fsize}\selectfont}%
  \ifx\svgwidth\undefined%
    \setlength{\unitlength}{705bp}%
    \ifx\svgscale\undefined%
      \relax%
    \else%
      \setlength{\unitlength}{\unitlength * \real{\svgscale}}%
    \fi%
  \else%
    \setlength{\unitlength}{\svgwidth}%
  \fi%
  \global\let\svgwidth\undefined%
  \global\let\svgscale\undefined%
  \makeatother%
  \begin{picture}(1,0.60106383)%
    \lineheight{1}%
    \setlength\tabcolsep{0pt}%
    \put(0,0){\includegraphics[width=\unitlength,page=1]{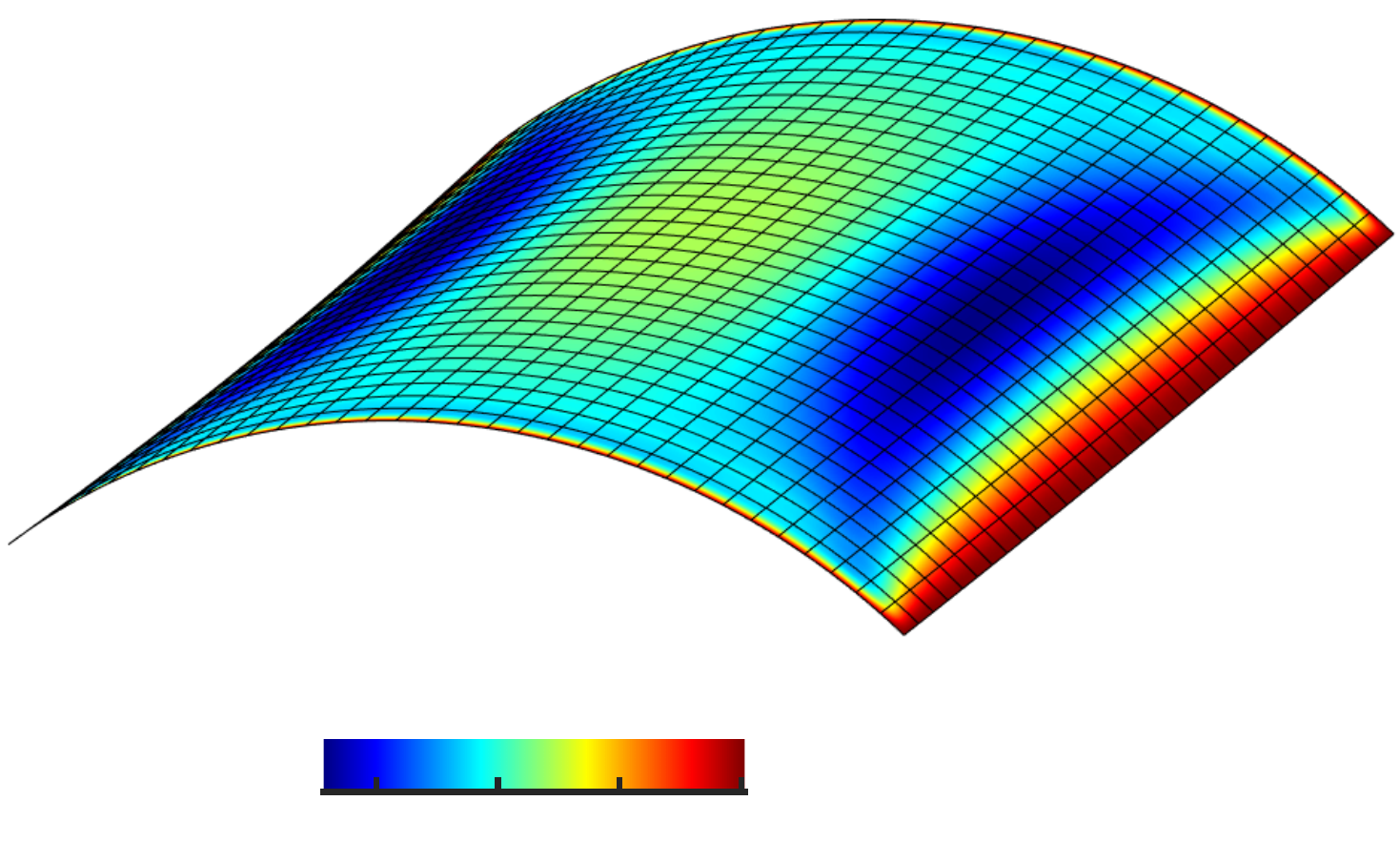}}%
    \put(0.54002038,0.05882767){\color[rgb]{0,0,0}\makebox(0,0)[lt]{\lineheight{1.25}\smash{\begin{tabular}[t]{l}$n^{11}$\end{tabular}}}}%
    \put(0.52113929,0.00013862){\color[rgb]{0,0,0}\makebox(0,0)[lt]{\lineheight{1.25}\smash{\begin{tabular}[t]{l}0\end{tabular}}}}%
    \put(0.41855161,0.0002037){\color[rgb]{0,0,0}\makebox(0,0)[lt]{\lineheight{1.25}\smash{\begin{tabular}[t]{l}-10\end{tabular}}}}%
    \put(0.33478518,0.0002037){\color[rgb]{0,0,0}\makebox(0,0)[lt]{\lineheight{1.25}\smash{\begin{tabular}[t]{l}-20\end{tabular}}}}%
    \put(0.25015732,0.00026884){\color[rgb]{0,0,0}\makebox(0,0)[lt]{\lineheight{1.25}\smash{\begin{tabular}[t]{l}-30\end{tabular}}}}%
  \end{picture}%
\endgroup%

        }}

    \caption{Scordelis-Lo roof: the membrane stress resultant $n^{11}$ for a slenderness ratio of $R/t = 1,000$, computed with our approach on a mesh of $16 \times 16$ elements. To enable a comparison with the results given in \cite{Sauer_locking2024}, we scale the self-weight with $d^2$, that is, we apply a self-weight of $1.44\cdot10^3\,d^2$ per unit area.} \label{Scord1}
\end{figure}

\begin{figure}[t!]
    \centering
    \captionsetup[subfloat]{labelfont=scriptsize,textfont=scriptsize}
    \subfloat[Quadratic splines - $p=2$]{{
        \def\svgwidth{0.43\textwidth}
\begingroup%
  \makeatletter%
  \providecommand\color[2][]{%
    \errmessage{(Inkscape) Color is used for the text in Inkscape, but the package 'color.sty' is not loaded}%
    \renewcommand\color[2][]{}%
  }%
  \providecommand\transparent[1]{%
    \errmessage{(Inkscape) Transparency is used (non-zero) for the text in Inkscape, but the package 'transparent.sty' is not loaded}%
    \renewcommand\transparent[1]{}%
  }%
  \providecommand\rotatebox[2]{#2}%
  \newcommand*\fsize{\dimexpr\f@size pt\relax}%
  \newcommand*\lineheight[1]{\fontsize{\fsize}{#1\fsize}\selectfont}%
  \ifx\svgwidth\undefined%
    \setlength{\unitlength}{705bp}%
    \ifx\svgscale\undefined%
      \relax%
    \else%
      \setlength{\unitlength}{\unitlength * \real{\svgscale}}%
    \fi%
  \else%
    \setlength{\unitlength}{\svgwidth}%
  \fi%
  \global\let\svgwidth\undefined%
  \global\let\svgscale\undefined%
  \makeatother%
  \begin{picture}(1,0.60106383)%
    \lineheight{1}%
    \setlength\tabcolsep{0pt}%
    \put(0,0){\includegraphics[width=\unitlength,page=1]{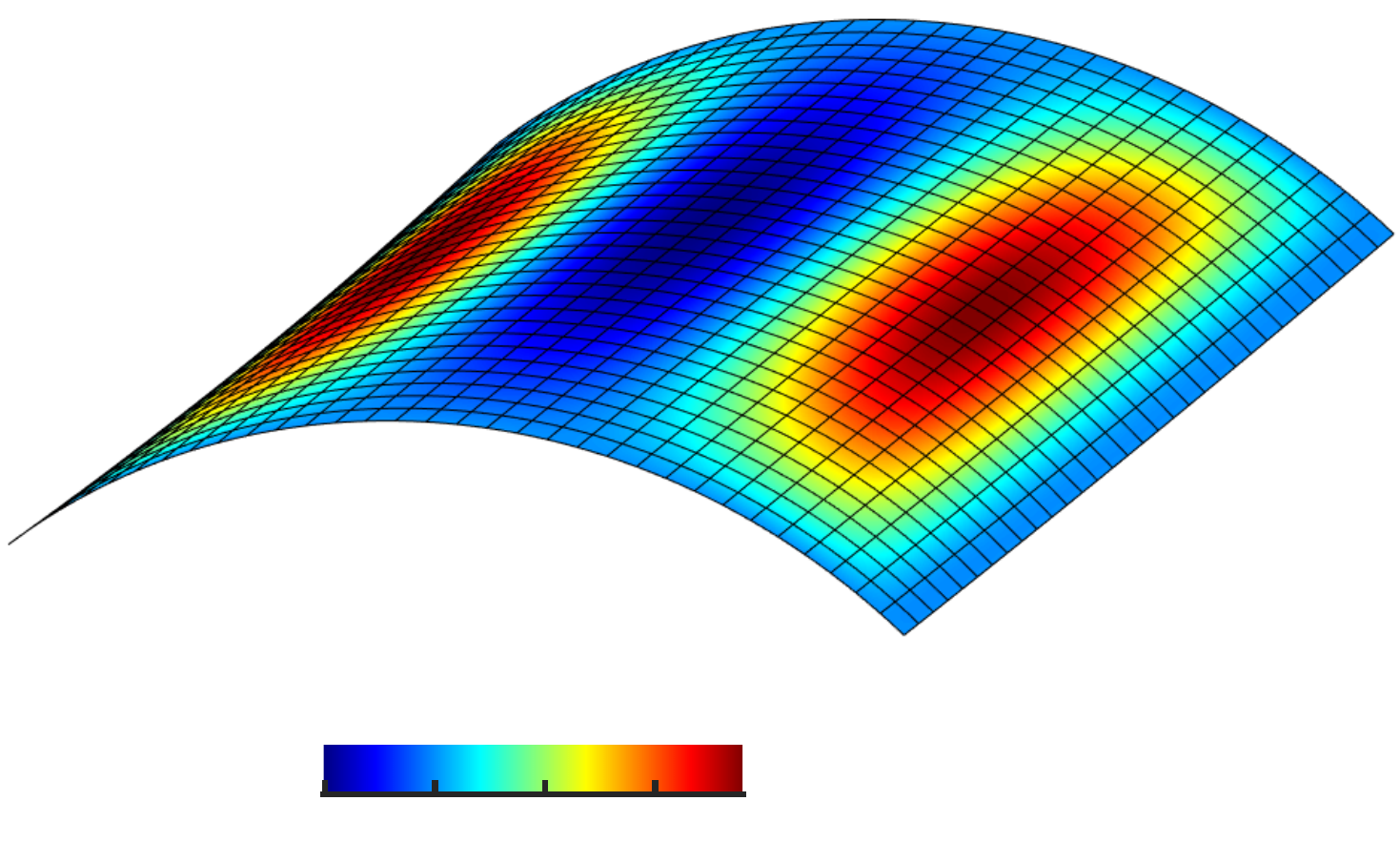}}%
    \put(0.53950193,0.04313231){\color[rgb]{0,0,0}\makebox(0,0)[lt]{\lineheight{1.25}\smash{\begin{tabular}[t]{l}$m^{11}$\end{tabular}}}}%
    \put(0.38370154,0.00190497){\color[rgb]{0,0,0}\makebox(0,0)[lt]{\lineheight{1.25}\smash{\begin{tabular}[t]{l}$2$\end{tabular}}}}%
    \put(0.21408098,0.00175999){\color[rgb]{0,0,0}\makebox(0,0)[lt]{\lineheight{1.25}\smash{\begin{tabular}[t]{l}$-2$\end{tabular}}}}%
    \put(0.3046175,0.00168749){\color[rgb]{0,0,0}\makebox(0,0)[lt]{\lineheight{1.25}\smash{\begin{tabular}[t]{l}$0$\end{tabular}}}}%
    \put(0.46168217,0.00173586){\color[rgb]{0,0,0}\makebox(0,0)[lt]{\lineheight{1.25}\smash{\begin{tabular}[t]{l}$4$\end{tabular}}}}%
  \end{picture}%
\endgroup%

        }} \hspace{1cm}
    \subfloat[Cubic splines - $p=3$]{{
        \def\svgwidth{0.43\textwidth}
\begingroup%
  \makeatletter%
  \providecommand\color[2][]{%
    \errmessage{(Inkscape) Color is used for the text in Inkscape, but the package 'color.sty' is not loaded}%
    \renewcommand\color[2][]{}%
  }%
  \providecommand\transparent[1]{%
    \errmessage{(Inkscape) Transparency is used (non-zero) for the text in Inkscape, but the package 'transparent.sty' is not loaded}%
    \renewcommand\transparent[1]{}%
  }%
  \providecommand\rotatebox[2]{#2}%
  \newcommand*\fsize{\dimexpr\f@size pt\relax}%
  \newcommand*\lineheight[1]{\fontsize{\fsize}{#1\fsize}\selectfont}%
  \ifx\svgwidth\undefined%
    \setlength{\unitlength}{705bp}%
    \ifx\svgscale\undefined%
      \relax%
    \else%
      \setlength{\unitlength}{\unitlength * \real{\svgscale}}%
    \fi%
  \else%
    \setlength{\unitlength}{\svgwidth}%
  \fi%
  \global\let\svgwidth\undefined%
  \global\let\svgscale\undefined%
  \makeatother%
  \begin{picture}(1,0.60106383)%
    \lineheight{1}%
    \setlength\tabcolsep{0pt}%
    \put(0,0){\includegraphics[width=\unitlength,page=1]{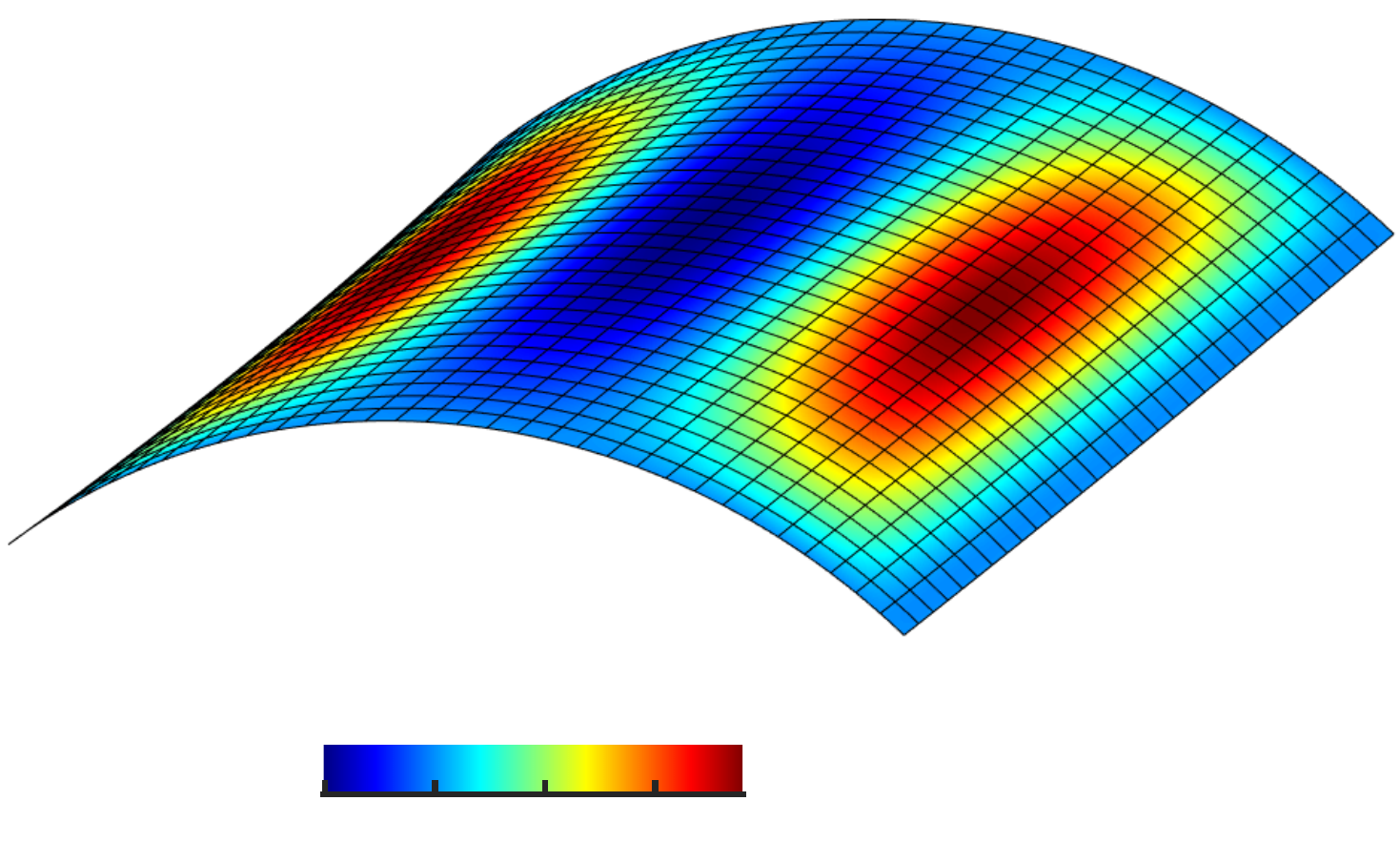}}%
    \put(0.53950193,0.04313231){\color[rgb]{0,0,0}\makebox(0,0)[lt]{\lineheight{1.25}\smash{\begin{tabular}[t]{l}$m^{11}$\end{tabular}}}}%
    \put(0.38370153,0.00190498){\color[rgb]{0,0,0}\makebox(0,0)[lt]{\lineheight{1.25}\smash{\begin{tabular}[t]{l}$2$\end{tabular}}}}%
    \put(0.21408097,0.00175999){\color[rgb]{0,0,0}\makebox(0,0)[lt]{\lineheight{1.25}\smash{\begin{tabular}[t]{l}$-2$\end{tabular}}}}%
    \put(0.3046175,0.00168749){\color[rgb]{0,0,0}\makebox(0,0)[lt]{\lineheight{1.25}\smash{\begin{tabular}[t]{l}$0$\end{tabular}}}}%
    \put(0.46168217,0.00173586){\color[rgb]{0,0,0}\makebox(0,0)[lt]{\lineheight{1.25}\smash{\begin{tabular}[t]{l}$4$\end{tabular}}}}%
  \end{picture}%
\endgroup%

        }}

    \caption{Scordelis-Lo roof: the bending moment $m^{11}$ for a slenderness ratio of $R/t = 1,000$, computed with our approach on a mesh of $16 \times 16$ elements. We again apply a self-weight of $1.44\cdot10^3\,d^2$ per unit area.} \label{Scord2}
\end{figure}

\begin{figure}
    \centering
    \captionsetup[subfloat]{labelfont=scriptsize,textfont=scriptsize}
    \subfloat[$n^{12}$]{{
        \def\svgwidth{0.43\textwidth}
\begingroup%
  \makeatletter%
  \providecommand\color[2][]{%
    \errmessage{(Inkscape) Color is used for the text in Inkscape, but the package 'color.sty' is not loaded}%
    \renewcommand\color[2][]{}%
  }%
  \providecommand\transparent[1]{%
    \errmessage{(Inkscape) Transparency is used (non-zero) for the text in Inkscape, but the package 'transparent.sty' is not loaded}%
    \renewcommand\transparent[1]{}%
  }%
  \providecommand\rotatebox[2]{#2}%
  \newcommand*\fsize{\dimexpr\f@size pt\relax}%
  \newcommand*\lineheight[1]{\fontsize{\fsize}{#1\fsize}\selectfont}%
  \ifx\svgwidth\undefined%
    \setlength{\unitlength}{622.5bp}%
    \ifx\svgscale\undefined%
      \relax%
    \else%
      \setlength{\unitlength}{\unitlength * \real{\svgscale}}%
    \fi%
  \else%
    \setlength{\unitlength}{\svgwidth}%
  \fi%
  \global\let\svgwidth\undefined%
  \global\let\svgscale\undefined%
  \makeatother%
  \begin{picture}(1,0.93975904)%
    \lineheight{1}%
    \setlength\tabcolsep{0pt}%
    \put(0,0){\includegraphics[width=\unitlength,page=1]{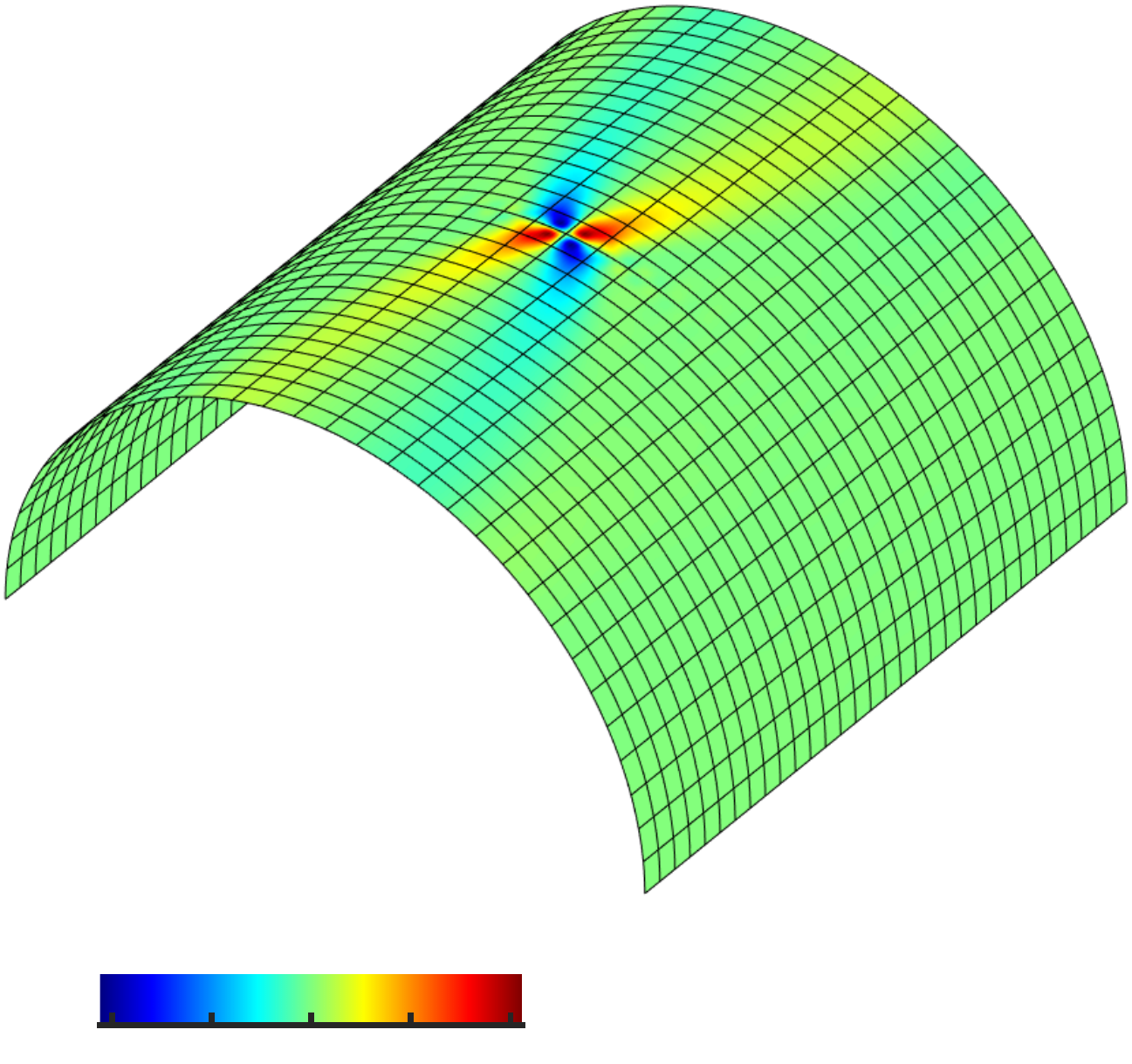}}%
    \put(0.47602571,0.07676062){\color[rgb]{0,0,0}\makebox(0,0)[lt]{\lineheight{1.25}\smash{\begin{tabular}[t]{l}$n^{12}$\end{tabular}}}}%
    \put(0.47570389,0.03706974){\color[rgb]{0,0,0}\makebox(0,0)[lt]{\lineheight{1.25}\smash{\begin{tabular}[t]{l}$\times 10^{-2}$\end{tabular}}}}%
    \put(0.35223546,0.00025466){\color[rgb]{0,0,0}\makebox(0,0)[lt]{\lineheight{1.25}\smash{\begin{tabular}[t]{l}8\end{tabular}}}}%
    \put(0.26097965,0.00032432){\color[rgb]{0,0,0}\makebox(0,0)[lt]{\lineheight{1.25}\smash{\begin{tabular}[t]{l}0\end{tabular}}}}%
    \put(0.16196278,0.00032432){\color[rgb]{0,0,0}\makebox(0,0)[lt]{\lineheight{1.25}\smash{\begin{tabular}[t]{l}-8\end{tabular}}}}%
    \put(0.0598358,0.00014838){\color[rgb]{0,0,0}\makebox(0,0)[lt]{\lineheight{1.25}\smash{\begin{tabular}[t]{l}-16\end{tabular}}}}%
    \put(0.43417091,0.00025466){\color[rgb]{0,0,0}\makebox(0,0)[lt]{\lineheight{1.25}\smash{\begin{tabular}[t]{l}16\end{tabular}}}}%
  \end{picture}%
\endgroup%

        }}\hspace{0.4cm}
    \subfloat[$m^{11}$]{{
        \def\svgwidth{0.43\textwidth}
\begingroup%
  \makeatletter%
  \providecommand\color[2][]{%
    \errmessage{(Inkscape) Color is used for the text in Inkscape, but the package 'color.sty' is not loaded}%
    \renewcommand\color[2][]{}%
  }%
  \providecommand\transparent[1]{%
    \errmessage{(Inkscape) Transparency is used (non-zero) for the text in Inkscape, but the package 'transparent.sty' is not loaded}%
    \renewcommand\transparent[1]{}%
  }%
  \providecommand\rotatebox[2]{#2}%
  \newcommand*\fsize{\dimexpr\f@size pt\relax}%
  \newcommand*\lineheight[1]{\fontsize{\fsize}{#1\fsize}\selectfont}%
  \ifx\svgwidth\undefined%
    \setlength{\unitlength}{622.5bp}%
    \ifx\svgscale\undefined%
      \relax%
    \else%
      \setlength{\unitlength}{\unitlength * \real{\svgscale}}%
    \fi%
  \else%
    \setlength{\unitlength}{\svgwidth}%
  \fi%
  \global\let\svgwidth\undefined%
  \global\let\svgscale\undefined%
  \makeatother%
  \begin{picture}(1,0.93975904)%
    \lineheight{1}%
    \setlength\tabcolsep{0pt}%
    \put(0,0){\includegraphics[width=\unitlength,page=1]{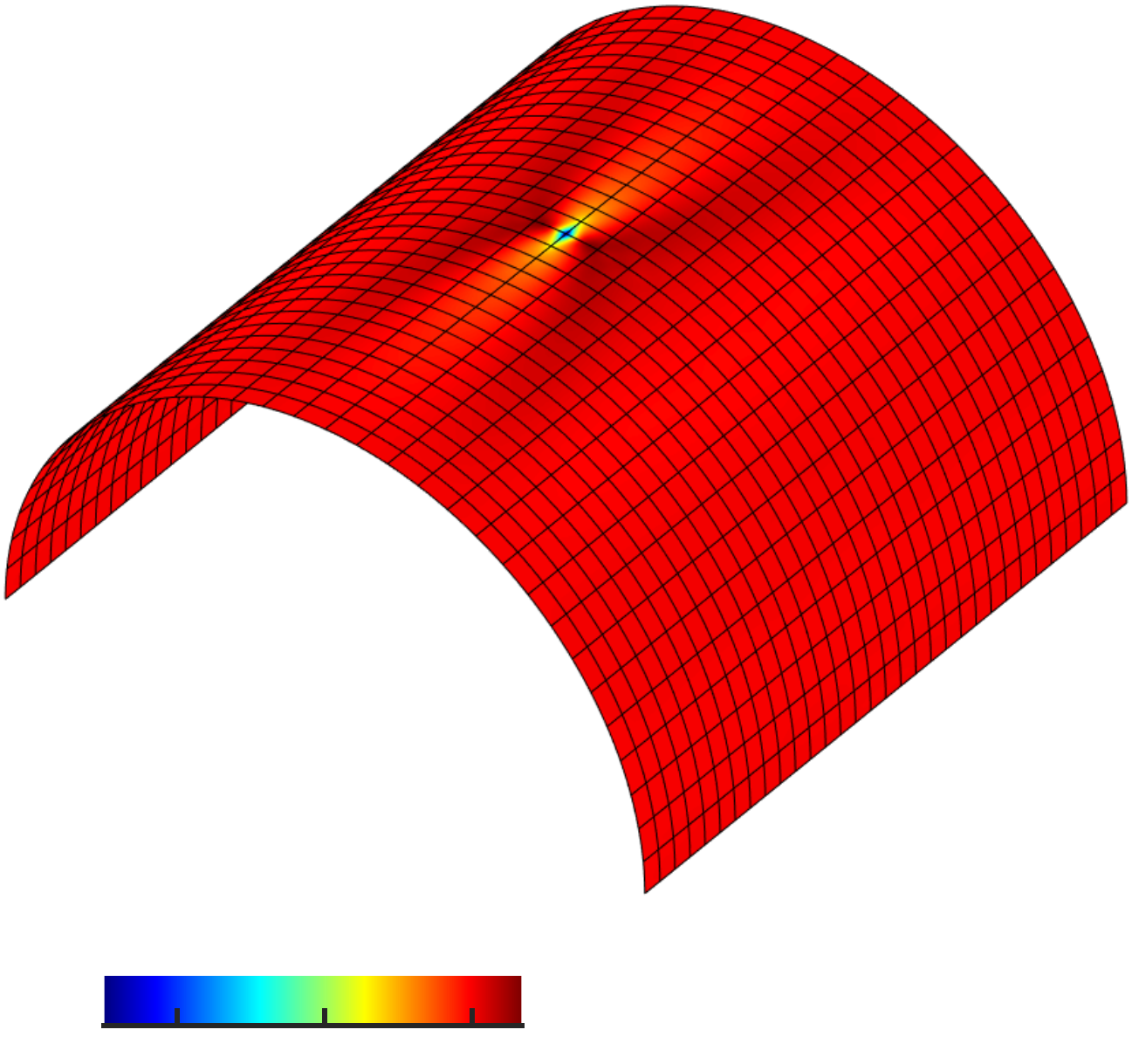}}%
    \put(0.47327203,0.07358199){\color[rgb]{0,0,0}\makebox(0,0)[lt]{\lineheight{1.25}\smash{\begin{tabular}[t]{l}$m^{11}$\end{tabular}}}}%
    \put(0.47327203,0.03095476){\color[rgb]{0,0,0}\makebox(0,0)[lt]{\lineheight{1.25}\smash{\begin{tabular}[t]{l}$\times 10^{-2}$\end{tabular}}}}%
    \put(0.27588136,0.00012621){\color[rgb]{0,0,0}\makebox(0,0)[lt]{\lineheight{1.25}\smash{\begin{tabular}[t]{l}-8\end{tabular}}}}%
    \put(0.13694102,0.00029368){\color[rgb]{0,0,0}\makebox(0,0)[lt]{\lineheight{1.25}\smash{\begin{tabular}[t]{l}-16\end{tabular}}}}%
    \put(0.41381839,0.00015824){\color[rgb]{0,0,0}\makebox(0,0)[lt]{\lineheight{1.25}\smash{\begin{tabular}[t]{l}0\end{tabular}}}}%
  \end{picture}%
\endgroup%

        }}

    \caption{Pinched cylinder: the membrane stress resultant $n^{12}$ and the bending moment $m^{11}$ for a slenderness ratio of $R/t = 1,000$, computed with our approach and cubic splines on a mesh of $16 \times 16$ elements. To enable a comparison with the results given in \cite{Sauer_locking2024}, we scale the pinching forces with $d^3$.} \label{Pinch1}
\end{figure}

\begin{figure}
    \centering
    \captionsetup[subfloat]{labelfont=scriptsize,textfont=scriptsize}
    \subfloat[$|\vect{u}|$]{{
        \def\svgwidth{0.43\textwidth}
        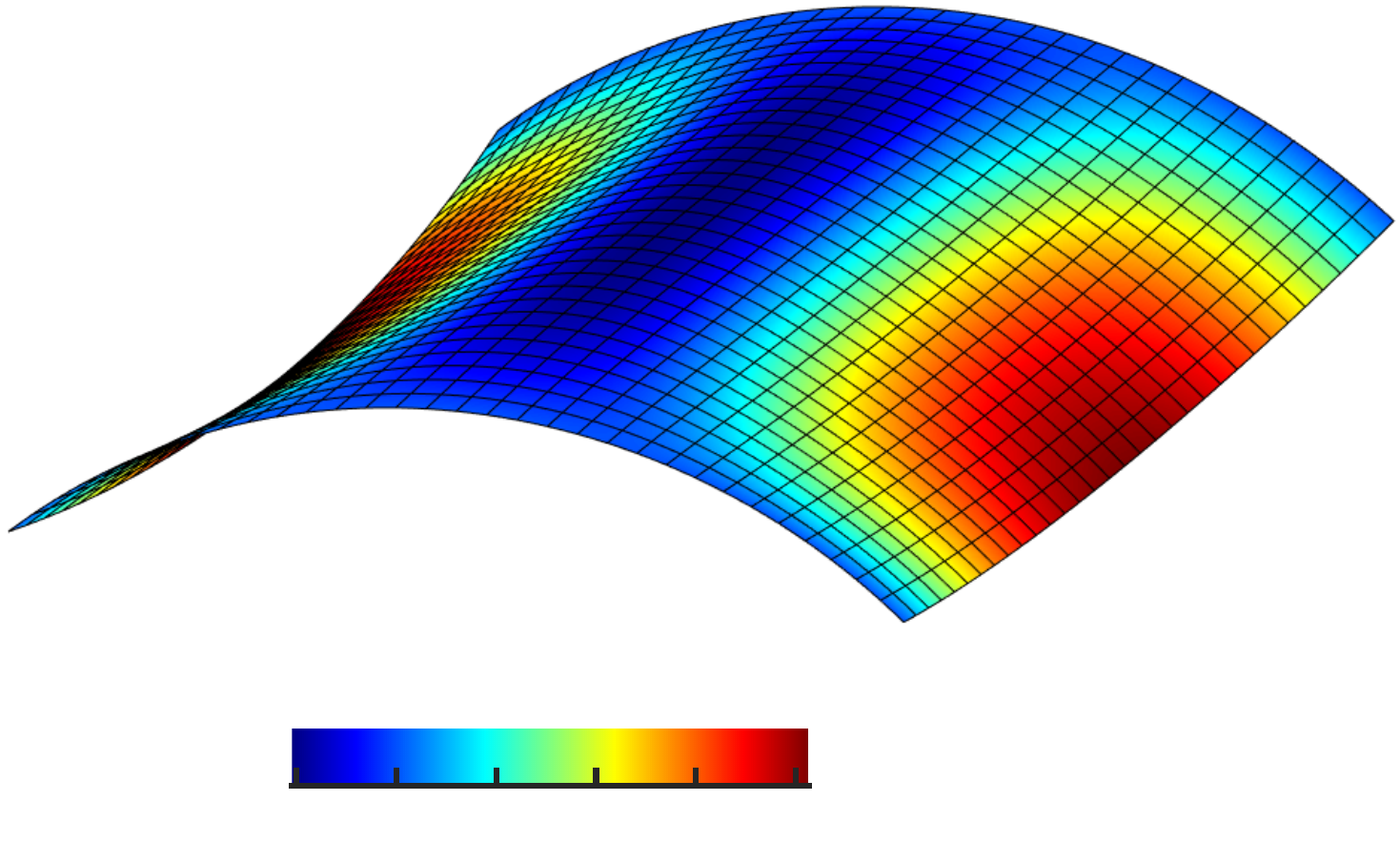        
        }}
    \subfloat[$n^{11}$]{{
        \def\svgwidth{0.43\textwidth}
\begingroup%
  \makeatletter%
  \providecommand\color[2][]{%
    \errmessage{(Inkscape) Color is used for the text in Inkscape, but the package 'color.sty' is not loaded}%
    \renewcommand\color[2][]{}%
  }%
  \providecommand\transparent[1]{%
    \errmessage{(Inkscape) Transparency is used (non-zero) for the text in Inkscape, but the package 'transparent.sty' is not loaded}%
    \renewcommand\transparent[1]{}%
  }%
  \providecommand\rotatebox[2]{#2}%
  \newcommand*\fsize{\dimexpr\f@size pt\relax}%
  \newcommand*\lineheight[1]{\fontsize{\fsize}{#1\fsize}\selectfont}%
  \ifx\svgwidth\undefined%
    \setlength{\unitlength}{705bp}%
    \ifx\svgscale\undefined%
      \relax%
    \else%
      \setlength{\unitlength}{\unitlength * \real{\svgscale}}%
    \fi%
  \else%
    \setlength{\unitlength}{\svgwidth}%
  \fi%
  \global\let\svgwidth\undefined%
  \global\let\svgscale\undefined%
  \makeatother%
  \begin{picture}(1,0.60106383)%
    \lineheight{1}%
    \setlength\tabcolsep{0pt}%
    \put(0,0){\includegraphics[width=\unitlength,page=1]{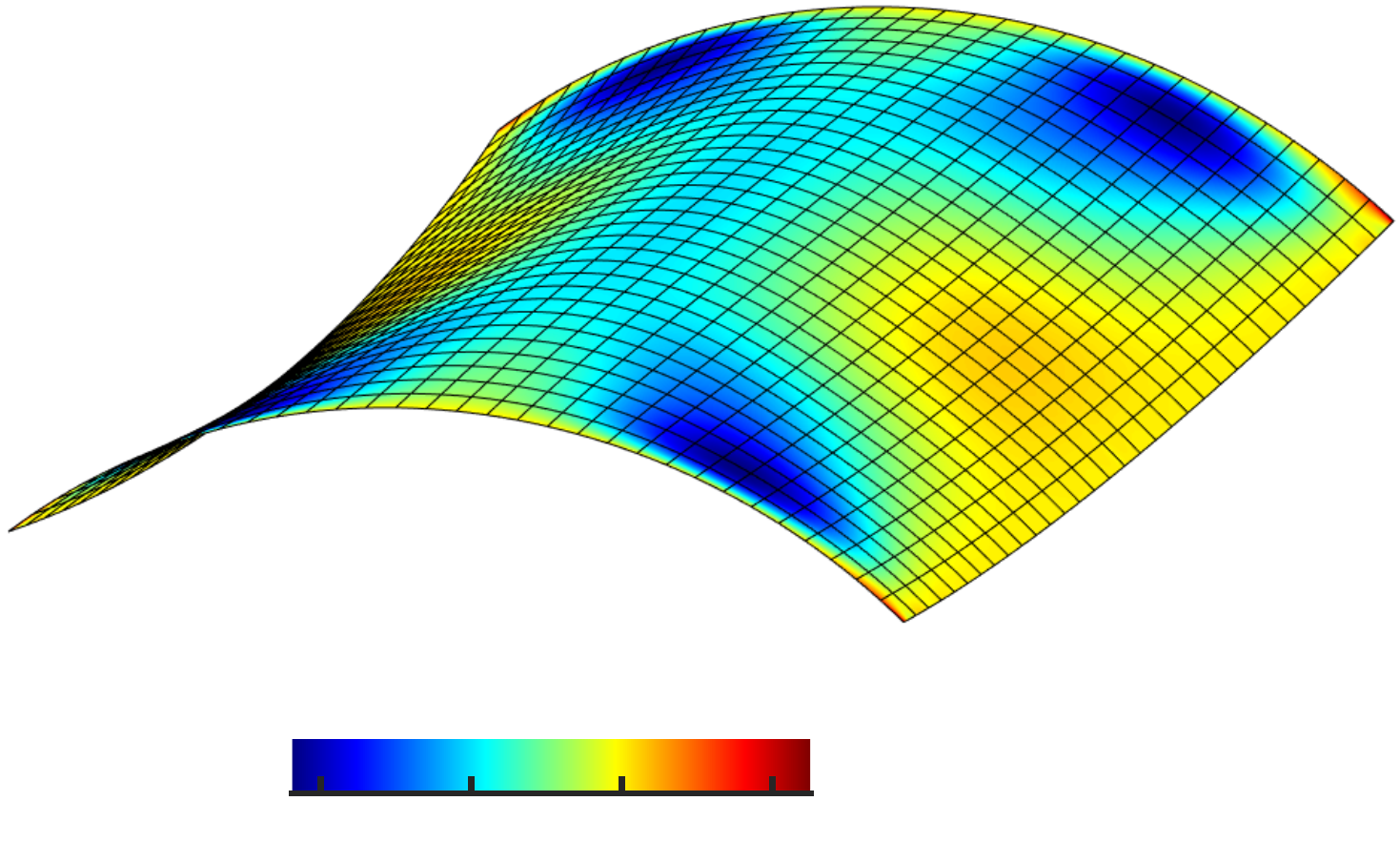}}%
    \put(0.43318812,0.00196829){\color[rgb]{0,0,0}\makebox(0,0)[lt]{\lineheight{1.25}\smash{\begin{tabular}[t]{l}$0$\end{tabular}}}}%
    \put(0.28259985,0.00196829){\color[rgb]{0,0,0}\makebox(0,0)[lt]{\lineheight{1.25}\smash{\begin{tabular}[t]{l}$-300$\end{tabular}}}}%
    \put(0.16983404,0.00196829){\color[rgb]{0,0,0}\makebox(0,0)[lt]{\lineheight{1.25}\smash{\begin{tabular}[t]{l}$-600$\end{tabular}}}}%
    \put(0.52124597,0.00196829){\color[rgb]{0,0,0}\makebox(0,0)[lt]{\lineheight{1.25}\smash{\begin{tabular}[t]{l}$300$\end{tabular}}}}%
    \put(0.58551001,0.05761176){\color[rgb]{0,0,0}\makebox(0,0)[lt]{\lineheight{1.25}\smash{\begin{tabular}[t]{l}$n^{11}$\end{tabular}}}}%
  \end{picture}%
\endgroup%

        }}
    
    \caption{Scordelis-Lo roof with large deformations: total displacements and stress resultant $n^{11}$ at maximal load value, plotted on the deformed configuration and computed with our formulation using cubic splines on a mesh of $16 \times 16$ elements.}\label{fig:roof_nonlinear_u_n11}.
\end{figure}

\begin{figure}
\centering
\captionsetup[subfloat]{labelfont=scriptsize,textfont=scriptsize}
\subfloat[$p=2$, $u_{\text{midpoint}}$]{{ \def\svgwidth{0.43\textwidth}
    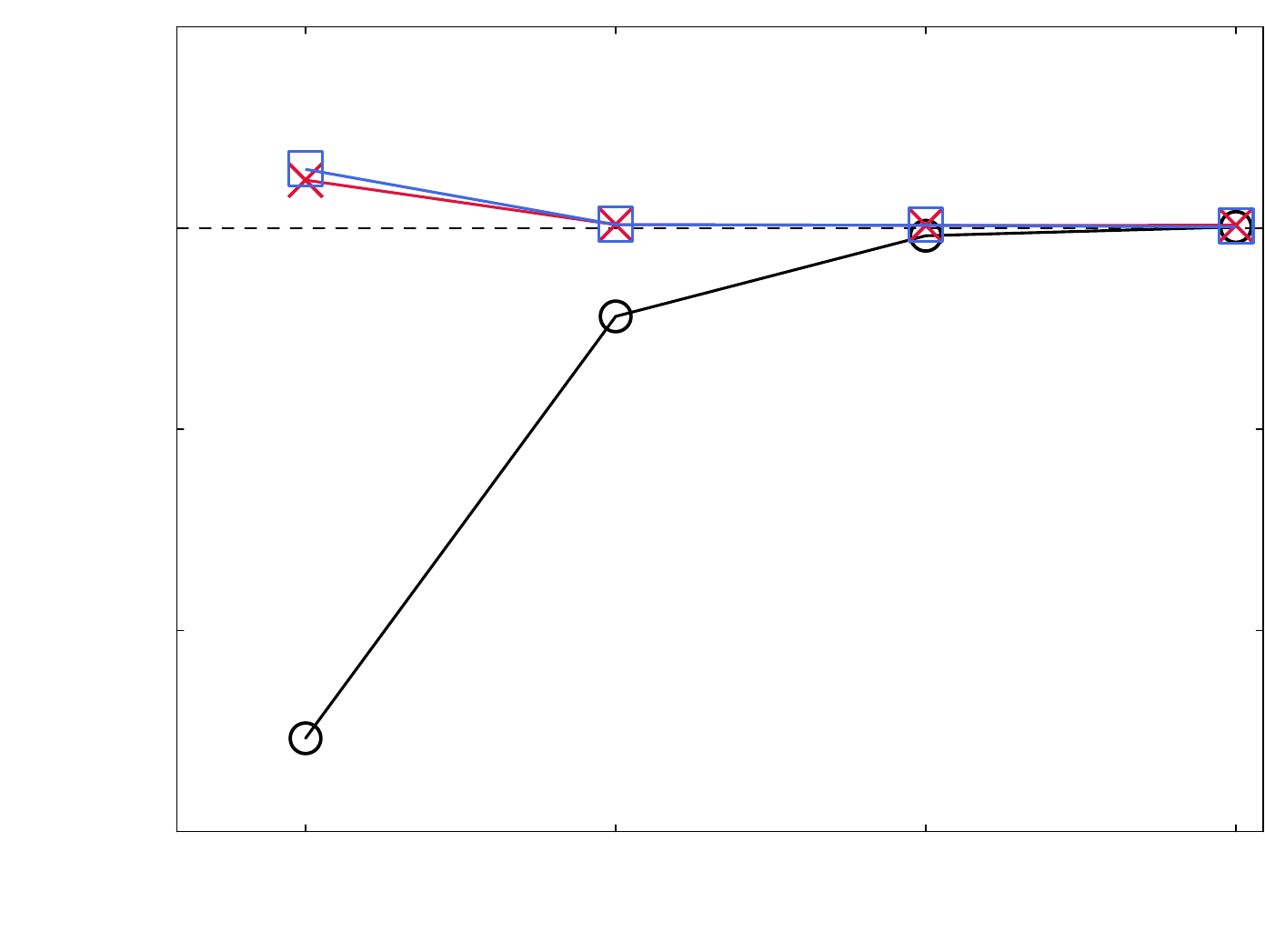 }}
\subfloat[$p=3$, $u_{\text{midpoint}}$]{{ \def\svgwidth{0.43\textwidth}
    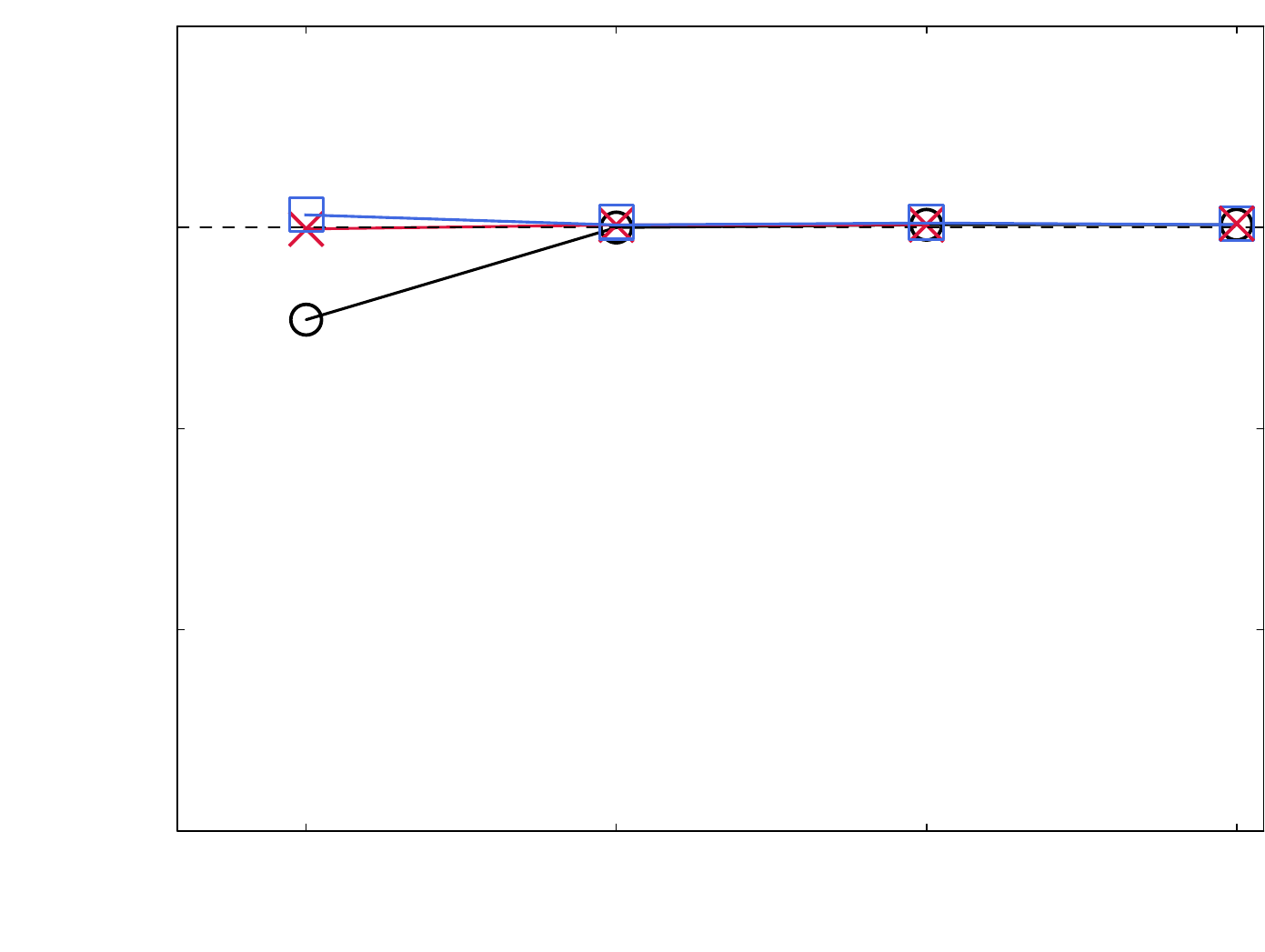 }}

\subfloat[$p=2$, $n^{11}_{\text{min}}$]{{ \def\svgwidth{0.43\textwidth}
    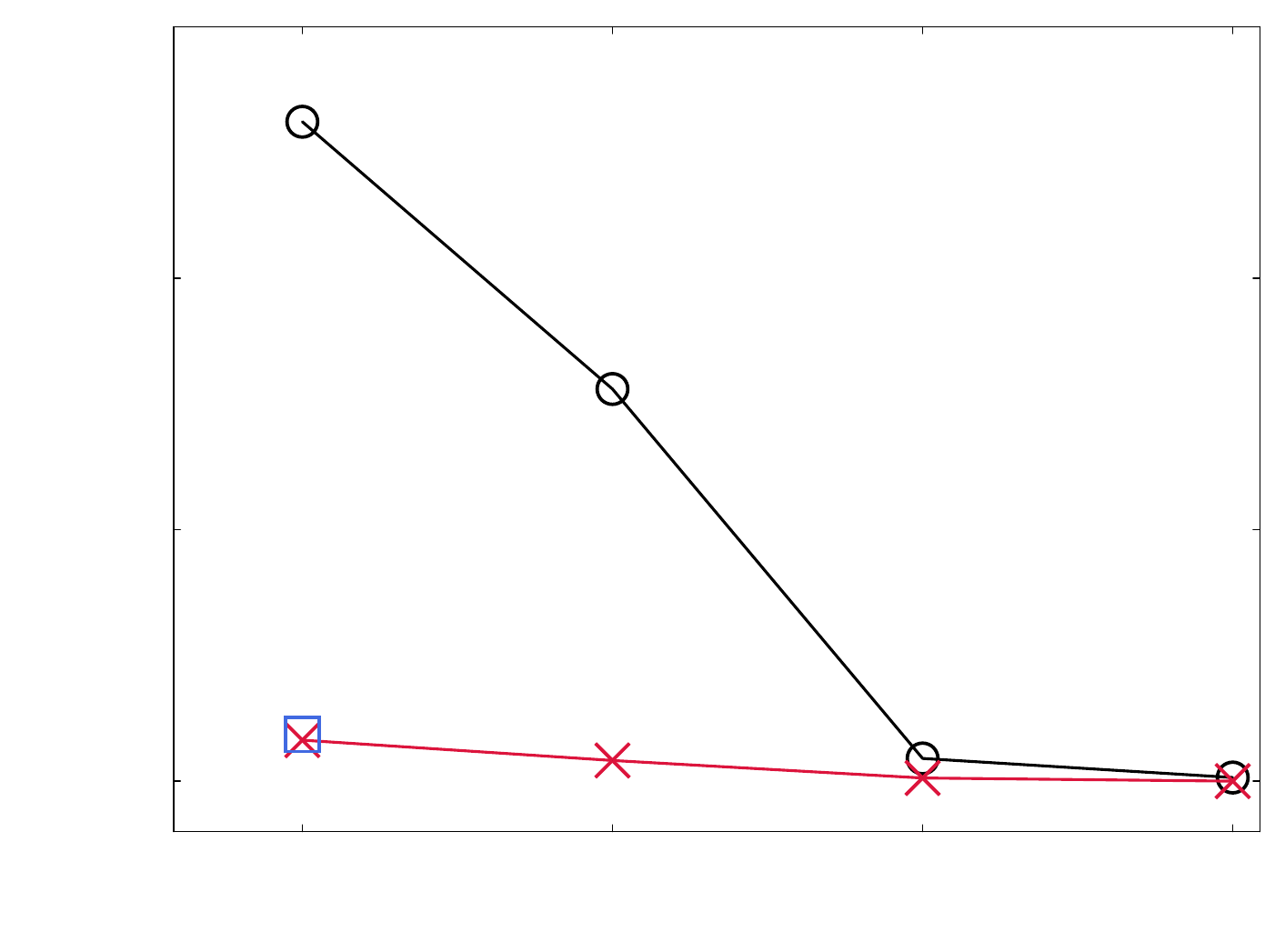 }}
\subfloat[$p=3$, $n^{11}_{\text{min}}$]{{ \def\svgwidth{0.43\textwidth}
    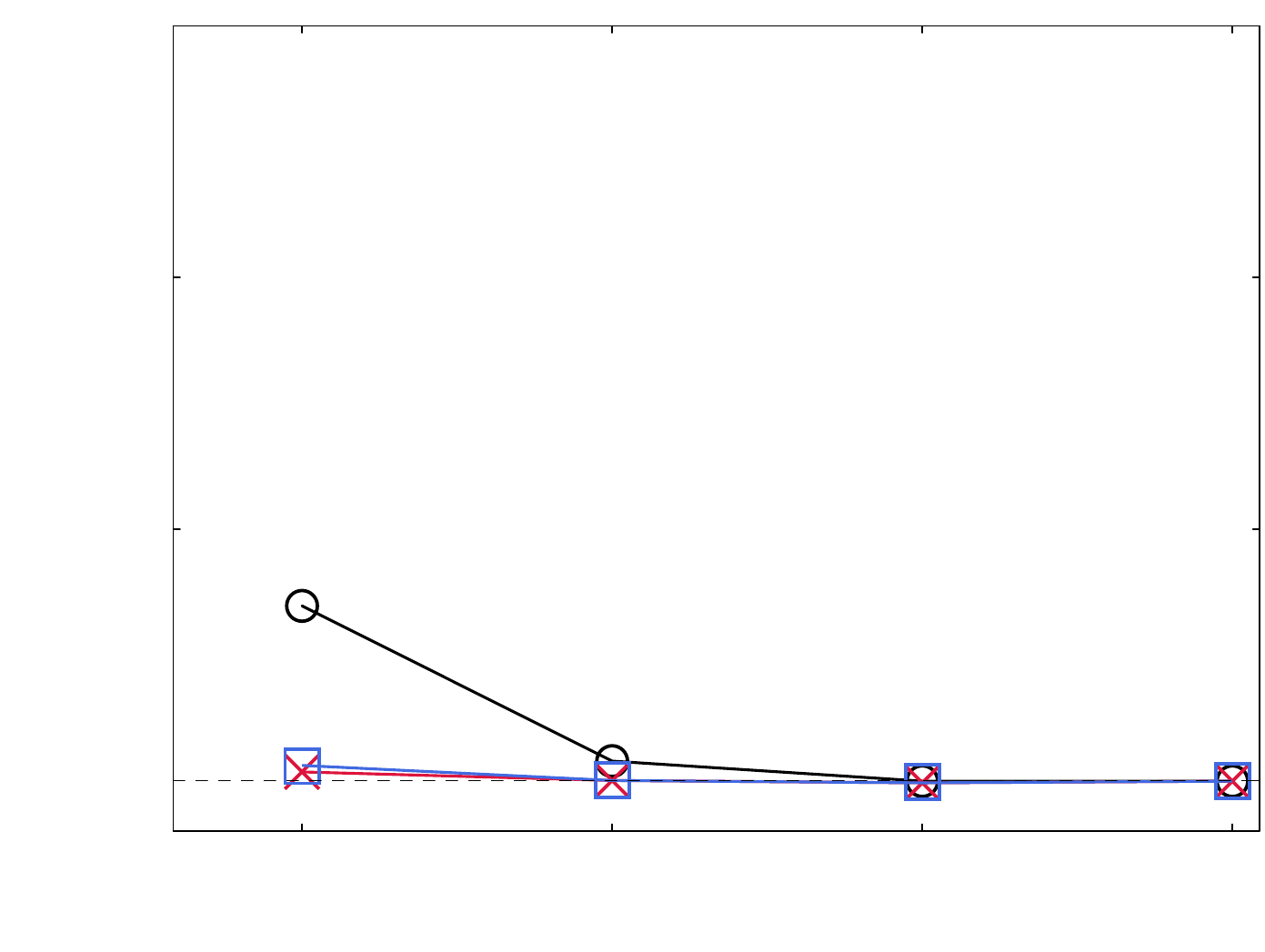 }}

\vspace{0.2cm}
{\begin{tikzpicture}
    \filldraw[black,line width=1pt, solid] (0.0,0) -- (0.6,0);
    \filldraw[black,line width=1pt] (0.3,0) [fill=none] circle (3pt);
    \filldraw[black,line width=1pt] (0.7,0) node[right]{\scriptsize Displacement-based formulation};
    \filldraw[red1,line width=1pt, solid] (7.5,0) -- (8.1,0);
    \filldraw[red1,line width=1pt] (7.5,0) node[right]{\scriptsize $\boldsymbol{\bigtimes}$};
    \filldraw[red1,line width=1pt] (8.2,0) node[right]{\scriptsize Mixed formulation, consistent strain projection};
\end{tikzpicture}

\begin{tikzpicture}
    \filldraw[blue1,line width=1pt, solid] (4.5,0.05) -- (5.1,0.05);
    \filldraw[blue1,line width=1pt] (4.7,-0.08) [fill=none] rectangle ++(0.25,0.25);
    \filldraw[blue1,line width=1pt] (5.2,0) node[right]{\scriptsize Mixed formulation, lumped strain projection, approximate duals};
\end{tikzpicture}}
\caption{Scordelis-Lo roof with large deformations: convergence of the normalized vertical displacement at the midpoint of the free edge and the minimum value of the membrane stress resultant $n^{11}$, computed with different formulations.}\label{fig:roof_nonlinear_convergence}
\end{figure}

At the request of an anonymous reviewer, we provide the following additional stress resultant plots for the shell obstacle course, enabling a full comparison with the plots shown in \cite{Sauer_locking2024}. In addition to the plots shown in the main body of the paper, these include plots of the membrane stress resultant $n^{11}$ and the bending moment $m^{11}$ for the slenderness ratio $R/t=1,000$ for the Scordelis-Lo roof in Figures \ref{Scord1} and \ref{Scord2}, and for the pinched cylinder in Figure \ref{Pinch1}. We observe that in our formulation, all stress resultant fields are indeed locking-free.

To obtain geometrically nonlinear examples, we follow \cite{Sauer_locking2024} and reduce our original definition of Young's modulus by a factor of 15. We apply the load in 80 load steps, using the standard Newton-Raphson method to keep the solution on the equilibrium path. In Figure \ref{fig:roof_nonlinear_u_n11}, we plot the total displacement field and the membrane stress resultant $n^{11}$ for the slenderness ratio $R/t=100$ for the geometrically nonlinear variant of the Scordelis-Lo roof for the maximal load value, computed with our formulation and cubic splines on a $16 \times 16$ mesh. In addition, Figure \ref{fig:roof_nonlinear_convergence} plots the convergence of the normalized displacement at the midpoint of the free edge and the minimal value of $n^{11}$ due to maximal loading, for which a reference solution is given in \cite{Sauer_locking2024}. We observe that in our formulation, all stress resultant fields are indeed locking-free.


\section{Equivalence of Galerkin projections with and without explicit material matrix}\label{sec:equivalentProj}

At the request of an anonymous reviewer, we address the following question concerning the Galerkin projection of strain-like quantities in \eqref{eq:HR-mixed-form-KL}: whether the explicit presence of the material tensor \( \mathbf{C} \) in the discrete projection equations -- in \eqref{eq51} and \eqref{eq54}, where \( \mathbf{C} \) appears explicitly, versus \eqref{neardiag}, where \( \mathbf{C} \) does not appear explicitly -- affects the result of the projection. 

In the following, we will demonstrate rigorously that the Galerkin projection yields the same function regardless of whether \( \mathbf{C} \) appears explicitly, provided the test functions are consistently defined. In particular, what may appear as a ``dropping'' of \( \mathbf{C} \) is, in fact, the result of a variationally consistent change of basis.

\subsection{Variational model problem}

To this end, we consider the following variational model problem: Let \( \Omega \subset \mathbb{R}^d \) be the spatial domain, and let \( \mathbf{C} \in \mathbb{R}^{n \times n} \) denote a symmetric positive definite (SPD) matrix, constant over \( \Omega \). We consider the projection of a given vector function \( \varepsilon \in [L^2(\Omega)]^n \) onto a finite-dimensional subspace \( V_h \subset [L^2(\Omega)]^n \) with respect to the inner product induced by \( \mathbf{C} \).

We seek \( e_h \in V_h \) such that
\begin{equation}
\int_\Omega \delta e^T \mathbf{C} \, (\varepsilon - e_h) \; dx = 0 \quad \forall \; \delta e \in V_h.
\label{eq:galerkin_energy}
\end{equation}
This defines the Galerkin projection of \( \varepsilon \) onto \( V_h \) in the energy inner product \( \langle u, v \rangle_\mathbf{C} := \int_\Omega u^\top \mathbf{C} v \, dx \).

\subsection{Finite element representation}

We represent the unknown vector field \( e_h \in V_h \) using a specific finite element basis, while test functions \( \delta e_h \in V_h \) are represented using an arbitrary (and potentially different) basis for the same space \( V_h \). 
Hence let \( \{ \phi_i(x) \}_{i=1}^{n_e} \) be a basis for the trial space \( V_h \), and \( \{ \tilde{\phi}_j(x) \}_{j=1}^{n_e} \) be a basis for the test space \( V_h \).
We express the finite element approximation in compact matrix notation as:
\begin{align}
e_h(x) = N(x) \, \hat{e}, \quad \delta e_h(x) = \tilde{N}(x) \, \delta \hat{e}, \quad
\end{align}
The array $N(x) =
\begin{bmatrix}
\phi_1(x) I_{n} & \phi_2(x) I_{n} & \cdots & \phi_{n_e}(x) I_{n}
\end{bmatrix}
\in \mathbb{R}^{n \times (n n_e)},$ collects the trial basis functions, and analogously $\tilde{N}(x) := \begin{bmatrix}
\tilde{\phi}_1(x) I_{n} & \tilde{\phi}_2(x) I_{n} & \cdots & \tilde{\phi}_{n_e}(x) I_{n}
\end{bmatrix}$ for the test basis functions, with $I_{n}$ being the identity matrix of the corresponding size. The global coefficient vector $\hat{e} \in \mathbb{R}^{n n_e}$ (and analogously $\delta\hat{e}$)  is arranged in blocks $\hat{e}_i \in \mathbb{R}^{n}$ so that all $n$ components of the vector field associated with the same basis function $\phi_i(x)$ are stored together.

Inserting the finite-dimensional representations into the variational equation \eqref{eq:galerkin_energy}, we obtain:
\begin{align}
\delta \hat{e}^T \left[ \int_\Omega \tilde{N}^T \mathbf{C} \left( \varepsilon - N \hat{e} \right) dx \right] = 0 \quad \forall \; \delta \hat{e} \in \mathbb{R}^{nn_e} \; . \label{D19}
\end{align}

This yields the discrete linear system, $\mathbf{M} \, \hat{e} = \mathbf{b}$, with
\begin{align}
\mathbf{M} & = \int_\Omega \tilde{N}^T \mathbf{C} \, N \; dx,\\
\mathbf{b} & = \int_\Omega \tilde{N}^T \mathbf{C} \, \varepsilon \; dx.
\end{align}

This formulation is valid for any (reasonable) choice of test basis \( \tilde{N}(x) \). The matrix \( \mathbf{C} \) appears explicitly in the system matrix because the projection is performed with respect to the energy inner product induced by \( \mathbf{C} \). In the canonical form of the Galerkin method, the test and trial spaces are represented by the same basis, i.e.\ \( \tilde{\phi}_j = \phi_j \), or equivalently \( \tilde{N} = N \). This corresponds directly to \eqref{eq51} and \eqref{eq54} in the main text. We emphasize that from a rigorous mathematical viewpoint, the discrete form \eqref{D19} is always a Galerkin method, as long as the trial and test spaces are equivalent, irrespective of their basis representation.

\begin{figure}[ht!]
	\centering
    \captionsetup[subfloat]{labelfont=scriptsize,textfont=scriptsize}
    \subfloat[$p=2$]{{ \def\svgwidth{0.48\textwidth}
    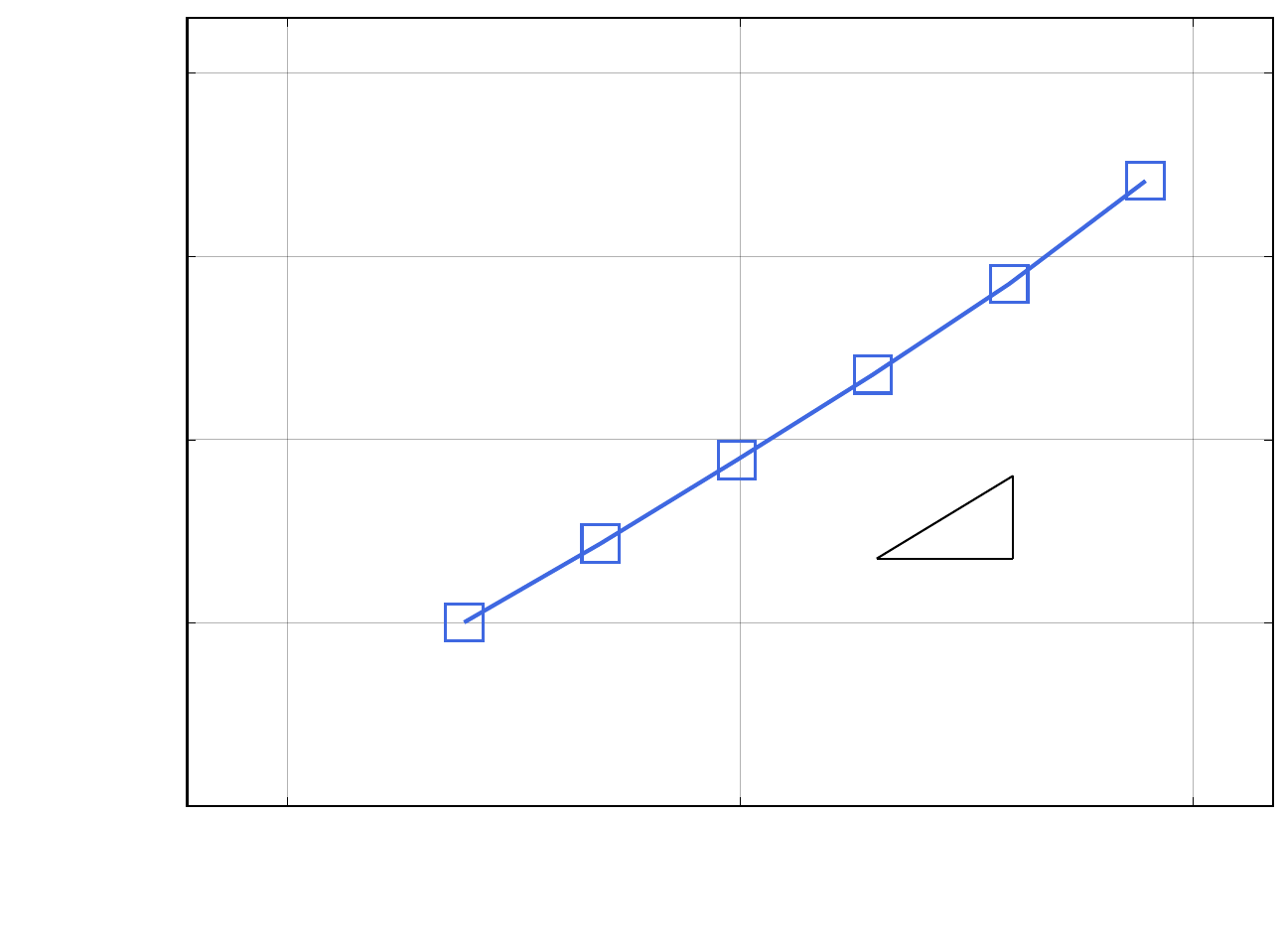 }}
    \subfloat[$p=3$]{{ \def\svgwidth{0.48\textwidth}
    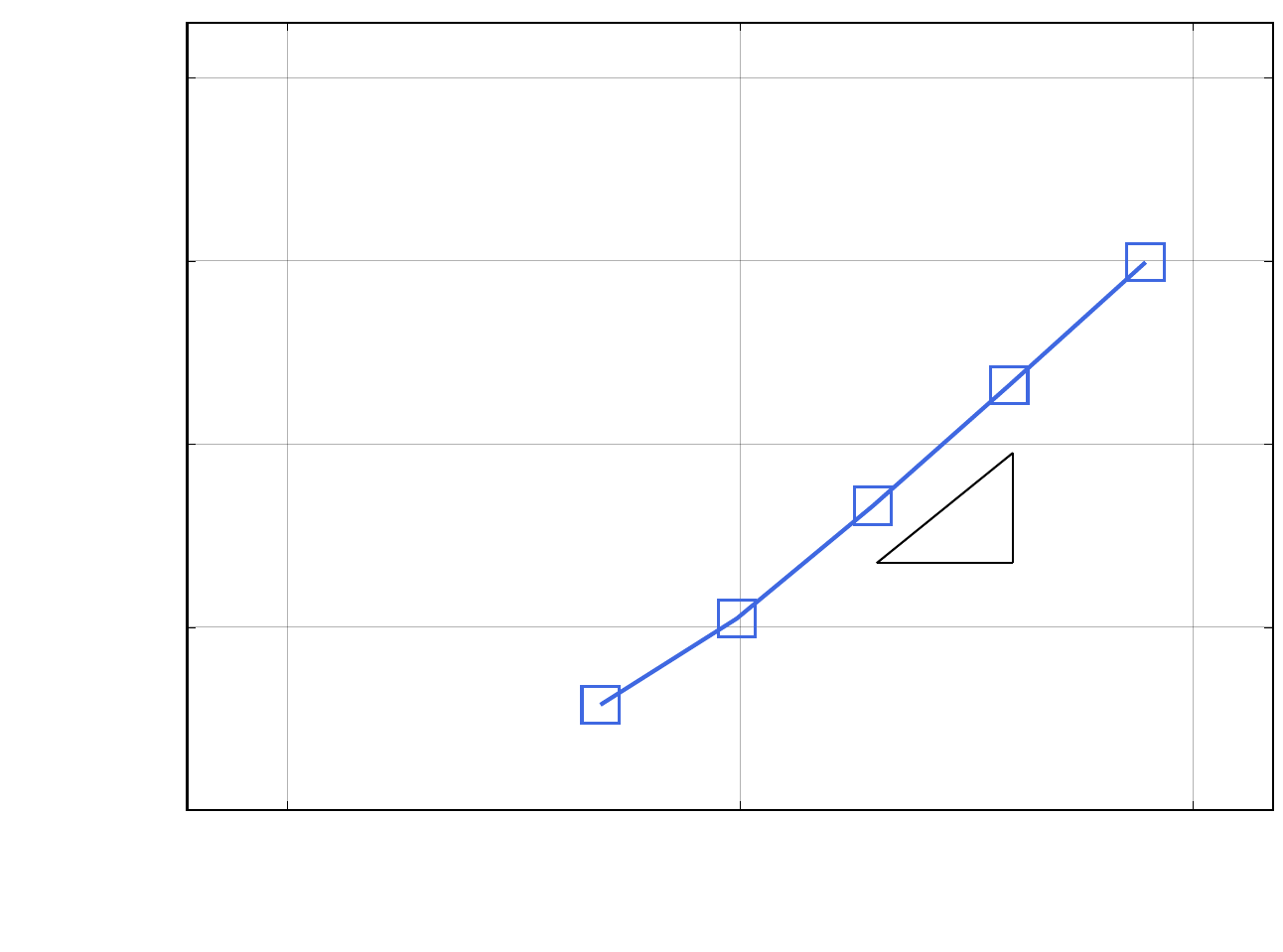 }}
    
    \vspace{0.2cm}
    \begin{tikzpicture}
    \filldraw[blue1,line width=1pt, solid] (0.2,0.0) -- (0.8,0.0);
    \filldraw[blue1,line width=1pt] (0.2,0.0) node[right]{$\boldsymbol{\Delta}$};
    \filldraw[blue1,line width=1pt] (0.9,0.0) node[right]{No scaling with $\mathbf{C}^{-1}$};
    \filldraw[blue1,line width=1pt, solid] (5.2,0.05) -- (5.9,0.05);
    \filldraw[blue1,line width=1pt] (5.4,-0.08) [fill=none] rectangle ++(0.25,0.25);
    \filldraw[blue1,line width=1pt] (6.0,0) node[right]{Scaling with $\mathbf{C}^{-1}$};
\end{tikzpicture}
    \caption{Curved Euler-Bernoulli beam: relative error in the $L^2$-norm of the displacement field, computed with and without scaling the test functions in the membrane and bending strain projection by $\mathbf{C}^{-1}$.}\label{fig:beam_scalingC}

\vspace{0.2cm}

    \captionsetup[subfloat]{labelfont=scriptsize,textfont=scriptsize}
    \subfloat[$p=2$]{{ \def\svgwidth{0.47\textwidth}
    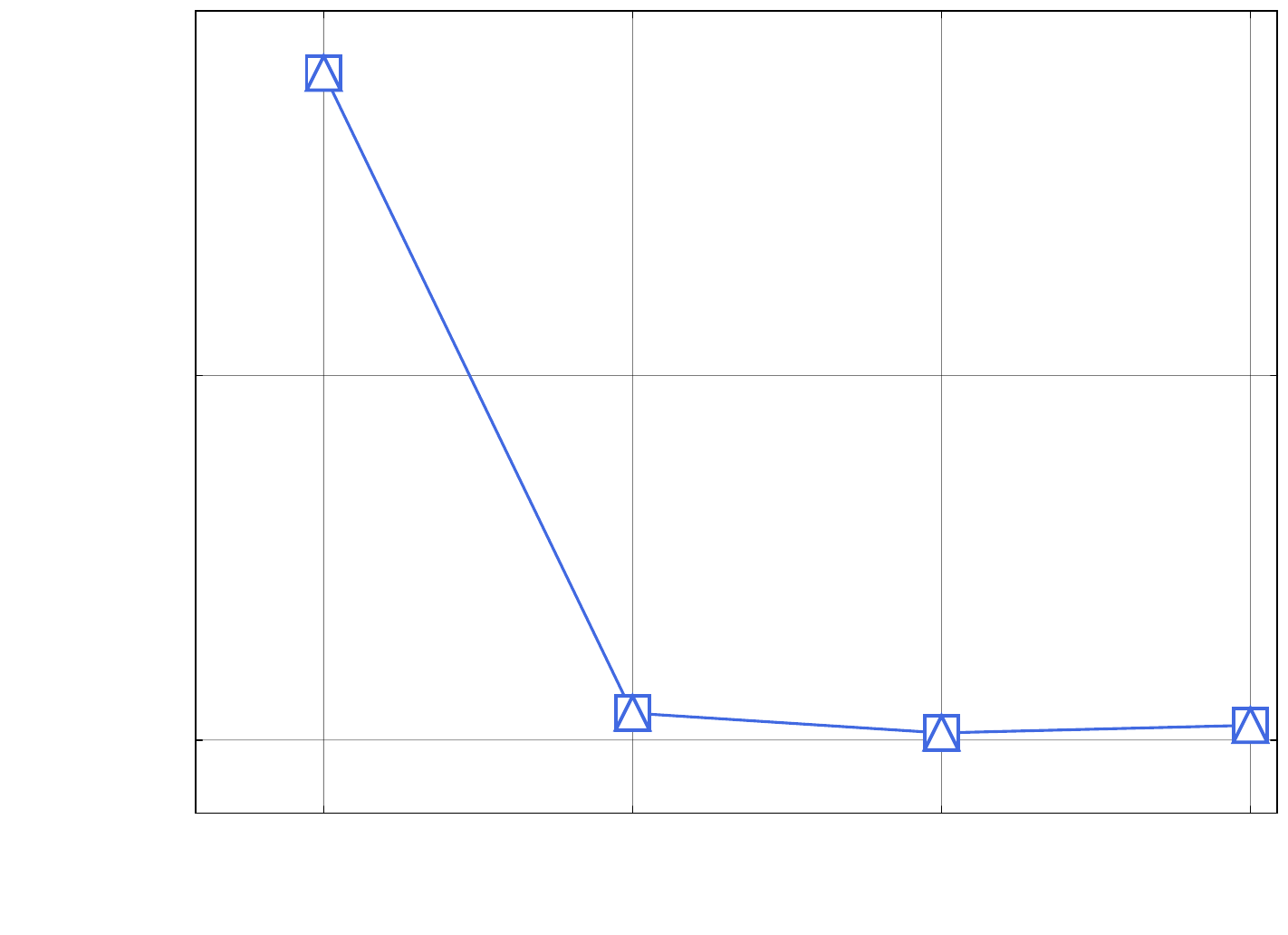 }}
    \subfloat[$p=3$]{{ \def\svgwidth{0.47\textwidth}
    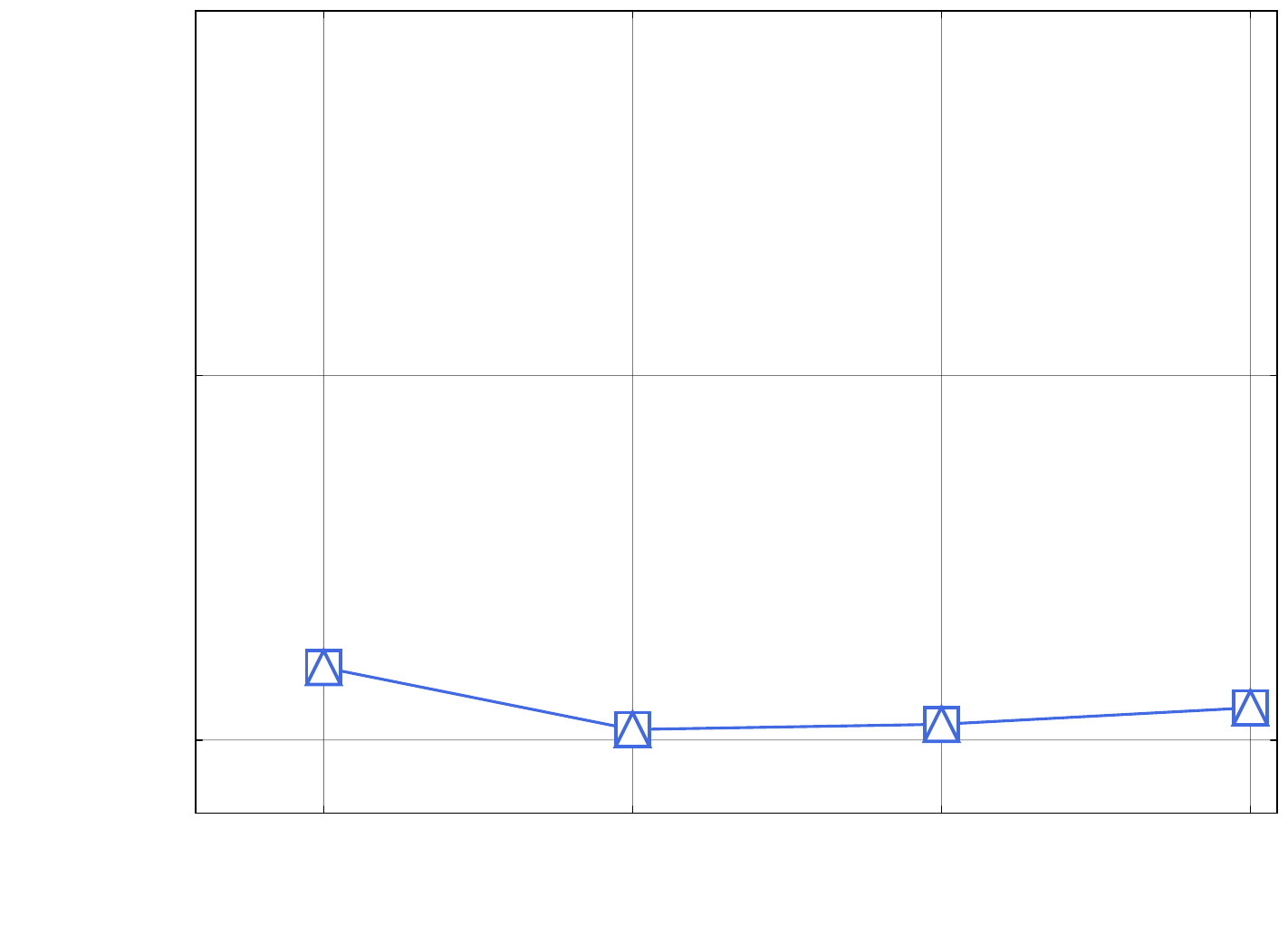 }}
    
    \vspace{0.2cm}
    \begin{tikzpicture}
    \filldraw[blue1,line width=1pt, solid] (0.2,0.0) -- (0.8,0.0);
    \filldraw[blue1,line width=1pt] (0.2,0.0) node[right]{$\boldsymbol{\Delta}$};
    \filldraw[blue1,line width=1pt] (0.9,0.0) node[right]{No scaling with $\mathbf{C}^{-1}$};
    \filldraw[blue1,line width=1pt, solid] (5.2,0.05) -- (5.9,0.05);
    \filldraw[blue1,line width=1pt] (5.4,-0.08) [fill=none] rectangle ++(0.25,0.25);
    \filldraw[blue1,line width=1pt] (6.0,0) node[right]{Scaling with $\mathbf{C}^{-1}$};
\end{tikzpicture}
    \caption{Scordelis-Lo roof ($R/t=100$): convergence of the normalized vertical displacement at the midpoint of the free edge, computed with and without scaling the test functions in the membrane and bending strain projection by $\mathbf{C}^{-1}$.}\label{fig:roof_scalingC}
\end{figure}

\subsection{Change of test basis: absorbing \( \mathbf{C} \)}

Suppose the test functions are redefined as:
\begin{align}
\tilde{N}(x) := \mathbf{C}^{-1} N(x) \label{D22},
\end{align}
i.e., the test functions are scaled pointwise by \( \mathbf{C}^{-1} \). This definition corresponds precisely to the definitions \eqref{scal1} and \eqref{scal2} in the main text. Substituting into \eqref{D19}, we obtain:
\begin{align}
\mathbf{M} & = \int_\Omega (\mathbf{C}^{-1} N)^T \mathbf{C} \, N \, dx = \int_\Omega N^T \mathbf{C}^{-T} \mathbf{C} \, N \, dx = \int_\Omega N^T N \, dx,  \label{D23}\\
\mathbf{b} & = \int_\Omega (\mathbf{C}^{-1} N)^T \mathbf{C} \, \varepsilon \, dx = \int_\Omega N^T \mathbf{C}^{-T} \mathbf{C} \, \varepsilon \, dx = \int_\Omega N^T \varepsilon \, dx.
\end{align}

We observe that for the choice \eqref{D22}, our system matrix does not involve \( \mathbf{C} \) explicitly. However, scaling (and re-arranging) the test functions by the inverse material matrix $\mathbf{C}^{-1}$ in \eqref{D22} does not change the underlying variational problem or the Galerkin method. Instead, it constitutes a change of basis within the same test space \( V_h \), designed to absorb the material matrix $\mathbf{C}$ into the test functions.

As a result, the bilinear form no longer shows $\mathbf{C}$ explicitly in the discrete system, but the projection remains consistent with the energy norm defined by $\mathbf{C}$, and the projected function \( e_h(x) = N(x) \hat{e} \in V_h \) remains unchanged. Therefore, \eqref{D22} is not an approximation or a simplification -- it is a variationally consistent reformulation, where algebraic simplifications in the projection matrix \eqref{D22} (and in \eqref{neardiag} in the main text) arise naturally from the chosen basis.

Therefore, what may appear as a ``dropping'' of \( \mathbf{C} \) is, in fact, a variationally consistent change of basis. The discrete projection remains mathematically equivalent and physically meaningful, preserving the energy norm structure. As requested by an anonymous reviewer, we ``validate'' this statement for the examples of the curved Euler-Bernoulli beam and the Scordelis-Lo roof. The corresponding plots in Figures \ref{fig:beam_scalingC} and \ref{fig:roof_scalingC} confirm that using a test basis with or without \( \mathbf{C}^{-1} \) scaling yields exactly the same results (as dictated by the shown mathematical structure).

\bibliography{wileyNJD-AMA.bib}




\end{document}